%% file: main.tex
\documentstyle[12pt,twoside,singlespace,bibcontents]{mitthesis}

\input epsf.tex

\begin{document}

\newcommand{\vecmu}{{\mbox{\boldmath $\mu$}}}
\newcommand{\ethr}{{\epsilon_t}}
\newcommand{\ethrc}{{\epsilon_{t,c}}}
\newcommand{\kb}{{k_B}}
\newcommand{\zz}{{{\cal Z}}}
\newcommand{\zzinf}{{{\cal Z}_\infty}}
\newcommand{\veff}{{V_{eff}}}
\newcommand{\etal}{{\em et al.}}
\newcommand{\erf}{{\rm erf}}
\newcommand{\oneStwoS}{\mbox{$1S$-$2S$}}
\newcommand{\oneSoneS}{\mbox{$1S$-$1S$}}
\newcommand{\Hdown}{\mbox{$H\!\!\downarrow$}}
\newcommand{\Hup}{\mbox{$H\!\!\uparrow$}}
\newcommand{\Lalpha}{\mbox{Lyman-$\alpha$}}

\include{title}

\pagestyle{plain}
\include{contents}
\include{intro}
\include{statmech}

\include{rfcell}
\include{rfplay}

\include{condensates}

\include{conclusion}

\appendix
\input{statmech_MB_app}

\include{evap_collision}
\input{statmech_bose_app}

\input{evap_model}

\include{vel_distribs}
\input{cond_loss}
\input{trapshape}

\input{variable2}

\input{energy_units}

\bibliographystyle{ieeetr}
\bibliography{dgf}

\end{document}

%% file: title.tex
\title{Bose-Einstein Condensation of Atomic Hydrogen}
\author{Dale G. Fried}
\prevdegrees{B.S. Physics\linebreak Washington State University (1992)}
\department{Department of Physics}
\degree{Doctor of Philosophy}

\degreemonth{June}
\degreeyear{1999}
\thesisdate{March 31, 1999}

\supervisor{Daniel Kleppner}{Lester Wolfe Professor of Physics}
\supervisor{Thomas J. Greytak}{Professor of Physics}

\chairman{Thomas J. Greytak}
{Chairman, Department of Physics Graduate Committee}

\maketitle

%
%

\pagestyle{empty}
\cleardoublepage
\typeout{Clearing double page}
\pagestyle{empty}
\setcounter{savepage}{\thepage}
\begin{abstractpage}
\input{abstract}


\end{abstractpage}


\pagestyle{empty}
\cleardoublepage
\typeout{Clearing double page}

\section*{Acknowledgments}

\input{acknowledgments}

%% file: abstract.tex

This thesis describes the observation and study of Bose-Einstein
condensation (BEC) of magnetically trapped atomic hydrogen.  The
sample is cooled by magnetic saddlepoint and radio frequency
evaporation and is studied by laser spectroscopy of the \oneStwoS\
transition in both the Doppler-free and Doppler-sensitive
configuration.  A cold collision frequency shift is exploited to infer
the density of both the condensate and the non-condensed fraction of
the sample.  Condensates containing $10^9$ atoms are observed in
trapped samples containing $5\times10^{10}$ atoms.  The small
equilibrium condensate fractions are understood to arise from the very
small repulsive interaction energy among the condensate atoms and the
low evaporative cooling rate, both related to hydrogen's anomalously
small ground state $s$-wave scattering length.  Loss from the
condensate by dipolar spin-relaxation is counteracted by replenishment
from the non-condensed portion of the sample, allowing condensates to
exist more than $15$~s.  A simple computer model of the degenerate
system agrees well with the data.  The large condensates and much
larger thermal reservoirs should be very useful for the creation of
bright coherent atomic beams.  Several experiments to improve and
utilize the condensates are suggested.

Attainment of BEC in hydrogen required application of the rf
evaporation technique in order to overcome inefficiencies associated
with one-dimensional evaporation over the magnetic field saddlepoint
which confines the sample axially.  The cryogenic apparatus (100~mK)
had to be redesigned to accomodate the rf fields of many milligauss
strength.  This required the removal of good electrical conductors
from the cell, and the use of a superfluid liquid helium jacket for
heat transport.  Measurements of heat transport and rf field strength
are presented.

The rf fields in the apparatus allow rf ejection spectroscopy to be
used to measure the trap minimum as well as the temperature of the
sample.

%% file: acknowledgments.tex

This thesis is dedicated to Jesus Christ, who I understand to be the
One who created quantum mechanics, and indeed all of physics and the
whole of the entire universe.  The extremely rich and subtle
complexity of the physical world, and that of quantum systems such as
dilute gases at low temperatures and high densities in particular,
give me reason to worship Him as an amazing intelligence with
creativity and complexity much farther beyond the grasp of the human
mind than the physics described in this thesis is beyond the grasp of
my sister-in-law's golden retriever.  As I understand it, our
Creator's deep love for us has caused Him to initiate relationship
with us, a relationship that has implications far beyond our
activities here.  For this, too, I worship Him and give to Him my
loyalty.  He is the King of the Universe.

Having gratefully acknowledged the Inventor of the physics studied in
this thesis, I also gratefully acknowledge my parents, Ray and Twyla,
and my brother, Glenn, who patiently but thoroughly instilled in me a
pleasure in working hard, a curiosity about the world, a
self-confidence that has allowed me to take risks.  Their
encouragement and interest during my graduate work has been
unwavering, a crucial help when I myself was wavering.

As Daniel Kleppner says, one of the important roles of a thesis
advisor is to ask a good question.  He and my other advisor, Thomas
Greytak, have done this superbly.  I am grateful to both of them for
allowing me to work in their research group and for supporting me
financially.  Their intellectual commitment to the research and their
personal commitment to me as a student have made my years at MIT a
pleasure.

Much of a graduate education comes through one's co-workers, and much
of the pleasure comes through the camaraderie and friendship in the
research group.  I am fortunate to have worked with many world-class
colleagues during my time at MIT.

Mike Yoo, my first office mate, helped me learn practical skills, such
as programming ``makefiles'', and also helped me shoulder responsibilities
with confidence.  His cheerful insights into graduate student life
helped take the edge off all night problem sets.

John Doyle, a postdoc when I started in the group, taught me the power
of back of the envelope calculations and modeled a systematic, bold
approach to solving problems.  He has been a continuing encouragement
during my graduate career.

Jon Sandberg's helpful ability to teach me the basics of the
experiment planted the seed in my head that one day I would be able to
run the experiment, too.  His confidence in the ``younger generation''
has helped me push through times of discouragement.

Albert Yu's cheerily optimistic ``there is no problem which cannot be
solved'' has been my rallying cry more than once.  Albert's friendship
is typified by generous hospitality, including one night when he met me
after I was mugged on the subway.

Claudio Cesar contributed a contagious creativity to the experiment,
and a fun demeanor to life in the lab.  His optimism and encouragement
helped me move toward more of a leadership position.  His friendship
has helped me maintain perspective.

Adam Polcyn joined the experiment the same year I did.  Beginning with
the finding that our tastes in used books were compatible, I have
enjoyed my friendship with him immensely.  I often rely on him for
reality checks both in my physics and in life.

Thomas Killian has been my primary colleague over the most recent
several years of work.  I learned an intellectual thoroughness from
his agile and deep approach to physics.  His excitement and engagement
with the experiment has made long nights in the lab fruitful and
enjoyable.

David Landhuis joined the group more recently.  His work to rigorously
understand physics is very welcome, as is his cheerful willingness to
carry out thankless lab chores such as being the safety officer.
Dave's friendship makes work in the lab a pleasure.

Stephen Moss, my new office mate, also brings a tenacious quest for
deep understanding to the experiment.  His sense of humor, humility,
and flexibility are much appreciated.  His friendship significantly
adds to the fun of working in this research group.

Lorenz Willmann, who joined the group two years ago as a postdoc, has
taught me about experiment documentation and data analysis.  His calm
manner has brought a broader perspective to problems that loomed
large in my thinking.  

I am confident of a bright future for this experiment in the capable
hands of Dave, Stephen, and Lorenz.

Finally, I wish to thank my wife, Diana, for marrying me and giving me
her love.  I met her three years ago at a French abbey on Easter,
while traveling after a physics conference at Les Houches in the
French Alps.  She has made the last three years of my life 
meaningful and joyful.  I am deeply grateful for her patience and sacrifice
during the sometimes long days and nights I have spent in the lab.

%% file: contents.tex

\tableofcontents
\listoffigures
\listoftables

%% file: intro.tex

\chapter{Introduction}

\section{The Lure of Bose-Einstein Condensation}

Bose-Einstein condensation (BEC) of a dilute gas has been a very
important goal since the beginning of experimental research on
spin-polarized atomic hydrogen.  The original intent \cite{sn76} was
to study quantum phase transition behavior and search for
superfluidity in this dilute gas, a weakly interacting system that is
much more theoretically tractable than strongly interacting degenerate
quantum systems such as superfluid liquid $^4$He.  This phase
transition is a purely quantum mechanical effect, and, unlike all
other phase transitions, it occurs in the {\em absence} of any
interparticle interactions.

Although the search for BEC in dilute gases began in spin-polarized
hydrogen, the first observations of the effect used dilute gases of
the alkali metal atoms Rb \cite{aem95}, Na \cite{dma95}, and Li
\cite{bst95,bsh97} in the summer of 1995.  The properties of hydrogen
that were so attractive in the beginning turned out, in the end, to be
irrelevant or even a hindrance.  Nevertheless, we have finally
observed BEC in hydrogen \cite{fkw98,kfw98}.  In several ways our
hydrogen condensates are rather different from those of the alkalis,
and we use different techniques to probe the sample.  Hydrogen thus
contributes significantly to the extraordinary flurry of experimental
activity creating and probing Bose condensates, and applying them to
intriguing new areas.

The formation of a Bose condensate out of a gas can be studied
as the occurrence of a quantum mechanical phase transition, but perhaps
even more interesting are the properties of the condensate itself.
Centered in the middle of a trapped gas, this collection of atoms
exhibits such bizarre effects as a vanishingly small kinetic energy,
long range coherence across the condensate \cite{atm97}, a
single macroscopic quantum state, immiscibilty of two components of a
gas \cite{sis98,mss99}, and constructive/destructive
interference between atoms that have fallen many millimeters out of
the cloud \cite{ank98}.  These systems hold promise for
creating quantum entanglement of huge numbers of atoms for use in
quantum computing and vastly improved fundamental measurements.

\section{The Significance of the Work in this Thesis}

In this thesis we describe the culmination of a 22 year research
effort: the first observations of BEC in hydrogen\footnote{A two
dimensional analog of BEC has been observed very recently in atomic
hydrogen confined to a surface of liquid $^4$He \cite{svy98}}.
Furthermore, we demonstrate a new technique of probing Bose
condensates, optical spectroscopy.  This tool allows us to measure the
density and momentum distributions of the sample, and thus infer the
temperature, size of the condensate, and other properties.  We find
that the condensates we create contain nearly two orders of magnitude
more atoms than other BEC systems, and the condensates occupy a much
larger volume.  In addition, we find that nearly $10^{10}$ atoms are
condensed during the 10~s lifetime of the condensate.  This high flux
of condensate atoms implies that this system might be ideal for
creation of bright, coherent atomic beams.  Indeed, hydrogen has
proven to be a rich and promising atom for further studies of BEC.

The thesis is divided as follows.  Chapter \ref{statmech.chap} and the
associated appendices provide a detailed description of the behavior
of the trapped gas both in the classical and quantum mechanical
regimes.  Results are obtained to guide the reader's intuition
throughout the remainder of the thesis.  Chapter \ref{rfcellchapter}
describes changes to the apparatus that allowed BEC to be achieved.
Chapter \ref{rfplay.chap} details use of the new capabilities of the
improved apparatus.  The most important contributions described in
this thesis are found in chapter \ref{results.chap} where studies of
BEC are presented.  Finally, conclusions and suggestions for future
work are made in chapter \ref{conclusions.chap}.

Much of the work described here was done in close collaboration with a
group of people, as shown by the several names on our papers.  In
particular, this thesis should be considered a companion thesis to
that of Thomas Killian \cite{killian99}.  Both should be read to
obtain a complete picture of the whole experiment.  The work builds on
earlier work by Claudio Cesar, whose thesis \cite{cesar95} describes
the first \oneStwoS\ spectroscopy of trapped atomic hydrogen; Albert
Yu, whose thesis \cite{yu93} on universal quantum reflection includes
important insights into non-ergodic atomic motion in the trap; Jon
Sandberg, whose thesis \cite{san93} describes the physics of
\oneStwoS\ spectroscopy in a trap and the development of the uv laser
system we use; and John Doyle, whose thesis \cite{doyle91} describes
trapping and cooling of spin-polarized hydrogen and the magnet system
we use to trap the gas.

\section{The Basics of Trapping and Cooling Hydrogen}
\label{discussion.of.loading}

There is a long and interesting history of spin-polarized hydrogen in
the laboratory, summarized well by Greytak \cite{gre95}.  Here we
briefly summarize our method of creating spin-polarized hydrogen,
loading it into a trap, and cooling it to BEC (quantum degeneracy).  A
broader description is in the thesis of Killian \cite{killian99}.

\begin{figure}[tb]
\centering \epsfxsize=5in\epsfbox{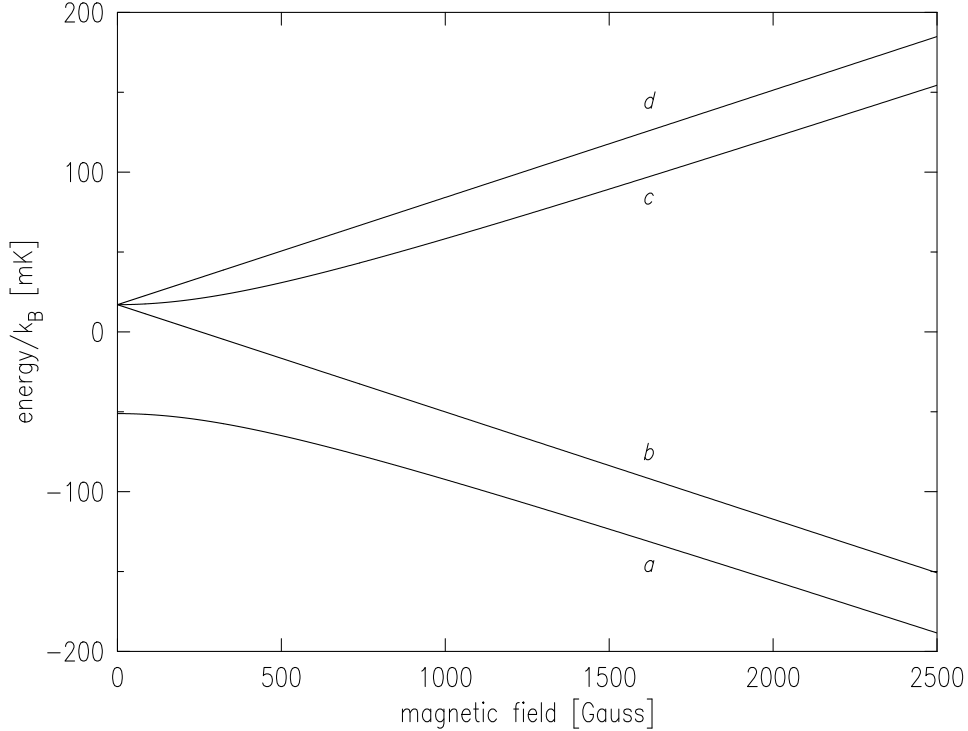}
\caption[hyperfine-Zeeman diagram of atomic hydrogen]{Hyperfine-Zeeman
diagram of ground state atomic hydrogen.  The eigenstates of the
combined hyperfine and Zeeman Hamiltonian are described in section
\ref{hyperfine.zeeman.sec}.  Atoms in states $c$ and $d$ can be
trapped in a magnetic field minimum. }
\label{intro.zeeman.fig}
\end{figure}

Hydrogen is composed of a proton and electron, whose spins can couple
in four ways: the four hyperfine states are labeled $a$-$d$ (these
states are derived in section \ref{hyperfine.derivation.sec}).  The
high-field seeking states ($a$ and $b$) are pulled toward regions of
high magnetic field, and the low-field seeking states ($c$ and $d$)
are expelled from high field regions.  Figure \ref{intro.zeeman.fig}
shows the energies of the four states as a function of field strength.
Our experiments use the pure, doubly-polarized $d$-state.

\begin{figure}[t]
\centering \epsfysize=6.65in \epsfbox{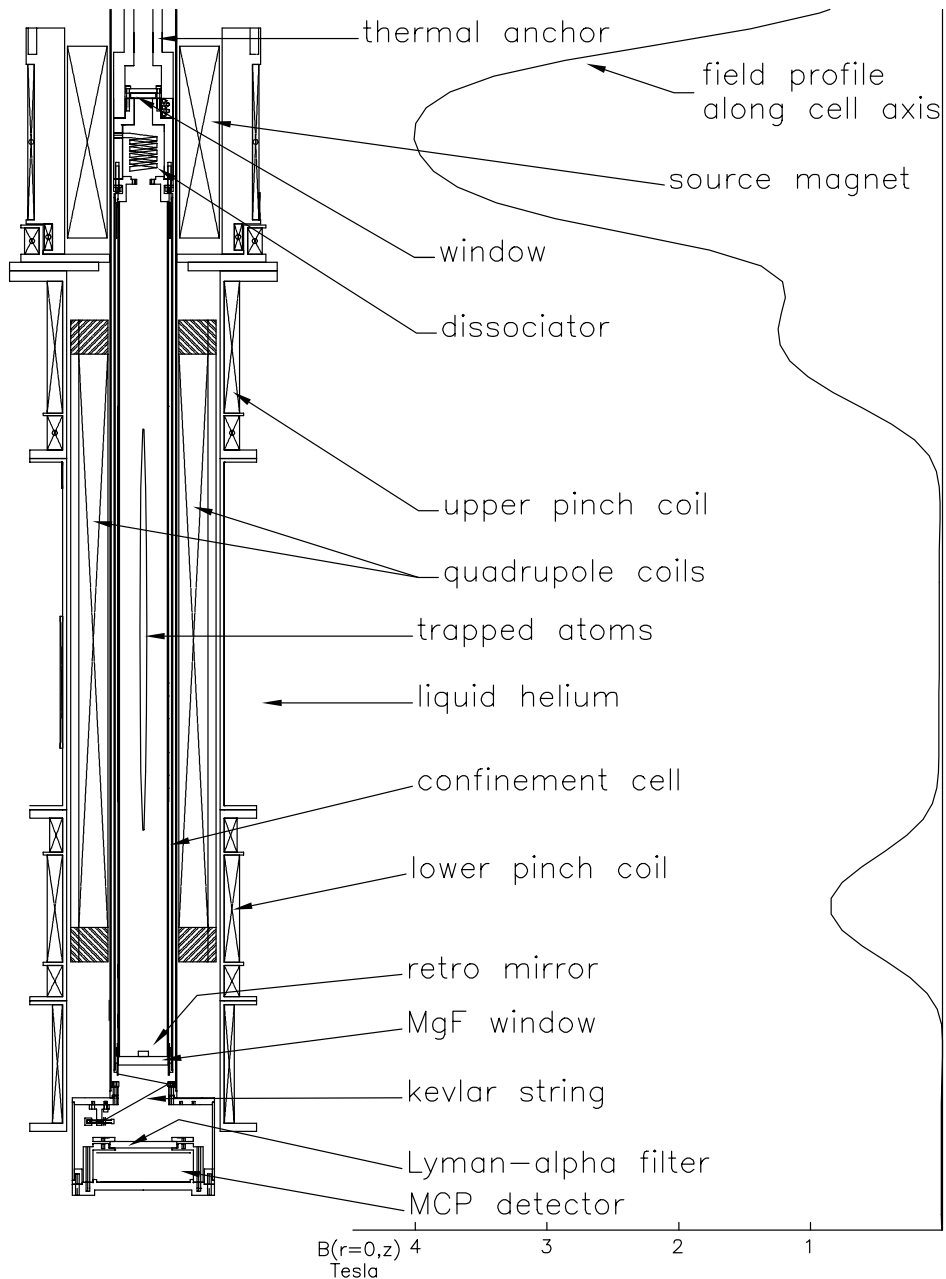}
\caption[cutaway view of apparatus]{Cutaway view of apparatus.
Molecular hydrogen is dissociated in an rf discharge in a 4~tesla region.
The atoms move into the confinement cell which has superimposed on it a
trap, created by the quadrupole coils (radial confinement) and the
pinch coils (axial confinement).  The coils are bathed in 4.2~K liquid
helium.  The cell is centered in the vacuum can with three Kevlar
strings.  It is thermally anchored to the mixing chamber of a dilution
refrigerator (not shown).  The laser beam passes through the upper
window, retroreflects on the mirror at the bottom of the cell, and
exits  through the upper window.  Atomic fluorescence photons
pass through the MgF window and Lyman-$\alpha$ filter, and are counted
on the microchannel plate detector.  }
\label{trap.apparatus.fig}
\end{figure}

At the beginning of an experimental run molecular hydrogen is loaded
into the cryogenic apparatus by blowing a mixture of H$_2$ and $^4$He
into a cold ($T\simeq 1$~K) can, called the dissociator, where it
freezes.  Then, for each loading of the trap, H$_2$ molecules are
dissociated by pulsing an rf discharge (the dissociator is in a region
of high magnetic field, 4~T).  The low field seekers are blown into a
confinement cell (4~cm diameter, 60~cm length), over which the
inhomogeneous trapping field is superimposed.  The trap fields are
created by currents in superconducting coils, described by Doyle
\cite{doyle91} and Sandberg \cite{san93}.  The coils create a trap
with maximum depth 0.82~T, which, for hydrogen's magnetic moment of 1
Bohr magneton, corresponds to an energy $\ethr/\kb=550$~mK. (See
appendix \ref{energy.unit.app} for conversions between various
manifestations of energy in this experiment.)  The trap depth is the
difference between the field at the walls of the containment cell and
the minimum field in the middle of the cell.  There are seventeen
independently controlled coils in the apparatus which are used to
adjust the trap shape.  Figure \ref{trap.apparatus.fig} is a cutaway
diagram of the system.
Expanded views of the top and bottom of the cell are in figures
\ref{celltop.mechanical.fig} and \ref{cellbottom.mechanical.fig}.  The
trap shape used for capturing the atoms is basically a long trough.
The field increases linearly away from the $z$ axis of the trap (the
potential exhibits near cylindrical symmetry about the $z$ axis); the
potential is small and roughly uniform for about 20~cm along the $z$
axis.  ``Pinch'' coils at each end provide the axial confinement.  The
field profile indicated in figure \ref{trap.apparatus.fig} is along
the $z$ axis of the cell.

The trap depth sets an energy scale: for atoms to be trapped their
total energy must be less than the trap depth.  Two techniques are
used to cool the atoms into the trap.  First, after H$_2$ molecules
are dissociated \cite{doyle91,kfr_tobepublished} the hot atoms are
thermalized on the walls of the cell, held at 275~mK.  In order to
prevent the atoms from sticking tightly to the cold surfaces, a
superfluid $^4$He film covers the walls and reduces the binding energy
of the H atoms to a manageable $1.0$~K \cite{mjb81EB,jmb82,hhc87}.
When the atoms leave the wall their total energy ($3\kb T/2$ of
kinetic energy and $\ethr$ of potential energy) is still greater than
the trap depth.  The second stage of cooling into the trap involves
collisions among the atoms that are crossing the trapping region.
Sometimes these collisions result in one atom having low enough energy
to become trapped.  The partner atom in the collision goes to the wall
and is thermalized.  The gas densities expected in the cell correspond
to a collision length many times greater than the cell diameter, so
atoms pass through the trapping region many times before becoming
trapped.  A recent study of the trap loading process will be published
elsewhere \cite{kfr_tobepublished}.

After loading the trap, the cell walls are quickly cooled to below
150~mK to thermally disconnect the sample from the confinement cell.
At these lower temperatures the residence time of an atom on the
surface of the cell becomes much longer than the characteristic
H+H$\rightarrow$ H$_2$ recombination time, and so the surface is
``sticky''; all the atoms that reach the surface recombine before
having a chance to leave the surface.  Thermal disconnect between the
cryogenic cell and the trapped sample is achieved because no warm
particles can leave the walls and carry energy to the trapped gas.
Atoms can leave the trapped gas and reach the wall, however, if they
have large enough total energy to climb the magnetic hill.  As these
very energetic atoms irreversibly leave the trapped sample, it cools.  In fact,
this preferential removal of the most energetic atoms, called
evaporation, is very useful.  It is the cooling mechanism we exploit
to create samples as cold as $20~\mu$K, as described in chapter
\ref{results.chap}.  A thorough description of the heating and cooling
processes which set the temperature of the system is found in appendix
\ref{classical.ideal.gas.app}.

We catch both the $d$ and $c$ low field seeking states in the trap.
However, inelastic collision processes involving two $c$ state atoms
quickly deplete the $c$-state population, and the remaining atoms
constitute a doubly polarized sample (both electron and proton spins
are polarized).  Because the spin state is pure, the spin-relaxation
rate is rather small.  For peak densities in the normal gas of
$n\simeq10^{14}~{\rm cm}^{-3}$ (nearly the largest for these
experiments) the characteristic decay time is 40~s.

The trapped gas is probed using various techniques.  The simplest
method involves quickly lowering the magnetic confinement barrier at
one end of the trap, and measuring the flux of escaping atoms as a
function of barrier height \cite{doyle91,dsm89}.  In a limited regime
this technique can be used to infer the gas temperature and density.
This technique, and its limitations, is described in chapter
\ref{rfplay.chap}.  Another technique is rf ejection spectroscopy,
described in chapter \ref{rfplay.chap}.  The third probe technique is
laser spectroscopy of the $1S$-$2S$ transition, envisioned
\cite{kleppner89,san93}, realized \cite{cesar95,cfk96}, and perfected
\cite{killian99} in our lab for a trapped gas.  This diagnostic tool
can give the density and temperature of the gas over a wider range of
operating conditions than the other techniques, and has proven crucial
for studies of BEC.  The laser probe will be described in more detail
in chapter \ref{results.chap}, and is a principle subject of the
companion thesis by Killian \cite{killian99}.

%% file: statmech.tex

\chapter{Theoretical Considerations}
\label{statmech.chap}

\input{statmech_MB}
\input{statmech_bose}

%% file: statmech_MB.tex

This chapter addresses two important topics from a theoretical
perspective.  First, we explain why previous experiments could not
bring trapped hydrogen into the quantum degenerate regime.  This
understanding allowed us to solve the problem and achieve BEC.  The
second topic is related to the small fraction (only a few percent) of
atoms we are able to put into the condensate.  This situation differs
markedly from that observed in Rb and Na BEC experiments.  An
explanation of this difference and others is the second topic we
explore theoretically.

The starting point for understanding the behavior of the trapped gas
is classical statistical mechanics.  Good explanations have been
developed elsewhere \cite{lrw96,wal94,doyle91,bpk87}.  For
completeness, we present a development in appendix
\ref{classical.ideal.gas.app}.  Since the gas exists in a trap of
finite depth $\ethr$, we use a truncated Maxwell-Boltzmann occupation
function.  Knowledge of the trap shape allows us to calculate the
population, total energy, density, etc.  The effects of truncation of
the energy distribution at the trap depth are included in these
derivations.  We see that for deep traps ($\eta\equiv\ethr/\kb T>7$)
the truncation does not significantly alter the properties of the
system from estimates based on an untruncated distribution.  Appendix
\ref{classical.ideal.gas.app} also explains the density of states
functions (see section \ref{density.of.states.sec}) used in the
remainder of the thesis and summarizes evaporative cooling (see
section \ref{collision.sec}).

\section{Dimensionality of Evaporation}
\label{evaporation.dimensionality.sec}

The temperature of the gas is set by a balance between heating and
cooling processes, as described in section \ref{equilibrium.temp.sec}.
Previous attempts to attain BEC in hydrogen \cite{dsy91,pmw98} failed
because the cooling process became bottlenecked by the slow rate at
which energetic atoms could escape, thus reducing the effective
cooling rate.  To understand this bottleneck we must first consider
the details of the trap shape.  We then study the motion of the
particles that have enough energy to escape.

The trap shape used to confine samples at $T<200~\mu$K is often called
the ``Ioffe-Pritchard'' \cite{pri83,lrw96} type (labeled ``IP'').
Using axial coordinate $z$ and radial coordinate $\rho$, the potential
has the form
\begin{equation}
V_{IP}(\rho,z)=\sqrt{(\alpha\rho)^2 + (\beta z^2 + \theta)^2}-\theta
\end{equation}
with radial potential energy gradient $\alpha$ (units energy/length),
axial potential energy curvature $2\beta$ (units of
energy/length$^2$), and bias potential energy $\theta$.  See section
\ref{density.of.states.sec} for a summary of the density of states
functions for this trap.

In the limit of $\rho\ll\theta/\alpha$, the Ioffe-Pritchard potential
is harmonic in the radial coordinate, as may be seen by expanding the
potential in powers of $\alpha \rho/(\beta z^2 + \theta)$:
\begin{equation}V_{IP}(\rho,z) = \beta z^2 + \frac{1}{2} 
	\frac{\alpha^2}{\beta z^2 + \theta}\: \rho^2 \; + \; 
	\frac{1}{8}\:\frac{\alpha^3}{(\beta z^2 + \theta)^2}\: \rho^3\; + 
	\cdots.
\label{IPharmonic.expansion.eqn}
\end{equation}
The trap is harmonic in the radial direction when the third term 
is much smaller than the second term.  This is true for radial
coordinates $\rho\ll\rho_{anharm}$, where we have defined the
``anharmonic radius'' at which the second term matches the third term:
\begin{equation}
\rho_{anharm}\equiv 4\:\frac{\beta z^2 + \theta}{\alpha}.
\end{equation}
The trap appears
harmonic in all three directions to short samples for which the radial
oscillation frequency is essentially uniform along the length of the
sample.  This occurs for temperatures $T\ll 4\theta/\kb$.  In the
harmonic regime, the axial oscillation frequency is
\begin{equation}
\omega_z=\sqrt{\frac{2\beta}{m}},
\label{axial.osc.freq.defn.eq}
\end{equation}
and the radial oscillation frequency is
\begin{equation}
\omega_\rho=\alpha\sqrt{\frac{m}{\beta z^2+\theta}}.
\label{radial.osc.freq.defn.eq}
\end{equation}

For the evaporation to proceed efficiently, atoms with energy greater
than the trap depth $\ethr$ (called ``energetic atoms'') must leave
the trap promptly, before having a collision.  As shown in appendix
\ref{evapcoll:appendix}, in the vast majority of cases an energetic
atom that collides with a trapped atom will become trapped again.  The
rare collision that produced the energetic atom will be ``wasted''.
It is essential to understand the details of the particle removal
process if maximum evaporation efficiency is to be achieved.

Previous attempts to cool hydrogen to BEC utilized ``saddlepoint
evaporation'', in which energetic atoms escape over a saddlepoint in
the magnetic field barrier at one end of the trap.  To escape, the
atom must have energy in the axial degree of freedom ($z$) that is
greater than $\ethr$.  This atom removal technique is inherently one
dimensional.  The collisions which drive evaporation produce many
atoms with high energy in the radial degrees of freedom, and in order
for these to escape the energy must be transferred to the axial degree
of freedom.  This energy transfer process was analyzed theoretically
by Surkov, Walraven, and Shlyapnikov \cite{sws96}, and we follow their
analysis.

In a harmonic trap the potential is separable, and the
particle motion is completely regular; no energy exchange occurs.  In
the Ioffe-Pritchard trap, however, energy exchange can occur because
the potential is not separable; the radial oscillation frequency
depends on the axial coordinate, $z$, and so radial motion can couple
to axial motion.  (See equation \ref{radial.osc.freq.defn.eq}.)

This energy mixing can be understood by considering how rapidly the radial
oscillation frequency changes as an atom moves along the $z$
axis.  If the frequency changes slowly (``adiabatically''), then the
energy will not mix among the degrees of freedom.  The adiabaticity
parameter is the fractional change of the radial oscillation frequency
in one oscillation period as the atom moves axially through the trap.
Strong mixing occurs when
\begin{equation}
\frac{\dot{\omega}_\rho}{\omega_\rho^2} \sim 1.
\end{equation}
Here $\dot{\omega}_\rho=(d\omega_\rho/dz)(dz/dt)$. For a
Ioffe-Pritchard trap with a bias that is large compared to $\kb T$,
$\omega_\rho=\alpha/\sqrt{(\beta z^2 + \theta)m}$.  We have used the
expansion of $V_{IP}$ from equation \ref{IPharmonic.expansion.eqn},
which is valid if $\kb T \sim \alpha \rho \ll \theta$.  Given that
$\kb T \ll \theta$, the adiabaticity parameter is
\begin{equation}
\frac{\dot{\omega}_\rho}{\omega^2_\rho}=v_z\: \frac{\beta z
	\sqrt{m}}{\alpha \sqrt{\beta z^2 + \theta}}.
\end{equation}
We see that several factors contribute to good mixing: large axial
velocity $v_z$ (which occurs at high temperature), small radial
gradient $\alpha$, small bias field $\theta$, and large axial
curvature $\beta$.  In practice, however, achieving BEC requires low
temperatures ($v_z$ small) and high densities (obtained with large
compressions, and thus large $\alpha$).  Consequently, the degrees of
freedom do not mix and evaporation becomes essentially one
dimensional.  Typical values for our experiment are
$\alpha/\kb=16~{\rm mK}/{\rm cm}$, $\beta/\kb=25~\mu{\rm K}/{\rm
cm^2}$, $\theta/\kb=30~\mu{\rm K}$, $T=100~\mu{\rm K}$, $z\sim 2~{\rm
cm}$, and $v_z=140~{\rm cm}/{\rm s}$, so that
$\dot{\omega_\rho}/\omega_\rho^2 \sim 10^{-3}$.  For these conditions
it takes about $10^3$ oscillations to transfer energy, but there are
only about $\omega_\rho/2\pi\Gamma_{col}\sim 200$ oscillations per
collision for a peak sample density $n_0=10^{14}~{\rm cm}^{-3}$
($\Gamma_{col}$ is the elastic collision rate,
p\pageref{Gamma.def.page}).  There is not enough time to transfer the
radial energy to axial energy before the particle has a collision. The
energy mixing is thus very weak and the evaporation is one
dimensional.  Surkov \etal\ \cite{sws96} pointed out the consequences
of one-dimensional evaporation.  They estimated that the efficiency is
reduced by a factor of $4\eta$ compared to that of full 3D
evaporation.  The evaporative cooling power thus drops dramatically.
Experiments in our laboratory (unpublished) have confirmed that phase
space compression ceases near $100~\mu$K\@.  These results were
duplicated and studied in more depth by Pinkse \etal\ \cite{pmw98}.

In order to maintain the evaporation efficiency a technique is
required that quickly removes {\em all} particles with energy greater
than the trap depth.  To this end we implemented rf evaporation, as
described in detail in chapter \ref{rfcellchapter}.

%% file: statmech_bose.tex

\section{Degenerate Bose Gas}

\subsection{Bose Distribution}

The statistical mechanical description of a classical gas, presented
in appendix \ref{classical.ideal.gas.app}, is not correct when quantum
effects are important.  A simple way to understand the crossover to
the quantum regime is to recall that particles are characterized by
wavepackets whose size is related to their momentum by the Heisenberg
momentum-position uncertainty relation.  As a gas is cooled, the
particle momenta decrease, and the wavepackets enlarge.  The classical
(point-particle) description of the system breaks down when these
wavepackets begin to overlap.  The quantum treatment correctly deals
with the effect of particle indistinguishability.  There are many
excellent treatments of quantum statistical mechanics \cite{hua63}.
Here we review the basic features that are important for our
experiment.

A collection of identical particles with integer (half-integer) spins
must have a total wavefunction that is symmetric (anti-symmetric) when
two particles are exchanged.  The connection between spin and symmetry
has been explained using relativistic quantum field theory
\cite{pau40}.  Particles with integer (half-integer) spin are called
bosons (fermions).  In contrast to fermions, multiple bosons may
occupy a single quantum state.

\begin{figure}[tb]
\centering \epsfxsize=5in \epsfbox{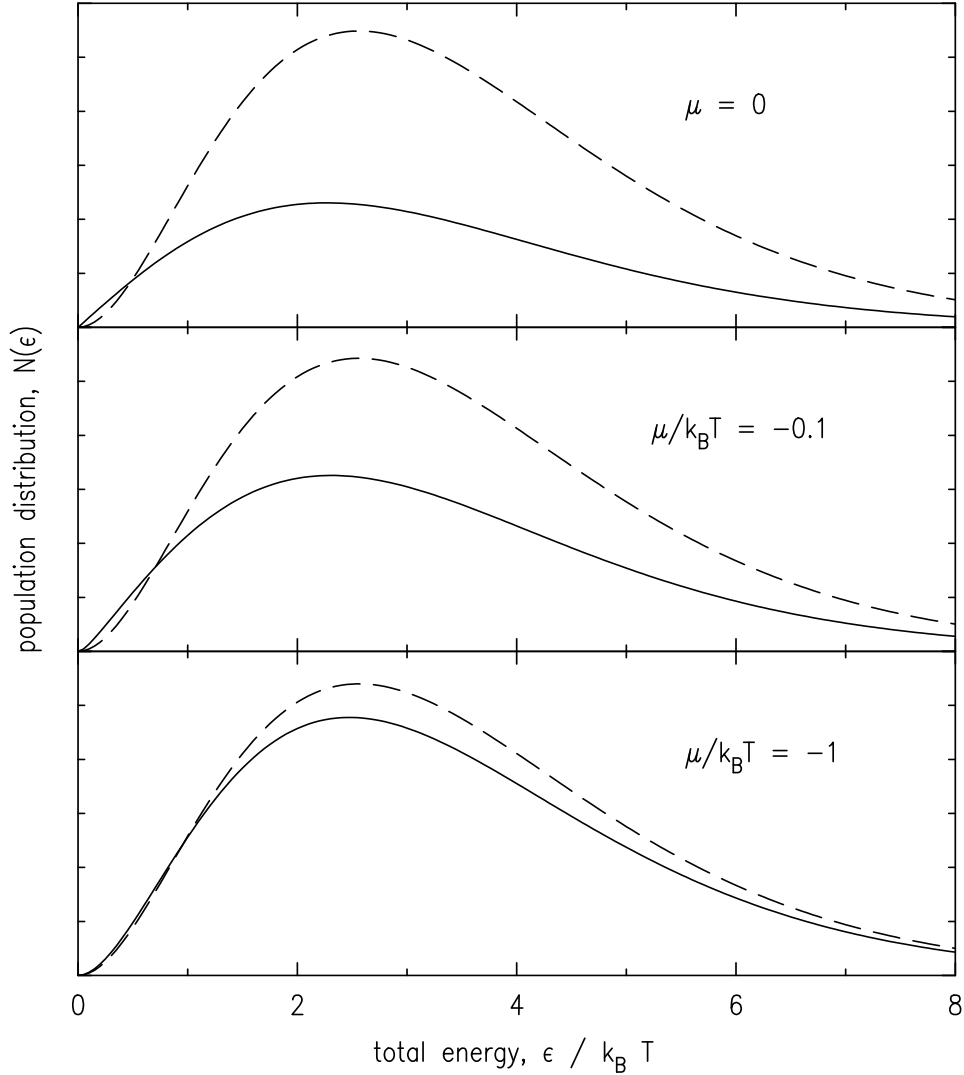}
\caption[Maxwell-Boltzmann and Bose-Einstein population
distributions]{ Comparison of Maxwell-Boltzmann (dashed line) and
Bose-Einstein (solid line) distributions of particle energies, plotted
for various chemical potentials.  The peak density, a quantity related
to the onset of BEC and to evaporation and decay rates, is identical
for each pair of curves.  The vertical axis is the number of atoms
with the given energy.  This is proportional to the occupation
function weighted by the total energy density of states for a
Ioffe-Pritchard trap, $\propto \epsilon^3+2\theta\epsilon^2$.  Here
$\theta/\kb T=1$.  }
\label{MB.BE.population.fig}
\end{figure}

The occupation function for a gas of $N$ identical bosons in a box of
volume $V$, and in the limit of $N\rightarrow\infty$ and
$V\rightarrow\infty$ but constant $N/V$, is called the Bose-Einstein
occupation function \cite{hua63}
\begin{equation} 
f_{BE}(\epsilon)=\frac{1}{\exp^{(\epsilon-\mu)/\kb T}-1} ,
\label{BE.distrib.func.eqn}
\end{equation}
where $\mu$ and $T$\label{muT.def.page} are Lagrange multipliers which constrain the
system to exhibit the correct population and total energy
through the conditions
\begin{equation}
\frac{N}{V}=\int d\epsilon \;\frac{\rho(\epsilon)}{V}\:f_{BE}(\epsilon)
\end{equation} 
and
\begin{equation}\frac{E}{V}=\int d\epsilon \;\epsilon \:\frac{\rho(\epsilon)}{V}\:
f_{BE}(\epsilon).
\end{equation} 
The physical interpretation of these parameters is that $\mu$ is the
chemical potential and $T$ is the temperature.  Figure
\ref{MB.BE.population.fig} 
shows the energy distribution of the population in the trap for a
classical gas described by the Maxwell-Boltzmann (MB) distribution and
for a quantum gas described by the Bose-Einstein (BE) distribution.
The two functions shown describe systems with identical peak
densities, a parameter relevant to the creation of a condensate.  The
MB distribution requires many more particles than the BE to create the
same peak density because the atoms are distributed at higher
energies, further from the center of the trap.

The difference between the BE and MB distributions originates in
the different assumptions about particle distinguishability.  To
intuitively understand this, we consider the outcome of these
differing assumptions for sparsely and densely occupied sets of energy
states.  
In the classical treatment there are many ways to arrange the
distinguishable particles among sparsely occupied energy states
because interchanging any pair of particles leads to a different
arrangement.
For very densely occupied states, interchanges of distinguishable
particles often result in the same arrangement because the particles
often do not change energy level.  For a collection of distinguishable
particles with a given total energy, there are ``extra''
arrangements in which the particles are sparsely distributed.
In contrast, for indistinguishable particles any interchange of two
particles results in the same overall arrangement of particles,
regardless of whether the states are sparsely or densely occupied.
There are thus no ``extra'' sparse arrangements of particles.

In a gas near the quantum degenerate regime the occupation of the
energy levels ranges from dense ($\sim 1$ at low energies) to sparse
($\ll 1$ at high energies).  Using the assumption of equal {\em a
priori} probabilities \cite{hua63}, distinguishable particles will
most likely be found in those arrangements with the more sparse level
occupations (higher overall energies) because there are so many more
of these arrangements.  Indistinguishable particles give no rise to
such ``extra'' arrangements, and so the most likely arrangements will
involve more dense level occupations (lower overall energies) than
predicted by classical theory.  Figure \ref{MB.BE.population.fig}
demonstrates this effect.  For each pair of curves the peak density,
$n_0$, is identical.  The peak density is the experimentally
observable quantity.  Since the MB occupation function favors
arrangements with higher energy (and thus atoms are distributed over a
larger volume in the trap), more atoms are required in the sample to
produce the same peak density.  The curves thus do not exhibit equal
area.

Bose-Einstein condensation occurs when the chemical potential goes to
zero and the occupation of the lowest energy state diverges.  This
occurs at the critical density \cite{hua63}
\begin{equation}
n_c=g_{3/2}(1)/\Lambda^3(T)
\label{bec.density.eqn}
\end{equation}
where $g_{n}(z)\equiv\sum_{l=1}^\infty
z^l/l^n$\label{boseg.def};
$g_{3/2}(1)=2.612$.  A gas that has undergone the Bose-Einstein phase
transition is said to be in the quantum degenerate regime because a
macroscopic fraction of the particles are in an identical quantum
state.

Although a hydrogen atom consists of two fermions, it behaves like a
composite boson for the studies in this thesis because the collision
interaction energies are extremely small compared to the
electron-proton binding energy \cite{eho31}.  The two fermions act as
a unit except in high energy collisions where electron exchange is
possible.  The typical interaction energy during low temperature
collisions is $\sim 1$~mK, which corresponds to $10^{-7}$~eV, $10^8$
times smaller than the binding energy.

To analyze the behavior of the degenerate Bose gas we shall separately
treat the condensed and non-condensed portions of the system.  The
condensate is treated in section \ref{condensate.qm.sec}.  The
non-condensed fraction, which we call the ``normal gas'' or ``thermal
gas'', is treated in appendix \ref{normal.degenerate.app}.  The
results derived here will be used to understand how the hydrogen
system is different from other gases that have been Bose condensed
(section \ref{hbec.props.sec}), and to interpret the experimental
results in chapter \ref{results.chap} (note that truncation effects
are important because of the shallow traps used for those experiments,
$4<\eta<6$).

\subsection{Description of the Condensate}
\label{condensate.qm.sec}

\subsubsection{Gross-Pitaevskii Equation}
When Bose-Einstein condensation occurs, a macroscopic fraction of the
particles occupy the lowest energy quantum state of the system, and
thus have the same wavefunction.  For a non-interacting Bose gas, that
wavefunction is simply the lowest harmonic oscillator wavefunction for
the trap.  Interactions become important when many particles occupy
this region of space and the local density increases.  In this case
the wavefunction spreads out due to repulsion among the atoms.  There
is a large body of literature on Bose-Einstein condensates (see
\cite{dgp98} for a recent review), and so we simply quote
here the most important results.

The Schroedinger equation for the
interacting condensate is called the Gross-Pitaevskii equation
\cite{pit61,gro61,gro63}, and has the form
\begin{equation}
-\frac{\hbar^2}{2m}\nabla^2\psi({\bf r}) 
+ V({\bf r})\psi({\bf r})
+ U_0 \left|\psi({\bf r})\right|^2 \psi({\bf r}) 
= \mu \psi({\bf r})
\label{gp.eqn}
\end{equation}
where $\psi({\bf r})$ is the condensate wavefunction to be determined.
The eigenenergy of the wavefunction is $\mu$, which is the total
energy of each condensate atom.  The quantity $U_0\equiv 4\pi\hbar^2
a/m$\label{U0.def} parameterizes the mean field energy, which is the
energy of interaction among the atoms per unit density, and is
repulsive for $s$-wave scattering lengths $a>0$.  For hydrogen in the
ground state, $a=0.648~{\rm \AA}$\cite{jdk95}, and $U_0/\kb
=3.92\times 10^{-16}~\mu{\rm K~cm}^{3}$.  The mean field energy
augments the trap potential by an amount proportional to the local
condensate density $n_{cond}({\bf r})=\left|\psi({\bf
r})\right|^2$\label{ncond.def.page}.  Note that interactions between
the condensate and non-condensed atoms are neglected here.  For the
experiments in this thesis, this interaction energy is small because
the density of non-condensed atoms (called the ``thermal gas'' or
``normal gas'') is small.  Furthermore, the interaction energy with
the thermal gas is essentially uniform across the condensate because
the density of non-condensed atoms varies only weakly; the thermal
energy of the normal gas ($T_c\sim60~\mu$K) is much larger than the
condensate mean field energy ($\mu/\kb\sim 2~\mu$K), and thus the
density of the normal gas does not change appreciably on the length
scale of the condensate.

We identify the eigenenergy $\mu$ in equation \ref{gp.eqn} with the
chemical potential of the system in equilibrium.  The chemical
potential is the energy required to add a particle to the system.
When a condensate is present, the normal gas is ``saturated'', and any
particles added to the system go into the condensate.  The energy
required to add the last atom to the condensate is $\mu$, the
eigenenergy, and so we  link $\mu$ with the chemical potential.  In
practice we measure $\mu$ spectroscopically through the peak density at
the center of the condensate, which is in turn measured through the
cold-collision frequency shift.  This is explained in section
\ref{peak.cond.dens.sec}.

\subsubsection{Thomas-Fermi Approximation}

If the condensate density is large enough so that the mean field
energy is much greater than the kinetic energy of the wavefunction,
then the kinetic energy term (also called the quantum pressure term)
in equation \ref{gp.eqn} may be neglected. Using this ``Thomas-Fermi''
approximation we obtain the condensate density profile
\begin{equation} n_{cond}({\bf r})= n_p-V({\bf r})/U_0
\label{tf.cond.profile.eqn}
\end{equation}
where $n_p\equiv\mu/U_0$ \label{np.def.page} is the peak condensate
density (the largest density in the condensate, found at the bottom of
the trap).  The Thomas-Fermi approximation is valid over most of the
volume of the condensate, but near the edges the condensate density
approaches zero, and the kinetic energy term should be included.  We
ignore this small correction in the calculations which follow because
it is minor for interpretation of the experiments described in this
thesis.  See \cite{pes97} for a good description of various ways to go
beyond the Thomas-Fermi approximation.

The condensate density profile may be obtained without the
Gross-Pitaevskii equation by assuming the condensate is stationary and
its particles are at rest, and then balancing hydrodynamic forces
\cite{gk82}.  A condensate particle in a region of potential energy
$\varepsilon$ has total energy $E=\varepsilon + n_{cond}(\varepsilon)U_0$.
Since there must be no net force on the particle, $E={\rm const}\equiv
n_p U_0$ and $n_{cond}(\varepsilon)=n_p-\varepsilon/U_0$.

\subsubsection{Population and Loss Rate}
The population of the condensate is easily computed using the
Thomas-Fermi wavefunction in the bottom of the IP trap, which is
parabolic for the condensate sizes and trap parameters of interest in
this thesis.  We approximate the potential energy density of states
(see p\pageref{varrho.def.page}) as $\varrho(\varepsilon)\approx {\cal
A}_{IP} \sqrt{\varepsilon}\theta$ and obtain
\begin{eqnarray}
N_c & = & \int_0^{U_0n_p}d\varepsilon\; {\cal A}_{IP}\: \theta \sqrt{\varepsilon} 
(n_p-\varepsilon/U_0) \nonumber \\
 &=&\frac{4}{15}{\cal A}_{IP} \:\theta\: U_0^{3/2} n_p^{5/2} .
\label{Ncond.eqn} 
\end{eqnarray}
As noted above, the total energy of each condensate atom is the
chemical potential $\mu=n_p U_0$, so $E_c=\mu N_c$\label{Ec.def.page}.

The loss rate from the condensate due to two-body dipolar relaxation is found
by integrating the square of the density over the volume of the
condensate:
\begin{eqnarray}
\dot{N}_{2,c} &=& -\int_0^{U_0n_p}d\varepsilon\; {\cal A}_{IP}\: \theta 
\sqrt{\varepsilon}\:(g/2!)\: (n_p-\varepsilon/U_0)^2 \nonumber \\
 &=& -\frac{16}{105}\:\frac{g}{2!}{\cal A}_{IP} \theta \:U_0^{3/2} n_p^{7/2}
\label{Ndot.condensate.eqn}
\end{eqnarray}
where the $2!$ accounts for correlation properties of the condensate
\cite{wam94,bgm97}.  Here the dipolar decay rate constant $g$ is
slightly different from that given in equation
\ref{dipolar.decay.const.eqn} because the rate constant
$G_{dd\rightarrow cd}$ should be multiplied by 2; the energy $\theta$
liberated in this process is large compared to $\mu$.  This term is
small, so $g$ is not effected much.
The energy loss rate is $\dot{E}_{2,c}=\mu \dot{N}_{2,c}$\label{Edot2c.def}.
The characteristic condensate decay rate in a parabolic trap is
\begin{equation}
-\frac{\dot{N}_{2,c}}{N_c} = \frac{2}{7}g n_p,
\end{equation}
which is $1.6~{\rm s}^{-1}$ for a typical peak density
$n_p=5\times10^{15}~{\rm cm}^{-3}$.

\section{Properties of a Bose-Condensed Gas of Hydrogen}
\label{hbec.props.sec}

Hydrogen differs in several ways from alkali metal atoms that
have been Bose condensed.  The principal differences are its small
mass and small $s$-wave scattering length.  How do these properties
influence the system?

First, the small mass implies that BEC occurs at a higher temperature
for a given peak density: from equation \ref{bec.density.eqn}, $T_c
\propto 1/m$.  We have observed the transition at temperatures roughly
50 times higher than in the other systems.

Further, as estimated by Hijmans \etal\ \cite{hks93} and will be
shown in the following sections, the maximum equilibrium condensate
fraction is small for hydrogen.  This will be explained by noting the
relatively high density of the condensate, as compared to the thermal
cloud.  This high density leads to high losses through dipolar decay,
which result in heating of the system.  This heating must be balanced
by evaporative cooling, which proceeds slowly in hydrogen because of
the small collision cross section.  The result of these factors is
that only small condensate fractions are possible in equilibrium.
Possible remedies will be noted.

Finally, hydrogen's small collision cross-section should allow
condensates of H to be produced by evaporative cooling that contain
many orders of magnitude more atoms than those possible in
alkali-metal species, as explained in the section
\ref{ultimate.condensate.pop.sec}.

\subsection{Relative Condensate Density}
\label{relative.condensate.density.sec}

For trapped hydrogen, the condensate density grows much greater than
the density of the non-condensed portion of the gas at even very small
condensate fractions \cite{gre95}, a distinct difference from other
species that have been Bose-condensed.  This is noteworthy for
possible future studies of behavior near the phase transition.  As
shown in section \ref{achievable.condensate.density.sec}, it also has
implications for the maximum condensate fraction that may be achieved
in hydrogen.  In this section we calculate the ratio of the densities
as a function of the condensate fraction.  We compare hydrogen to Li,
Na, and Rb.

The condensate fraction is
\begin{equation}
F\equiv \frac{N_c}{N_c + N_t}
\label{condensate.fraction.def.eqn}
\end{equation}
where the population of the thermal cloud, $N_t$, is found using
equation \ref{thermal.population.eqn} and the population of the
condensate is given by equation \ref{Ncond.eqn}.  The fraction can be
written in terms of the more convenient population ratio, $f\equiv
N_c/N_t$\label{f.def}, as $F=f/(1+f)$.  For a given peak condensate density, sample
temperature, and set of trap parameters, the population ratio is
\begin{equation}
f=
\underbrace{\left(\frac{2^{7/2}h^6a^{3/2}}{\pi^{7/2}m^3\kb^3}\right)}
 _{\textstyle f_0}
\underbrace{\left(\frac{\phi}{\pi^4+60\zeta(3)\phi}\right)}
_{\textstyle f_{trap}(\phi)}
\underbrace{\left(\frac{n_p^{5/2}}{T_c^3 A_0(\eta,\phi)}\right)}
_{\textstyle f_{sample}}
\label{condensate.fraction.from.density.eqn}
\end{equation}
where $f_0=7.37\times 10^{-34}~({\rm cm}^3)^{5/2} \mu{\rm K}^3$ for H;
$f_{trap}$ indicates whether the trap shape experienced by the thermal
cloud is predominantly harmonic (large $\phi$) or predominantly linear
(small $\phi$) in the radial direction ($\phi\equiv\theta/\kb T$ is a
unitless measure of the trap bias energy); and $f_{sample}$ carries
the details of the thermal cloud and condensate occupation.

The ratio of the peak condensate density to the critical BEC density,
$R\equiv n_p/n_c(T_c)$, can be expressed in terms of the
occupation ratio as
\begin{equation}
R=
\underbrace{\left(\frac{1}{2^{29/10} g_{3/2}(1)\pi^{1/10}}\right)}
 _{\textstyle R_0}
\underbrace{\left(\frac{h^2}{m \kb T_c a^2}\right)^{3/10}}
	_{\textstyle R_{atom}}
\left(\frac{f A_0(\eta,\phi)}{f_{trap}(\phi)}\right)^{2/5}
\label{condensate.density.ratio.eqn}
\end{equation}
The prefactor is $R_0=0.0457364$.
Table \ref{condensate.density.ratio.table} 
\begin{table}[tb]
\centering\begin{tabular}{|c|c|c|c|c|c|c|c|}
\hline
species & $m$  & $a$ & $T_c$ & $R_{atom}$ &
$\phi$ & $f_{trap}(\phi)$ & 
$R_0 R_{atom}/f_{trap}(\phi)^{2/5}$ \\
\hline
H & 1 & 0.648 \cite{jdk95} & 60 & 230 & 0.6  & 0.0043 & 94 \\
Li \cite{bsh97} & 7 & -14.4 \cite{ams95} & 0.30 & 98 & $2.2\times 10^5$ & 0.012 & 25  \\
Na \cite{mad96} & 23 & 27.5 \cite{twj96} & 2.0 & 26 & 30 & 0.013 & 6.8  \\
Rb \cite{bgm97} & 87 & 57.1 \cite{vtf97} & 0.67 & 16 & 160 & 0.011 & 4.4  \\
\hline
\end{tabular}
\caption[comparison of density ratios] {Comparison of the parameters
governing the ratio of condensate density to thermal gas density for
the various BEC experiments.  The units are amu, \AA, $\mu$K.  The
density ratio (for small condensate fractions) is the last column
multiplied by the condensate fraction to the 2/5 power.  This ratio is
much larger for H than for Na or Rb, indicating that two-body loss
from the condensate becomes important at much lower $f$.  For Li,
hydrodynamic collapse of the condensate occurs before the two-body
loss process studied here becomes important. }
\label{condensate.density.ratio.table}
\end{table}
gives the parameters appearing in equation
\ref{condensate.density.ratio.eqn} for the various BEC experiments.
For even a small occupation ratio of $f=0.05$, the H condensate will
be 28 times more dense than the thermal gas, which can be compared to
a Rb condensate that will be only be 1.3 times more dense.  The loss
rates from the condensate are thus very different for the two systems,
as shown in the next section.  Note that the trap oscillation
frequencies play no role in these results.  The only assumptions are
that the trap is of the IP form with bias field\footnote{When $\phi$
is small, the condensate experiences a harmonic potential while the
thermal gas experiences a predominantly linear potential in the radial
direction.  When $\phi$ is large, both condensate and thermal gas
experience the same potential functional form.}  $\phi$, that the
condensate is in the Thomas-Fermi regime, and that $\mu\ll\kb T$ so
that mean-field interaction energy of the condensate with the thermal
cloud can be neglected.

\subsection{Achievable Condensate Fractions}
\label{achievable.condensate.density.sec}

The temperature of a trapped sample, and thus its condensate fraction,
is set by a competition between heating and cooling, as described in
section \ref{evap.cooling.rate.sec}.  Hydrogen's anomalously low
scattering length means that the elastic collision rate is small, and
thus evaporation proceeds slowly.  The evaporative cooling power is
small, and the heating-cooling balance favors higher temperatures
and lower condensate fractions than would occur if hydrogen had a larger
collision cross section.  Higher temperatures are also favored by the
large density of the condensate, which places an extra heat load on
the system due to dipolar relaxation.  This extra heat load can easily
be much larger than the dipolar relaxation heat load of the thermal
gas.  In this section we examine each of these factors in more detail.

\begin{figure}[tb]
\centering \epsfxsize=4in \epsfbox{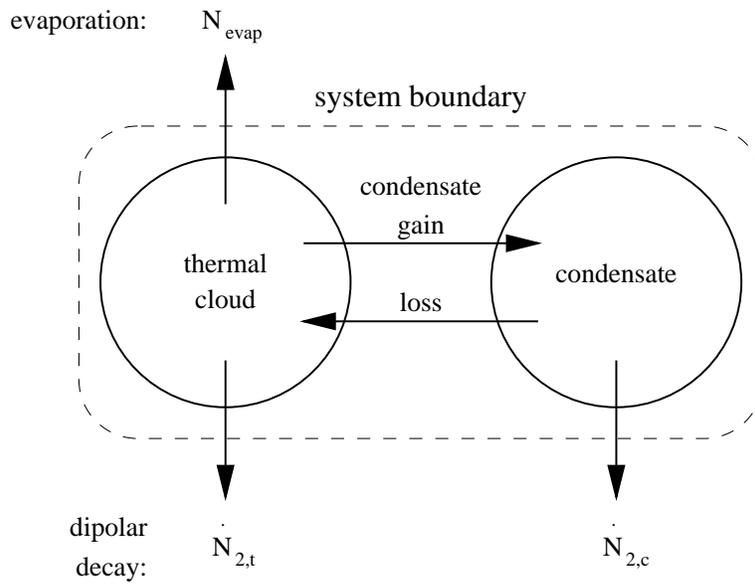}
\caption[diagram of losses from trapped gas system]{ Diagram of system
consisting of thermal cloud and condensate.  We assume the particle
and energy exchange rates between the condensate and thermal cloud are
fast compared to the loss rates.  The dynamics of the gas are set by
evaporation and dipolar decay from the normal gas when the condensate
is small.  When the condensate is large, however,
$\dot{N}_{2,c}\gg\dot{N}_{2,t}$, and the dynamics are dominated by the
dipolar decay rate from the condensate in balance with the evaporation
rate.}
\label{system.cartoon.fig}
\end{figure}

We model the system as two components, the condensate and thermal gas,
that are strongly linked.  As indicated in figure
\ref{system.cartoon.fig}, losses from the system occur through dipolar
decay in the thermal gas, dipolar decay in the condensate, and
evaporation from the thermal cloud.  
We assume that particles and energy are exchanged between the
condensate and thermal gas quickly compared to all loss rates.  The
system is in a dynamic equilibrium.  In section
\ref{condensate.feeding.rate.sec} we examine the validity of the
assumption of fast feeding of the condensate.

\subsubsection{Dipolar Heating Rate}
\label{cond.dipolar.heating.sec}
Here we study the rate of heating the system due to dipolar decay in
the condensate and in the thermal gas.  We define heating as the
removal of atoms from the sample which carry less than the average
amount of energy per particle.  We find that for hydrogen the
condensate heat load exceeds the thermal gas heat load at condensate
fractions of 0.3\%.  We then compare this ratio for H to that of the
atomic species in other BEC experiments.  While illustrative, this
comparison is not completely appropriate because the loss mechanisms
in the other systems are different.  Hydrodynamic collapse of the Li
condensate prevents it from growing larger than about 1000 atoms in
the experiments of Hulet \etal\ \cite{bsh97,ssh98}.  Three-body loss
process are dominant over two-body dipolar decay in the Rb and Na
experiments.  The comparison made in this section therefore serves
simply to indicate how different hydrogen is from the other species.

The heating rate for a process (labeled $\sigma$), with energy and particle
loss rates $\dot{E}_\sigma$ and $\dot{N}_\sigma$, is the difference
between the average energy per particle in the system and the energy
per particle that is removed, multiplied by the rate at which
particles are removed:
\begin{equation}
H_\sigma= \dot{N}_\sigma
\left(\frac{\dot{E}_\sigma}{\dot{N}_\sigma}-\frac{E}{N}\right)
\label{H.def.eqn}
\end{equation}
where $E/N$ is the average energy per particle for the whole system.
In this analysis $N=N_c+N_t$.  The characteristic rate for this
process is $H_\sigma/E$.

We wish to find the ratio of the heating rates due to dipolar decay
from the condensate ($H_c$) and from the thermal cloud ($H_t$) as a
function of the occupation fraction $f$.  Losses from the condensate
play a significant role in the trap dynamics when
$H_c \geq H_t$.

\begin{figure}[tb]
\centering \epsfxsize=5in \epsfbox{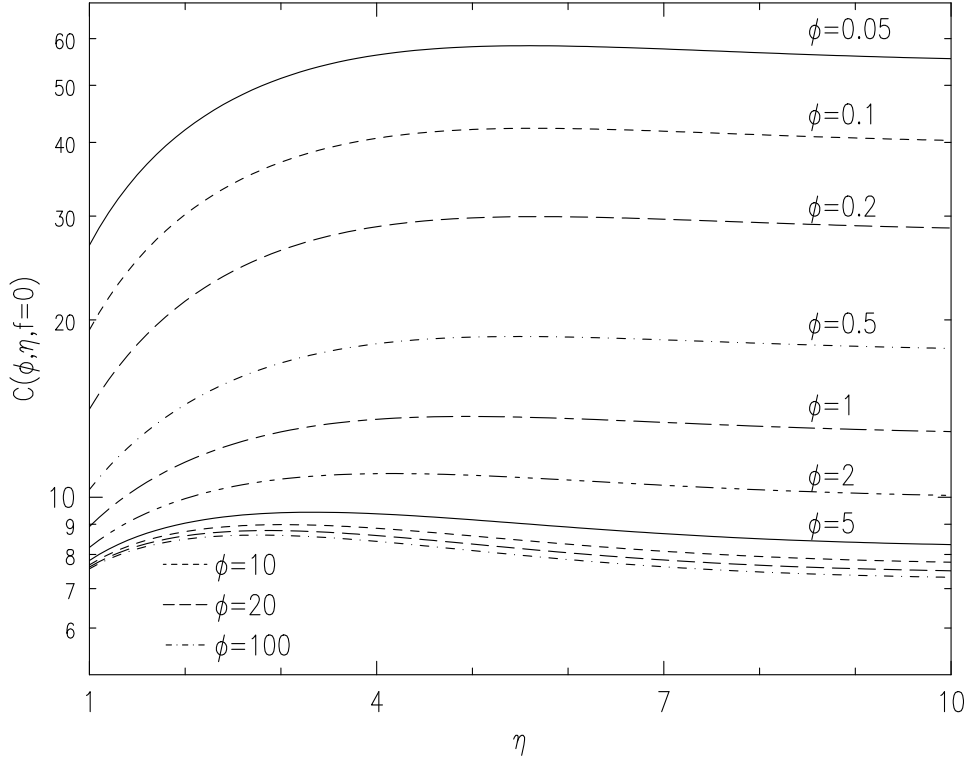}
\caption[behavior of $C(\eta,\phi,f=0)$]{ Behavior of function
$C(\phi,\eta,f=0)$, plotted for various trap bias fields
$\phi=\theta/\kb T$.  The asymptotic behavior for large $\phi$ can be
seen.}
\label{cfunc.fig}
\end{figure}

By inverting equation \ref{Ncond.eqn} to get $n_p$, and using the
definition of $f$, we obtain $\dot{N}_{2,c}$ (from equation
\ref{Ndot.condensate.eqn}) in terms of $f$ and $N_t$.  For a small
condensate (i.e. $\mu\ll \overline{E}_t$), straightforward algebra obtains
\begin{equation}
\frac{H_c}{H_t} = f^{7/5}\;
\underbrace{\left(\frac{h^2}{m \kb T_c a^2}\right)^{3/10}}
	_{\textstyle R_{atom}}
\;
\underbrace{\frac{\pi^{2/5}}{2^{21/10} \;105}\;
	\frac{(\pi^4 + 60 \zeta(3)\phi)^{7/5} A_0(\eta,\phi)^{7/5}}
		{\phi^{2/5} Q_1(\phi,\eta)\left(1-
	\frac{Q_2(\phi,\eta)A_0(\eta,\phi)(1+f)}
	{Q_1(\phi,\eta)A_1(\eta,\phi) B(\phi)}\right)}}
_{\textstyle  C(\phi,\eta,f)}
\label{heating.ratio.eqn}
\end{equation}
where $B(\phi)$ was defined in equation \ref{b.phi.defn.eqn}.  The
function $C(\phi,\eta,f)$ has only weak dependence on $f$ for $f\ll
1$,  so the dominant dependence of the heating ratio on the
occupation fraction is $f^{7/5}$.  The function $C(\phi,\eta,f=0)$ is
plotted in figure \ref{cfunc.fig}.

For the experiments described in this thesis $\phi\sim 0.6$ and
$\eta\sim 5$.  Then $C\sim 17$, and  for small $f$
\begin{equation}
\frac{H_c}{H_t} = f^{7/5}\;3.9\times10^3.
\end{equation}
The heating rates are equal when $f=0.27\%$.  This means that heating
of the sample due to losses from the condensate becomes a very
significant problem even when the condensate is still quite small.  It
is difficult to hold the system in equilibrium for large condensates
because the heat load that must be balanced by evaporation quickly
grows too large.  To obtain a large condensate fraction, the normal
gas could, of course, be removed, and then $E/N$ would be $\mu$.  The
heating rate would be zero.  However, the condensate would decay
quickly and there would be no reservoir from which to replenish it.

The experiments on hydrogen may be compared to other BEC experiments
using Na, Rb, and Li.  Relevant values for the quantities in
equation~\ref{heating.ratio.eqn} are tabulated in table
\ref{heatingratiotable}.  These values help us understand why high
equilibrium condensate fractions can be created in Na and Rb
systems.  Note that three-body loss processes, while insignificant in
hydrogen, are the limiting factor which precludes larger condensate
densities in Na and Rb.

\subsubsection{Evaporative Cooling Rate}
In this section we compare the evaporative cooling rates of hydrogen
and other species.  We use expressions derived using the MB energy
distribution for the thermal gas because analytic expressions exist
and should yield rates reasonably close to those obtained using the BE
distribution and the quantum Boltzmann transport equation.  We
postulate this because evaporation is driven preferentially by atoms
with high energy, which are in regions where the energy distribution
is essentially classical.  Furthermore, we examine here only the
scaling of the evaporation rate with the atomic properties $m$, $a$,
and the experimentally chosen temperature $T$; exact rates are not of
interest.  The correct particle loss rate may be smaller by a factor
of order unity from the results presented here, but the energy carried
away by evaporating atoms should be very similar for the two
distributions.  The escaping atoms must have energy greater than the
trap depth whether the gas is classical or quantum.

As shown in section \ref{evap.cooling.rate.sec}, the energy loss rate
due to evaporation is
\begin{equation}
\frac{\dot{E}_{evap}}{\dot{N}_{evap}} \simeq \kb T (\eta + 1)
\end{equation}
for $\eta>6$.  The cooling rate is
\begin{equation}
H_{evap} = \dot{N}_{evap}\left(	\frac{\dot{E}_{evap}}{\dot{N}_{evap}}
-\frac{E}{N}\right)
\end{equation}
which, for large $\eta$, is $\dot{N}_{evap}\kb T$ times a factor
which depends linearly on $\eta$.  To compare the cooling rates of
gases of different species, it is thus sufficient to consider
$\dot{N}_{evap}/N \sim n_0\sigma \bar{v} e^{-\eta}(\eta+{\rm const})$.
When $n_0=n_c(T_c)$, the ratio is, for fixed $\eta$,
\begin{eqnarray}
\frac{\dot{N}_{evap}}{N} & \propto & \frac{(m\kb T_c)^{3/2}}{h^3}\: 
8 \pi a^2 \sqrt{\frac{8\kb T_c}{\pi m}} \nonumber \\
& \propto & m a^2 T_c^2 
\end{eqnarray}
which is listed in table \ref{heatingratiotable}.  
\begin{table}[tb]
\centering\begin{tabular}{|c|c|c|c|}
\hline
species & $m a^2 T_c^2$ & $C(\phi,\eta=\infty,f=0)$ & $f_=$ \\
\hline
H & 1000 & 16 & 0.28 \% \\
Li \cite{bsh97} & 130 & 7.2 & 0.91 \% \\
Na \cite{mad96} & 70,000 & 7.4 & 2.3 \% \\
Rb \cite{bgm97} & 130,000 & 8.3 & 3.1 \% \\
\hline
\end{tabular}
\caption[comparison of heating-cooling parameters] {Comparison of the
parameters entering the heating/cooling balance for the various BEC
experiments.  The units are amu, \AA, $\mu$K. The second column,
$ma^2T_c^2$, is proportional to the cooling rate.  The third column
carries the trap shape dependence of the dipolar decay process in the
thermal gas.  The forth column,
$f_==(R_{atom}C(\phi,\infty,0))^{-5/7}$, is the occupation fraction at
which the heating rate due to dipolar decay in the thermal cloud
matches that due to the condensate.}
\label{heatingratiotable}
\end{table}
We see that, for a
similar $\eta$, the cooling rate for H is much smaller than that for
Na and Rb.  To increase the cooling rate in H, the parameter $\eta$
must be reduced, which reduces the cooling efficiency.  

We conclude that the equilibrium condensate fraction expected in H is
much smaller than that in Na and Rb for two reasons.  First, loss from
the condensate significantly influences the thermodynamics of the
system at much lower condensate fractions in H experiments since $f_=$ is
so small.  Furthermore, the cooling rate for a given $\eta$ is much
less in the H experiment.  It is thus easier to achieve large
condensate fractions in Rb and Na than in H.  Note that the condensate
fraction for Li experiments is limited by hydrodynamic collapse of the
condensate \cite{ssh98}.

A possible remedy for the low evaporation rate in H experiments has
been suggested by Kleppner \etal\ \cite{kgk99}.  A small admixture of
impurity atoms in the trap could act as ``collision moderators'' since
the collision cross section with H would most likely not be
anomalously low; it should be on the order of $10^3$ times larger than
that of H-H collisions.  The increased evaporation rate should allow
more atoms to be condensed, and with a much faster experimental cycle
time.

\subsection{Condensate Feeding Rate}
\label{condensate.feeding.rate.sec}

As the condensate decays by dipolar relaxation it must be fed from the
thermal cloud (which is assumed to be at the critical BEC density,
$n_c$).  This feeding process may be bottlenecked by the small
collision cross section, $\sigma$, of hydrogen.  To estimate the
maximum condensate replenishment rate, $R_{max}$, we consider the
event rate for collisions between two thermal atoms assuming that the
collisions occur inside the condensate.  Ignoring stimulated
scattering for the moment, this rate is the collision rate per atom
times the number of thermal atoms in the region of the condensate,
\begin{equation}
\dot{N}_{col,feed}=(n_c \sigma \bar{v}_{BE} \sqrt{2})\; (V_{cond}\; n_c)
\end{equation}
where
\begin{equation}
\bar{v}_{BE}=\frac{g_2(e^{\mu/\kb T})}{g_{3/2}(e^{\mu/\kb T})}\:
\sqrt{8\kb T/\pi m}
\end{equation}
is the mean particle speed in a Bose gas
(ignoring truncation; $g_2(1)/g_{3/2}(1)=0.6297$).  The condensate volume (for samples in the
Thomas-Fermi regime) is, using equation \ref{tf.cond.profile.eqn},
\begin{equation}
V_{cond}= 2{\cal A}_{IP}\;\mu^{3/2}
\left\{\frac{\mu}{5}+\frac{\theta}{3}\right\}.
\label{Vcond.def.eqn}
\end{equation}
For small chemical potentials we approximate $V_{cond}=2{\cal
A}_{IP}\theta\mu^{3/2}/3$.  We assume that some fraction $\zeta$ of
the collisions result in an atom being added to the condensate.  One
might expect $\zeta$ to be less than unity because only a fraction of
collisions involve atoms with initial momentum and energy consistent
with one atom going into the BEC wavefunction, which has nearly zero
momentum and energy.  On the other hand, the Bose enhancement factor
very strongly favors population of the condensate.  Calculations by
Jaksch \etal \cite{jgg98} (equation 19a) indicate that for a
condensate near its equilibrium population, and with $\mu\ll\kb T$,
the fraction is $\zeta\sim 1/5$.  For the moment we leave the
parameter free.  The maximum condensate replenishment rate is $\zeta
\dot{N}_{col,feed}$.

In equilibrium the condensate population is varying slowly compared to
the feeding and loss rates.  We may therefore equate the maximum
replenishment rate to the condensate dipolar decay loss rate (from
equation \ref{Ndot.condensate.eqn}):
\begin{eqnarray}
-\dot{N}_{2,c} &=& \zeta \dot{N}_{col,feed} \nonumber \\
\frac{16}{105}\:\frac{g}{2!}{\cal A}_{IP} \theta \:U_0^{3/2} n_p^{7/2}
& = &
\frac{2\sqrt{2}}{3}\zeta n_c^2 \sigma \bar{v}_{BE} 
{\cal A}_{IP} U_0^{3/2} n_p^{3/2}
\label{feed.loss.rate.eqn}
\end{eqnarray}
We find that the loss rate matches the maximum feeding rate when the
ratio of peak condensate density to thermal density is
\begin{equation}
\frac{n_p}{n_c} = a\;\sqrt{\frac{280\zeta}{g\sqrt{\pi}}}\;
\left(\frac{\kb T}{m}\right)^{1/4}.
\end{equation}
For hydrogen at $T=60~\mu$K and $\zeta=1$ the ratio is $n_p/n_c=16$.
The ratio is strongly influenced by the scattering length, and is
about 30 times larger for the Rb experiment listed in table
\ref{heatingratiotable}, assuming a decay rate $g=10^{-16}~{\rm
cm^3/s}$ \cite{bgm97}.  In the experiments described in chapter
\ref{results.chap} the largest observed value of this ratio is about
20.

A detailed understanding of the maximum condensate feeding rate is
clearly desirable, but is beyond the scope of this thesis.
Nevertheless, it is clear that the finite rate of replenishing the
condensate from the thermal cloud can bottleneck growth of the
condensate.

\subsection{Ultimate Condensate Population}
\label{ultimate.condensate.pop.sec}

One figure of merit for a Bose condensate is the total number of
condensed atoms, which is related to the number of atoms in the trap
just prior to condensation.  We investigate here one limit to this
quantity.

For a trap of effective length $L$ and effective radius
$r$ which contains a density $n$, the atomic population is 
\begin{equation}
N\simeq\pi r^2 L n.
\end{equation} 
If evaporative cooling is used for phase space compression, one must
consider that for efficient cooling to occur, energetic atoms must be
able to reach a pumping surface\footnote{A pumping surface is
basically a one-way valve: atoms which reach the surface leave the
trap and never return.}  before having a collision.  For a
cigar-shaped trap this sets a maximum radius given by the collision
length
\begin{equation}
r_{max}=\bar{\ell}=\frac{1}{n \sigma}.
\end{equation} 
For a density-limited sample size the maximum trap
population is
\begin{equation}
N_{max}=\frac{\pi L}{n\sigma^2}
\end{equation}
which exhibits a strong dependence on the scattering cross section.
Spin-polarized atomic hydrogen features a cross section more than
$10^{3}$ times smaller than that of the alkalis, indicating that
samples with $10^6$ times more atoms are possible.  Of course, the
alkali traps can work at lower densities, but this necessitates slower
forced evaporation, and hence increases the loss due to background gas
collisions.  Use of a ``pancake'' instead of ``cigar'' geometry would
mean the sample size is only limited in one dimension instead of two.
However, the dimensionality of evaporation would also be reduced from
two to one dimensions, and evaporation would thus become significantly
less efficient.

Table \ref{ultimate:evap:table} summarizes this fundamental limit on
condensate population for thermal cloud densities of $n\simeq
1\times10^{14}~{\rm cm}^{-3}$.
\begin{table}[tb]
\centering\begin{tabular}{|c|c|c|c|} \hline
species & $\sigma$ & $r_{max}$ & $N_{max}/L$ \\
 & ${\rm cm}^2$ & cm & ${\rm cm}^{-1}$ \\ \hline
H & $1.1 \times 10^{-15}$ & 10 & $3\times 10^{16}$ \\
Na & $2 \times 10^{-12}$ & $5 \times 10^{-3}$ & $9\times 10^{9}$ \\
Rb & $8 \times 10^{-12}$ & $1.2 \times 10^{-3}$ & $5\times 10^{8}$ \\
\hline
\end{tabular}
\caption[comparison of limits on number of trapped atoms] {Limits on
number of trapped atoms when evaporative cooling is the sole cooling
mechanism.  We assume a peak density $n_0\simeq 1\times10^{14}~{\rm
cm}^{-3}$.}
\label{ultimate:evap:table}
\end{table}
Perhaps more important that producing huge condensates is the
sustained production rate of cold, coherent atoms for an atom laser.
This production rate should scale as $N_{max}$ times the evaporation
rate, which scales as $ma^2T_c^2$.  As shown in table
\ref{heatingratiotable}, the evaporation rate for Na and Rb is about
$10^2$ higher than for H.  We therefore expect the production rate of
coherent atoms to be $10^{6-2}=10^4$ times bigger for H than for Na
and Rb.

%% file: rfcell.tex

\chapter{Implementing RF Evaporation}
\label{rfcellchapter}

Previous attempts to realize BEC in hydrogen were thwarted by
inefficient evaporation (see section
\ref{evaporation.dimensionality.sec}).  This chapter discusses the
solution to this problem and the hardware required to implement this
solution.

In the improved apparatus an rf magnetic field couples the Zeeman
sublevel of the trapped atoms to an untrapped level, but only in a
thin shell around the trap called the resonance region.  Resonance
occurs where the energy of the rf photons, $\hbar \omega_{rf}$,
matches the energy splitting between hyperfine states, proportional to
the strength of the trapping field.  Only those atoms with enough
energy to reach the high potential energy of the resonance region are
ejected.  The resonance region thus constitutes a ``surface of death''
which surrounds the sample, and so atoms with high energy in any
direction are able to quickly escape.

The rf evaporation scheme needed to be incorporated into an existing
cryogenic experiment.  The basic geometry of the trapping cell is
constrained by the superconducting magnets which create the trap
potential.  A vertical bore of 5~cm diameter is available for the cell
along the cylindrical symmetry axis of the magnets, as shown in figure
\ref{trap.apparatus.fig}.  Into this bore fits the ${\rm H_2}$
dissociator and the plastic tube which contains the gas as it is being
trapped.  The length of the trapping region is about 60~cm.  The cell
is thermally and mechanically anchored to the dissociator, which is
anchored to the mixing chamber of an Oxford Model 2000 dilution
refrigerator.  The mixing chamber is situated above the magnets.

A complete redesign of the cryogenic trapping cell was required in
order to incorporate the rf magnetic field.  Any good electrical
conductors in the vicinity of the rf fields and thermally connected to
the cell would absorb power from the rf field via rf eddy currents,
causing the cell to heat up. Metals, some of which had supplied
thermal conductivity, thus needed to be eliminated.  A superfluid
helium jacket was employed to provide the heat transport formerly
supplied by these metals.  In addition, the coils which drive the
field were compensated so that the field is very weak far away from
the atoms, where the large pieces of copper that comprise the
dissociator and mixing chamber are located.  In this chapter we discuss
these and other design considerations, provide construction notes, and
describe performance tests of the apparatus.

\section{Magnetic Hyperfine Resonance}
\label{hyperfine.zeeman.sec}

In this section we determine the magnitude and frequency of the rf
magnetic field required to drive the evaporation efficiently.  We
summarize the ground states of hydrogen, beginning with the hyperfine
interaction, and then adding the Zeeman effect.  We obtain the four
ground states and their energies.  We then derive the matrix elements
for rf transitions between these states.  After making a series of
simplifying approximations, we obtain the Rabi frequency for these
transitions.  An analytic theory for transition probabilities in
$N$-level systems is then adapted to our situation.  We then calculate
the field strength required to drive evaporation efficiently for the
trap and sample parameters of immediate interest in this experiment.

\subsection{H in a Static Magnetic Field}
\label{hyperfine.derivation.sec}
The ground states of hydrogen in a magnetic field are influenced by
the hyperfine interaction (interaction between electron and proton
magnetic moments) and the Zeeman interaction (the proton and
electron magnetic moments interacting with the applied magnetic
field).  Here we summarize Cohen-Tannoudji \etal\ \cite{ctdl77} (we
use SI units).  

We start in the zero magnetic field regime.  The hyperfine interaction
Hamiltonian is
\begin{equation}
{\cal H}_{hf}=A {\bf I}\cdot{\bf S}
\end{equation}
where ${\bf I}$ is the nuclear spin operator, ${\bf S}$ is the
electron spin operator, and $A\hbar = 2\pi\times \nu_{hf}$.  The
hydrogen ground state hyperfine frequency is $\nu_{hf}=1.420~$GHz.
The total angular momentum is ${\bf F}={\bf I}+{\bf S}$.  The
eigenstates $\left|F,m_F\right\rangle$ can be written in the
$\left|m_S,m_I\right\rangle$ basis:
\begin{eqnarray}
\left|1,1\right\rangle & = & \left|+,+\right\rangle \\
\left|1,0\right\rangle & = & \frac{1}{\sqrt{2}}
\left(\left|+,-\right\rangle + \left|-,+\right\rangle\right) \\
\left|1,-1\right\rangle & = & \left|-,-\right\rangle \\
\left|0,0\right\rangle & = & \frac{1}{\sqrt{2}}
\left(\left|-,+\right\rangle - \left|+,-\right\rangle\right).
\end{eqnarray}

\begin{figure}[tb]
\centering \epsfxsize=5in \epsfbox{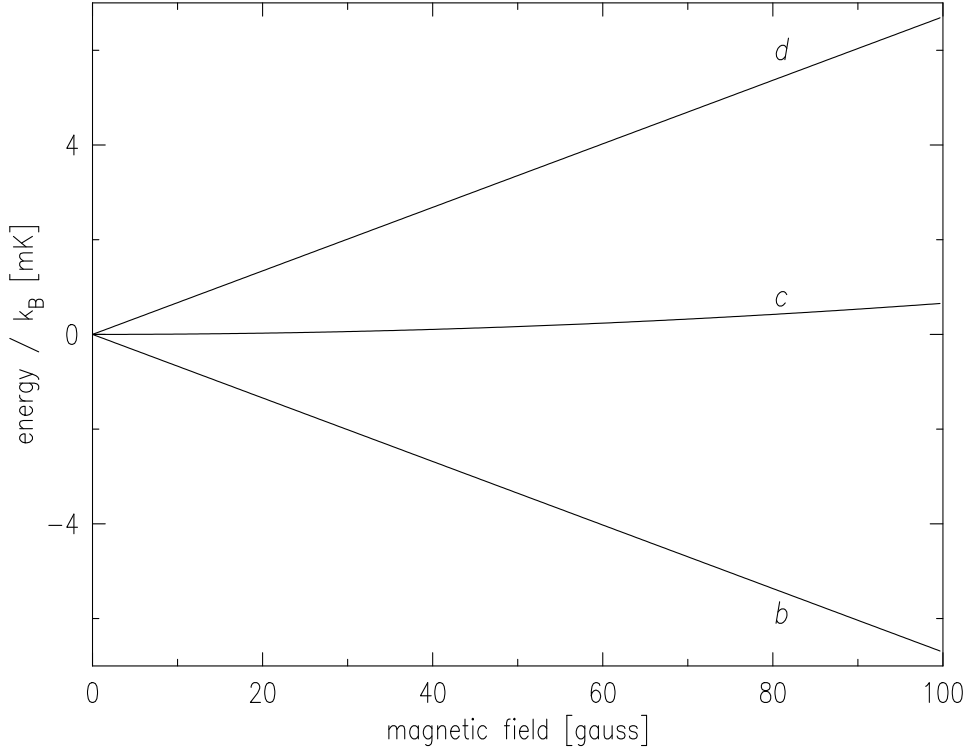}
\caption[low-field Zeeman diagram of hydrogen]{ Zeeman diagram of
hydrogen $F=1$ states in low fields.  State $d$ is trapped, state $c$
is untrapped, and state $b$ is ``antitrapped'', expelled from the
trapping region.  At high fields the $c$ state is trapped, as shown in
figure \ref{intro.zeeman.fig}.}
\label{hyperfine.lowfield.fig}
\end{figure}

In an applied magnetic field, ${\bf B}=B_0\hat{z}$, the Hamiltonian
includes the Zeeman term,
\begin{equation}
{\cal H}_Z=-{\bf B}\cdot({\bf M}_S + {\bf M}_I),
\end{equation}
where ${\bf M}_S=(q g_e/2 m_e){\bf S}$, ${\bf M}_I=(-q g_p/2m_p){\bf
I}$, $q$ is the electron charge ($q<0$), $g_e\simeq 2.0023$ is the electron
gyromagnetic ratio, and $g_p\simeq 5.585$ is the proton gyromagnetic
ratio.  The eigenstates of the combined Hamiltonian ${\cal H}={\cal H}_{hf}+{\cal H}_Z$ are then
\begin{eqnarray}
\left|d\right\rangle & = & \left|1,1\right\rangle \\
\left|c\right\rangle & = & \cos\theta\left|1,0\right\rangle + \sin\theta\left|0,0\right\rangle \\
\left|b\right\rangle & = & \left|1,-1\right\rangle \\
\left|a\right\rangle & = & \sin\theta\left|1,0\right\rangle - \cos\theta\left|0,0\right\rangle  \\
 & & \nonumber \\
\tan\theta & = & \frac{-1 + \sqrt{1+x^2}}{x}
\end{eqnarray}
where $x=-B_0/B_{hf}$, $B_{hf}=A\hbar^2/\mu_B g_e (1+\varphi)=506.65~$G,
$\mu_B=|q|\hbar/2m_e$ is the Bohr magneton\label{muB.def.page}, and $\varphi=g_p m_e/g_e
m_p\simeq1.5\times10^{-3}$ is a measure of the relative contribution
of the proton spin to the energy.  The energies are, in order of
decreasing energy,
\begin{eqnarray}
E_d & = & \mu_B B_0 (1-\varphi) \\
E_c & = & -\frac{A\hbar^2}{2}\left(1-\sqrt{1+x^2}\right) \\
E_b & = & -\mu_B B_0 (1-\varphi) \\
E_a & = & -\frac{A\hbar^2}{2}\left(1+\sqrt{1+x^2}\right)
\label{hfz.energies.eqn}
\end{eqnarray}
Figure
\ref{hyperfine.lowfield.fig} 
shows these energy levels.  Confining our attention to the the low
fields which are of interest for trapped samples near the BEC
transition and to the $F=1$ manifold, we see that atoms in the $d$
state are trapped, those in the $c$ state are untrapped, and those in
the $b$ state are antitrapped (expelled).

\subsection{H in an RF Magnetic Field}
\label{rf.transitions.theory.sec}
Atoms may leave the trap when their state is switched from $d$ to $a$,
$c$, or $b$.  We drive transitions between Zeeman sublevels by
applying an rf magnetic field perpendicular to the static
field\footnote{We consider the polarization of the rf field in more
detail below}, ${\bf B}_{rf}=B_1 \hat{x}\cos\omega_{rf} t$ which
couples states $d$ and $b$ to states $c$ and $a$.  Transitions between
the states are induced by the interaction ${\cal H}_{rf}= -{\bf
B}_{rf}\cdot({\bf M}_S+{\bf M}_I)$ with matrix elements
$H_{\alpha,\beta} \cos\omega_{rf}t \equiv\left\langle\alpha\right|{\cal H}_{rf}
\left|\beta\right\rangle$.  The non-vanishing terms are
\begin{eqnarray}
H_{d,c}\sqrt{2}/\mu_B B_1 &=&(1-\varphi)\cos\theta+(1+\varphi)\sin\theta \\
H_{d,a}\sqrt{2}/\mu_B B_1 &=&(1-\varphi)\sin\theta-(1+\varphi)\cos\theta \\
H_{b,c}\sqrt{2}/\mu_B B_1 &=&(1-\varphi)\cos\theta-(1+\varphi)\sin\theta \\
H_{b,a}\sqrt{2}/\mu_B B_1 &=&(1+\varphi)\cos\theta+(1-\varphi)\sin\theta.
\end{eqnarray}
In the limit of low trapping field $B_0$, $\theta\simeq -B_0/B_{hf}$.
We make the approximations $\cos\theta=1$, $\sin\theta=0$, and
$\varphi=0$, obtaining
\begin{equation}
H_{d,c}=H_{b,c} = \hbar \Omega_R
\end{equation}
where $\Omega_R=\mu_B B_1/\hbar\sqrt{2}$ is the Rabi frequency.  We
ignore the $a$-state since it is far separated in energy from the
$F=1$ manifold.  With these approximations we have an effective three
level system with equal energy separation $\mu_B B_0$.

As the atoms move through the varying magnetic field, resonant
transitions occur in a region of thickness $\Delta
r=\hbar\Omega_R/\alpha$ which is given by the resonance's energy
width, $\hbar \Omega_R$, scaled by the local potential energy
gradient, $\alpha$.  In the vicinity of the resonance region, the
Zeeman sub-levels can be assumed to vary linearly in space.  For an atom
whose speed $v$ changes only slightly as it crosses the resonance region,
the energy levels may then be considered to vary linearly in time.
The probability of changing Zeeman sub-levels as the atom crosses the
resonance region is calculated using a variant of the Landau-Zener
formalism \cite{rkl81} for avoided level crossings.  Vitanov and
Suominen \cite{vsu97} have analytically calculated the dynamics for
avoided level crossings in systems with an arbitrary number of levels.
They show that, for an atom initially in the $d$ state, the
probability of emerging from the resonance region in the $d$, $c$, or
$b$ state is
\begin{eqnarray}
P_d &=& (1-p)^2 \\
P_c & = & 2(1-p)p \\
P_b &=& p^2
\label{spin.flip.probability.eqn}
\end{eqnarray}
where $p=1-\exp(-\zeta)$ is the Landau-Zener transition probability
for a two-state system, and \label{zeta.definition.sec}
\begin{equation}
\zeta=2\pi\hbar\Omega_R^2/\alpha v
\label{zeta.definition.eqn} 
\end{equation}
is the adiabaticity parameter for a particle of
speed $v$ in a magnetic field with magnetic field gradient
$\alpha/\mu_B$.  For large $\zeta$ the atom absorbs two rf photons
while in the resonance region, and emerges in the $b$ state,
antitrapped.  For small $\zeta$ (the diabatic case of interest here)
there is only a small probability of leaving the trapped $d$ state,
and the vast majority of atoms that do leave this state will emerge in
the $c$ state, untrapped.  They are pulled by gravity out of the
trapping region along the trap axis.

\subsection{RF Field Amplitude Requirement}
\label{rf.evap.power.requirements.sec}

In order to realize the improved evaporation efficiency we seek, the
characteristic time required to eject an energetic atom must be less
than the characteristic collision time of these atoms \cite{ked96}.
This condition sets a minimum rf field amplitude.  

Here we consider an atom traversing the trap radially, passing through
the center.  Atoms in different trajectories will be effected
differently by the rf field and sample density.  However, in order to
have a significantly different orbit (while still having enough radial
energy to reach the resonance region), the atom would need high energy
in both the radial and azimuthal degrees of freedom.  Since this is
rare (see appendix \ref{evapcoll:appendix}), we ignore these orbits
and focus on the simple case of purely radial motion.

The characteristic time to eject an energetic atom, $\tau_{eject}$, is
the oscillation period $T$ divided by the probability of ejecting the
atom per period, $4P_c$.   (We ignore coherence between resonance
crossings).   The atom encounters a resonance
region four times per period since there is a resonance region on both
sides of the trap, and the atom passes through each region twice per
oscillation (assuming the turning point is outside the surface of
death).   The oscillation period of a particle of
mass $m$ moving in a linear potential with gradient $\alpha$ and with
an initial energy $\epsilon_t$ is $T=4\sqrt{2m\epsilon_t}/\alpha$.
For the parameters of interest ($\epsilon_t/\kb\sim 500~\mu$K,
$\alpha/\kb\sim 16~$mK/cm used near the BEC transition), we obtain a period
$T\sim 1~$msec.  Thus $\tau_{eject}\sim 1~{\rm msec}/4P_c$.

To find $P_c$ and the minimum rf field strength, we require that the
ejection time be less than the characteristic collision time
$\tau_{col}=1/n_0\sigma\bar{v}$ (A rigorous analysis includes an
average of the quantity $n(\rho)v(\rho)$ over the atom's trajectory,
which produces a correction factor of about 3).  For atomic hydrogen
near the onset of BEC at $T_c\sim 50~\mu$K, $\tau_{col}\sim 100~$msec.
The particles thus sample the resonance region roughly 400 times
before having a collision.  The ejection probability $P_c$ must be at
least $P_{min}=T/4\tau_{col}\simeq 1/400$.  The minimum adiabaticity
parameter is then $\zeta_{min}\simeq 10^{-3}$.  For the gradient
$\alpha$ and typical average velocity $v\simeq100$~cm/s, we obtain a
minimum Rabi frequency $\Omega_{min}=\sqrt{\alpha v
\zeta_{min}/h}=2\pi\times 1$~kHz.  The minimum rf field is
$B_{1,min}=\hbar\Omega_{min}\sqrt{2}/\mu_B=1$~mG.  The thickness of
the resonance region is $\Delta r=30$~nm, so the approximation of a
linearly varying trapping field made above is valid.  We note that for
atoms with more complicated hyperfine structure ($F>1$) the rf field
must be larger because an initially trapped atom must be taken
through one or more intermediate states before arriving at an
untrapped state.  For a given $g$-factor and hyperfine state the
required Rabi frequency and rf field scale as $\sqrt{ma^2T_c^3}$
independent of trap parameters (for linear and harmonic traps).

A potential problem is trap loss and heating through dipolar decay as
atoms in the $c$ state pass through the cloud of trapped $d$ atoms
\cite{ked96}.  This turns out to not be a problem for trapped H.  The
time constant for this process is $\tau_{dip}=1/G_{cd}n_c$ where
$G_{cd}$ is the total event rate constant for dipolar decay in
collisions between $c$ and $d$ atoms.  For H, the total event rate
constant is $G_{cd}\simeq 10^{-15} ~{\rm cm}^3/{\rm s}$ which is about
the same as $G_{dd}$, the total event rate constant for dipolar decay
through $d$-$d$ collisions \cite{lsv86}.  The density of $c$ atoms is
much less than $n_0$ since they are free to fall out of the trap
($\tau_{fall}\sim 100$~msec).  We conclude that $\tau_{dip}$ is long
compared to the lifetime of the sample as set by $d$-$d$ dipolar
relaxation.  Therefore, we need not be concerned with dipolar
relaxation events between the $c$ and $d$ atoms.

\subsection{RF Coil Design}

Here we describe the coil design considerations for creating an rf
magnetic field that will drive evaporation efficiently in the
cryogenic apparatus.

\begin{figure}[tb]
\centering \epsfxsize=5in 
\epsfbox{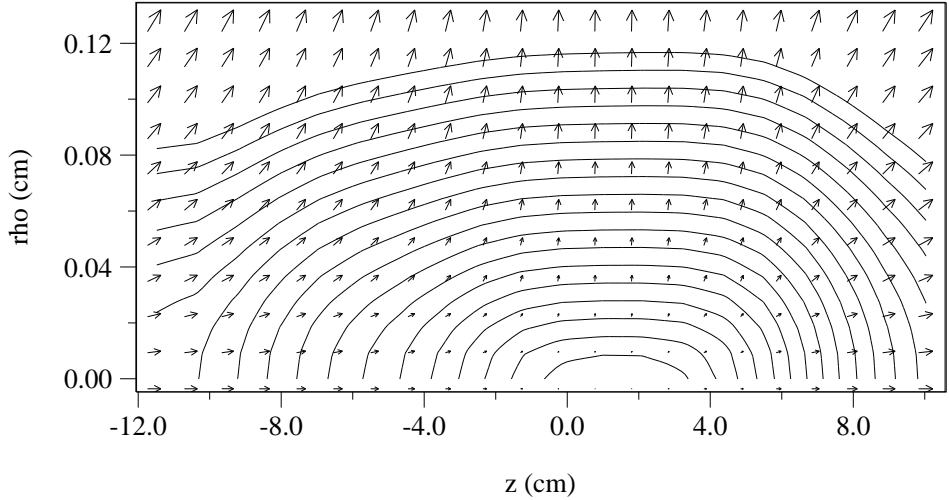}
\caption[direction of trapping field]{ Static magnetic field which
confines the gas, viewed in a plane which cuts through the center of
the trap.  The vertical axis of the figure is the distance away from
the central axis (symmetry axis) of the trap, in cm.  The horizontal
axis is the long axis of the trap.  The atoms are confined to the
bottom of the trap near $\rho=0$ and $z=0$.  The contours are spaced
by $100~\mu$K.  The arrows indicate the direction of the field.  Note
that the axes are very different lengths.  This trap is used to
confine samples below $T=130~\mu$K.  In order to eject energetic
atoms, the rf magnetic field must be perpendicular to this trapping
field.  The saddlepoint in the trapping field is on the left.  The top
of the cell and dissociator are to the right.}
\label{trap.field.direction.fig}
\end{figure}

The rf magnetic field must have a component in a direction
perpendicular to the trapping field, as discussed above.  Figure
\ref{trap.field.direction.fig} 
indicates the direction of the trapping field in a plane that cuts
through the trap.  This field is for a typical trap used for rf
evaporation experiments.  Along the sides of the trap the rf field
must point in the $\hat{z}$ direction.  This field is generated by
coils wrapped around the the cylindrical confinement cell.  These
coils are termed the ``axial coils''.

\begin{figure}[tb]
\centering \epsfysize=3.5in
\epsfbox{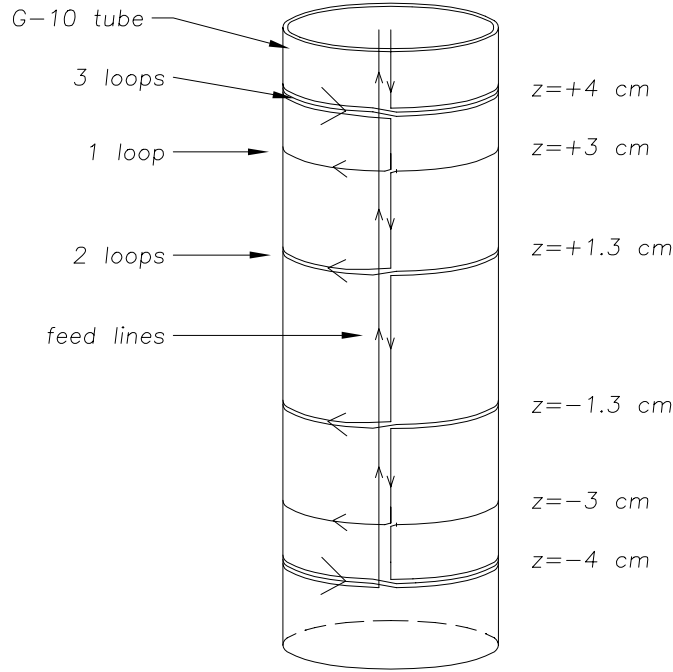}
\caption[winding pattern of axial rf coil]{ Winding pattern of axial
rf coil, shown wrapped on a section of the cell inner G-10 tube
(3.8~cm OD).  The coil is fed by the two wires extending out the top
of the section.  The loops are wound as close together as possible,
and the feed wires are laid tightly next to each other.  The point
$z=0$ is very near the laser beam focus, and also the minimum of the
trapping potential.}
\label{axial.winding.pattern.fig}
\end{figure}

It is important that the rf field be weak enough at the top of the
cell to avoid heating the large copper pieces which make up the cell
top, discharge and mixing chamber (heating rates are outlined in
section \ref{rf.heating.calc.sec}).  The field generated by a loop of
diameter $a$ decays as $(a/z)^{3}$ at large distances from the loop
along the loop axis.  However, the field can be made to decay as
$(a/z)^{5}$ if the loop is compensated appropriately.  In the
far-field the arrangement appears as a quadrupole instead of a dipole.
The heating rate is proportional to the squared magnitude of the
field, so the heating rate falls as $(a/z)^{10}$ for a compensated
loop.  The geometry of the axial coils was chosen to provide large
fields over the length of the cold samples and to produce fields which
decay rapidly outside the trapping region.  The coils are wrapped
around the outside of the inner tube of the cell (3.8~cm diameter).
They consist of three loops in the ``forward'' direction at
$z=\pm4$~cm, 1 loop ``backward'' at $z=\pm3$~cm, and two loops
``backward'' at $z=\pm1.2$~cm.  The winding pattern is shown in figure
\ref{axial.winding.pattern.fig}.

\begin{figure}[tb]
\centering \epsfxsize=5in 
\epsfbox{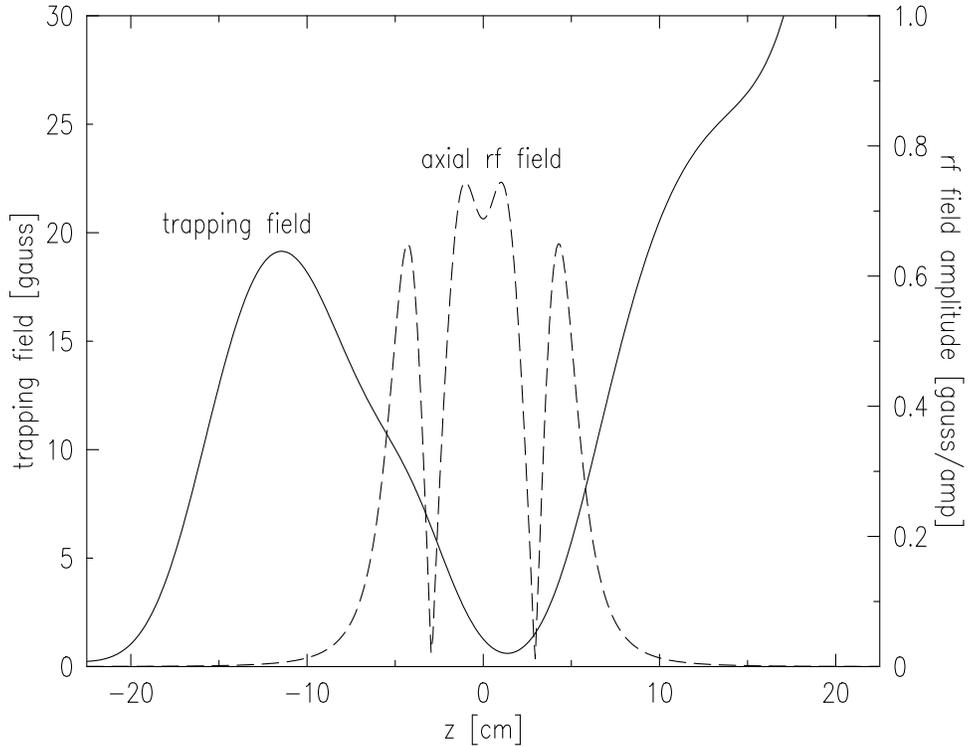}
\caption[rf and trapping fields]{ Profiles of the rf and dc magnetic
fields along the axis of the trap.  The dashed line is the rf field
magnitude, and the solid line is the trapping field for the trap used
to achieve BEC. }
\label{rf.field.geom.fig}
\end{figure}

\begin{figure}[tb]
\centering \epsfxsize=5in 
\epsfbox{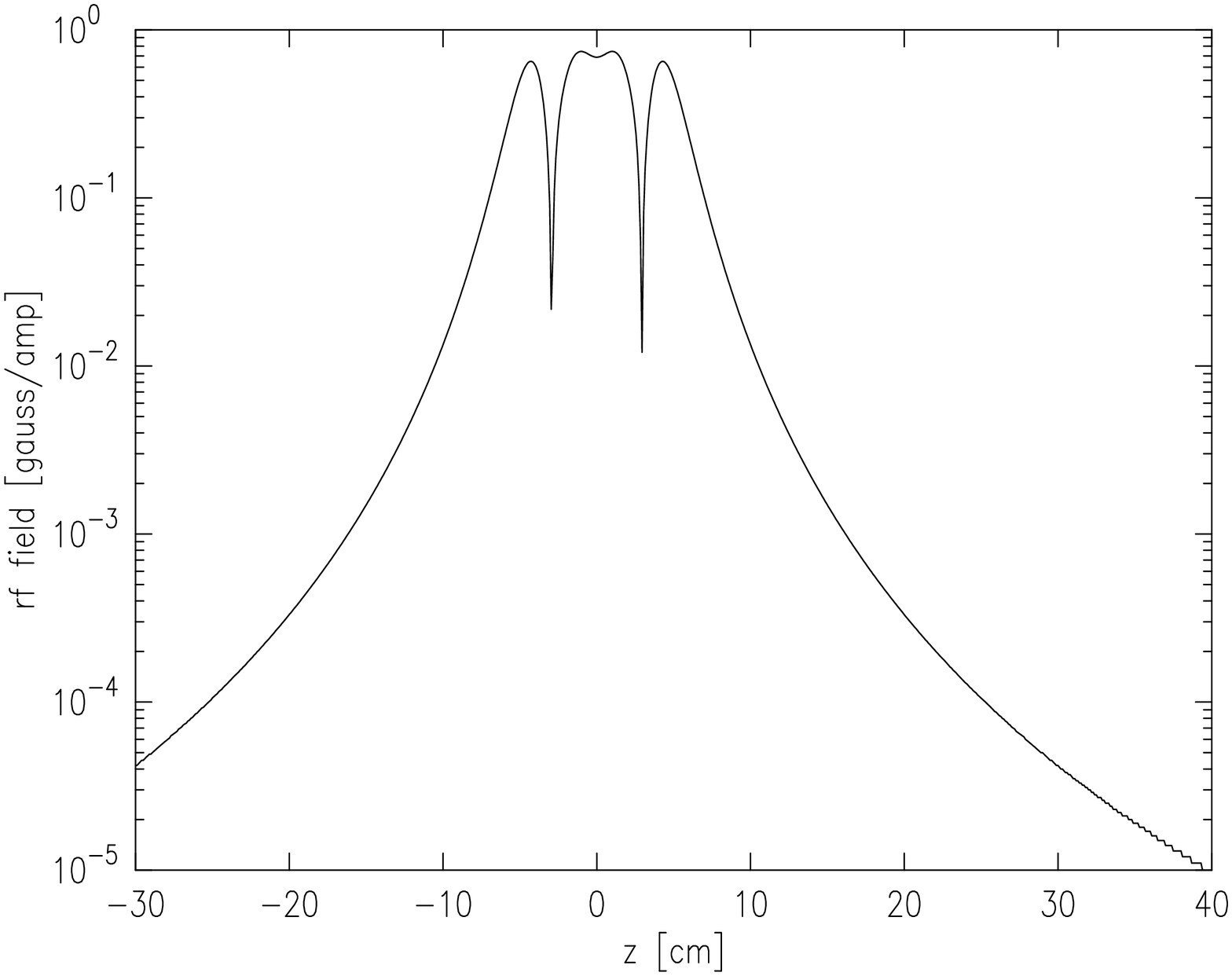}
\caption[rf field strength along the trap axis]{ Profile of the rf
magnetic field along the axis of the trap showing the field strength
at $z=40$~cm where the nearest pieces of cold metal are located.}
\label{rf.field.log.fig}
\end{figure}

Figure \ref{rf.field.geom.fig} shows a calculation\footnote{Since the
wavelength of this radiation ($\lambda=3$~m at 100~MHz) is much larger
than the length scale of interest we disregard propagation effects and
calculate the dc field.}  of the magnitude of the rf field generated
by the axial coils along the trap axis.
Figure \ref{rf.field.log.fig} shows how the field decays along the
trap axis.  
The nearest big pieces of metal that must be kept cold are at
$z=40$~cm.  Because the rf field is compensated well, the dominant
heating of the top of the cell, dissociator, and mixing chamber arises
from other mechanisms, described in section \ref{rf.heating.sec}.

\begin{figure}[tb]
\centering \epsfysize=3.5in
\epsfbox{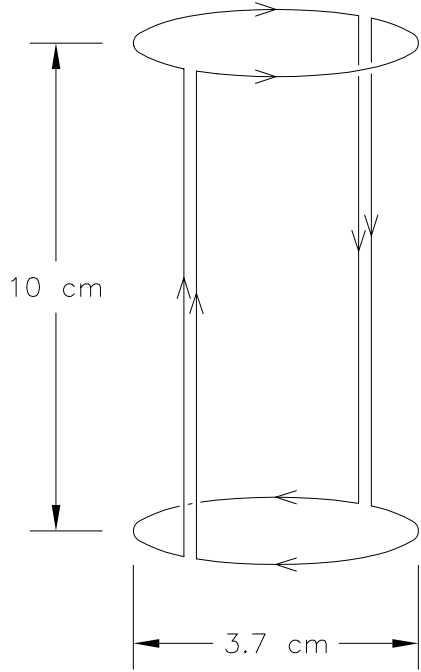}
\caption[winding pattern of transverse rf coil]{ Winding pattern of
transverse rf coils.  The circular ends wrap around the cell, and the
straight sections run parallel to the symmetry axis of the cell.  The
coil is centered at $z=0$.  }
\label{trans.winding.pattern.fig}
\end{figure}

The rf fields are primarily used to remove the most energetic atoms,
but it is also sometimes desirable to remove the atoms at the very
bottom of the trap where the trapping field is aligned along the
${z}$ axis.  Another set of rf coils, called the ``transverse
coils'', generate a field predominantly perpendicular to $\hat{z}$.
The winding pattern is shown in figure
\ref{trans.winding.pattern.fig}.

\section{Mechanical Design of the Cell}

The designs of the cells used previously in the experiment are
described by Doyle \cite{doyle91} and Sandberg \cite{san93}.  The
principle difference between those cells and the cell used for this
experiment is the presence of the rf magnetic fields.  Previous
designs relied on metals to provide thermal conductance along the
length of the cell, but metals must be excluded in the new design.
They would allow rf eddy currents to flow, which would heat the cell
and which would screen the atoms from the applied rf fields.  In
section \ref{rf.heating.calc.sec} we examine the problem of rf heating.  In
section \ref{superfluid.jacket.sec} we explain the design of a jacket
of superfluid $^4$He which surrounds the cell and provides the
required thermal conductance.

\subsection{Exclusion of Good Electrical Conductors}
\label{rf.heating.calc.sec}

RF eddy currents in electrical conductors will lead to heating of the
cell.  This heating during rf evaporation must be limited.
The cell must be kept at a temperature below $\sim 150$~mK to ensure
that walls are sticky for H atoms and to ensure that the vapor
pressure of the superfluid $^4$He film, which is crucial for loading the
trap, does not rise high enough to create a significant background gas
density.  A failure in either regard would allow high energy particles
($T\sim100$~mK) to knock cold atoms out of the trap.  A technical
limitation also exists: it is desirable to analyze the trapped sample
very soon after rf evaporation, and this analysis requires the cell to
be below 90~mK.  If the cell is heated too much during the rf
evaporation process it will take too long to cool below this
temperature.

The typical thermal conductivity along the length of
the cell is $\sim 1~\mu$W/mK so that heat loads of only 50~$\mu$W are
tolerable for a mixing chamber temperature of 100~mK.  We
therefore must beware of power deposited via rf eddy currents in any
metals thermally connected to either the refrigerator or the cell.  In
previous designs, good thermal conductivity along the length of the
cell (a tube about 4~cm in diameter and 60~cm long) was supplied by
about 100 Cu wires \cite{doyle91}.  These wires, and many other
materials, needed to be replaced in the new design in order to
eliminate rf eddy current heating.

\begin{figure}[tb]
\centering \epsfysize=3in 
\epsfbox{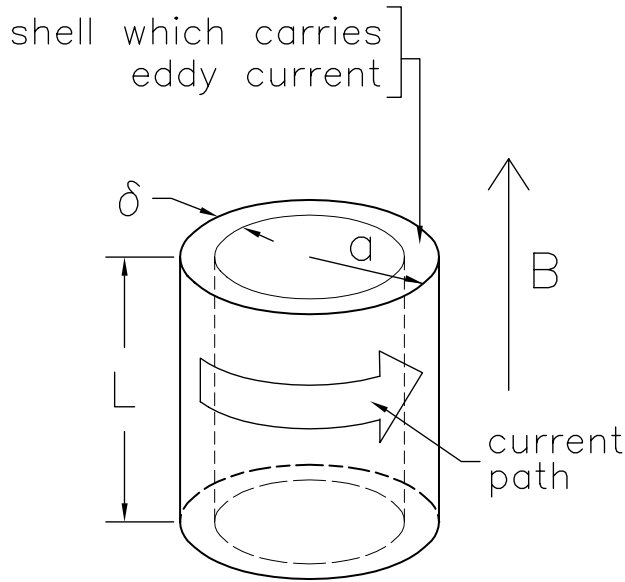}
\caption[geometry for calculation of rf eddy currents]{ geometry for
calculation of rf eddy currents in a wire of radius $a$ and length $L$
immersed in a uniform magnetic field ${\bf B}$ aligned along the axis
of the wire.  The eddy current flows around the cylindrical shell of
thickness $\delta$, height $L$, and perimeter $2\pi a$.  The effective
area for magnetic induction is $2\pi a \delta$.  }
\label{eddy.current.fig}
\end{figure}

To justify the effort required to remove good electrical conductors
from the cell, we here estimate the heat per unit length deposited in
a wire by eddy currents.  Consider a cylindrical conductor of radius
$a$, length $L$, and electrical conductivity $\sigma$ immersed in a
uniform magnetic field $B$, aligned along the axis of the wire, that
is oscillating at angular frequency $\omega$, as shown in figure
\ref{eddy.current.fig}.
For a good conductor the skin depth is roughly
$\delta=\sqrt{2/\sigma\mu\omega}$ \cite{lcl88.skindepth} where $\mu$
is the permeability; for the non-magnetic materials of interest here,
we take $\mu=\mu_0=4\pi\times 10^{-7}$~H/m, the permeability of free
space.  

For frequencies larger than $\omega_s=2/\sigma\mu a^2$ the skin depth
is smaller than the wire radius.  For $\omega\gg\omega_s$ we may
assume that all the eddy current flows in a shell of thickness
$\delta$, and that the magnetic field is extinguished in the wire
everywhere except in this shell.  We thus simplify the problem to a
shell carrying current in the azimuthal direction through a
cross-sectional area $\delta L$ and around a length $2\pi a$. The
conductance around the shell is then $S=\sigma\delta L/2\pi a$.  The
shell is pierced by the magnetic field in an area $A=2\pi a \delta$.
The power deposited (per unit length of wire) is $P/L=S V^2/L$ where
$V=AB\omega$ is the induced voltage around the shell.  We obtain
\begin{equation}
\frac{P}{L}=\frac{2^{5/2} \pi a B^2 \omega^{1/2}}{\sqrt{\sigma\mu}}
\end{equation}  
Typical values are $B=10^{-2}$~G, $\omega=2\pi\times10^8$~Hz.  The Cu
wire formerly used for thermal connection along the length of the cell
has radius $a=0.5$~mm and conductivity $\sigma\sim 10^{10}~(\Omega~{\rm
m})^{-1}$ \cite{pobell}, so that $P/L=20~$nW/cm.  It is clear that for
100 wires of length $L=10$~cm this heating rate is non-negligible.
Since this heating rate is so close to the limit, an alternative
mechanism for heat transport was necessary.  A further problem with
the Cu wires is that they shield the interior of the cell from the rf
field.

\subsection{Design of the Superfluid Jacket}
\label{superfluid.jacket.sec}

\begin{figure}[tb]
\centering \epsfysize=4in 
\epsfbox{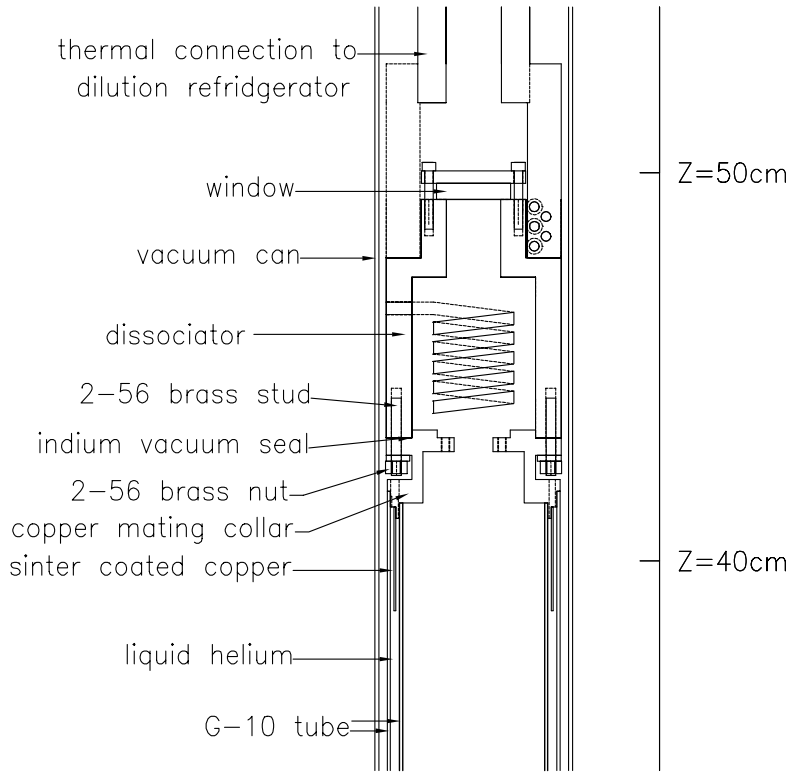}
\caption[cross-section through top of cell]{ Cross-section through
dissociator and upper end of cell.  The origin of the scale on the
right hand side is at the center of the trap.  The cell and
dissociator are inside a brass vacuum can which keeps 4~K liquid
helium out.  The dissociator couples to the cell via a copper mating
collar and an indium o-ring.  A thin copper fin, covered with silver
sinter, makes thermal contact with the superfluid helium shell that is
confined by two concentric G-10 tubes.  The window at the top of the
dissociator provides optical access to the trapped atoms for laser
diagnostics.  Not shown are electrical feedthrough tubes which are
soldered into the copper mating collar and lie in grooves on the
outside of the dissociator,}
\label{celltop.mechanical.fig}
\end{figure}

In order to obtain good thermal conductance along the length of the
cell we exploited the extraordinary heat transport properties of
superfluid helium.  The new cell design involves two concentric G-10
tubes \cite{spaulding.composites} with a layer of helium between them,
about 2.2~mm thick.  Figure \ref{celltop.mechanical.fig} is a
cross-section through the top of the cell.
There are several issues that were addressed in
the design of the cell.  These are discussed in this section.

\subsubsection{Thermal Conductivity}
The superfluid jacket must be thick enough to carry heat sufficiently
well from the bottom to the top of the cell where it is transferred to
the refrigerator.  To calculate the quantity of helium necessary, we
must estimate the required thermal conductance and understand heat
flow in the superfluid.

There are three temperature regimes which demand good conductance.
During loading of the trap, when $T=250$~mK, the conductivity must be
at least $S=0.1$~mW/20~mK so that $S/T^3=0.3$~W/${\rm K}^4$.  During
laser spectroscopy, when $T=140$~mK, the conductivity must be at least
$S=0.2$~mW/20~mK so that $S/T^3=4$~W/${\rm K}^4$.  Finally, during
bolometric detection, when $T=80$~mK, the conductivity must be at
least $S=10~\mu$W/10~mK so that $S/T^3=2 ~{\rm W}/{\rm K}^4$.

In metals heat is carried by electrons, which typically scatter off
impurities on a length scale which is short compared to the dimensions
of the object.  In contrast to this effectively diffusive heat flow
behavior, heat transport in superfluid is essentially ballistic
because there is no impurity scattering.  Furthermore, the heat is
carried by phonons which have a different excitation spectrum than
electrons.  The notable implications are that the effective
conductance scales with temperature as $T^3$ in He but only as $T^1$
in metals, and that the conductivity of a column of superfluid depends
on the diameter of the column; phonons scatter when they hit the
walls.  The empirical conductivity relation for a tube of superfluid
is \cite{pobell} $\kappa = (20~{\rm W}/ {\rm K}^4~{\rm cm}^2)\:d\:
T^3$ \label{superfluid.conductivity.relation.page} where $d$ is the
diameter of the vessel and $T$ is the temperature.  In our design the
outer radius of the inner tube is $a=1.92$~cm and the inner radius of
the outer tube is $b=2.14$~cm.  We take $d=b-a=2.2$~mm.  The
conductance along the cell is
\begin{equation}
S=\frac{\dot{Q}}{\Delta T}=\frac{\kappa A}{L}=\frac{\kappa \pi
(b^2-a^2)}{L}
\end{equation}
where $L=66$~cm is the length of the cell.  For our geometry
$S/T^3=0.2~{\rm W}/{\rm K}^4$.  The thickness of the shell causes
scattering on a length scale $d$ in the radial dimension, but the
scattering length is significantly longer in the azimuthal direction.
We thus expect the conductivity to be larger by a factor of two or
three.  Furthermore, the empirical formula for $\kappa$ above applies
to surfaces from which the phonons scatter diffusely.  If specular
reflection occurs, then the effective scattering length is much longer
\cite{awh68}.  The characteristic phonon wavelength at 100~mK is
300~nm \cite{fan74}; since the surfaces of the coated G-10 tubes
exhibit specular reflection of optical wavelengths ($\lambda\sim
500$~nm, observed by shining a lamp on the surface), the superfluid
phonons should also reflect specularly.  As will be explained in
section \ref{spec.diffuse.exp.sec} below, we have observed the
superfluid conductivity to be 5 times higher in tubes which exhibit
specular optical reflection, as compared to tubes for which the
optical reflection is diffuse.  Putting all these factors together, we
expect a conductance $S/T^3=3~{\rm W}/{\rm K}^4$.  There is not much
margin for error in the design.  Measurements of the thermal
conductivity are detailed in section \ref{cell.measurements.sec}.

\subsubsection{Thermal Link from Superfluid to Refrigerator}
Once heat has been transferred along the length of the cell, it must
be transferred from the liquid into metal links which lead to the
refrigerator.  This process is frustrated by the large boundary
resistance between the bulk superfluid and the metal.  This
resistance, called the Kapitza resistance \cite{pobell}, arises in two
ways from the factor of 25 mismatch in the speed of sound in the two
materials.  First, for a phonon to propagate into the metal it must
approach the surface at very nearly normal incidence; phonons with
wavevectors outside a small cone of acceptance angles experience total
internal reflection.  Furthermore, the very different speeds of sound
give rise to a significant impedance mismatch, and so most of the
phonons within the acceptance cone will be reflected.  The empirical
Kapitza resistance equation is $R_{K}=(0.02~{\rm
m^2~K^4/W})/AT^3$ \cite{pobell} for a superfluid-Cu boundary of area A
and temperature $T$ near 100~mK.  For a modest maximum resistance
$R_K=10~{\rm mK}/100~\mu$W at 140~mK a boundary area of at least
$A=700~{\rm cm}^2$ is required.  This is far too large to fit
conveniently in the apparatus.

One way to create an effective surface area that is much greater than
the simple geometric surface area is to apply sintered silver to the
surface \cite{pobell}.  The network of voids between the grains of
the powdered metal creates a huge surface area.  It is possible to
create an effective surface area of $10^3~{\rm cm}^2$ for a coating of
silver sinter on an area of copper that is straightforward to
incorporate into the apparatus, $A=25~{\rm cm^2}$.

\subsubsection{Heat Capacity}

After loading the trap at $T=250$~mK it is important to cool the cell
quickly to below 150~mK to achieve thermal isolation between the
trapped sample and the cell walls, as explained in section
\ref{discussion.of.loading}.  Thus, it is important to ensure that the
heat capacity of the superfluid is low.

The total heat that must be extracted from the superfluid is the
integral of the heat capacity between the initial and final
temperatures.  The heat capacity is $C=C_0 T^3$ with
$C_0=8\times10^{-2}~{\rm J/mol~K^4}$ \cite{pobell}.  The quantity of
heat to remove from $N$ moles of liquid cooled from $T_2$ to $T_1$ is
$U=C_0 N (T_2^4-T_1^4)/4$.  From the 6 moles of superfluid in the
jacket we must thus remove 0.4~mJ.  The refrigerator cooling power is about
1~mJ/s, and so the time constant is less than a second.  This is
acceptable.

\subsubsection{Filling and Emptying the Jacket}

Since superfluid is such a good heat conductor, we must be concerned
about heat links from the coldest part of the refrigerator to warmer
regions through the fill line which leads to the jacket.  Heat may
flow through two mechanisms.  When a tube is filled with superfluid
the thermal conductivity is given by the empirical relation quoted
above (p\pageref{superfluid.conductivity.relation.page}).  In this
case the total heat conductance is proportional to $d^3/L$.  Smaller
diameter tubes are obviously much better.  To limit the heat flow to
less than $10~\mu$W between the 1~K and 0.1~K stages of the
refrigerator requires $d^3/L<6\times10^{-7}~{\rm cm^2}$; for $L=50~$cm
we obtain $d<0.3$~mm.  The other heat conduction mechanism occurs in
tubes with only a superfluid film; the film flows to a warm region,
evaporates, and sends gas back to colder regions.  As it recondenses
it deposits heat.  The upper limit on this process is given by the
superfluid film flow rate.  A saturated film of thickness $t\sim30$~nm
flowing at the critical velocity $v_c=45~$cm/s up a tube with limiting
perimeter $\pi d$ will evaporate, and then recondense depositing power
$P_{flux}=Lv_c t \pi d=(0.9~{\rm mW/cm})d$ where $L=2~{\rm J/cm^3}$ is
the latent heat.  To deposit less than $10~\mu$W at the cold end of
the tube we require $d<0.1$~mm.  This is an overestimate of the power
transport because the warm vapor does not make it all the way to the
cold end of the tube before condensing.  To minimize heat flow through
the fill line we chose a segmented design; 40~cm sections of 0.2~mm ID
tubing connect heat exchangers which are thermally anchored to
progressively colder parts of the refrigerator.  In this way heat is
deposited as far as possible from the coldest part of the
refrigerator.

If the cell is warmed much above 4~K, the vapor pressure of the liquid
helium rises far above atmospheric pressure.  The tiny fill tube is a
huge impedance to the gas flowing out of the jacket.  Effectively we
have created a big bomb.  A pressure relief valve was developed to
bleed off excess pressure in an emergency, thus protecting the cell
from bursting.  More details are given in section
\ref{pressure.relief.valve.sec}.

\section{Construction Details}

In this section we discuss some of the techniques that were crucial
for the construction of the cell.

\subsection{Materials and Sealing Techniques}
It is a significant accomplishment to create joints between different
materials that are leak-tight to superfluid flow at low temperatures.
The cell is a composite unit consisting of a ${\rm MgF}_2$ window, two
concentric convolute wound G-10 tubes, an OFHC copper collar that
seals the superfluid jacket at the upper end and mates the cell to the
dissociator, two rf feedthroughs, and many low frequency electrical
leads.  The sealing techniques described here proved reliable.

\begin{figure}[tb]
\centering \epsfxsize=5in 
\epsfbox{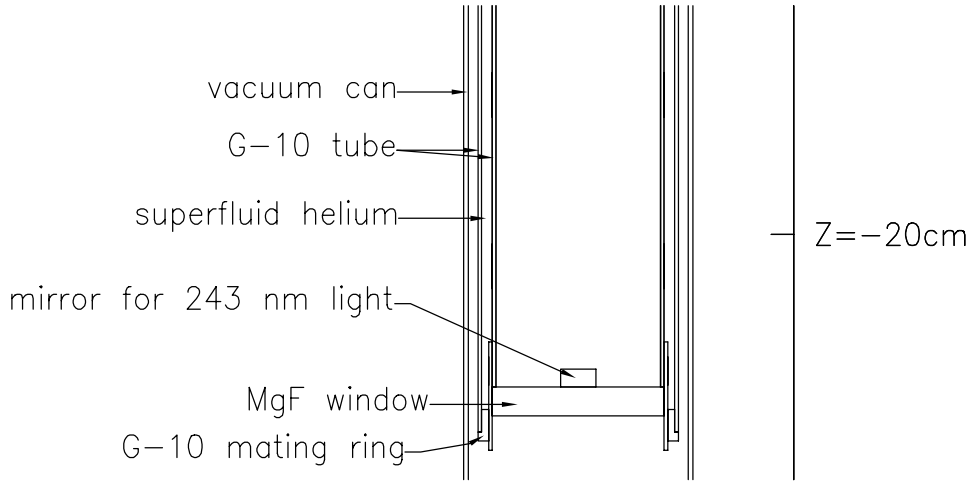}
\caption[cross-section through bottom of cell]{ Cross-section through
the bottom end of the cell.  The origin of the scale on the right hand
side is at the center of the trap.  The mating ring at the bottom
joins the two concentric tubes.  The MgF$_2$ window is set against the
bottom of the inner tube.  A uv mirror retroreflects the laser beam
which passes along the axis of the trap.  The bolometer is placed
about 1~cm above the window.  }
\label{cellbottom.mechanical.fig}
\end{figure}

The G-10 tubes \cite{spaulding.composites} came from the factory with
rough surfaces.  To make them leak-tight and to create the specularly
reflecting surface crucial for high thermal conductivity we coated
them with an epoxy mixture of 100 parts by weight Epon-828 \cite{epon}
to 32 parts Jeffamine D-230 \cite{jeffamine} catalyst.  The tubes were
hung vertically while the epoxy cured and was then heat set at 60 C
for 2 hours.  The rf coils were wrapped onto the tube before the
coating process to allow the epoxy coating to produce a smooth surface
for specular reflection of the superfluid phonons.  The tubes were
joined at the bottom to a G-10 mating ring with Stycast 2850FT epoxy,
cured with 24LV catalyst \cite{1266}.  See figure
\ref{cellbottom.mechanical.fig}.
The MgF$_2$ window was also glued in place with Stycast 2850FT epoxy.
In order to prevent the glue from contaminating the inside surface of
the window (and blocking Lyman-$\alpha$ transmission) the following
gluing procedure was used: the window was placed on a pedestal, and
the outer edge of the window was coated with glue.  The cell was then
held above the window and slid gently downward.  If necessary,
addition glue was dabbed to create a fillet around the bottom of the
window against the G-10 tube.  The cell was held vertically while the
glue cured.  Because 122~nm light must pass through this window, the
transmission of a window glued this way was tested in the ultraviolet,
using easily accessible 243~nm radiation.  No degradation was observed.
Windows glued in this way have been thermally cycled to 1.4~K many
times without ever leaking.  Thermal shock is a problem, though, so
the windows were never brought in direct contact with liquid nitrogen
during the leak checks.

Dual thermometers and a heater are placed at each end of the cell.
Other wires lead into the superfluid jacket space and connect to pads
or meshes used to produce the electric field required for rapid
quenching of the metastable atomic $2S$ state.  All of these leads are
brought into the jacket space through a simple epoxy feedthrough; the
wires pass through a 1.67~mm ID tube which is filled with 2850 epoxy
for several centimeters.  Each rf coil is driven by a twisted pair of
superconducting leads (Cu clad NbTi wire with a single 33~$\mu$m
diameter core) The twisted pair for the axial coils is connected to a
semirigid coax line (2.16~mm OD Cu \cite{microcoax}) which passes
through the Cu mating block at the top of the cell.  The other end of
the line is soft-soldered to a brass cylinder containing a hermetic
SMA connector \cite{macomSMA} soldered in place.  This sealing
technique works reliably.  The twisted pair which feeds the transverse
rf coil passes through an epoxy feedthrough and is connected to an SMA
connector in the vacuum space that surrounds the cell.

The semirigid coax cables which deliver the rf power to the cell are
composite units that optimize the compromise between low thermal
conductivity and high electrical conductivity.  Cables with high
electrical and thermal resistivity (Lakeshore model CC-SR-10
\cite{lakeshore}) carry the rf power to the 1.6~K pumped $^4$He pot,
where the cables are thermally anchored.  The ohmic heating of these
lines is too much of a heat load for the mixing chamber, however, so
lower resistivity lines (Lakeshore type C) carry the rf power to the
cell (which is thermally connected to the mixing chamber).  To reduce
the thermal conductivity of these lines, a 1~cm section of the high
resistivity cable is spliced in.  The thermal conductance is
dramatically reduced, but the electrical conductance is high enough to
eliminate excessive heating.

\subsection{Sintering}

Effective heat transfer from superfluid to copper is difficult to
achieve, as explained above.  A layer of sintered silver was fused
onto each side of the Cu fin, shown in figure
\ref{celltop.mechanical.fig}, to increase the surface area for heat
transfer and reduce the overall thermal resistance.  The fin is high
purity copper \cite{puratronic}, 0.5~mm thick,
and is brazed \cite{handy.harmon} to the Cu mating
collar.  The sintering process consists of several steps
\cite{bcg84,kw84}.  First, the surface of the fin is cleaned by
immersion in 1 molar nitric acid for 30~s, then rinsed with deionized water.
The surface is then plated with silver to increase the bonding of the
sinter to the Cu fin.  The surface is immersed in a plating solution
(electroless silver cyanide \cite{transene}) for 15 minutes at 95~C.
The surface is then rinsed again with deionized water, and the
assembly is baked at 120~C for 35 minutes in an air atmosphere.  The
baking was suggested to ``diffusion weld'' the plating to the Cu
substrate.  The Ag sintering powder (nominally 700~${\rm \AA}$
particle size) \cite{sinter} is first cleaned by baking at 50~C for
30~min in an ${\rm H_2}$ atmosphere.  The powder is then compressed
onto the Cu fin in a specially designed curved press which matches the
curved surface of the Cu fin.  The fin was sintered sequentially in
eight radial segments at a pressure of $5\times 10^5$~Pa.  The press
apparatus and mating collar were heated to about 100~C during the
operation.  About 7~g of powder were bonded to the fin, producing an
estimated $10~{\rm m^2}$ surface area.

\subsection{Bolometer}

The bolometer is used as a diagnostic of the trapped gas
\cite{doyle91}.  As atoms are released from the trap, their flux is
measured by detecting the molecular recombination heat, 4.6~eV per
event.  A resistive element serves as both a thermometer and a heater;
the electrical power required to maintain the bolometer at a constant
temperature is recorded.  The bolometer used in these experiments is
similar to previous designs.  It is constructed from a quartz plate
1~cm $\times$ 1~cm $\times$ 60~$\mu$m.  The plate is suspended by
nylon threads from a G-10 cradle which is glued into the cell.
Electrical contact is made through superconducting filaments extracted
from multifilament superconducting wire by dissolution of its Cu
matrix in nitric acid.  The wires are contacted to evaporated Au pads
on the quartz by conducting epoxy \cite{silver.epoxy} overlaid with
Stycast 1266 epoxy.  The Au pads touch graphite resistive elements
which were formed by dabbing Aerodag \cite{aerodag} onto the surface.
The differences in the present bolometer design from previous designs
arise from efforts to reduce rf eddy current heating.  Large gold
contact pads have been replaced by pads of a few ${\rm mm}^2$ surface
area.  A redundant design was used consisting of four contact pads in
a row with three resistive elements between them.  The pads are spaced
very closely so that the resistive elements could have resistances of
a few k$\Omega$ without large mass.  This bolometer has been reliable.


\subsection{Pressure Relief Valve}
\label{pressure.relief.valve.sec}

The vapor pressure of the six moles of liquid helium in the
superfluid jacket of the cell can cause the cell to explode if the
temperature is raised too high before the liquid is removed.  This
conjecture has been tested.  Recall that the fill line for the jacket
has a very low gas conductance, and so it takes many hours to remove
the liquid.  In the case of an accidental overheating, it is desirable
that the liquid in the jacket have an escape path with a high
conductance.

\begin{figure}[tb]
\centering \epsfysize=3in 
\epsfbox{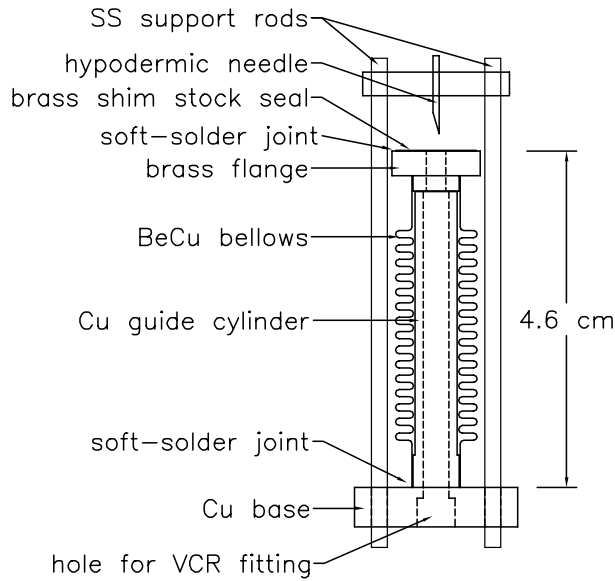}
\caption[pressure relief valve]{ Cross-section through the pressure
relief valve.  A VCR fitting \cite{vcr.fittings} brazed into the base
makes a superfluid tight vacuum connection to the superfluid jacket
through 2~mm ID tubes.  The BeCu bellows tube fits around a Cu guide
cylinder that prevents buckling and keeps the hole in the brass flange
aligned with the hypodermic needle.  A thin brass sheet is
soft-soldered to the brass flange to provide a vacuum seal that is
easy for the needle to puncture. }
\label{pressure.relief.valve.fig}
\end{figure}

We constructed a reusable pressure relief valve which provides this
high conductance escape path.  The reusable valve is at the same
temperature as the superfluid jacket so that large diameter tubing can
connect the two; there are no heat leak concerns which dictate small
tubing.  The design of the relief valve is conceptually divided into a
pressure-sensing part and a seal-breaking part.  The pressure sensor
is a BeCu bellows tube \cite{bellows} which lengthens as the pressure
inside increases.  The seal to be broken is a 25~$\mu$m thick piece of
brass shim stock which is soft-soldered to a brass flange, and is
replaceable.  When the seal is broken the gas escapes into the large
vacuum space which surrounds the cell.  The flange is brazed to the
bellows and moves toward a stainless steel hypodermic needle as the
pressure increases (The bellows are pre-stretched during installation.
They do not start to move until there is a pressure differential of
about 0.5~bar).  A touch sensor makes electrical contact with the
brass shim stock before it meets the needle, alerting the operator
that seal rupture is imminent.  The relief valve is shown in
figure~\ref{pressure.relief.valve.fig}.
The relief valve has been tested.  It is simple and reliable.

\section{Measurements of  Cell Properties}
\label{cell.measurements.sec}

Measurements have been made of several key parameters, such as thermal
conductivity, heat capacity, rf field strength, and rf heating rates.
In this section we outline these results.

\subsection{Measurements of RF Field Strength}
\label{rfantenna.bode.sec}

The strength of the rf field generated by the axial and transverse
coils was measured by inserting a pick-up coil inside the inner G-10
tube during construction of the cell.  The absolute field strength and
field profile along the cell axis were found to agree well with
calculations for frequencies up to about 5~MHz, where the inductive
impedance matched the $50~\Omega$ driving impedance and the response
began to drop.  When the cell is cooled to 100~mK the response could
be different because of the lower resistance of the superconducting
wire which forms the coils.

\begin{figure}[tb]
\centering \epsfxsize=5in 
\epsfbox{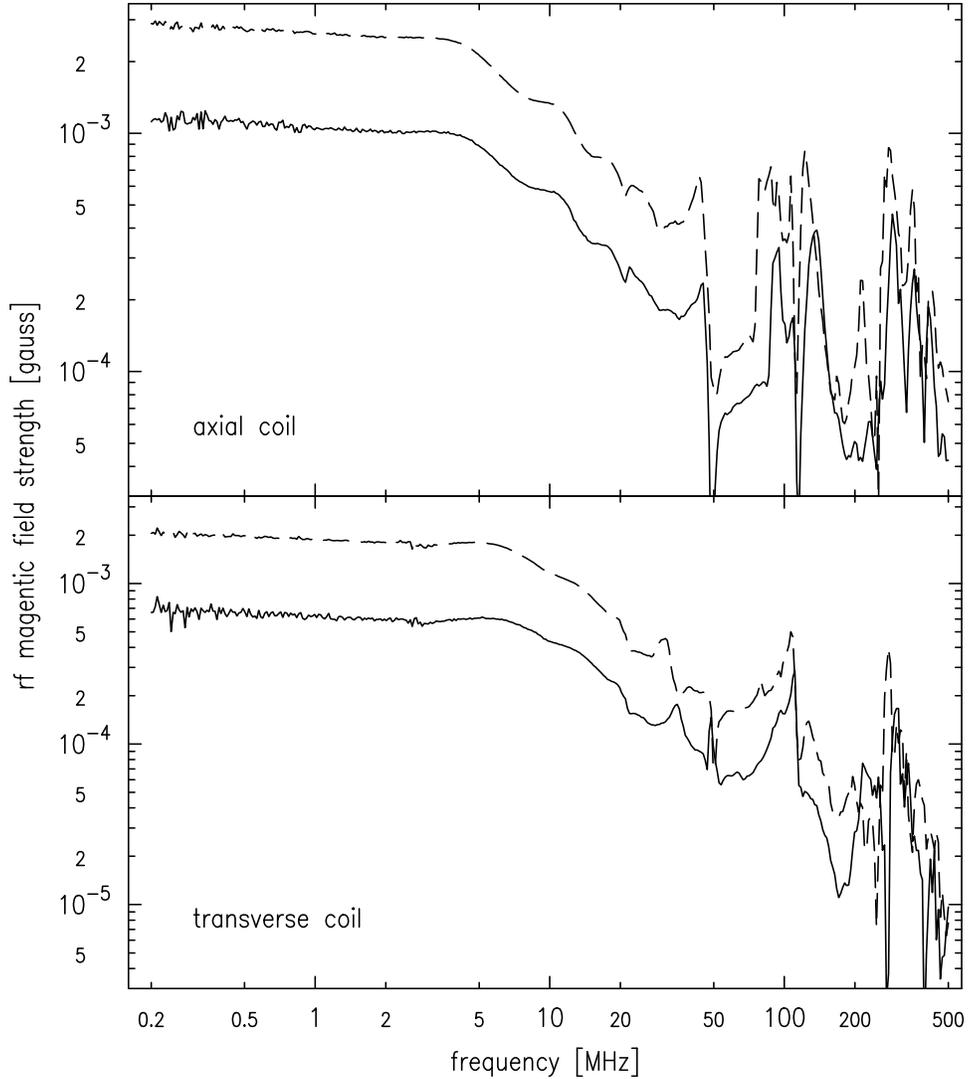}
\caption[frequency response of rf coils]{
Measured frequency response of axial and transverse rf coil.  The
solid line is the field when the cell is surrounded by the brass
vacuum can.  The dashed line corresponds to no conductors in the
vicinity of the cell.  The coil is driven through the coax lines in
the cryostat.}
\label{coil.bode.fig}
\end{figure}

The axial field was measured with a 1.6~cm diameter pickup loop
positioned in the center of the cell (the test was done before
attaching the bottom window onto the cell).  To match the impedance of
the loop to that of coax signal cable, an op amp \cite{ad811} was connected
directly to the loop.  The output of the unity gain amplifier drove
the signal cable through a 50~$\Omega$ resistor.  The frequency response of
the amplifier was measured, and so the pickup loop probe has a
calibrated response to 200~MHz.  Knowledge of the geometry of the loop
and of the gains in the system allow us to measure the field with an
accuracy of about 20\%.  An HP4195A network analyzer was used both to
calibrate the probe and to measure the fields.  The coils were driven
through the high thermal impedance microminiature coax cable
\cite{lakeshore} used to convey rf power from room temperature through
the refrigerator to the cell.  The response of the coils is shown in figure
\ref{coil.bode.fig}.
The structure above 50~MHz is most likely due to resonances in the
feed line and coil structure.

The cell resides inside a brass vacuum can in the experiment, and this
conductor, near the coils, can influence the field; a counter current
is generated which tends to partially cancel the field.  This effect
has been modeled, and the reductions shown in figure
\ref{coil.bode.fig} are consistent.

Note that as the apparatus is cooled to operating temperature the
resistance of the axial (transverse) coil changes from 13~$\Omega$
($26~\Omega$) to zero.  We thus expect a 30\% (50\%) increase in field
at low frequencies, and the corner frequency should move higher by
30\% (50\%).

\subsection{Thermal Conductivity}

The thermal conductivity along the length of the cell has been
measured in the temperature range 50~mK to 450~mK.  As described
above, the superfluid jacket is a shell between two concentric
cylinders.  The outer tube has an inside diameter of 4.27~cm, and the
inner tube has an outside diameter of 3.83~cm.  The thickness of the
shell is thus 2.2~mm.  The vertical distance between the thermometers
used for the following measurements is about 59~cm.  The jacket space
is cluttered by 5~mil diameter manganin wires which connect to
thermometers, heaters, the bolometer, and plates used to apply
electric fields.  The thermal conductivity is reduced since phonons
scatter off these structures.

\begin{figure}[tb]
\centering \epsfxsize=5in 
\epsfbox{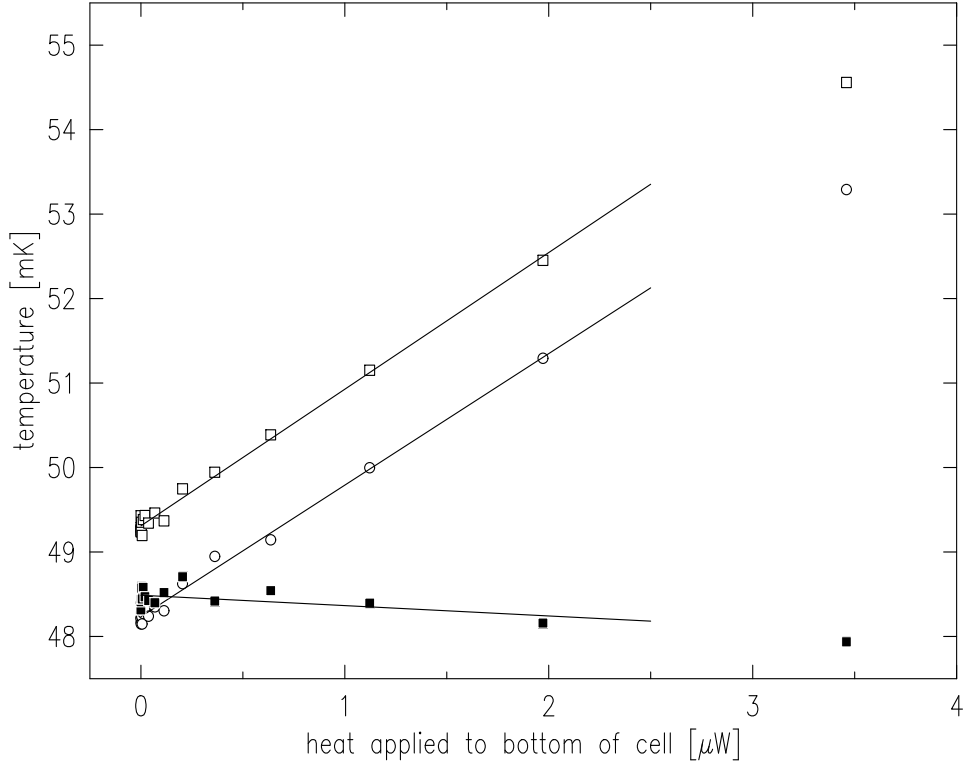}
\caption[response of cell to heat applied at the bottom]{ Typical
measurements used to extract thermal conductance of cell at one
temperature.  The temperature of the bottom (open symbols) of the cell
increases linearly as heat is applied to the bottom.  The temperature
at the top (solid symbols) stays roughly constant since it is
temperature controlled (using a different thermometer as the sensor).
The lines are linear fits for the range of powers from 0 to
2.5~$\mu$W.  The slopes of the two upper lines give the thermal
resistance of the cell at this temperature.  The resistances measured
by the two thermometers are $1.62\pm0.04~{\rm mK/\mu W}$ and
$1.56\pm0.04~{\rm mK/\mu W}$.  The 1~mK offset between the two bottom
thermometers is due to miscalibration.}
\label{cell.heat.response.fig}
\end{figure}

\begin{figure}[tb]
\centering \epsfxsize=5in 
\epsfbox{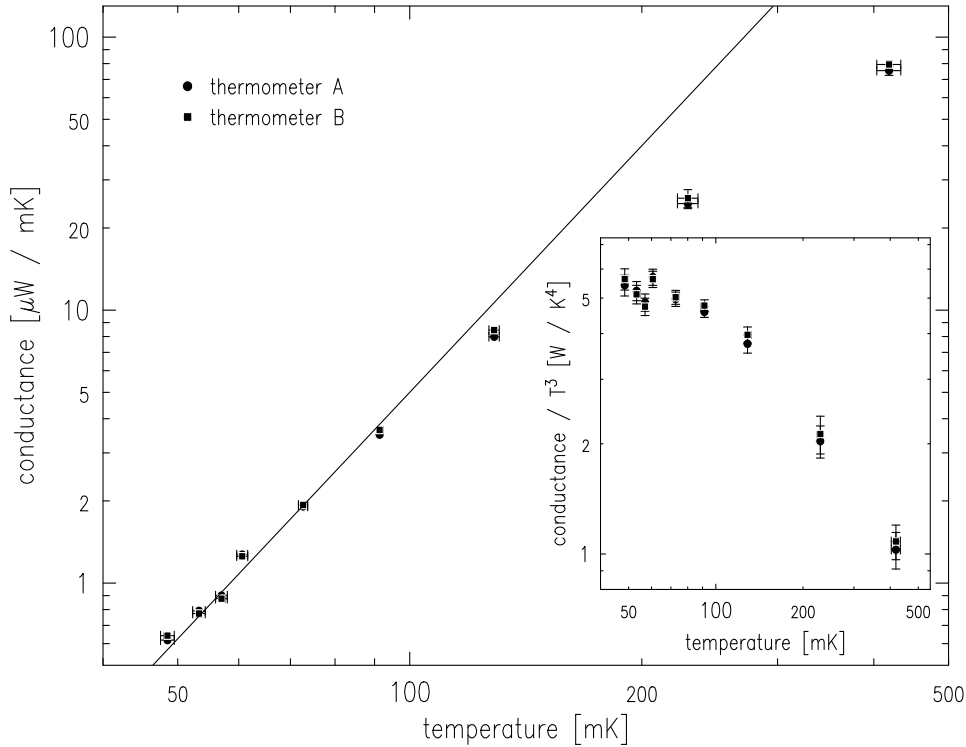}
\caption[thermal conductance of the cell]{
Thermal conductance along the length of the cell.  The two data sets
are from the two thermometers at the bottom of the cell.  Approaching
low temperatures one expects the conductance to fall like $T^3$, as
indicated by the line.  The inset shows the conductance with this
$T^3$ dependence factored out.  The error bars are the statistical fit
uncertainties, and do not include systematics. The vertical error bars
in the inset are dominated by the temperature uncertainty.  Data
without error bars have uncertainties smaller than the symbols.  }
\label{cell.thermal.cond.fig}
\end{figure}

Thermal conductivity measurements were made by controlling the
temperature at the top of the cell, applying heat at the bottom of the
cell, and measuring the temperatures.  The two resistance thermometers
at the bottom of the cell allowed consistency checks against each
other; one of the thermometers at the top of the cell verified the
operation of the temperature controller, which used the other cell top
thermometer.  The thermometers consist of a RuO paste on a ceramic
substrate, manufactured for surface mounting on printed circuit boards
\cite{Dale.chips}.  The four thermometers used in the cell were
calibrated in a previous cooldown against another RuO resistance
thermometer, which was itself calibrated by Oxford Instruments.  The
heaters at the top and bottom of the cell were each made of two
$5~{\rm k}\Omega$ metal film resistors.  The two resistors for each
heater were positioned on opposite sides of the jacket, but at the
same vertical ($\hat{z}$) position.  The heaters were at about the
same vertical position as the thermometers.  The temperature of the
top of the cell was controlled by a proportional-integral controller
(1~s proportional time constant, 3~s integrator time constant).
The thermal conductance at a given temperature was extracted from a
series of measurements such as those in figure
\ref{cell.heat.response.fig}.
Conductances were measured at several cell temperatures.  They are compiled
in figure \ref{cell.thermal.cond.fig}. All temperature rises were less
than 10\%.
The conductances measured here are adequate for the experiments we
wish to perform.

\begin{figure}[tb]
\centering \epsfxsize=5in 
\epsfbox{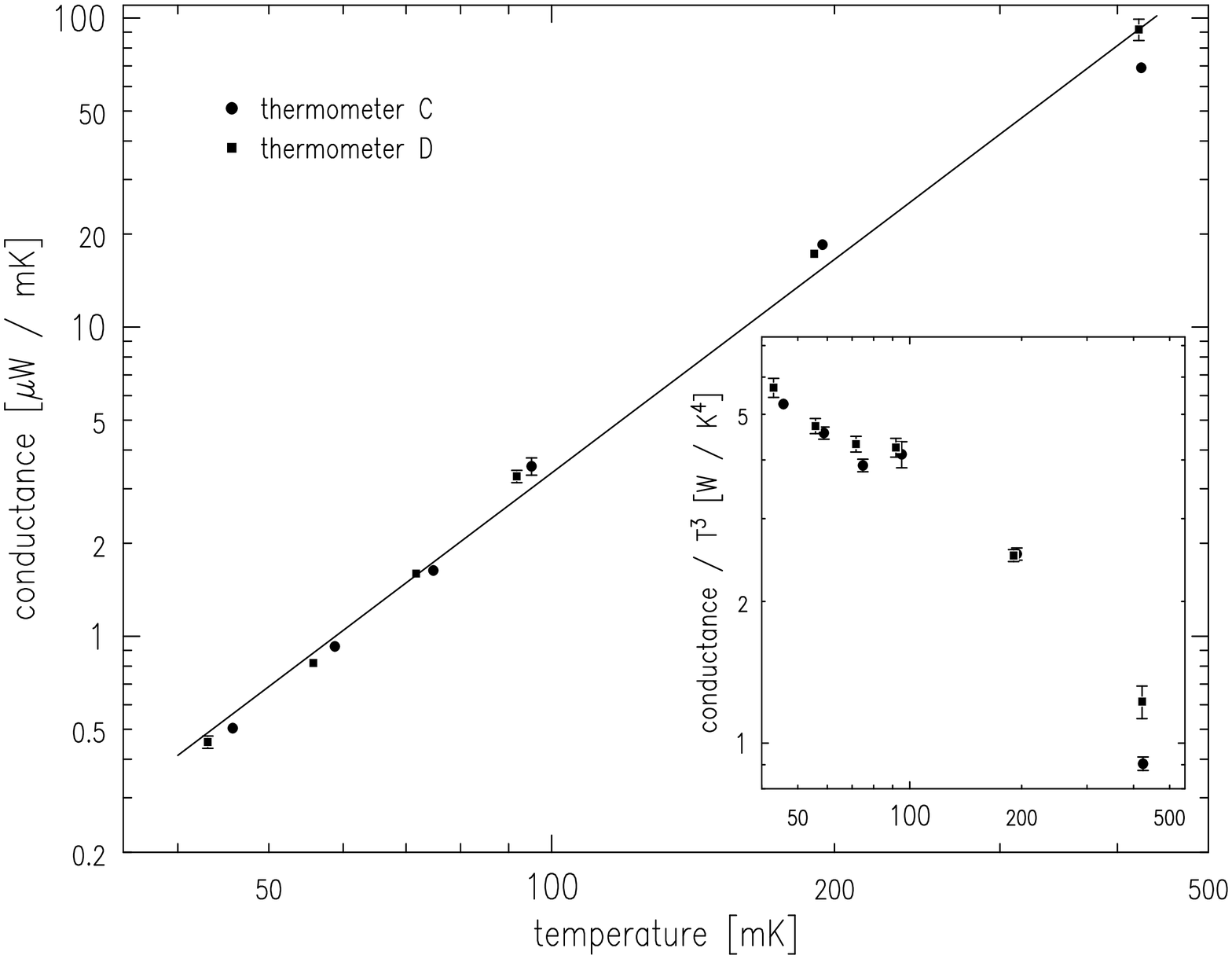}
\caption[thermal conductance of the Ag sinter]{
Thermal conductance across the Ag sinter, from the superfluid jacket
to the metal that comprises the dissociator.  The two data sets are
from the two thermometers at the top of the cell.  Simple acoustic
mismatch theory predicts a $T^3$ dependence.  This is factored out in
the inset.  The line, a fit to the data points, is $S=(8.6\times
10^{-5} ~{\rm \mu W/mK^{3.3}}) \times T^{2.3}$.  The error bars
reflect the statistical fit uncertainties.  Data without error bars
have uncertainties smaller than the symbols.  }
\label{sinter.conductivity.fig}
\end{figure}

The thermal conductance across the sinter plates, where heat is
transferred from the superfluid to metal and eventually to the
refrigerator, may be measured in an analogous way.  The temperature of
the dissociator was controlled with a thermometer and heater near the
top of the dissociator (see figure \ref{celltop.mechanical.fig}).  The
thermal gradient measured thus includes the thermal resistance across
the indium o-ring vacuum seal (superconductivity quenched by 4~T
field) and along the length of the dissociator.  The data from the two
thermometers at the top of the cell were fit to straight lines,
similarly to the data in figure \ref{cell.heat.response.fig}.  Two
temperatures and conductivities were thus obtained for each
temperature control point.  These two data sets are shown in figure
\ref{sinter.conductivity.fig}.


The conductance of the cell is similar to the conductance across the
sinter.  Any significant improvements to the thermal conductivity of
the system would need to address both of these segments of the heat
transport channel.

\label{spec.diffuse.exp.sec}
In a separate experiment the effects of specular and diffuse phonon
scattering were studied.  A cell similar to that described above was
made.  The surfaces of the bottom 23~cm of both the inner and outer
tubes were sanded with 120 grit paper to create an optically dull
finish.  The remaining 42~cm exhibited the shiny surface
characteristic of the epoxy coating.  The thermal conductance of the
two sections was measured in the range 50~mK to 300~mK.  The section
with diffuse optical reflection had a conductance three times lower
than the other section.  We thus conclude that diffuse reflection
reduces the conductivity by a factor of $3\times42/23\simeq5$.

\subsection{Heat Capacity}

\begin{figure}[tb]
\centering \epsfxsize=5in 
\epsfbox{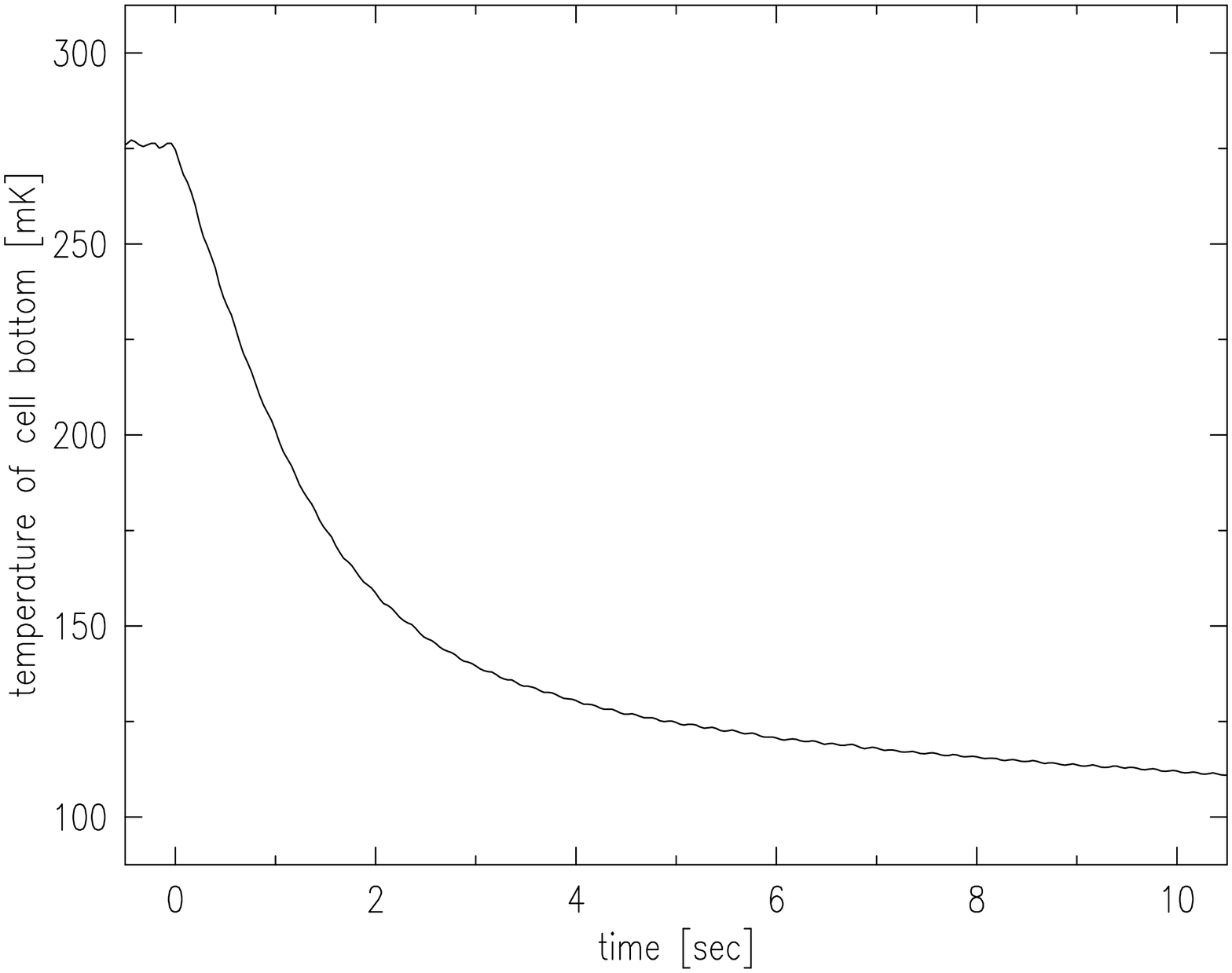}
\caption[cooling rate of cell]{
Temperature of the bottom of the cell while cooling from 280~mK.  The
cooling rate is dominated by the heat capacity of other metal parts,
not the superfluid.}
\label{cell.cooling.fig}
\end{figure}

The heat capacity of the cell and dissociator have been measured
together in a typical trap loading sequence.  The dissociator was held
by a temperature controller at 280~mK for 7~sec while pulsing an rf
discharge inside.  The heater and discharge where then turned off, and
the cell plus dissociator assembly was cooled.  The temperature of the
bottom of the cell is shown in figure \ref{cell.cooling.fig}.
The cooling rate is expected to be dominated by the large heat
capacities of the large copper pieces which comprise the dissociator
and mixing chamber.  Note that the molar specific heats of
copper and $^4$He are nearly identical at 100~mK, and there is much
more Cu than $^4$He.

\subsection{RF Heating}
\label{rf.heating.sec}
When rf power is introduced into the cryostat the cell and mixing
chamber are warmed.  Ohmic heating occurs in the coax lines that deliver
the power to the cell, and rf eddy currents cause heating of the
metals in the vicinity of the coils.  Measurements of these heating
rates is complicated by imprecise knowledge of where the power is
deposited and by disruption of the thermometer readout circuits by
the rf fields.  Nevertheless, some information can be obtained.

The most reliable piece of information relates to the ejection of
atoms from the trap.  If the cell becomes too warm, $^4$He atoms
evaporate and careen through the trapped sample, knocking atoms out of
the trap.  In many cases the maximum rf power is set by this
criterion.  A test was done of this ejection process by simply ramping
the rf frequency across the range of interest (23~MHz to 2~MHz) at
various rf powers, and noting how many atoms were ejected.  To prevent
resonant ejection a large bias field was applied so that the minimum
resonance frequency of the trapped atoms was always above the applied
rf frequency.  No ejection was noted for powers up to +27~dBm outside
the cryostat (corresponding to about 30~mG fields).  The cell
thermometers indicated 220~mK (bottom) and 180~mK (top).  The maximum
rf powers used for the experiments in this thesis were +15~dBm, so
ejection of atoms from the trap was not a problem.

Another piece of reliable data is that the cell will stay below 160~mK
when rf fields of roughly 5~mG are applied at frequencies of 23~MHz
and below.  This temperature is low enough to allow the cell to cool
reasonably quickly to temperatures at which atoms may be dumped out of
the trap and analyzed.

We note that there are a few resonances at which significant heating
occurs.  These are above 25~MHz.  One possible rf leak might occur
where the  superconducting twisted pair is joined to the coax.  Further
investigations of where heating occurs should be pursued.  Methods could
include a consideration of how fast objects cool when the rf is turned
off and the time required for heat to reach various thermometers.


%% file: rfplay.tex

\chapter{Manipulating Cold Hydrogen by RF Resonance}
\label{rfplay.chap}

Resonant RF ejection of atoms from the trap is a tool for manipulating
and studying the hydrogen gas.  This tool has not previously been
applied to trapped hydrogen, and so the peculiarities of this system
have not been explored.  We find---as anticipated---that rf ejection
can efficiently remove energetic atoms that would otherwise not find
their way out of the trap over the magnetic saddlepoint barrier.
Experiments are described in section \ref{rfevap.sec}.  Furthermore,
we find that rf ejection spectroscopy can constitute a valuable
diagnostic technique near BEC, where the trap dump technique used
previously fails.  RF ejection spectroscopy is also an essential tool
for measuring the magnetic field at the bottom of the trapping
potential.  These topics are covered in section
\ref{rf.spectroscopy.sec}.

\section{RF Evaporation}
\label{rfevap.sec}

\subsection{Need for RF Evaporation: Orbits with Long Escape Times}

In section \ref{evaporation.dimensionality.sec} we explained that, for
low temperature samples in tightly compressed traps, the saddlepoint
evaporation technique leads to one dimensional evaporation.  Recall
that atoms with high radial energy are unable to escape before having
a collision.  In this section we demonstrate this problem experimentally.

\begin{figure}[tb]
\centering \epsfxsize=5in 
\epsfbox{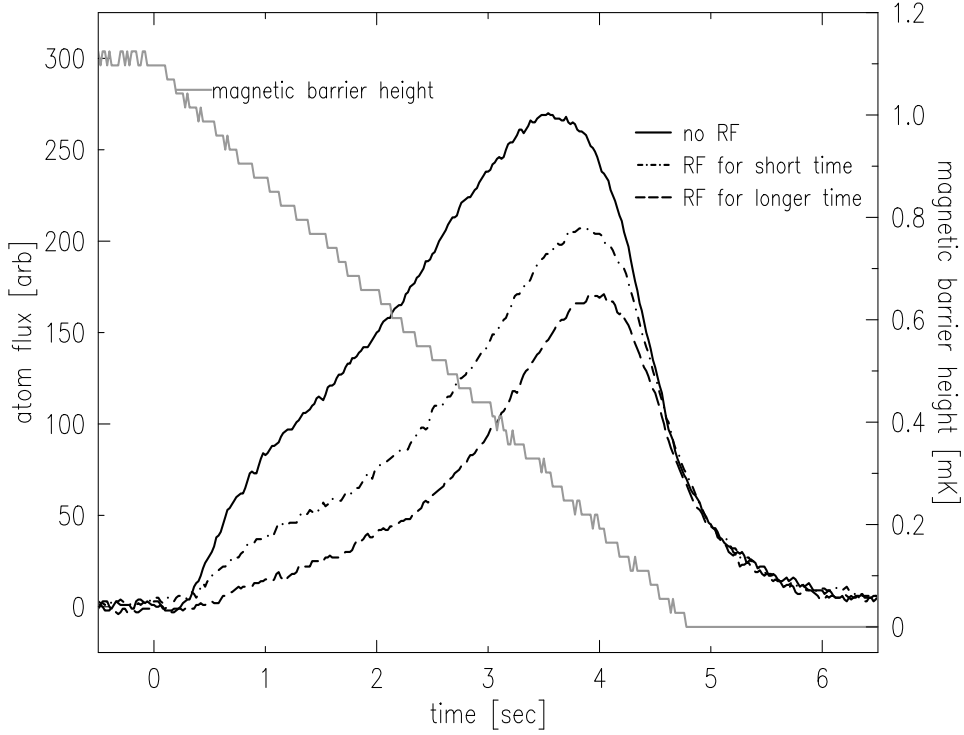}
\caption[demonstration of long escape times]{ Comparison of dump
signals from samples with and without rf ejection.  The solid sloping
line shows the trap depth as it is reduced from 1.1~mK to zero.  The
curves indicate the atom flux as the trap depth is reduced.  The
dashed curves correspond to a samples that were exposed to rf tuned to
the trap depth for 9~s and 23~s.  The solid curve corresponds to a
sample with no rf exposure.  The rf fields eject a sizeable fraction
of the sample and change the shape of the profile.  The response time
of the detection process is seen at the end of the ramp, at t=5~s. }
\label{rfcleanup.fig}
\end{figure}

Three successive samples are identically prepared by cooling into a
trap with depth $\epsilon_t/\kb =1.1$~mK, set by a magnetic
saddlepoint.  The trap radial gradient is $\alpha/\kb = 4.4$~mK/cm.
After the evaporation cycle is completed, the sample is held for many
seconds ($\sim 20$), on the order of three characteristic collision times.  The
saddlepoint barrier is then reduced to zero in 5~s while the atom flux
is recorded as a function of barrier height by sensing the heat of
molecular recombination on a sensitive bolometer (section
\ref{bolo.detection.sec}).  In its simplest interpretation, this flux
profile is the energy distribution of the trapped cloud.  (Corrections
to this picture will be described in section
\ref{dump.shape.ambiguities.sec}).  Figure \ref{rfcleanup.fig} shows
atom flux profiles for the three samples.
The largest profile is for a sample (sample 1) that was held for 10~s,
and then dumped from the trap.  For the smaller two profiles, a weak
rf field resonant at the trap depth ($\ethr=h\nu_{rf}-\theta$) was applied
for 9~s (sample 2) and 23~s (sample 3) at the point in the sequence
where sample 1 was held.  The samples were then dumped out of the trap
in the same way as sample 1.  We see that a significant fraction of
the atoms have been ejected from samples 2 and 3.  To be ejected,
these atoms must have had a total energy that was above the magnetic
saddlepoint energy.  Since these atoms did not leave the trap in
sample 1, they were apparently in orbits with escape times of at least
ten seconds.  This experiment demonstrates the long times required for
energetic atoms to escape when the energy barrier is set only by a
magnetic field saddlepoint at one end of the trap, and the reduction
in escape time when the atoms can leave in any direction.
Furthermore, the decrease in trap population between samples 2 and 3
indicates the long thermalization time of the system as it
equilibrates to the lower effective trap depth for the radial degrees
of freedom.  This trap had a radial confinement that was a factor of
$\sim3$ weaker than those used for attaining BEC as described in
section \ref{bec.sec}, and so these experiments place a lower bound on
the problem.

\subsection{Mixing of Energy}
\label{energy.mixing.sec}

It is postulated that there is only very weak coupling among the
motional degrees of freedom of a particle moving in the trap
potential.  This weak coupling between transverse and axial degrees of
freedom is the impetus for driving evaporation with rf fields.  In
this section we describe a simple experiment which demonstrates this
very weak coupling.

Four identically prepared samples are each cooled into a trap with
depth $\ethr=1.1$~mK and radial gradient $\alpha=4.5$~mK/cm.  The trap
is significantly longer than the one shown in figure
\ref{rf.field.geom.fig}, with the sample extending from roughly
$z=-6$~cm to $z=+9$~cm.  The sample is adiabatically expanded by
opening the trap slowly, reducing the radial gradient to
$\alpha=1.3$~mK/cm, but leaving the length constant.  The peak sample
density is approximately $n_0=4\times10^{12}~{\rm cm^{-3}}$.  The
thermalization time is thus on the order of 10~s, and the dipolar
decay time is on the order of $10^3$~s.  

\begin{figure}[tb]
\centering\epsfxsize=5in\epsfbox{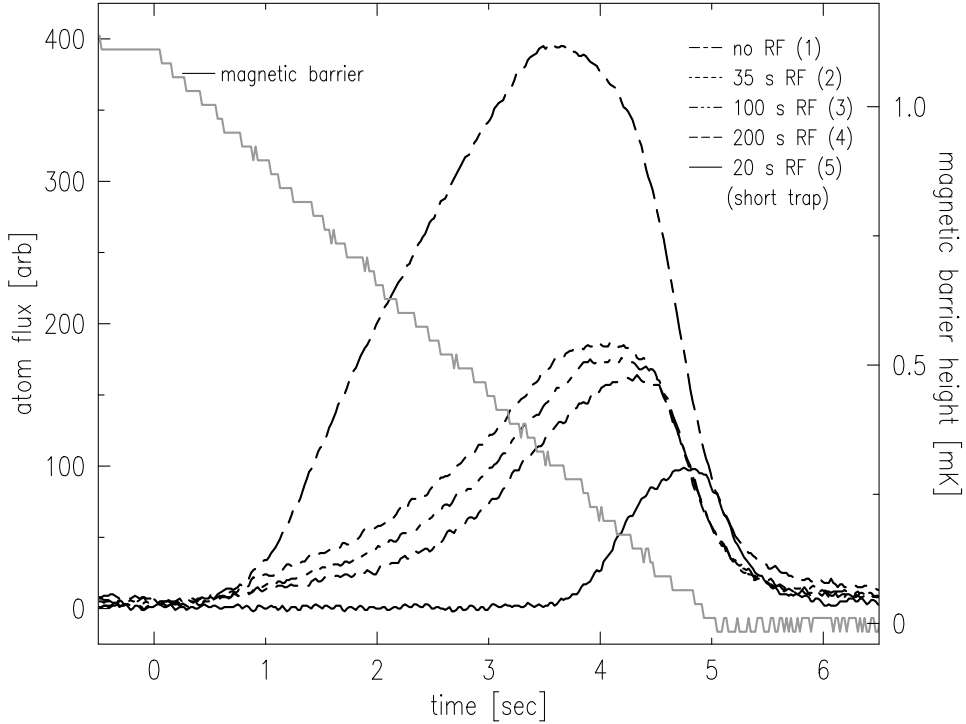}
\caption[slow mixing of energy from axial to transverse degrees of
freedom]{ Demonstration of slow transfer of energy from axial to
transverse motional degrees of freedom.  The five curves are the flux
of atoms escaping the trap as the magnetic energy barrier at one end
of the trap is slowly ramped to zero (linearly sloping line), for each
of five samples.  The samples labeled 1-4 correspond to various
durations of exposure of the sample to 10~MHz rf fields with strengths
of a few milligauss.  Sample 5 was confined in a short trap, and was
exposed to an rf field at 10~MHz for 20~s.  The field had ten times
more power than for samples 1-4.  }
\label{energy.mixing.fig}
\end{figure}

Figure \ref{energy.mixing.fig} shows the flux of atoms escaping the
trap as the magnetic confinement field at one end of the trap is
linearly ramped to zero.  
As explained above, this atom flux can be interpreted as an indication
of the distribution of energies of the trapped atoms.  The four traces
labeled 1-4 correspond to the four samples.  The first sample was
simply held in the trap for 20~s after preparation, and then dumped.
The second, third, and fourth samples had an rf field (a few
milligauss strength) applied for various times---35~s, 100~s, and
200~s.  The frequency (10~MHz) corresponds to slightly less than half
the trap depth set by the magnetic field barrier at the end of the
trap.  Even though the rf field is produced by both the axial and
transverse coils, the rf fields have significant strength (and
appropriate polarity) only along the sides of the trap, and not at the
ends.  We expect the rf fields to only eject atoms with energy in the
transverse (not axial) directions.

Comparing sample 1 to sample 2, we see that about two-thirds of the
atoms are ejected when the rf is applied for 35~s, and that the energy
distribution is narrowed, a sign of cooling.  As the rf exposure is
continued (samples 3 and 4), a few more atoms are ejected, and the
sample continues to slowly cool.  Two time scales thus appear.  The fast
time scale ($\sim10$~s) must correspond to removal of atoms with high
energy in the transverse degrees of freedom.  These atoms interact
directly with the rf fields.  The slow time scale ($\sim100$~s) is
for removal of atoms that originally had high energy in the axial
degree of freedom, but low energy in the transverse direction.  It
apparently takes on the order of 100~s for enough energy to transfer
from axial motion to transverse motion to allow the atoms to reach the
resonance region.

This interpretation is supported by the data shown in figure
\ref{energy.mixing.fig} for sample 5.  This sample was prepared
identically to samples 1-4, but the trap was short.  There was
significant field strength (and appropriate polarity) at the ends as
well as the sides of the trap, so that atoms with high energy in
either the axial or the transverse direction could pass through the
resonance region (again, both transverse and axial coils are used, as
is done for the routine rf evaporation in chapter \ref{results.chap}).
The resonance region wraps completely around the sample, in contrast
to samples 1-4 for which the resonance region is essentially a tube
around half the length of the sample.  We observe that the energy
distribution is narrowed very significantly with only short
application of the rf field.  This happens because energy does not
need to mix among degrees of freedom before ejection can occur.

\subsection{RF Field Strength Required for Evaporation}

As discussed in section \ref{rf.evap.power.requirements.sec}, the rf
field strength used during evaporation must be strong enough to eject
energetic atoms from the trap before they have a collision.  We have
experimentally determined this power requirement for a typical sample
density and trap shape.  A series of samples are identically prepared
at $T\sim 70~\mu$K, and then evaporatively cooled to a given trap
depth using rf fields of varying power.  After this rf cooling cycle
the atoms are dumped from the trap, and the total number of trapped
atoms is recorded.  For very low rf powers essentially no rf
evaporation occurs, and the sample is not depleted.  For higher powers
the sample is significantly depleted, up to an rf power that saturates
the process.  At this high power, energetic atoms are promptly removed
before having a collision; no further sample depletion is possible
with increasing power because the atom ejection rate is no longer
bottlenecked by the rf ejection rate, but rather by the rate at which
collisions in the gas create energetic atoms.

\begin{figure}[tb]
\centering \epsfxsize=5in 
\epsfbox{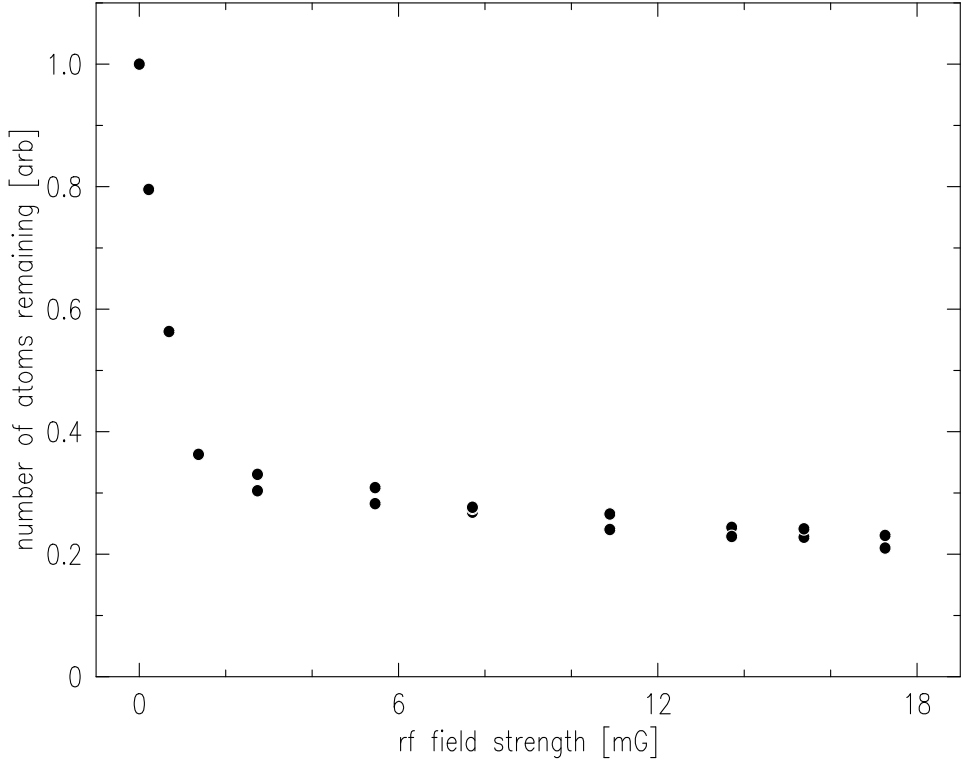}
\caption[rf power required for evaporation]{Determination of the field strength
required for rf evaporation.  Atoms are depleted from the sample by
fields of various strengths during an rf evaporation cycle.  The peak
sample density is $n_0\sim 7\times 10^{13}~{\rm cm}^{-3}$ and the trap
radial gradient is $\alpha/\kb = 4.5$~mK/cm.  The horizontal scale is
the peak rf field strength, as estimated from room temperature
measurements.  There is a factor of two scale uncertainty which arises
from imprecise knowledge of the temperature dependence of the
impedances and of the details of the field geometry.  }
\label{rf.evap.depletion.fig}
\end{figure}

Figure \ref{rf.evap.depletion.fig} shows one such saturation measurement.  
The rf frequency is swept from 17~MHz to 9.7~MHz at various rf field
strengths.  Two thirds of the atoms are removed when the field
strength is above 2~mG\@.  This agrees well with the calculation in
section \ref{rf.evap.power.requirements.sec}.  Note, however, the
large scale uncertainty on the horizontal axis.  Ideally we would plot
an effective field strength which is the appropriate average of the
ejection rate over the resonance surface (see section
\ref{rf.spec.theory.sec}); ignorance of the details of the rf field
direction and strength makes this average difficult.  We have
therefore plotted on the horizontal axis the strength of the field at
low frequencies, as measured at one point in the cell at room
temperature for a given power applied outside the cryostat.  A
correction was made for the frequency response measured in section
\ref{rf.eject.effic.sec}, assuming a driving frequency of 10~MHz.  The
number of remaining atoms continues to slowly drop as a function of
field strength even for fields much greater than the saturation field
of 2~mG.  This behavior is not understood.  It might indicate the
presence of atoms in high angular momentum orbits which only very
rarely cross the resonance region.

%

\section{RF Ejection Spectroscopy}
\label{rf.spectroscopy.sec}

The simple technique of rf ejection spectroscopy is useful as a
diagnostic of the temperature and density of the trapped gas near BEC.
It has significant advantages over the trap dump technique used
previously, which becomes completely ambiguous near the BEC threshold.
For probing the condensate, the technique is inferior to optical
spectroscopy (described in section \ref{1s2s.spec.theory.sec}), but it
is much simpler to use.  In this section we explain the problems which
make the trap dump technique ambiguous.  We then develop a theory that
explains the rf ejection measurements.  After an explanation of the
bolometric detection process employed in the rf ejection spectroscopy,
we show how the technique can be used to measure the magnetic field at
the bottom of the trap and to measure the temperature of a cold
sample.  

\subsection{Ambiguities of the Trap Dump Technique}
\label{dump.shape.ambiguities.sec}
In previous experiments the sample was studied by dumping it from
the trap and measuring its energy distribution (see section 2.5 and
2.6 in \cite{doyle91}); the distribution was inferred by measuring the flux of
escaping atoms as the axial trap barrier was reduced (see figure
\ref{rfcleanup.fig} for example).  Ambiguities arise, however, because
of conflicting timescales.  In order to measure an energy
distribution, the flux detector must respond in a time $\tau_{det}$
which is short compared to the duration over which the trap energy
threshold is reduced to zero, $\tau_{dump}$.  Furthermore, the
characteristic sample equilibration time, $\tau_{equil}$, must be long
compared to $\tau_{dump}$ so that the energy distribution does not
change during the dump process.  For the dense samples studied near
the BEC transition the equilibration time is $\tau_{equil}\sim300$~ms.  The detector response time is  $\tau_{det}\sim200$~ms.  
A further crippling complication is that the
characteristic escape time, $\tau_{esc}$, for an atom with total
energy greater than the trap depth must be short compared to
$\tau_{dump}$ so that all of the atoms in a given energy slice leave
the trap before atoms from the next slice begin to come out; in the
tightly compressed traps used near the BEC transition this escape time
has been observed to be many tens of seconds, as demonstrated by
figure \ref{rfcleanup.fig}.  The timescale hierarchy,
\begin{equation}
\tau_{det}, \tau_{esc} \ll \tau_{dump} \ll \tau_{equil}
\end{equation}
is not achievable, and the trap dump technique fails to providing
reliable measurements of the energy distribution of samples near the
BEC regime.  There is also a practical problem.  As the axial
confinement magnetic field is lowered, the trap shape changes.  Often
a magnetic field zero forms, further complicating the measurements.

RF ejection spectroscopy, in which a resonance region is tuned through
the trap to eject atoms with various potential energies, addresses all
these problems.  The trap shape remains unchanged, so no zeros are
introduced.  The particle escape time can be short because atoms are
coupled out in all directions.  Finally, the energy distribution of
the sample does not change significantly during the measurement
process because only a small fraction of the atoms are removed.

\subsection{Theory of RF Ejection Spectroscopy of a Trapped Gas}
\label{rf.spec.theory.sec}

In this section we use the results of the Landau-Zener theory
discussed in section \ref{rf.transitions.theory.sec} to explain the rf
ejection spectrum expected for a trapped gas, i.e. the flux of atoms
ejected as a function of the frequency of the rf magnetic field.  This
expression is used to interpret the spectra presented later in the
chapter.  

Consider a sample trapped in a magnetic potential $V({\bf r})=\mu_B
\left|{\bf B}_{trap}({\bf r})\right|$ described by the potential
energy density of states function $\varrho(\varepsilon)$.  The bottom
of the trap is at a magnetic field $B_0=\theta/\mu_B$.  An rf magnetic
field, oscillating at frequency $\nu_{rf}$ with strength ${\bf
B}_{rf}$, is superimposed on the trap, and resonantly interacts with
atoms crossing a surface of constant potential
$\varepsilon_{res}=h\nu_{rf}-\theta$.  For the analysis that follows
we note that this surface area and its differential thickness is
related to $\varrho(\varepsilon)\;d\varepsilon$, which is the
differential volume in the trap with potential energy in a range
$d\varepsilon$ around $\varepsilon$.  This volume may be written as an
integral, over the area of the surface of constant $\varepsilon$, of
the differential ``thickness'' $d\varepsilon/|\alpha({\bf r})|$ of the
surface:
\begin{equation}
\varrho(\varepsilon)\;d\varepsilon
=\int dA\frac{d\varepsilon}{|\alpha({\bf r})|}
\label{area.thickness.eqn}
\end{equation}
where $\alpha({\bf r})$ is the gradient of the potential at position ${\bf
r}$.  This result will be used to derive equation \ref{b.rf.eff.eqn}.

We wish to find the rate at which atoms leave the trap as a result of
Zeeman sublevel transitions in the resonance region.  In a patch of
resonance surface of differential area $dA$ at position ${\bf r}$, the
differential atom loss rate is $d\dot{N}=FP\;dA$, where $F$ is the
flux passing through the surface from either direction and $P$ is the
probability of the atom emerging from the resonance surface in an
untrapped Zeeman sublevel.  Equations \ref{spin.flip.probability.eqn}
and \ref{zeta.definition.eqn} indicate that, in the limit of low rf
field strength, the vast majority of transitions are made from the $d$
to $c$ Zeeman sublevels.  The transition probability is
$P=P_c=2(1-p)p$, where $p=1-\exp(-\zeta)$ and $\zeta({\bf r},v)=\pi
\mu_B^2 B_{rf}^2({\bf r})/\alpha({\bf r})v\hbar$ is the adiabaticity
parameter for an atom with velocity $v$ passing through resonance at a
position ${\bf r}$.  Here $B_{rf}$ is the component of the rf field
that is perpendicular to the dc trapping field.  In the weak field
approximation, $P\simeq 2p$ and $p\simeq\zeta$.  The transition
probability exhibits a velocity dependence.  The flux of atoms with
velocity $v$ passing through the surface is $F(v)=n({\bf r}) f_v(v)v$
where $f_v(v)$ is a normalized one dimensional velocity distribution
function, such as derived in appendix \ref{be.velocity.distrib.app},
and $n({\bf r})$ is the density at position ${\bf r}$ (either the
Bose-Einstein or Maxwell-Boltzmann density, as appropriate).  Since
both $P$ and $F$ exhibit a velocity dependence, the differential atom
loss rate must be derived from an integral over velocities
\begin{eqnarray}
d\dot{N} & = & dA \int dv \; n({\bf r})f(v)v\:2\zeta({\bf r},v) \nonumber \\
& = & dA\; \frac{2 \pi \mu_B^2 B_{rf}^2({\bf r})n({\bf r})}{|\alpha({\bf r})|\hbar}.
\label{Ndot.per.area.eqn}
\end{eqnarray}

The total atom loss rate ${\cal F}(\varepsilon)$, through a resonance
surface at potential energy $\varepsilon$, is
\begin{eqnarray}
{\cal F}(\varepsilon) & = & \int d\dot{N} \nonumber \\
& = & \frac{2 \pi \mu_B^2 n(\varepsilon)}{\hbar}\;
\int dA \frac{B_{rf}^2({\bf r})}{|\alpha({\bf r})|} \\
& = & \frac{2 \pi \mu_B^2}{\hbar}\;  n(\varepsilon)\: 
B_{eff}^2(\varepsilon) \varrho(\varepsilon).
\label{rf.flux.eqn}
\end{eqnarray}
Here we have defined $B_{eff}$ as an effective rf field strength
obtained by a weighted average, over the resonance surface, of the
magnitude of the component of the rf field that is orthogonal to the
trapping field, ${\bf B}_{trap}$.  The weight function is the
thickness of the surface.  Using equation \ref{area.thickness.eqn}
\begin{equation}
B_{eff}(\varepsilon)\equiv \left(\frac{1}{\varrho(\varepsilon)}
\int_{V({\bf r})=\varepsilon} dA\;
\frac{\left({\bf B}_{rf}({\bf r})\times{\bf B}_{trap}({\bf r})\right)^2}
{|\alpha({\bf r})| B_{trap}^2({\bf r})}\right)^{1/2}.
\label{b.rf.eff.eqn}
\end{equation}
If the polarization of ${\bf B}_{rf}$ is randomized all over the
resonance surface, then the cross product in equation
\ref{b.rf.eff.eqn} is essentially a scalar product, and
$B_{eff}(\varepsilon)$ is the average, over the equipotential surface,
of the square of the rf field strength.  This approximation is similar
to a random phase approximation.  Furthermore, if the length scale for
variations in the field strength is significantly different from the
size of the trap, then the field averaged over the surface will be
independent of the equipotential chosen.  Certainly if the field is
uniform over the trap (characteristic length long compared to size of
trap), then the average of the field over the equipotential surface
does not depend on which surface is chosen.  In the other limit, if
the characteristic length scale of the fluctuations in the field
strength is short compared to the size of the trap, then the
variations average out over the surface regardless of the surface
chosen.  The result in either case is that $B_{eff}$ is independent of
$\varepsilon$.  The field generated in the cell by a given rf driving
power, measured outside the cryostat, {\em can} vary as a function of
frequency; this ``transfer function'' is an artifact of the reactance
of the coils and the drive line impedance, and is measured in section
\ref{rf.eject.effic.sec}.  This dependence is not included here.

In our experiment these two approximations (random orientation and
incommensurate length scale) are not so bad; the rf field
is a superposition of the fields generated by both the transverse and
axial coils. The strength of the axial field varies on a length scale
much shorter than the trap length, but much longer than the trap
diameter, due to the field compensation scheme employed.  Figure
\ref{rf.field.geom.fig} shows the strength of the axial field along
the symmetry axis of the trap.  The field of the transverse coil is
essentially uniform all across the trap.  Since the relative strengths
of the rf field components vary all over the trap, the direction of
the field also varies.  Tests of this simplifying assumption are
outlined in section \ref{rf.eject.effic.sec}.  The relative phase of
the axial and transverse field is assumed to be zero, independent of
frequency and position.  Violations of this assumption have not been
considered.

If $B_{eff}$ has no strong dependence on $\varepsilon$, then the flux
of atoms ejected from the trap for a given $\nu_{rf}$ is proportional
to the number of atoms with potential energy $\varepsilon$,
$n(\varepsilon)\varrho(\varepsilon)$, as found in equation
\ref{b.rf.eff.eqn}.  A distribution of potential energies may thus be
obtained by sweeping the resonance surface through the trap.  This
distribution can be used to infer the sample temperature.

In the derivation of equation \ref{Ndot.per.area.eqn}, we made the
approximation that the rf power was low.  We assumed that $\zeta\ll
1$.  Inverting equation \ref{zeta.definition.sec}, this requires that
$B_{rf}\ll \sqrt{\alpha v\hbar/\pi \mu_B^2}$.  For typical parameters
$mv^2/2=(50~\mu{\rm K})\kb/10$ and $\alpha/\kb=(16~$mK/cm) the
requirement is that $B_{rf}\ll\sim 20$~mG\@.  Near the bottom of the
trap where the potential flattens the small gradient implies that the
experimenter must use very low rf field strength to remain in the low
rf power regime.

A further approximation used in the interpretation of the spectra is
that the sample's energy distribution does not change while the
spectrum is being measured.  This condition could be fulfilled by
measuring the spectrum quickly compared to the rethermalization time,
but, as explained above, the response time of the detection process is
too slow.  Another way to realize this condition is to remove  only a
small fraction of the atoms while obtaining the spectrum.  The number
of atoms ejected from the trap in a
duration $\tau_{spec}$ is $N_{eject}=\int_0^{\tau_{spec}} dt\; {\cal
F}(\varepsilon(t))$.  For a linear ramp of the resonance from the
bottom to the top of the trap, we have
$\varepsilon(t)=(\varepsilon_t/\tau_{spec})t$ and
\begin{eqnarray}
N_{eject} & = & 
\frac{2 \pi \mu_B^2 B_{eff}^2\tau_{spec}}{\hbar \varepsilon_t}\;  
\int_0^{\varepsilon_t} n(\varepsilon)\: \varrho(\varepsilon) \: d\varepsilon
\nonumber \\
& = & \frac{2 \pi \mu_B^2 B_{eff}^2\tau_{spec}}{\hbar \varepsilon_t}\;  N
\label{N.eject.rf.eqn}
\end{eqnarray}
where $N$ is the trap population.  A spectrum should be obtained over
a duration, $\tau_{spec}$, much longer than the detector response
time; in practice we use $\tau_{spec}\sim5$~s.  For a trap depth of
1~mK we obtain the condition $B_{eff}\ll \sqrt{\hbar
\varepsilon_t/2\pi\tau_{spec}}/\mu_B\sim200~\mu$G\@.  This low
ejection rate condition can be difficult to fulfill while
simultaneously enjoying a large signal-to-noise ratio in the detector.

\subsection{Bolometric Detection}
\label{bolo.detection.sec}

Atoms coupled out of the trap by rf resonance are detected by sensing
a small fraction of the molecular recombination heat they liberate.
The detection process is explained in detail by Doyle \cite{doyle91}.
Atoms (in the $c$-state) leaving the trap stick to the cell wall.  A
fraction experience a spin flip to the $b$- or $a$-state, opening the
way for rapid recombination.  If the residence time on the wall is
long enough, and if the time associated with sliding along the wall
out of the detection region and into a high field region is long
enough, and if the surface density is high enough, then the vast
majority of the atoms will recombine.  The heat deposited on the
bolometer will be proportional to the number of atoms which escaped
from the trap.  The picture is modified slightly for the rf
spectroscopic studies presented here because the atoms are ejected in
the $c$-state instead of the doubly spin-polarized $d$-state treated
by Doyle.  State mixing should occur rapidly, however, induced by
magnetic impurities on the walls, and so the remaining considerations
are still valid.  They indicate that the cell walls should be cold
(80~mK) and that the atom flux should be high.  Both of these
conditions impose limitations on the techniques described here.  We
observe well resolved rf spectra with atom fluxes on the order of
$10^9$ atoms per second falling out of the trap toward the bolometer.

\begin{figure}[tb]
\centering \epsfxsize=5in \epsfbox{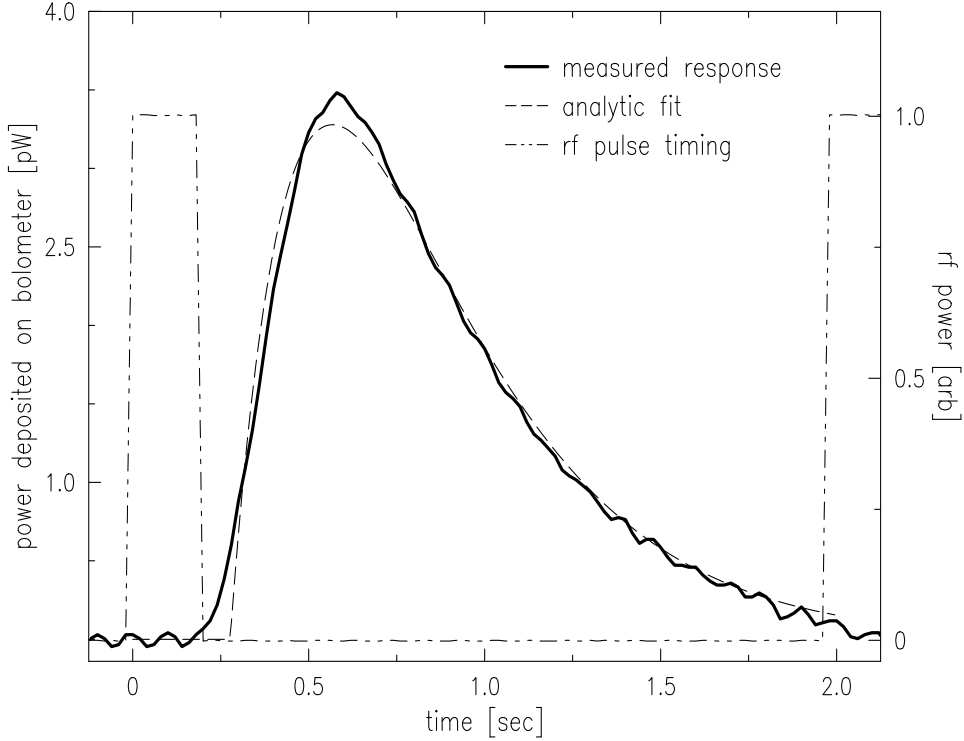}
\caption[impulse response of bolometric detection system]{Impulse response of
detection system.  An rf pulse is applied for 200~ms, and the
resulting bolometer signal is recorded.  For this pulse the fit delay
time is $\tau_{delay}=277$~ms, the fall time is $\tau_{fall}=340$~ms,
and the effective rise time is $\tau_{rise}=251$~ms.  The heat in this
pulse is 2.6~pJ, which corresponds to the heat from $3\times 10^6$
recombination events.  The quiescent power applied to the bolometer
was 100~pW, the bolometer temperature was 180~mK, and the cell
temperature was 90~mK.}
\label{rfpulse.fig}
\end{figure}

Crucial to the interpretation of the spectra presented is an
understanding of the time response of the bolometric detection
process, as suggested above.  The impulse response function of the
detection system was measured directly using a series of 200~ms rf
pulses at various frequencies for samples of two different
temperatures in the range of interest.  Figure \ref{rfpulse.fig} shows
a typical pulse
and an analytic fit of the form
\begin{equation}
g(t^\prime)=\left\{ \begin{array}{ll}
\frac{g_0}{\tau_{fall}-\tau_{rise}}
\left(e^{-t^\prime/\tau_{fall}} - 
	e^{-t^\prime/\tau_{rise}}\right) & t^\prime\geq 0 \\
0 & t^\prime <0
\end{array}
\right.
\end{equation}
where $t^\prime=t-\tau_{delay}$.  This form was chosen to 
model the time constants for an atom to stick to the wall and to find
a recombination partner.  The fit parameters were consistent to about
10\% for all the pulses, and the average values were
$\tau_{delay}=299$~ms, $\tau_{rise}=246$~ms, and $\tau_{fall}=337$~ms.
The dependence of these parameters on the film thickness has not been
studied systematically, but at low film thicknesses the H-He binding
energy increases and the atoms stick promptly; $\tau_{delay}$
decreases.  Because the rf pulse has finite duration, the measured
response is not, strictly speaking, the impulse response function.
Nevertheless, it is a good approximation, and the analytic fit
parameters quoted above were used in calculating the spectra presented
in the following sections.

The response time of the bolometer readout electronics was measured by
heating the bolometer with short pulses of light from a HeNe laser.
The response time was approximately $10^{-3}$~s, negligible compared
to the time constants for recombination.

\subsection{Measuring the Trap Bias Field}
\label{trap.bias.measurement.sec}

The bias field in the Ioffe-Pritchard trap, $\theta/\mu_B$, sets the
characteristic sample temperature at which the trap functional form
changes from predominantly linear ($T\gg\theta/\kb$) to predominantly
harmonic ($T\ll\theta/\kb$) in the two radial dimensions.  It is thus
an important quantity for analysis of Bose condensation experiments.
Furthermore, to know the trap depth $\epsilon_t=h\nu_{rf}-\theta$, one
must know both the trap energy barrier (set by the experimenter) and
the bias energy.

\begin{figure}[tb]
\centering \epsfxsize=5in \epsfbox{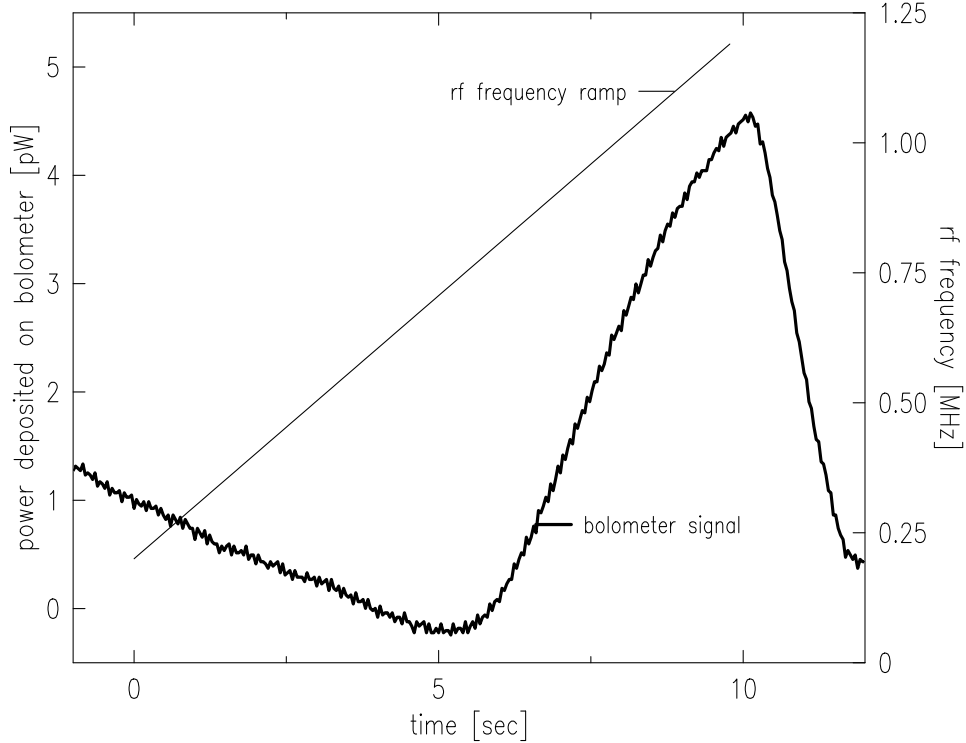}
\caption[measurement of $\theta$ by rf ejection]{Measurement of trap
bias using rf ejection.  The rf field is switched on at $t=0$ and
swept from 200~kHz to 1.2~MHz.  The atoms begin to leave the trap at
about 700~kHz.  The sloping baseline signal results from
atoms evaporating out of the trap as the sample continues to cool
after the end of the forced evaporative cooling cycle.  }
\label{rf.bias.fig}
\end{figure}

The bias is directly measured by sweeping up the frequency of the rf
field until atoms begin to be ejected from the trap.  Figure
\ref{rf.bias.fig} shows such a sweep. 
The atoms begin to be coupled out of the trap at about $t=5$~s,
corresponding to a frequency $\nu_\theta=700$~kHz; in this case we
find $\theta/\kb =\nu_\theta h/\kb = 33.6~\mu$K\@.  RF ejection is a
powerful and sensitive technique to quickly and accurately measure the
absolute trap bias.

The bias can also be inferred by deforming the trap potential until a
field zero is created; atoms begin to experience Majorana flips, leave
the bottom of the trap, and get detected by the bolometer.  The zero
is made by changing the current in a solenoid that creates a fairly
uniform field along the length of the trap.  Integration of the
Biot-Savart law gives the field strength per unit current.  This
method of finding the bias field is not as accurate as the rf ejection
method since the trap shape is deformed for the measurement.
Nevertheless, the two methods agree to within the experimental
uncertainties.

\subsection{Measuring the Effective RF Field Strength}
\label{rf.eject.effic.sec}

In order to analyze rf ejection spectra, one must know the effective
rf field strength, $B_{eff}$ given by equation \ref{b.rf.eff.eqn}.  It
may exhibit complicated polarization dependences if the random
orientation and incommensurate length scale approximations made in the
discussion of equation \ref{b.rf.eff.eqn} are invalid.  Furthermore,
there may be a frequency dependence in the field strength generated in
the cell by a given rf power applied outside the cryostat, as
discussed in section \ref{rfantenna.bode.sec}.  Clearly a measurement
of the efficiency is in order.

\begin{figure}[tb]
\centering \epsfxsize=5in \epsfbox{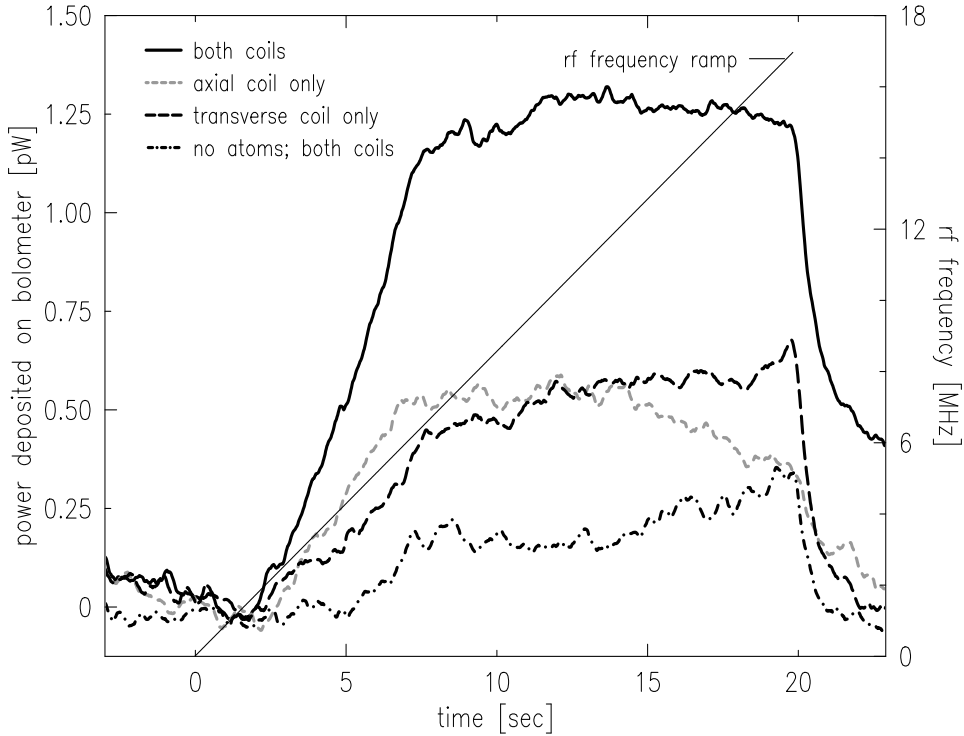}
\caption[ejection efficiency of rf coils]{Ejected atom flux as the
rf frequency is swept through the trap.  The lowest sweep indicates
the rf pickup and heating loads; no atoms were in the trap and both
coils were driven.  For the middle two sweeps only a single coil
was driven.  For the top sweep both coils were driven.  The
point at which atoms start to leave the trap, $t\sim 2$~s, indicates
the bias field.  Atoms stop leaving the trap when the rf field is
extinguished, $t\sim 20$~s.  At the lower frequencies the rf field
strength is $\sim 3\times 10^{-4}$~G\@.   Apparently the transverse coil is
responsible for most of the heating at high frequencies.}
\label{rf.freq.response.fig}
\end{figure}

RF spectra were measured to verify the ejection efficiency for
frequencies between 1 and 17~MHz.  The ejection rate at a given
frequency, given by equation \ref{rf.flux.eqn}, includes the density,
the potential energy density of states, and the effective rf field
strength.  In order to simplify the measurement of the efficiency, the density was kept
constant over the energy range of interest by using a warm sample,
$T\sim 1$~mK$>h\nu_{max}/\kb$\@.   The potential energy density of states functional form
is known well.  The quantity obtained, then, is the ejection
efficiency as a function of frequency.  The flux ejected as the
frequency of the rf field was swept upward is shown in figure
\ref{rf.freq.response.fig} for the axial coil alone, the transverse
coil alone, and both coils together.  
Another sweep, taken with no atoms in the trap but with both coils
driven, indicates rf pickup and heating effects.

\begin{figure}[tb]
\centering \epsfxsize=5in \epsfbox{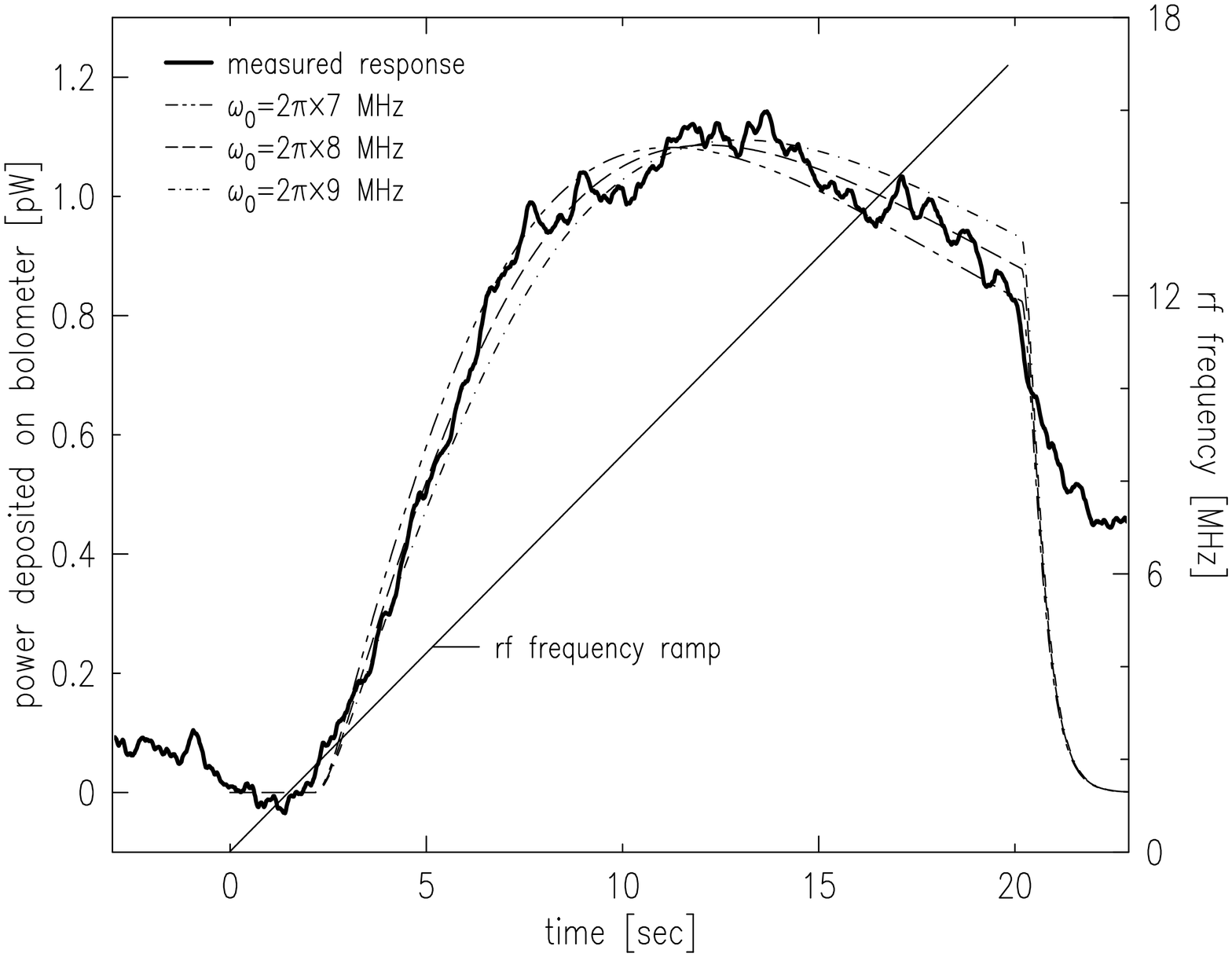}
\caption[frequency response of rf coils modeled]{Determination of the
frequency response of the coils by a comparison of calculated and
measured spectra.  The measured spectrum has the background, shown in
figure \ref{rf.freq.response.fig}, subtracted.  The only adjustable
parameters in the calculated spectra are the overall normalization,
the baseline, and $\omega_0$. }
\label{eject.eff.fit.fig}
\end{figure}

The simplest model of the frequency dependence of the ejection
efficiency ignores polarization effects and accounts only for
impedance matching (the voltage divider effects of the reactance of
the coils and the 50~$\Omega$ feed lines); we take
\begin{equation}
B_{eff}(\omega)\propto\left|\frac{1}{1+i\omega/\omega_0}\right|
\end{equation}
Figure \ref{eject.eff.fit.fig} indicates the expected spectra for
three values of $\omega_0$.  
Evidently $\omega_0\simeq2\pi\times 8$~MHz, which agrees reasonably
with the measured response of the coils below 20~MHz at room
temperature (figure \ref{coil.bode.fig}).  This value is used in the
calculated spectra presented next.

\subsection{Determining the Sample Temperature by RF Ejection Spectroscopy}

\begin{figure}[tb]
\centering \epsfxsize=5in\epsfbox{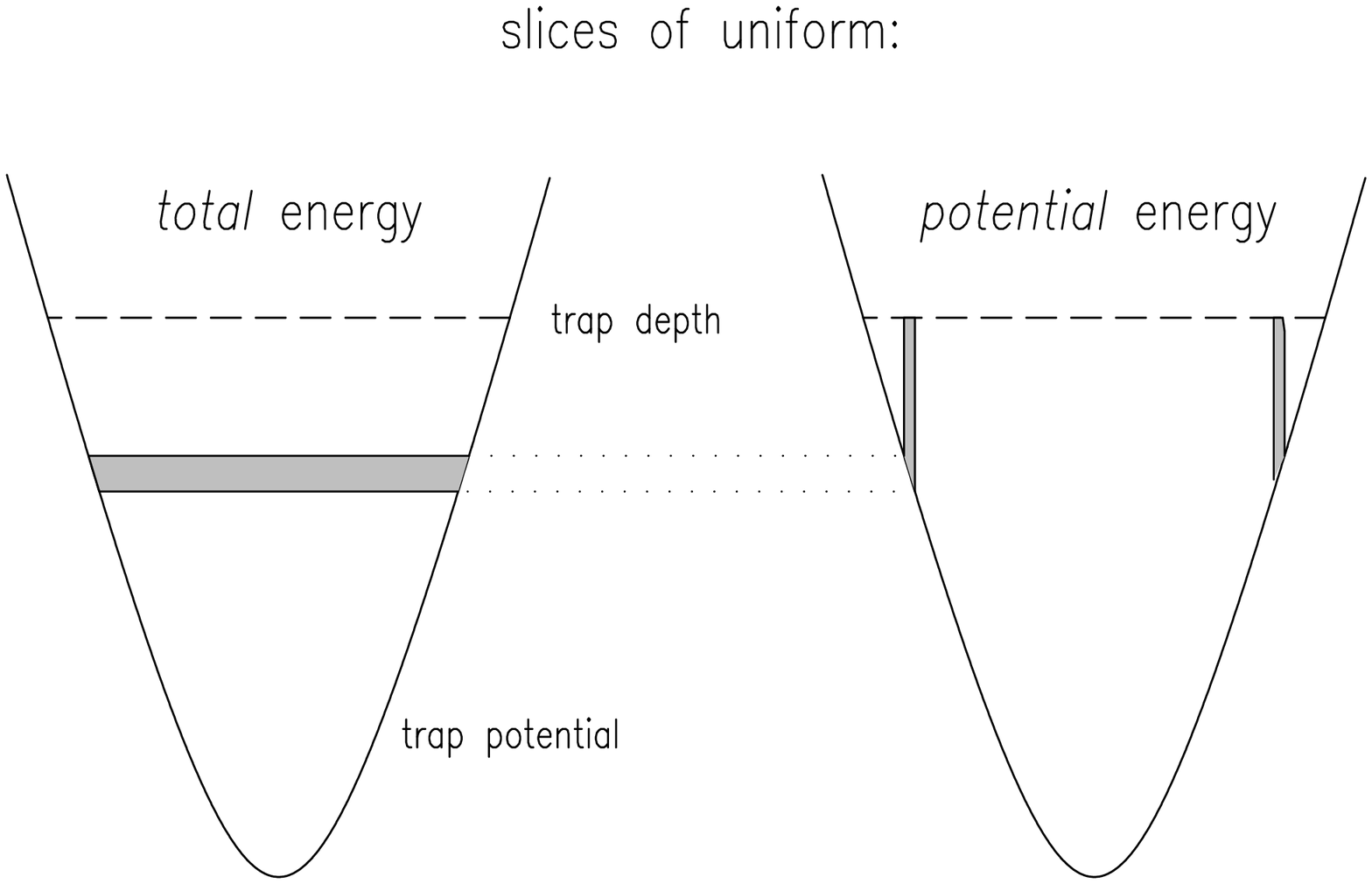}
\caption[probing distribution of total and potential energy]{Probing
the distribution of total energy and potential energy of a trapped
sample.  The total (potential) energy distribution
function gives the number of particles with a given total (potential)
energy.  The shaded area is the region of atoms probed by the
technique.  The solid lines indicate the trapping potential, and the
dashed line is the trap depth. }
\label{total.potential.distrib.fig}
\end{figure}

As suggested in equation \ref{rf.flux.eqn}, rf ejection spectroscopy
can give the temperature of the sample since it gives the 
{\em potential} energy distribution of the trapped atoms.  This spectrum is
different from the {\em total} energy distribution obtained by
dumping the atoms out of the trap, as done in previous hydrogen
experiments \cite{dsm89,doyle91}, or by saturating the rf transition
for the entire sample \cite{mhb88, phm88}.  
The difference between the {\em
total} and {\em potential} energy distributions is indicated in figure
\ref{total.potential.distrib.fig}.  
As outlined above, the rf ejection spectroscopy technique is free of
several complications implicit in the trap dump technique.  RF
ejection spectroscopy has been employed to measure the temperature of
the coldest samples of trapped H to date, a factor of 40 colder than
the photon recoil cooling limit.

\begin{figure}[tb]
\centering \epsfxsize=5in \epsfbox{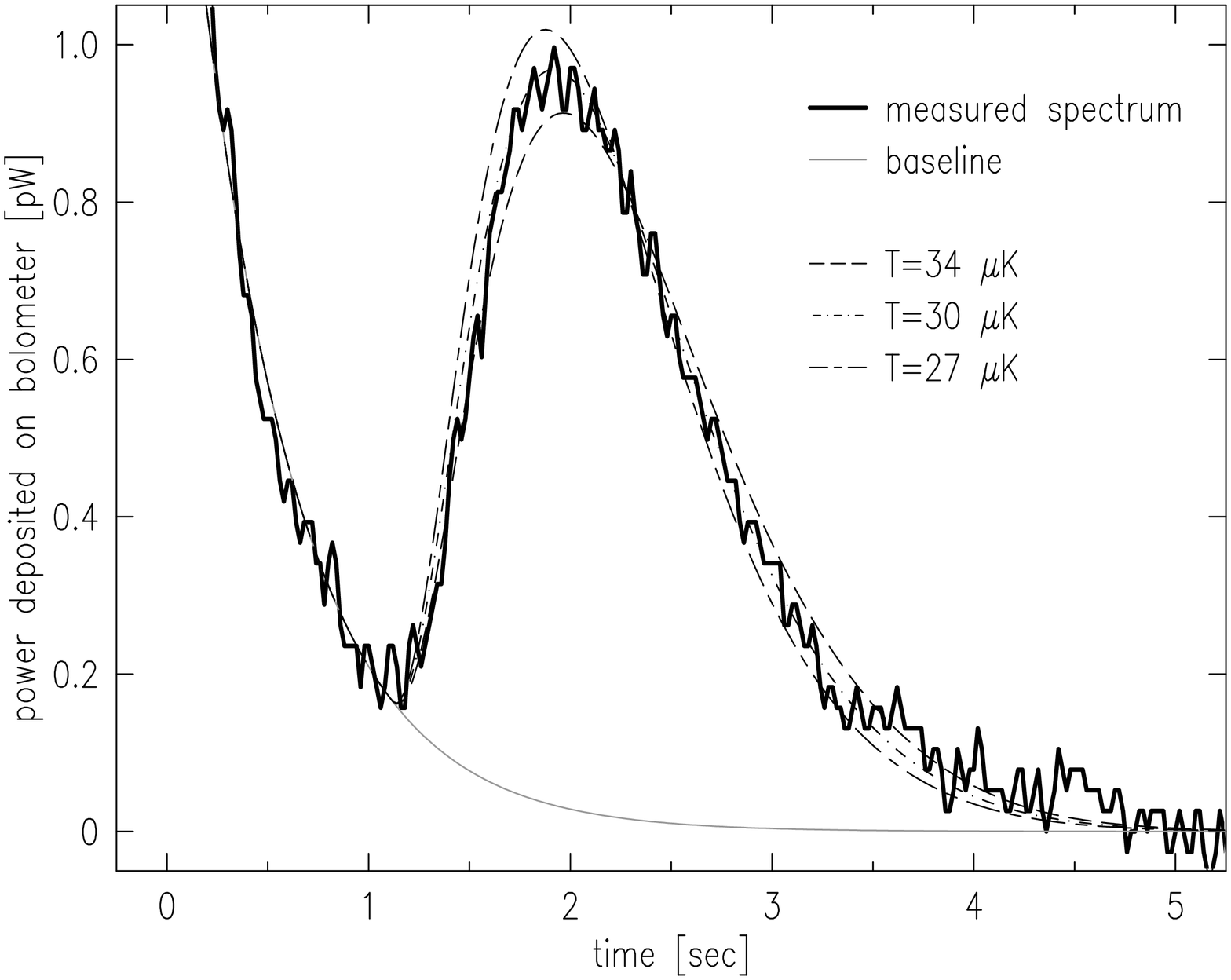}
\caption[rf spectrum of 30~$\mu$K sample]{RF spectrum of trapped gas
at $30~\mu$K\@.  The heavy line is the raw data.  The thin solid line
is the inferred background, whose functional form has been verified
independently.  The three dashed lines are calculations of the
spectrum for different temperatures assuming quantum degeneracy.  The
calculations include the time response of the detection system, the
measured trap depth, and the measured effective rf field strength.
The only remaining adjustable parameter is the normalization.  The rf
field strength used for this spectrum is $\sim 200~\mu$G\@.  }
\label{rf.trap.spec.fig}
\end{figure}

Figure \ref{rf.trap.spec.fig} shows the spectrum of a sample that was
cooled to $T=30~\mu$K by rf evaporation.  
This
sample should contain a small condensate, although it is not visible
here.  The three calculated spectra in figure \ref{rf.trap.spec.fig}
assume quantum degeneracy.  Furthermore, they include the measured
time response of the detection system, the measured trap bias, and the
measured effective rf field strength.  Possible
distortions, not considered in the calculations, arise from sample
depletion during the rf sweep and changing detection sensitivity due
to nonlinearities in the detection process at low atom fluxes.  This
second effect should be small, as suggested by the quality of the
calculated rf ejection spectrum, which ignored nonlinearity.  The
first effect can be shown to be small by taking a series of such
spectra at various rf powers, thus depleting the sample by various
(small) amounts.  Of course at lower powers the signal/noise ratio
becomes smaller; the power used for the spectrum in figure
\ref{rf.trap.spec.fig} involves a compromise between signal size and
distortion.  The apparent temperature is shifted no more than 10\%.

Determination of sample temperature through rf ejection spectroscopy
is limited by the signal rate, which is proportional to
$N_{eject}/\tau_{spec}$ in equation \ref{N.eject.rf.eqn}.  The
technique could be extended to much lower temperatures if the
detection efficiency is improved.

A practical limitation involves the requirement that the cell walls be
cold enough to support a long atom residence time, and thus high
recombination efficiency.  Because of rf heating effects, this
limitation usually requires the rf power to be reduced for several
seconds before the rf spectrum is obtained.  The sample can warm
during this time since the trap depth is not as well defined.




%% file: condensates.tex

\chapter{Observation of BEC}
\label{results.chap}

In this chapter we describe the first observations of Bose-Einstein
condensation of hydrogen.  

Our primary diagnostic technique is laser spectroscopy, which we
briefly describe here; detailed descriptions are given by Sandberg
\cite{san93} and Killian \cite{killian99} (Killian's thesis should be
considered a companion to this chapter).  With this probe we are able
to monitor the density and temperature of the gas as the temperature
is reduced by rf evaporation.  When the gas is cooled into the quantum
degenerate regime signatures of the Bose-Einstein condensate appear in
the spectrum; these can be analyzed to reveal the size and population
of the condensate.  As anticipated, the condensate is huge, containing
a factor of 50 more atoms than condensates in other experiments, and
having a length on the order of a centimeter.  We are able to study
the time evolution of the condensate, which is governed by feeding of
the condensate from the thermal gas.  We explore some of the
implications of these results for the production of high-flux coherent
atomic beams.


\section{Laser Spectroscopy of Trapped Hydrogen}

Laser spectroscopy of trapped atomic hydrogen has been pursued as both
a technique to study and manipulate a trapped {\em gas}, and as a
probe of the collection of individual {\em atoms} for metrology and
fundamental measurements \cite{kleppner89,san93,cesar95,cek99}.  We
focus here on spectroscopy as a tool to measure the density and
temperature of the trapped gas.  It has been suggested that the
coherence properties of Bose-condensed atoms could be exploited to
perform much improved fundamental measurements
\cite{bok97,bwi95}.

The principal transition in hydrogen, $1S$-$2P$, has a large
transition rate, but is difficult to use for sample diagnostics
because it has a large linewidth (100~MHz), it suffers from
complicated magnetic structure, and it is in the uv (121.6~nm).  The
transition has been used by the Amsterdam hydrogen trapping group as a
diagnostic for warm samples, and also for limited laser cooling and
laser induced evaporation \cite{lws93,swl93,lwr94}.  The Doppler
effect can be used to measure the sample temperature, but the short
$2P$ lifetime limits the technique to temperatures above a few mK\@.
In order to circumvent the short $2P$ lifetime, the Amsterdam group
employed two-photon spectroscopy to the $3S$ (160~ns) and $3D$ (16~ns)
states \cite{pmw97}.  To enhance the two-photon transition rate, one
photon was tuned near the \Lalpha\ resonance and the other near the
Balmer-$\alpha$.  In principle the $1S$-$3S$ transition could be used
for temperature measurement all the way into the sub-$\mu$K regime.

\subsection{Spectroscopy of the \oneStwoS\ Transition}
\label{1s2s.spec.theory.sec}

We employ the \oneStwoS\ transition and drive it with two photons
at 243~nm, twice the \Lalpha\ wavelength.  The $2S$ state is
metastable with lifetime $\tau_0=120$~ms, and the natural linewidth is
below 1~Hz.  The transition thus constitutes a very sensitive probe,
through shifts and broadenings, of the local environment of the atom.
As explained below, the transition can be used to measure the density
of the gas, in addition to the temperature.

To excite the transition the laser beam is retroreflected to
create a standing wave in the atom cloud.  In this configuration the
atom may absorb two photons from the same beam or one photon from each
beam.  Absorption of co-propagating photons involves the Doppler
shift, and the excitation is said to be ``Doppler-sensitive'' (DS).
In the absorption of counter-propagating photons the Doppler effect
cancels to first order, and the process is termed ``Doppler-free''
(DF).

The energy equation for two-photon excitation of an isolated atom from state
$i$ to $f$, with initial momentum ${\bf p}_i$ and final momentum ${\bf
p}_f = {\bf p}_i + \hbar({\bf k}_1 +{\bf k}_2)$, where ${\bf k}_1$ and
${\bf k}_2$ are the wave vectors of the laser beams, is
\begin{equation}
2 h\nu =
\sqrt{ p_f^2c^2 + (m c^2 + 2h\nu_o)^2} -
\sqrt{ p_i^2c^2 + (m c^2)^2} .
\end{equation}
\noindent
where the rest mass of the atom in the initial state is $m$ and
$2\nu_o=2.466\times10^{15}$~Hz \cite{uhg97} is the unperturbed
transition frequency.  Expanding, we obtain
\begin{equation}
\nu  = 
\nu_o
+\underbrace{\frac{({\bf k}_1+{\bf k}_2)\cdot{\bf p}_i}{4\pi m}
\left(1-\epsilon\right)}
	_{\textstyle \Delta\nu_{D1}}
+\underbrace{\frac{\hbar({\bf k}_1 + {\bf k}_2)^2}{8\pi m}
\left(1-\epsilon\right)}
	_{\textstyle \Delta\nu_{R}}
-\underbrace{\frac{\nu_o p_i^2}{2(mc)^2}}_{\textstyle \Delta\nu_{D2}}
+O(\epsilon^3).
\label{optical.energy.eqn}
\end{equation}
Here $\Delta\nu_{D1}$ and $\Delta\nu_{D2}$ are the first and second
order Doppler shifts, respectively, $\Delta\nu_R$ is the recoil shift,
and $\epsilon=2h\nu_o/mc^2=1.1\times 10^{-8}$ is a relativistic
correction which accounts for the mass change of the atom upon
absorbing energy $2h\nu_o$.  For hydrogen in the submillikelvin
regime, $\Delta\nu_{D2} \ll 1$~Hz and can be neglected.  In the
Doppler-sensitive configuration, ${\bf k}_1={\bf k}_2$ and
$\Delta\nu_R = 6.7$~MHz.  (All frequencies are referenced to the
243~nm laser source).  At a temperature of $50~\mu$K, $\Delta
\nu_{D1}\sim 2.6$~MHz, and thus the Doppler-sensitive peak is well
separated from the Doppler-free.  In the Doppler-free configuration,
${\bf k}_1=-{\bf k}_2$, and there is no recoil or first order Doppler
broadening.  A spectrum is shown in figure
\ref{overall.spec.nobec.fig}.

A complete discussion of the shape of the Doppler-free line in the
various regimes of density and temperature is presented in the thesis
of Thomas Killian \cite{killian99}.  We summarize his discussion by
noting that the narrow intrinsic linewidth allows the spectrum to be
used for measuring the density and temperature of the sample.  The
density is measured through the cold collision frequency shift, the low
temperature analog to the more familiar pressure shift.  In the
presence of a cloud of atoms of density $n$ in the $1S$ electronic
state, an atom in state $\sigma$ experiences a shift of its energy by
an amount $8\pi\hbar^2a_{\sigma-1S}n/m$, where $m$ is the atomic mass and
$a_{\sigma-1S}$ is the $s$-wave scattering length which parameterizes
collisions between atoms in state $\sigma$ and atoms in the $1S$
state.  Although both the $1S$ and $2S$ energy levels are perturbed by
surrounding atoms, $|a_{2S-1S}|\gg |a_{1S-1S}|$, so most of the
cold-collision frequency shift is due to perturbations of the $2S$ level.
The \oneStwoS\ resonance is shifted by an energy $\Delta E=h\chi n$,
where $h\chi=8\pi\hbar^2(a_{2S-1S}-a_{1S-1S})/m$ is the energy shift per unit
density.  The energy shift appears in the spectrum as a frequency
shift (at 243~nm) $\Delta_n=\chi n/2$.  A detailed
discussion of this mean-field picture of the energy shift has been presented
by Killian \cite{killian99}; here we merely emphasize that the process leads
to a frequency shift proportional to the sample density, and thus the spectroscopy is a valuable tool for measuring the sample density {\em in situ}.  We have
measured this shift to be \cite{kfw98}
\begin{equation}
\chi_m=-3.8\pm0.8\times10^{-10}~{\rm Hz~cm^3},
\label{chi.m.quote.eqn}
\end{equation}
which is in reasonable agreement with calculations \cite{jdd96}.  The
shift was measured for densities in the range $2-7\times10^{13}~{\rm
cm^{-3}}$ and for temperatures between 100 and 500 $\mu$K.  (For that
study the sample densities were measured using the known dipolar decay
rate constant, $g$, from equation \ref{dipolar.decay.const.eqn}.) The
frequency resolution of our laser system (currently 1~kHz) limits our
use of this technique to densities above a few times $10^{13}~{\rm
cm^{-3}}$.  The inhomogeneous sample density leads to a broadened
lineshape that must be interpreted; the interpretation process is
responsible for the majority of the $\sim$20\% uncertainty quoted for
that study.

At low densities the shape of the Doppler-free spectrum is governed by
the finite interaction time of the laser beam with the atom as the
atom traverses the beam (the cloud of atoms is much larger than the
beam diameter).  The characteristic linewidth (at 243~nm) is
$\delta_0=u/2\pi d_0$ where $u=\sqrt{2 \kb T/m}$ is the most probable
speed and $d_0$ is the waist diameter of the beam.  The expected
lineshape for a cloud of atoms having a Maxwell-Boltzmann energy
distribution, interacting with a laser beam with a Gaussian beam
profile, is the cusp-shaped $S(\delta)\propto
\exp(-|\delta|/\delta_0)$ for uv laser frequency detuning $\delta$
($\delta$ in Hz).  The linewidth is proportional to $\sqrt{T}$, making
this feature a useful probe of the sample temperature.  For the beam
waist diameter and laser frequency resolution achieved in these
experiments this technique is reliable above $100~\mu$K\@.  This
temperature probe suffers from systematic effects arising from
misalignment between the counterpropagating beams and divergence of
the beam over the length of the trap, effects which make the lineshape
difficult to interpret.  In an improved apparatus these effects could
be better characterized and controlled.


\subsection{Experimental Technique}

The \oneStwoS\ transition is excited by narrowband continuous-wave
radiation generated by frequency doubling the output of a 486~nm dye
laser.  This laser, which is frequency stabilized to an isolated Fabry-Perot
reference cavity, routinely exhibits a linewidth of 1~kHz and
output power of 480~mW.  After frequency doubling and spatial
filtering there is typically about 9~mW of uv power in a Gaussian mode
available at the window into the cryostat.

The trapped gas is illuminated by the uv beam as it traverses the
symmetry axis of the trap and is retroreflected by a spherical mirror.
The beam focus ($50~\mu$m waist radius) is positioned at the minimum
of the trapping potential so that maximum optical intensity overlaps
maximum sample density.  The atoms are excited during an illumination
interval $\tau_{exc}\sim 500~\mu$s.  Atoms in the metastable $2S$
state are held for a short delay $\tau_{delay}\sim 70~\mu$s while
background fluorescence dies away.  Then a 10~V/cm electric field is
applied for a short time $\tau_{quench}\sim 10~\mu$s; the $2S$ state
mixes with the $2P$, which quickly decays emitting a \Lalpha\ photon
which is counted on a microchannel plate detector.  Only photons which
hit the detector during the quenching interval are considered signal,
and so the background count rate is much suppressed.  The total
detection efficiency should be about $2\times 10^{-5}$, limited by
detection solid angle (0.13\%), transmission through a window (60\%)
and a \Lalpha\ filter (10\%), and detector quantum efficiency
(estimated at 25\%).  Actual detection efficiency appears to be about
10 times lower than this estimate, possibly due to contaminants on
windows \cite{killian99,lorenz.pc}.  The background count rate is
between 200 and 700~counts/sec, giving about 8 background counts per
second of laser exposure assuming the excitation and detection timing
sequence outlined above.  The minimum detectable signal thus
corresponds to $4\times 10^{6}$ atoms excited per second of laser
excitation.  Straightforward improvements to the optical access should
easily yield a factor of ten improvement in detection efficiency.

\subsection{The \oneStwoS\ Spectrum of a Non-Degenerate Gas}

\begin{figure}[tb]
\centering\epsfxsize=5in\epsfbox{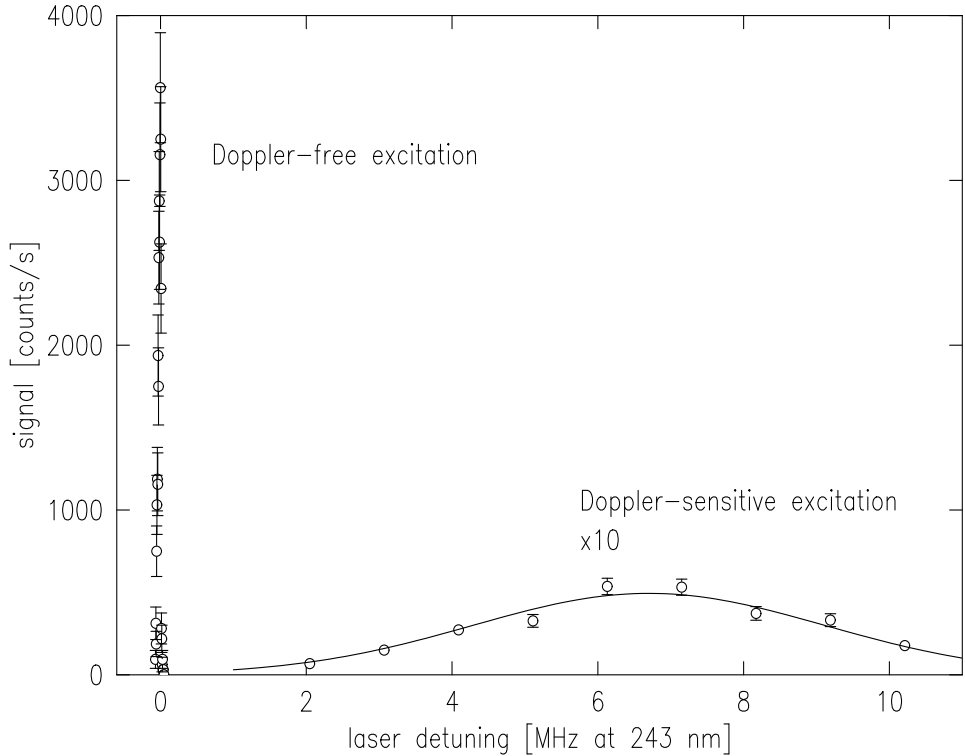}
\caption[spectrum of non-degenerate trapped hydrogen gas]{
Spectrum of non-degenerate gas.  The intense narrow peak on
the left results from Doppler-free excitation and the low broad peak
on the right results from Doppler-sensitive excitation which probes
the velocity distribution.  The solid line on the right is the
expected Doppler-sensitive spectrum for a nondegenerate sample at
42~$\mu$K.  At this low temperature the recoil-shifted
Doppler-sensitive line is clearly separated from the Doppler-free
line.  The error bars are statistical.  The ratio of peak heights
should be taken as only a rough indication of the transition rate
ratios because of variations in laser power and beam alignment.}
\label{overall.spec.nobec.fig}
\end{figure}

The spectrum of the trapped hydrogen gas slightly above the quantum
degenerate regime is shown in figure \ref{overall.spec.nobec.fig}.  
As described in section \ref{1s2s.spec.theory.sec}, there are two
components of the spectrum, corresponding to absorption of co- or
counter-propagating photons.  The wide, low feature on the right
(higher photon energy) is the Doppler-sensitive peak, and is offset by
the recoil shift $\Delta\nu_R$ from the intense, narrow Doppler-free
peak on the left.  The solid line is the Gaussian lineshape expected
for a Maxwell-Boltzmann distribution of kinetic energies in a sample
at $42~\mu$K.  The DS spectrum maps the velocity distribution through
the Doppler shift; only atoms with the proper velocity
$v=\lambda\Delta\nu$ are Doppler shifted into resonance for a given
laser detuning $\Delta\nu=\nu - (\nu_0+\Delta\nu_R)$ ($\nu$ is the optical
frequency of the laser).  Thus, only a tiny fraction of the sample is
resonant, and the excitation rate is correspondingly tiny.  The
fraction of atoms which are resonant is proportional to the ratio of
the excitation linewidth, $\delta_0$ from time-of-flight broadening,
to the Doppler linewidth $\Delta_D=\sqrt{2\kb T \log(2)/m \lambda^2}$;
in our apparatus this ratio is $\sim 5\times 10^{-4}$.  The very high
densities of our trapped samples have allowed us to measure the DS
\oneStwoS\ spectrum of hydrogen for the first time.

\begin{figure}[tb]
\centering \epsfxsize=5in \epsfbox{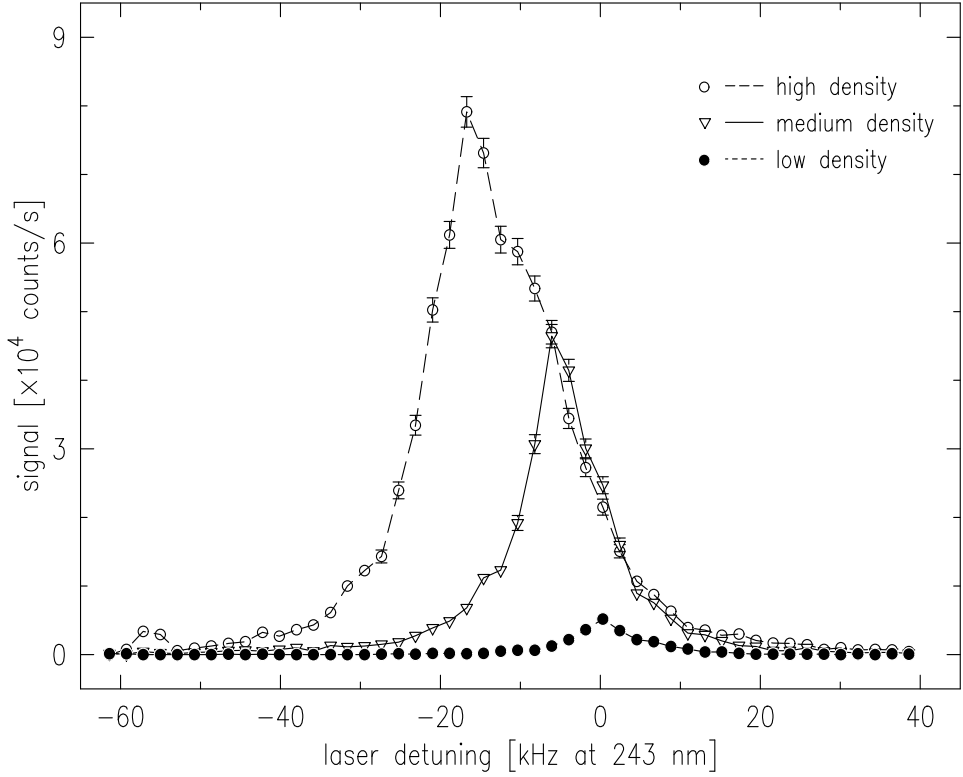}
\caption[Doppler-free spectra at various sample densities]{
Doppler-free spectra at various sample densities.  The trap depth is
570~$\mu$K, corresponding to a sample temperature $T\sim 70~\mu$K.
The spectra of the higher density samples are red-shifted and
broadened by the inhomogeneous density.  The lines connecting the
points are a guide to the eye.  The samples were confined by trap
shape A, described in table \ref{3trap.shape.summary.tab}.  The
highest density is about $10^{14}~{\rm cm^-3}$. }
\label{df.dens.fig}
\end{figure}

The Doppler-free spectrum gives information about the density
distribution and temperature of the gas, as described above.  Both of
these effects are shown by the spectra in figure \ref{df.dens.fig},
just three of many successive spectra that were measured as the
density of the sample slowly decayed.  The initial spectra exhibit a
large shift, arising from the cold-collision frequency shift at high
density.  The inhomogeneous density in the trap causes a broadening of
the spectrum as well.  As the sample density decreases, the shift and
broadening also decrease.  At low density, where the density-dependent
shift is negligible, the exponential lineshape characteristic of
transit time broadening is evident.  The frequency shift of the
spectra of high density samples is used to obtain the sample density,
and the width of the transit-time broadened line is related to the
sample temperature.

\section{Cooling into the Degenerate Regime}
\label{bec.sec}

\begin{figure}[tb]
\centering \epsfxsize=6in \epsfbox{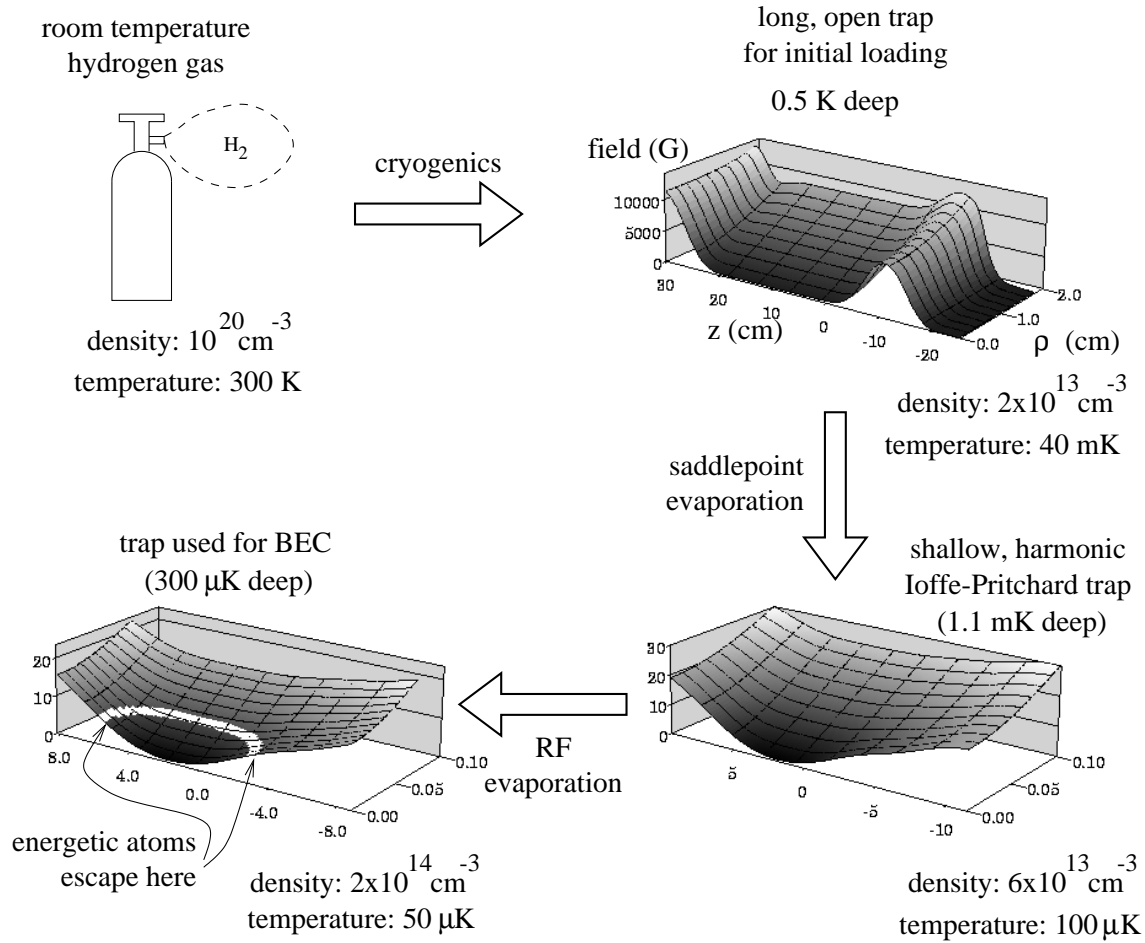}
\caption[schematic diagram of trap shape during cooling]{Schematic
diagram of trap shape as sample is loaded and evaporatively cooled to
BEC.  Room temperature molecular hydrogen (upper left diagram) is
cooled to $275~$mK by cryogenic techniques.  It is dissociated, and
the atoms are loaded into an open, long trap that is about 500~mK deep
(upper right diagram).  The magnetic field in the $\rho-z$ plane is
shown.  Note the very different scales for the radial coordinate
$\rho$ and the axial coordinate $z$.  During trap loading, the atoms
settle into the region of low field at the center of the trap.  The
sample is then cooled by magnetic saddlepoint evaporation.  It is
compressed radially and axially into a trap that is nearly harmonic
along the $z$ axis and predominantly linear along the radial axis
(lower right diagram).  It is this trap shape that is labeled ``A''
or ``B'' (see section \ref{cooling.paths.sec}).  RF evaporation is
used to further cool the sample while the trap shape remains fixed
(lower left diagram).  The entire cooling process takes about 5
minutes. }
\label{trap.progression.fig}
\end{figure}

Armed with an efficient cooling method (rf evaporation) and a
sensitive probe of sample density and temperature (spectroscopy), we
are equipped to push toward the quantum degenerate regime.  The trap
is loaded as described in section \ref{discussion.of.loading}, and the sample is cooled by
magnetic field saddlepoint evaporation.  The currents in the magnetic
coils that create the trapping potential are ramped to slowly change
the trap from the deep, long, open shape used for loading the atoms to
a short, tightly compressed, shallow trap from which rf evaporation is
commenced.  Figure \ref{trap.progression.fig} indicates the process.
In this section we discuss the path through density-temperature space
taken by the sample as it is cooled into the quantum degenerate
regime.

\subsection{Cooling Paths}
\label{cooling.paths.sec}

Previous attempts to achieve BEC in hydrogen were thwarted by
inefficient evaporation at the lowest temperatures.  In order to
verify that we have solved this problem, the progress of the sample
through phase space (density-temperature space) is monitored as the
sample is cooled.  We find that the sample is cooled efficiently.  We
also find and explore the phase boundary where the sample undergoes phase
separation into a condensate and normal gas.  This first evidence for
phase transition behavior is described below.

\begin{figure}[tb]
\centering \epsfxsize=5in \epsfbox{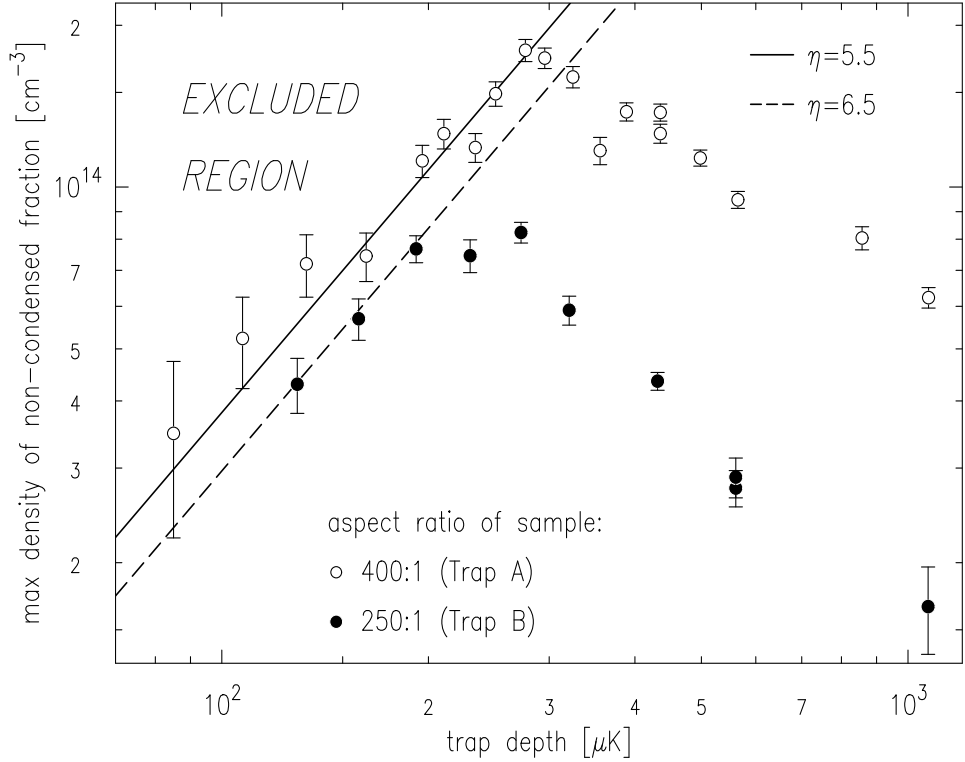}
\caption[path of sample through $n$-$T$ space as it is cooled into
degenerate regime]{ Path of non-condensed portion of the sample
through $n-T$ space as the sample is cooled into the degenerate
regime.  Paths are shown for samples in two different traps: open
circles --- trap A, closed circles --- trap B.  The trap depth is set
by the frequency of the rf magnetic field, and above degeneracy the
temperature is about $1/\eta\sim 1/8$th the trap depth.  The density
is measured through the cold-collision frequency shift.  The
non-condensed portion of the sample cannot have density greater than
the BEC critical density $n_c\propto T^{3/2}$, and thus the upper
left region of the phase diagram is not accessible.  The condensates
created in the two traps contain about the same number of atoms.}
\label{evap.trajectory.fig}
\end{figure}

A series of measurements were made to map the sample's trajectory
through phase space as it was cooled into the quantum regime.
Identical samples were cooled into a trap of depth
$\epsilon_t/\kb=1.1$~mK.  This process used magnetic saddlepoint
evaporation exclusively.  The trap shape was then held constant while
the trap depth was reduced by rf evaporation.  (Three trap shapes will
be discussed in this chapter, A, A$^\prime$, and B.  Their
characteristics are summarized in table \ref{3trap.shape.summary.tab}.
Here we refer to trap shape A.)
\begin{table}[tb]
\centering\begin{tabular}{|c||c|c|c|}
\hline \hline
parameter & trap A & trap A$^\prime$ & trap B \\ \hline \hline
$\alpha/\kb$ (mK/cm) & \multicolumn{2}{c|}{15.9} & 9.5 \\ \hline
$\beta/\kb$ ($\mu$K/cm$^2$) & \multicolumn{3}{c|}{25} \\ \hline
$\theta/\kb$ ($\mu$K) & $27\pm2$ & $35\pm 2$ & $34\pm 2$ \\ \hline
$n_0(T=120~\mu$K) ($\times 10^{13}$~cm$^{-3}$) & \multicolumn{2}{c|}{8} & 1.7 \\ \hline
\hline
\end{tabular}
\caption[summary of trap parameters]{ Summary of
parameters describing the three trap shapes used in this chapter.
The parameters $\alpha$, $\beta$, and $\theta$ describe the
Ioffe-Pritchard potential; $\alpha$ and $\beta$ are calculated, and
$\theta$ is measured. The peak density at $T=120~\mu$K is noted.
}
\label{3trap.shape.summary.tab}
\end{table}
The evaporation path (the trap depth as a function of time) was
identical for all samples, but the different samples were taken
various distances along the path.  The rf frequency $f(t)$ that sets
the trap depth $\ethr=hf(t)-\theta$ was ramped according to the
function\footnote{The evaporation path was not optimized before most
of the data presented in this chapter was obtained.  For future
studies an optimization of the functional form of the path is
desirable.}  $f(t)=f_{start}(f_{end}/f_{start})^{(t/d)^\xi}$ where the
start frequency is $f_{start}=23$~MHz, the end frequency is
$f_{end}=2$~MHz, the maximum ramp duration is $d=25$~s, and the shape
exponent $\xi=1.5$.  The ramp was executed for a duration $\tau_r$.
For $t\geq\tau_r$, the frequency was maintained at $f(\tau_r)$.
Evaporation could still occur, but forced evaporation halted.  After
rf evaporation was completed, the sample density was measured using
the cold-collision frequency shift of the DF spectrum as described
above.  The densities thus measured are shown in figure
\ref{evap.trajectory.fig}.
The density initially increases as the trap depth (and thus
temperature) is reduced because the atoms are confined to a smaller
volume closer to the bottom of the trap.  Some atoms are lost through
evaporation, but the good efficiency of evaporative cooling leads to
overall phase space compression.  The cooling efficiency parameter
$\gamma=d\log D/d\log N$ (the phase space density is $D\equiv
n_0\Lambda^3$) is $\gamma\simeq -1.5$ during this stage of the cooling.

At a trap depth of about $\epsilon_t/\kb =280~\mu$K the phase space
trajectory exhibits a dramatic kink.  Further cooling leads to
decreasing density in the normal gas; apparently population is being
rapidly removed.  The location of the kink and the slope of the
subsequent trajectory agree well with the BEC phase transition line,
$n_c=2.612 (2 \pi m \kb T)^{3/2}/h^3$ assuming a ratio of trap depth
to temperature of $\eta=\epsilon_t/\kb T\sim 6$, which is reasonable.
The ``excluded region'' above the line has the following
interpretation: as the sample is cooled, the thermal cloud cannot
support a peak density above $n_c$; extra atoms are transferred to the
condensate.  The condensate is not directly observed in these
measurements because its density is so large (as discussed in section
\ref{peak.cond.dens.sec}) that its spectrum is shifted far away from
the spectrum of the thermal cloud.

The magnetic trap shape used for these experiments (trap shape A) was
chosen as a compromise between high radial compression and adequate
magnetic saddlepoint evaporation efficiency.  High radial compression
is desirable because the correspondingly higher densities allow BEC to
be reached at a higher temperature.  As discussed in section
\ref{evaporation.strategy.sec}, evaporation and the approach to BEC
should be more efficient at higher temperatures because the ratio of
``good'' to ``bad'' collisions is more favorable.  On the other hand,
high radial compression is undesirable because the magnetic field
saddlepoint evaporation process (see figure
\ref{trap.progression.fig}) that cools and compresses the sample into
the tight trap is less efficient.  As discussed in section
\ref{evaporation.dimensionality.sec}, tight radial compression leads
to more regular motion of the trapped atoms, and thus to reduced
evaporation dimensionality.

A more open trap, trap B, has also been used.  A typical
phase space trajectory is shown by the filled circles in figure
\ref{evap.trajectory.fig} .  The evaporation was done more slowly,
with maximum duration $d=55$~s, but was otherwise identical to the
path described above.  The evaporation efficiency parameter is
$\gamma\sim -2.7$ for cooling above the degenerate regime.  The higher
evaporation efficiency has not been systematically studied, but could
arise simply because the trajectory duration $d$ was optimized for
trap B, but not trap A.

Computer models of the cooling process have been constructed, and are
described in appendix \ref{evap.model.app}.  The trajectories through
phase space calculated by them are in good agreement with
observations.

\subsection{Evaporation Strategies}
\label{evaporation.strategy.sec}

There are many paths that can be taken through phase space to
the BEC phase boundary.  In this section we discuss the experimental
limitations of the present apparatus and their implications for the
choice of evaporation path.  

As shown in section \ref{equilibrium.temp.sec}, evaporation is more
efficient in hydrogen at higher temperatures.  There the ratio of
``good'' collisions to ``bad'' collisions is higher.  ``Good''
collisions are those which lead to cooling, and they occur at a rate
$\Gamma_{evap}\propto\Gamma_{col}\propto n\sigma\bar{v}\propto
n\sqrt{T}$ for a sample of density $n$.  ``Bad'' collisions are those
which lead to dipolar relaxation and heating. They occur at a rate
$\Gamma_{dip}\propto g n$.  The collision ratio
$\Gamma_{evap}/\Gamma_{dip}\propto\sqrt{T}$ is higher at higher
temperatures.  More phase space compression can be accomplished per
atom lost from the trap.  This higher efficiency makes it desirable to
do most of the phase space compression at higher temperatures.

The competing consideration is that full evaporation dimensionality,
and thus efficiency, can only be realized using the rf evaporation
technique (instead of saddlepoint evaporation), and this technique is
only usable for trap depths less that about 1~mK.  Creation of deeper
traps requires higher frequency and higher power rf fields.  These
heat the current apparatus too much.

In an ideal apparatus the rf evaporation would be initiated at much
higher temperatures, phase space compression would lead to higher
sample densities, and the bulk of the cooling and compression would be
accomplished more quickly and more efficiently than in the present
experiments.  The sample could then be cooled at nearly constant phase
space density to cross the BEC phase transition line at the desired
density (one may desire smaller densities with correspondingly longer
lifetimes).  In the experiments described here more phase space
compression was accomplished at lower temperatures than is optimal
because of the heating effects of the rf power used for evaporation.
Improvements to the rf parts of the apparatus should allow larger
condensates to be created.

To test these ideas about efficiency we cooled to BEC in two traps,
A and B described above.  (These labels refer to the field
configuration at low temperature.  Identical trap shapes are used for
initial trap loading ($T=40$~mK, $n_0=2\times10^{13}~{\rm cm^{-3}}$)
and the first stage of saddlepoint evaporative cooling.  The trap
shape is converted into A or B during the last stage of the
saddlepoint evaporation.)  The evaporation occurred along different
paths for the two traps, but at the end the phase space density was
nearly identical; the condensate population was about $10^9$ atoms for
both traps.  One expects path B to have lower overall evaporation
efficiency (and thus fewer condensate atoms) because more of the phase
space compression is accomplished at lower temperatures (where the
evaporation efficiency is lower because the ratio of good to bad
collisions is less favorable).  On the other hand, if rf evaporation
is more efficient than saddlepoint evaporation, then one expects path
B to result in better evaporation efficiency, and thus more condensate
atoms, since more of the phase space compression is accomplished with
the more efficient rf evaporation.  Since the two paths result in the
same final phase space density, we conclude that the two effects
cancel.  This necessitates that rf evaporation be more efficient than
saddlepoint evaporation.  It would be advantageous to use rf
evaporation for more of the cooling.

\section{Spectroscopic Study of the Degenerate Gas}

When the sample is cooled into the quantum degenerate regime we
observe signatures of the condensate in the \oneStwoS\ spectrum of the
gas.  In this section we discuss two of these signatures from which we
extract the chemical potential and temperature.  Knowledge of these
thermodynamic parameters allows us to estimate several properties of
the system.

\begin{figure}[tb]
\centering \epsfxsize=5in \epsfbox{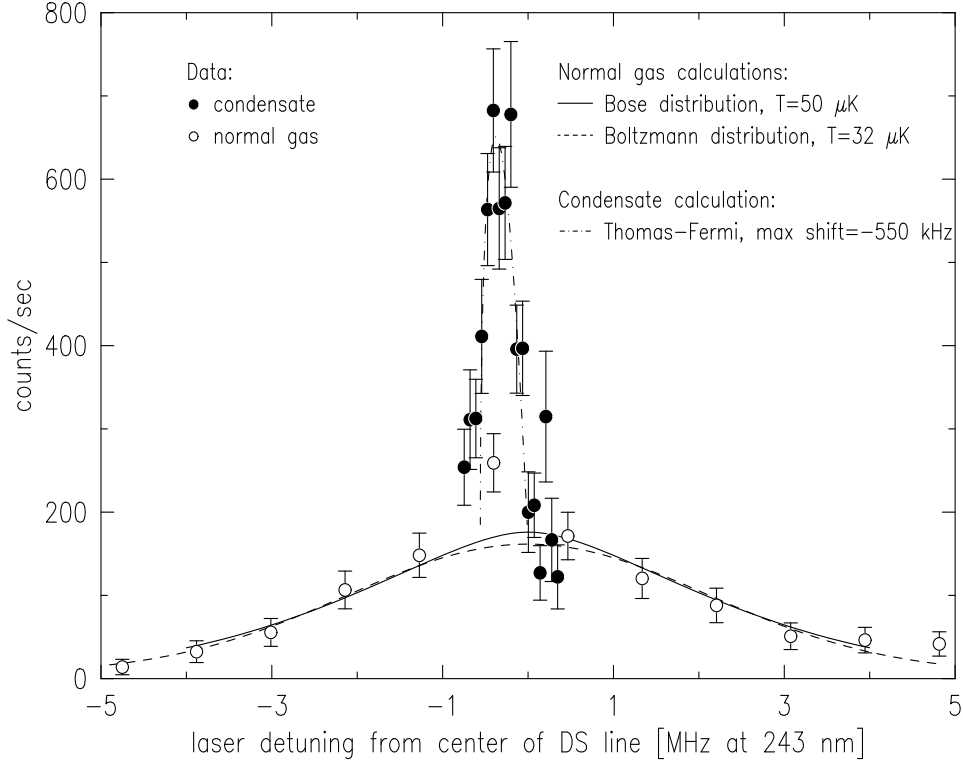}
\caption[Doppler-sensitive spectrum of degenerate
gas]{Doppler-sensitive spectrum of degenerate gas.  Zero detuning is
taken at center of the recoil-shifted spectrum, which is detuned
6.70~MHz blue of the Doppler-free resonance.  open circles---normal
gas, filled circles---spectrum focusing on condensate.  The dashed
line is a fit to the normal fraction data which assumes a
Maxwell-Boltzmann velocity distribution.  The solid line is a
calculation which assumes a Bose-Einstein distribution.  The
dot-dashed line is the condensate spectrum expected for a Thomas-Fermi
wavefunction in a harmonic trap, when the dominant spectral broadening
is the cold-collision frequency shift.  The temperature of the Bose
calculation was chosen to fit the observed spectrum.  The condensate
spectrum is a composite of data from the first sweep of each of three
loadings of the trap; the spectral sweep took 0.94~s.  The normal
spectrum is a composite of data from the first three sweeps of each of
two loadings of the trap; the spectral sweep took 0.94~s.  This data
is for trap shape B.}
\label{DS.degenerate.fig}
\end{figure}

Bose condensation involves the macroscopic occupation of the lowest
energy quantum state of the system.  In a trap this state is
concentrated at the minimum of the trapping potential, and it has very
small kinetic energy, given approximately by the Heisenberg
momentum-position uncertainty principle.  In the Doppler-sensitive
spectrum, which maps the momentum distribution, one would expect an
intense line at zero detuning from the recoil shift $\Delta\nu_R$,
rising above the background spectrum \cite{san93}.  Figure
\ref{DS.degenerate.fig} shows the DS spectrum of the normal gas and
the condensate together.

\begin{figure}[tb]
\centering \epsfxsize=5in \epsfbox{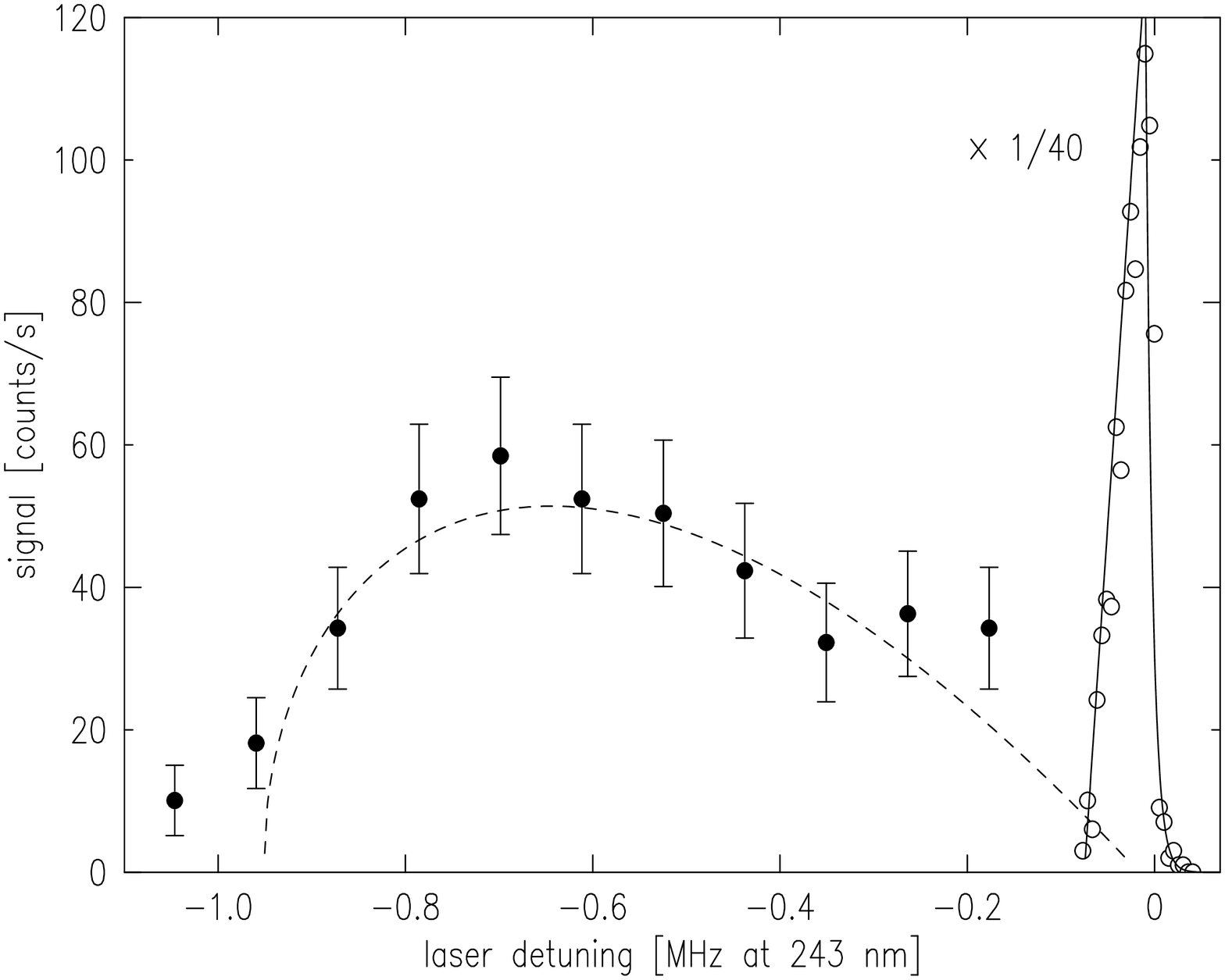}
\caption[Doppler-free spectrum of degenerate gas]{Doppler-free
spectrum of degenerate gas.  The intense, narrow feature on the right
is due to the normal gas, and is shrunk vertically by a factor of 40
to fit in the plot.  The wide, low, red-shifted feature is the
condensate.  The lineshape is that expected for a Thomas-Fermi
wavefunction in a harmonic trap.  The spectrum is a composite of the
first two spectral sweeps from each of 10 loadings of the trap; the
sweep took 0.67~s.  The normal gas spectrum is a single sweep from a
single loading of the trap.  This data is for trap shape A$^\prime$. }
\label{DF.condensate.spec.fig}
\end{figure}

Instead of appearing as a very narrow, intense peak centered at zero
detuning, however, the line is  red-shifted and broadened  by the very
high and inhomogeneous density of the condensate.  This same effect is
observed in the DF line, displayed in figure \ref{DF.condensate.spec.fig}.  
The condensate spectrum has a very different shape and shift than the
normal gas spectrum, implying very high densities and a density
distribution that is different from the normal gas.

The spectroscopic technique of probing the condensate differs
significantly from the spatial imaging techniques used by other
groups, providing somewhat complementary information.  The probe is
not limited to length scales larger than the spatial resolution of the
imaging optics, and so detection of very small features, such as
condensate ``droplets'' \cite{cok98} would be possible.  Also, we are
able to measure the momentum distribution of the normal gas {\em in
situ}.  Finally, the coherence length of the condensate can, in
principle, be obtained directly from the broadening of the spectrum,
as recently demonstrated in Na by Stenger \etal\ \cite{sic99}.
Because of the large condensate densities, we are only able to put a
very small lower bound on the coherence length for our H condensates.

\subsection{Peak Condensate Density}
\label{peak.cond.dens.sec}

The DF spectrum of the condensate, shown in figure
\ref{DF.condensate.spec.fig}, can be understood using a local density
approximation; atoms excited from different regions of the condensate
exhibit correspondingly different frequency shifts.  The signal size
$S$ at a given detuning $\Delta$ is proportional to the number of
condensate atoms on a surface of constant density
$n_{cond}=2\Delta/\chi_c$, where $\chi_c$ is the (two photon sum)
frequency shift per unit density for excitation out of the condensate.
(The relation of $\chi_c$ to $\chi_m$, the measured \cite{kfw98}
frequency shift per unit density for excitation out of the thermal gas
quoted in equation \ref{chi.m.quote.eqn}, will be addressed in section
\ref{question.of.chi.sec}.  Although there is significant ambiguity,
it is clear that either $\chi_c=\chi_m$ or $\chi_c=\chi_m/2$.)  The
spectrum calculated using this local density approximation in a
harmonic potential is parameterized only by the peak shift, $\Delta_p$,
and the overall scaling factor $S_0$.  The normalized spectrum is
\begin{equation}
S(\Delta)= \left\{ \begin{array}{lll}
 S_0\frac{-15\Delta}{4(-\Delta_p^{5/2})}\sqrt{\Delta-\Delta_p} & ;&
 \Delta_p\leq\Delta\leq0 \\ 
 0 & ;& {\rm otherwise}
\end{array}
\right. .
\label{TF.spec.eqn}
\end{equation}
(The negative signs are appropriate for red density shifts:
$\Delta_p<0$.)  A fit of this function to the measured spectrum is the
dashed line in figure \ref{DF.condensate.spec.fig}.  The peak shift is
$920\pm 70$~kHz, which indicates a peak condensate density
$n_p=2\Delta_p/\chi_c=4.8\pm0.4\times10^{15}~{\rm cm^{-3}}$ (assuming
$\chi_c=\chi_m$; the uncertainty reflects the statistical fitting
uncertainty, and ignores the 20\% uncertainty in $\chi_m$).  If we
assume $\chi_c=\chi_m/2$ the density is twice as big.  The quality of
the fit indicates the utility of the Thomas-Fermi wavefunction
approximation.

A quantum mechanical approach may also be used to understand this
spectrum, as developed in the thesis of Killian \cite{killian99}.
Before excitation, an atom is in the $1S$ electronic state, and its
motional quantum state is the condensate wavefunction.  After
excitation the atom is in the $2S$ electronic state; its motional
state has been projected onto the motional eigenstates of the modified
potential it experiences.  This new potential, a sum of the trap
potential and the mean-field interaction energy for a $2S$ atom in a
$1S$ cloud, is very different because the new mean-field energy has a
different sign and is much stronger.  The parabolic condensate density
profile, with density much larger than the background thermal cloud
density, gives rise to an attractive parabolic potential for the $2S$
atoms.  The motional quantum states the $2S$ atom finds available are
simple harmonic oscillator states.  The excitation spectrum is the
probability of exciting atoms from the condensate wavefunction to each
of these $2S$ motional states.  This probability is the wavefunction
overlap, which is straightforward to calculate.  Calculations by
Killian show very good agreement between the quantum and local density
approximation treatments of the spectrum.

The peak condensate density indicates the number of atoms in the
condensate, given a knowledge of the trap shape. (This data is for
trap shape A$^\prime$).  Using equation \ref{Ncond.eqn} we obtain
$N_c=1.2\pm0.2 \times10^9$.  This is larger by a factor of 50 than the
biggest condensate created so far in other labs \cite{sac98}.  Using
equation \ref{Ndot.condensate.eqn}, the characteristic dipolar decay
rate is $\Gamma_{2,c}=\dot{N}_{2,c}/N_c=1.7~{\rm s}^{-1}$.  The
chemical potential for this peak density is $\mu=n_pU_0=1.9~\mu$K.
The extent of the Thomas-Fermi wavefunction is determined from
equation \ref{tf.cond.profile.eqn} by setting the density to zero at
the edge of the condensate: $n_{cond} U_0=\mu-V_{IP}(\rho,z)=0$.  We
obtain
\begin{equation}
\rho_{max}=\frac{1}{\alpha}\sqrt{\mu^2+2\mu\theta}=7.3~\mu{\rm m}
\end{equation}
and
\begin{equation}
z_{max}=\sqrt{\frac{\mu}{\beta}}=2.8~{\rm mm}.
\end{equation}

For trap B, similar data has been obtained.  The peak shift of
$\Delta_p=620\pm 20$~kHz corresponds to a peak density
$3.26\pm0.10\times10^{15}~{\rm cm^{-3}}$ (assuming $\chi_c=\chi_m$;
the uncertainty reflects the statistical fitting uncertainty, and
ignores the 20\% uncertainty in $\chi_m$).  In this more open trap,
the condensate population is $N_c=1.19\pm0.10\times10^9$, the chemical
potential is $\mu/\kb=1.28\pm0.04~\mu$K, the diameter is
$2\rho_{max}=20~\mu$m, and the length is $2z_{max}=4.5$~mm.  If we
assume $\chi_c=\chi_m/2$, then the peak density is twice as high, and
the derived condensate parameters are correspondingly different.
These results are summarized in table \ref{trap.shape.summary.tab}.

\begin{table}[tb]
\centering\begin{tabular}{|c||c|c|c|c|}
\hline
parameter & \multicolumn{2}{c|}{trap A$^\prime$} & 
	\multicolumn{2}{c|}{trap B} \\ \hline
$\alpha/\kb$ (mK/cm) & \multicolumn{2}{c|}{15.9} & \multicolumn{2}{c|}{9.5} \\
$\beta/\kb$ ($\mu$K/cm$^2$) & \multicolumn{2}{c|}{25} & 
	\multicolumn{2}{c|}{25} \\
$\theta/\kb$ ($\mu$K) & \multicolumn{2}{c|}{$35\pm 2$} & 
	\multicolumn{2}{c|}{$34\pm 2$} \\
$\ethrc/\kb$ ($\mu$K) & \multicolumn{2}{c|}{280} & \multicolumn{2}{c|}{220} \\
$\ethr/\kb$  ($\mu$K) & \multicolumn{2}{c|}{272} & \multicolumn{2}{c|}{214} \\
$T_c$ ($\mu$K) &  \multicolumn{2}{c|}{$\sim 65$} & \multicolumn{2}{c|}{$\sim 50$} \\
$\Delta_p$ (kHz) & \multicolumn{2}{c|}{$920\pm 70$} & 
	\multicolumn{2}{c|}{$620\pm 20$} \\ 
\cline{2-5}
 & $\chi_c=\chi_m$ & $\chi_c=\chi_m/2$ & 
	$\chi_c=\chi_m$ & $\chi_c=\chi_m/2$ \\ 
$n_p$ ($\times 10^{15}$~cm$^{-3}$) & $4.8\pm 0.4\pm1$  & $9.7\pm0.7\pm2$ & 
	$3.3\pm0.1 \pm0.7$ & $6.5\pm0.2\pm1.3$  \\
$\mu/\kb$ ($\mu$K) & 1.9 & 3.8 & 1.3 & 2.6 \\
$N_c$ ($\times 10^9$) & $1.2\pm0.2(^{+0.7}_{-0.5})$ &
	$6.6\pm1.3(^{+4}_{-3})$ &
	$1.2\pm0.1(^{+0.7}_{-0.5})$ &
	$6.7\pm0.5(^{+4}_{-3})$ \\
$\Gamma_{2,c}$ (s$^{-1}$) & 1.7 & 3.3 & 1.1 & 2.2 \\
$2\rho_{max}$ ($\mu$m) & 15 & 21 & 20 & 28 \\
$2z_{max}$  (mm) & 5.5 & 7.8 & 4.5 & 6.4 \\
\hline
\end{tabular}
\caption[summary of trap and condensate properties]{ Summary of
parameters describing the two trap shapes used for achieving BEC, and
summary of the properties of the condensates.  The parameters
$\alpha$, $\beta$, and $\theta$ describe the Ioffe-Pritchard
potential; $\alpha$ and $\beta$ are calculated, and $\theta$ is
measured.  The trap depth at which BEC is first observed is $\ethrc$,
and the trap depth used for the observations listed here is $\ethr$.
The approximate temperature in the presence of the biggest condensates
is $T_c$.  The peak cold-collision frequency shift in the condensate
is $\Delta_p$.  The remainder of the table is divided to show the
implications of assuming $\chi_c=\chi_m$ or $\chi_c=\chi_m/2$ (recall
the sum frequency shift is $\chi_m=-3.8\pm0.8\times10^{-10}~{\rm
Hz~cm^3}$).  The peak condensate density is $n_p$, and the chemical
potential is $\mu$.  The number of condensate atoms is $N_c$.  The
characteristic condensate dipolar decay rate is
$\Gamma_{2,c}=\dot{N}_{2,c}/N_c$.  Finally, the length and diameter of
the condensates are given.  The uncertainties are divided into a
component which depends on the present experiment (first number), and
a component reflecting the 20\% uncertainty on $\chi_m$ (second
number).}
\label{trap.shape.summary.tab}
\end{table}

\subsection{Sample Temperature}
\label{degenerate.thermometry.sec}

The temperature and chemical potential of the trapped gas allows a
full thermodynamical description of the system, assuming a near
equilibrium situation.  The chemical potential was obtained above.
The temperature is extracted here using two methods, both of which
suffer from uncomfortably large uncertainties.  Nevertheless, in this
section we characterize the temperature with uncertainty on the 20\%
level.

The primary thermometry tool is the DS spectrum of the normal gas.  As
described in section \ref{1s2s.spec.theory.sec} above, the spectrum is
a map of the velocity distribution of the sample.  Far from quantum
degeneracy this (classical) velocity distribution is Gaussian,
and the normalized spectrum is
\begin{equation}
S(\Delta)=\frac{S_0}{\sigma \sqrt{2\pi}} 
\exp\left(-\frac{\Delta^2}{2\sigma^2}\right)
\end{equation}
where $\sigma=\sqrt{\kb T/m\lambda^2}=373.6 \sqrt{T/1~\mu{\rm K}}$~kHz
is the linewidth parameter and the detuning $\Delta$ is measured in
Hz.  We have not measured the DS spectrum of the gas far from quantum
degeneracy because the low sample density reduces the signal rate
below the background noise\footnote{The signal rate at zero detuning
is $S(0)\propto n_0/\sqrt{T}$, whereas the phase space density is
$D\propto n_0/T^{3/2}$.  The signal rate $S\propto DT$ falls less
rapidly than the phase space density as the temperature increases,
indicating the potential feasibility of obtaining DS spectra of
samples far from quantum degeneracy.}.  There is sufficient signal
near the degenerate regime, however.  For a degenerate Bose gas at a
given temperature, the velocity distribution is significantly narrower
than its classical counterpart because the particles are distributed
toward lower energies.  In a homogeneous potential the DS spectrum of
an ideal gas is cusp-shaped:
\begin{equation}
S(\Delta)\propto \log\left[1-\exp\left(
	\frac{\mu-\frac{\lambda^2\Delta^2m}{2}}{\kb T}\right)\right].
\end{equation}
In a trap, however, only the atoms at the very bottom of the trap are
in the quantum degenerate regime.  Atoms in regions of potential
energy $\varepsilon\geq\kb T$ experience a reduced phase space
density, parameterized by an effective chemical potential
$\mu^\prime=(\mu-\varepsilon) \leq -1\kb T$; their velocity
distribution is nearly classical.  The measured spectrum is an average
over the various velocity distributions in the gas.  Appendix
\ref{be.velocity.distrib.app} presents a detailed treatment of
Doppler-sensitive optical excitation in a trap, showing that if the
laser beam is small, then atoms in regions of high phase space density
are preferentially excited and the spectrum reflects the Bose velocity
distribution more dramatically. The shot noise on our measurements is
too big to allow a precise comparison between the shapes of the
classical and quantum spectra, but evidence of the effect is noted in
the spectrum below.

Although it is interesting to probe the Bose-Einstein velocity
distribution and compare it to the Maxwell-Boltzmann, the disadvantage
presented by the effect is the dependence of the spectrum on the
effective chemical potential, which leads to measurement
uncertainties.  The laser beam geometry and alignment must enter the
interpretation of the spectrum, and these factors are difficult to
measure directly.  The alignment of the beam with the sample is
verified by translating the trap radially relative to the laser beam
while exciting the condensate.  The condensate is used as a precise
alignment marker in the center of the trap.  With this technique we
can center the trap radially over the laser beam to within about a
half a beam radius (20~$\mu$m), small compared to the thermal radius
of the normal gas (100~$\mu$m).  Systematic axial adjustment is not
possible in the current apparatus.  The uncertainty in the axial
position of the beam focus is about 2~cm, about half the length of the
normal gas cloud.

\begin{figure}[tb]
\centering\epsfxsize=5in\epsfbox{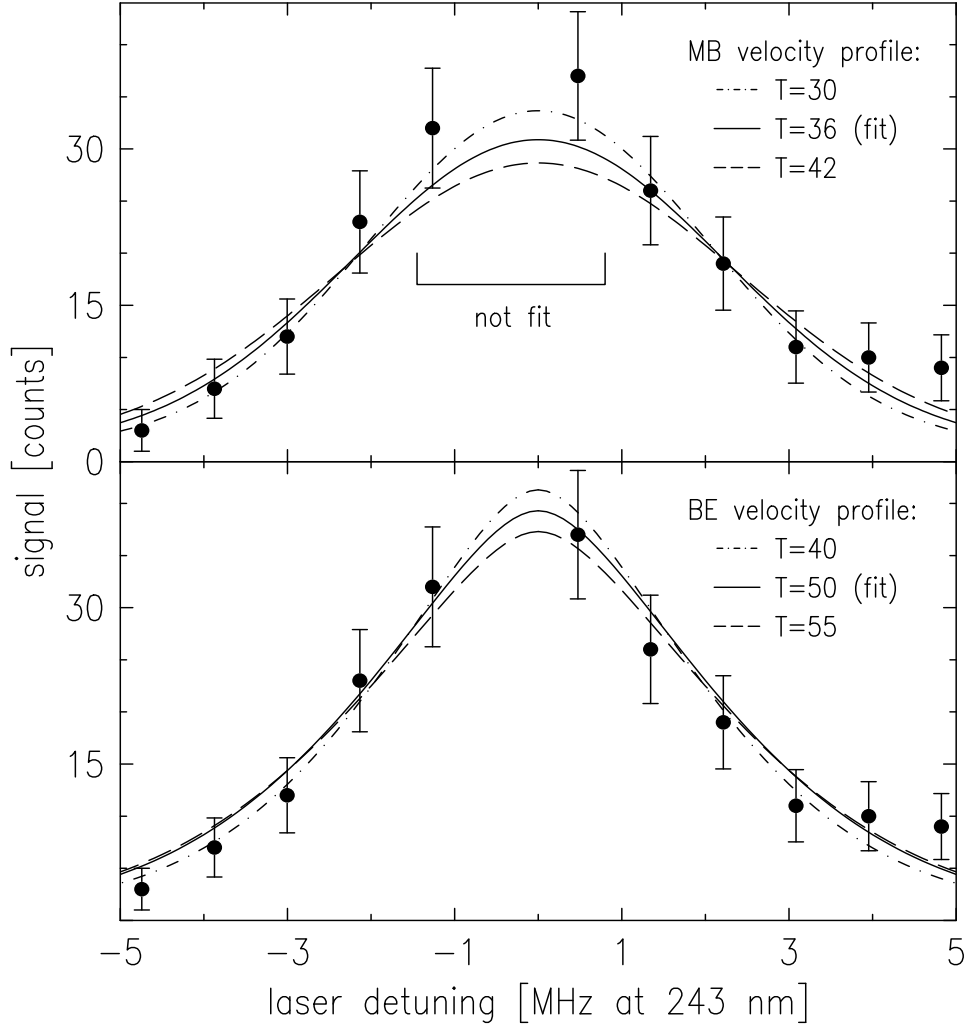}
\caption[extracting temperature from Doppler-sensitive spectrum of
normal gas]{ Doppler-sensitive spectrum of degenerate gas.  The
condensate signal (point at -0.5~MHz detuning) is off scale.  In the
upper panel the wings of the line (all points except center three) are
fit to a Gaussian, the expected lineshape for a Maxwell-Boltzmann (MB)
velocity profile.  The fit temperature is $T=36\pm6~\mu$K\@.  In the
lower panel the same data is compared to calculations of the spectrum
which assume a Bose-Einstein (BE) velocity distribution for a
degenerate gas in trap B.  Spectra calculated for three temperatures
are displayed.  We conclude that the sample temperature is
$T=50\pm10~\mu$K, corresponding to $\eta=4.3\pm0.9$.  The BE velocity
distribution fits the low velocity portion of the data better than the
MB distribution, indicating the presence of quantum effects in the
momentum distribution.  The data is compiled from the first three
sweeps of two samples; each sweep required 0.94~s.}
\label{ds.normal.tfit.fig}
\end{figure}

To demonstrate the technique we study here the DS spectrum of a sample
in trap B, in the quantum degenerate regime.  Plotted in figure
\ref{ds.normal.tfit.fig} is the DS spectrum, obtained immediately
after forced rf evaporation is completed; the trap depth is
$\ethr=214~\mu$K\@.
The first step in the analysis is to fit the wings of the spectrum with a Gaussian, an
analytic form that is simple to fit.  This fit is shown in the upper
panel along with the spectra expected for temperatures higher and
lower than the fit result by one standard deviation.  Next, spectra
are calculated assuming a truncated Bose-Einstein velocity
distribution, as described in appendix \ref{be.velocity.distrib.app}.
These calculations take into account the trap shape, trap depth, laser
beam waist, and alignment of laser beam relative to the center of the
trap (we choose ideal alignment here: beam focus at center of trap,
beam aligned with the $z$ axis).  The calculations are done for a
range of temperatures.  The calculated spectra are fit in the wings by
a Gaussian, and the spectrum with fit temperature closest to the fit
temperature of the data is chosen.  For the data in figure
\ref{ds.normal.tfit.fig}, the fit temperature for the data is
$36\pm6~\mu$K\@.  The calculated Bose spectrum with temperature
$T=50~\mu$K also had a fit temperature of 36~$\mu$K\@ (see below).
This spectrum, and those for $T=40~\mu$K and $55~\mu$K are plotted in
the lower panel of figure \ref{ds.normal.tfit.fig}.  We conclude that
the sample temperature is $T=50\pm10~\mu$K\@.

When a Gaussian is fit to a Bose velocity distribution, calculated for
temperature $T_{BE}$, the fitted temperature $T_{MB}$ is significantly
lower than $T_{BE}$ because of two effects.  First, the Bose
distribution concentrates population at the low energies corresponding
to the low velocity area near the center of the spectrum.  This effect
is tuned by changing the chemical potential from the quantum
degenerate regime ($|\mu|\ll\kb T$) to the classical regime
($|\mu|\gg\kb T$, $\mu<0$).  The fit temperature is also smaller than
$T_{BE}$ because the true energy distribution is truncated, and the
Gaussian profile disregards truncation.  This effect is tuned by
changing the trap depth parameter $\eta\equiv\ethr/\kb T_{BE}$ (in
practice $\ethr$ is fixed and $T_{BE}$ is varied).  For large $\eta$
truncation is unimportant, while for the small $\eta$ values used here
truncation effects cause significant deviations.  The calculation
procedure has been checked in these limits.

A systematic error can arise if the laser beam focus is not aligned in
the trap as expected.  The beam would be probing a region with
effective chemical potential different from expected.  Three beam
parameters affect the alignment: the beam waist, the radial
position of the beam, and the axial position of the beam.  We know the
waist radius to within 30\% because of the aperture constraints
encountered while transporting the beam into the cryostat and back
out.  The radial position of the beam is measured by translating the
trap, and is verified by the presence of the condensate spike in the
center of the spectrum.  The axial position of the beam is not as
important because the divergence length is on the order of the length
of the sample.  We conclude that uncertainties in the beam parameters
should not cause a major systematic effect compared to the shot noise
of the current measurements.

The other indication of sample temperature comes from the phase space
diagram, figure \ref{evap.trajectory.fig}.  For trap B, the phase
boundary appears at a ratio of trap depth to temperature of
$\eta\simeq 6.5$.  This would indicate that the temperature of the
sample in figure \ref{ds.normal.tfit.fig} should be 33~$\mu$K.  A
systematic error arises, however, in the extraction of the peak
density of the normal gas (plotted in figure
\ref{evap.trajectory.fig}) from the DF spectra.  The lineshape of the
spectrum changes significantly near the onset of quantum degeneracy,
as detailed in section \ref{asymetric.line.sec}.  This systematic
error could shift the apparent density higher by up to roughly 30\%
(measured density higher than actual density), translating to a
decrease in the indicated temperature by 20\% (measured temperature
lower than actual temperature).  This adjustment shifts the expected
sample temperature to $40~\mu$K, within the uncertainty of the DS
measurement.

We note that the DS spectrum does {\em not} constitute an absolute
measure of the sample temperature in the regimes of quantum degeneracy
and/or shallow traps (low $\eta$).  On the contrary, the details of
the trap shape and laser beam geometry are folded into the spectrum.
In spite of these difficulties, reliable temperatures can be
extracted.  In addition, we observe in figure \ref{ds.normal.tfit.fig}
hints of quantum perturbation of the momentum distribution of the
normal gas, a subject which could be studied with improved signal rate.

Our temperature derivation procedure could be complicated by a
non-uniform temperature in the sample.  Loss of atoms from the trap
occurs mostly in the condensate when the condensate density is high.
As explained in section \ref{cond.dipolar.heating.sec}, the effective
heating rate of the system is thus higher in the center 10\% of the
length of the trap where the condensate lives.  Near the ends of the
thermal gas the heat load is lower, and conceivably the temperature
could be lower.

In order for a temperature gradient to exist, the characteristic mean
free path must be less than the length of the sample.  At the highest
densities achieved, the thermal gas has peak density
$n_0=2\times10^{14}~{\rm cm^{-3}}$ corresponding to a mean free path
$l=1/n_0\sigma\simeq4$~cm.  This is about the characteristic length of
the sample.  While significant thermal gradients are not expected,
there is a danger of such effects.

\subsection{Condensate Fraction}

\begin{figure}[tb]
\centering\epsfxsize=5in\epsfbox{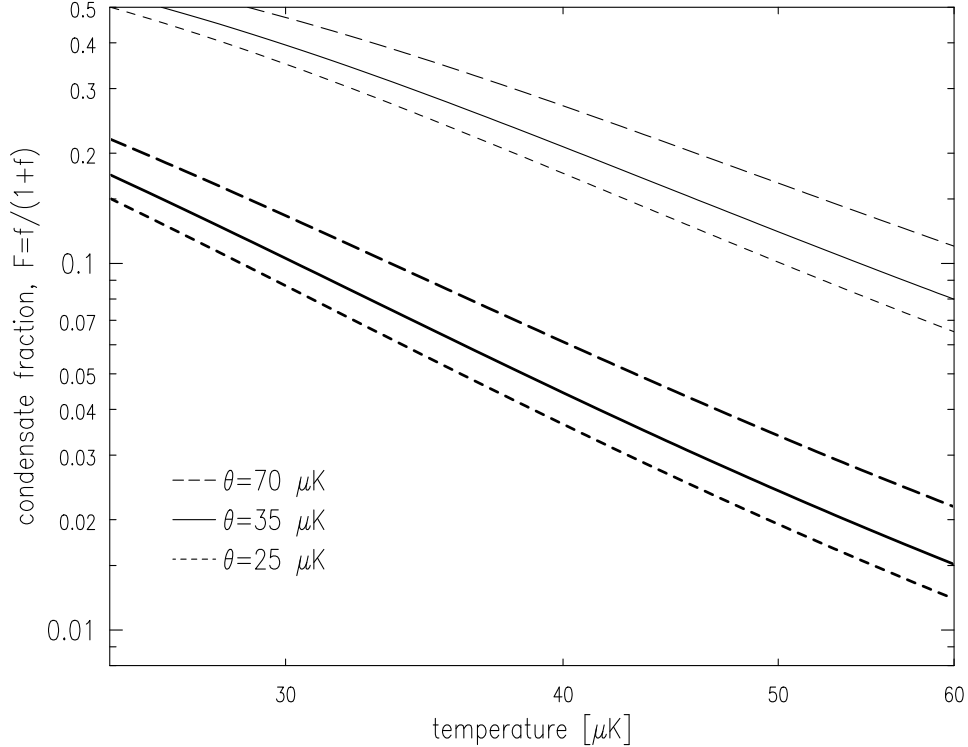}
\caption[calculated condensate fraction as a function of temperature]{
Condensate fraction $F=f/(1+f)$ as a function of sample temperature,
plotted for various bias energies.  The heavy lines correspond to peak
condensate density $n_p=3.3\times10^{15}~{\rm cm^{-3}}$, and the light
lines correspond to a peak density twice as big.  The trap depth
corresponds to trap B, as do the solid lines ($\theta/\kb=35~\mu$K).
A temperature of $50~\mu$K indicates a condensate fraction around 2\%
(12\%) for $\chi_c=\chi_m$ ($\chi_c=\chi_m/2$).  The calculations,
based on equation \ref{condensate.fraction.from.density.eqn}, properly
treat the finite trap depth.}
\label{condensate.frac.trapB.fig}
\end{figure}

The fraction of atoms in the condensate can be obtained using two
methods.  First, given knowledge of the trap shape and sample
temperature, we compute the population of the normal gas using
equation \ref{thermal.population.eqn} (which assumes thermal
equilibrium).  We then compare to the condensate population computed
above using equation \ref{Ncond.eqn}.  The trap shape parameters
$\alpha$ and $\beta$ cancel in the comparison, and the only quantities
that enter are measured: the temperature, peak condensate density
$n_p$, and trap bias energy $\theta$.  The population ratio
$f=N_c/N_t$ is given by equation
\ref{condensate.fraction.from.density.eqn}.  We use $f$ to plot the
condensate fraction, $F=f/(1+f)$, in figure
\ref{condensate.frac.trapB.fig} for the trap depth, bias energy, and
range of temperatures applicable to trap B.
The condensate fraction for trap B corresponding to $T=50\pm10~\mu$K is
$F=f/(1+f)=2.4(^{+2}_{-1})(^{+1.4}_{-1})$\% (uncertainties arising
from $T$ and $\chi_m$ listed separately).  This is in agreement
with estimates by Hijmans \etal\ \cite{hks93}.

We have found the condensate fraction just after crossing the BEC
phase boundary; condensate fractions could be determined at other
points on the phase diagram, but the high loss from the condensate
depletes the system as lower temperatures are attained, reducing the
achievable fraction.  Also, we are considering equilibrium condensate
fractions.  As explained in section \ref{cond.time.evol.sec}, the
condensate is fed from the thermal cloud.  An extremely high
condensate fraction could be created by simply removing all the
thermal atoms by rf ejection, but the resulting condensate would decay
rather quickly.

The other method of finding the condensate fraction exploits the
spectrum more directly; the population of the normal gas is
proportional to the integrated signal of the normal component of the
DF (or DS) resonance.  This integral is compared to the integral of
the condensate portion of the spectrum, and adjustments are made to
account for the partial illumination of the normal gas by the laser
beam.  (The condensate diameter is significantly smaller than the beam
diameter, but the normal gas diameter is significantly larger.) The
fraction computed with this technique \cite{killian99} is
$2.2^{+1}_{-0.5}$\%.  The two methods give consistent results; an
increased signal rate would reduce the large temperature uncertainties
and make the measurements useful.

\section{Time Evolution of the Degenerate Gas}
\label{cond.time.evol.sec}

Because the condensate fraction is small, there are enough atoms in
the normal gas to continuously replenish the condensate for many
seconds.  As a first look at this feeding process we study the time
evolution of the condensate.

Simple estimates of time scales reveal that the condensate lives much
longer than one might expect in the absence of replenishment.  The
large condensate density leads to fast condensate decay.  The
characteristic time an atom spends in the condensate before undergoing
an inelastic collision is $\tau_{2,c}=N_c/\dot{N}_{2,c}$, less than a
second for the condensates described in table
\ref{trap.shape.summary.tab}.  However, the condensate is
observed to live for more than $15\tau_{2,c}$, indicating that feeding
of the condensate by the thermal cloud occurs quickly.  In this
section we examine the time evolution of the condensate.  We find that
a simple equilibrium model reproduces the data.

\begin{figure}[tb]
\centering\epsfxsize=3.7in \epsfbox{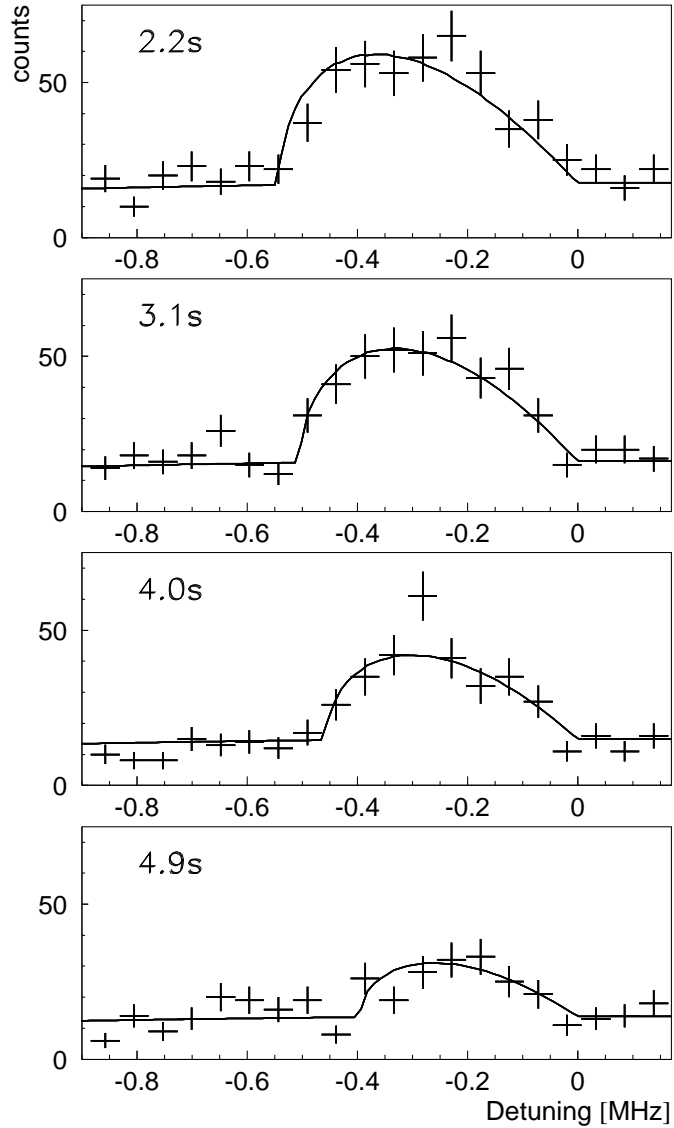}
\caption[time evolution of Doppler-sensitive spectrum of
condensate]{Time evolution of the Doppler-sensitive spectrum of the
condensate in trap B. Spectra are obtained sequentially after forced
rf evaporation is finished.  This data is compiled from five identical
loadings of the trap.  The solid line is a model which is fit to the
entire sequence of spectra, only four of which are shown here.
\cite{lorenz.pc}}
\label{cond.DS.spec.time.evol.fig}
\end{figure}

To measure the condensate dynamics many sequential condensate spectra
have been measured.  Figure \ref{cond.DS.spec.time.evol.fig} shows
part of a series of such spectra, obtained using DS absorption, for
trap B.
The solid curve is a fit to all the data in the series, with
adjustable parameters being the peak condensate density in each scan,
the overall amplitude scaling factors for the condensate and normal
gas, the temperature of the sample, and a time constant for the
temperature to change \cite{lorenz.pc}.  The peak condensate density
decreases only slowly in time, implying feeding of the condensate from
the normal gas.  For this feeding to occur, the normal gas must be
losing atoms, and thus density.  In order for the normal gas to remain
on the phase boundary, the temperature must therefore be decreasing.

\begin{figure}[tb]
\centering\epsfxsize=5in\epsfbox{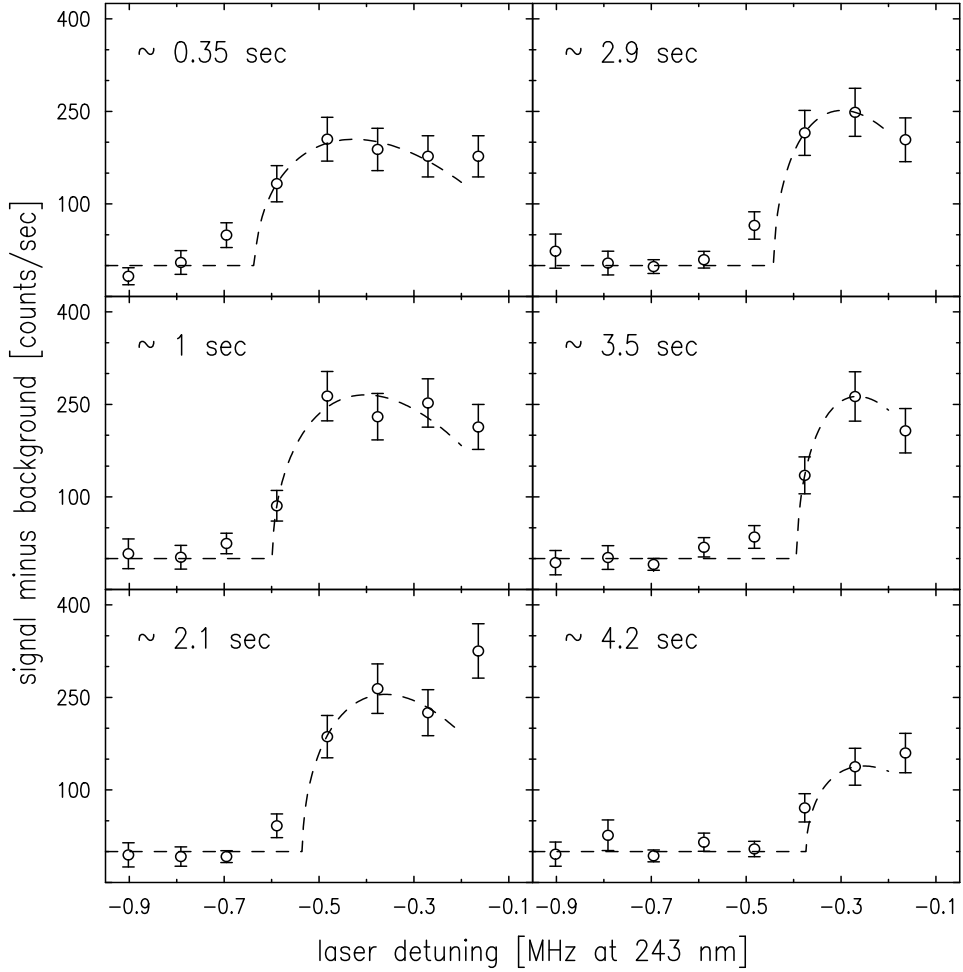}
\caption[time evolution of Doppler-free spectrum of condensate]{time
evolution of Doppler-free spectrum of condensate in trap B.  The noted
time is measured from the end of the forced evaporative cooling cycle
to the middle of the sweep.  Each spectrum is fit independently with a
two parameter model: the peak density $\Delta_p$ and an overall
scaling factor $S_0$; the background is estimated independently.  The
right side is not fit because of signal spillover from the very
intense DF normal line (not shown).  This data is a compilation of
sweeps from each of five identical loadings of the trap. }
\label{cond.DF.spec.time.evol.fig}
\end{figure}

The time evolution of the condensate can also be probed using the DF
spectrum, as shown in figure \ref{cond.DF.spec.time.evol.fig}.
Similar results are obtained.  Note the systematic misfit of points
just to the red of the peak shift, $\Delta_p$ in figures
\ref{cond.DF.spec.time.evol.fig} and \ref{DF.condensate.spec.fig}.
The spectra suggest that the density distribution of the condensate is
more complicated than that predicted by the Thomas-Fermi wavefunction
alone.  Unfortunately, the low signal rate precludes further study at
this time.

\begin{figure}[tb]
\centering\epsfxsize=4.5in\epsfbox{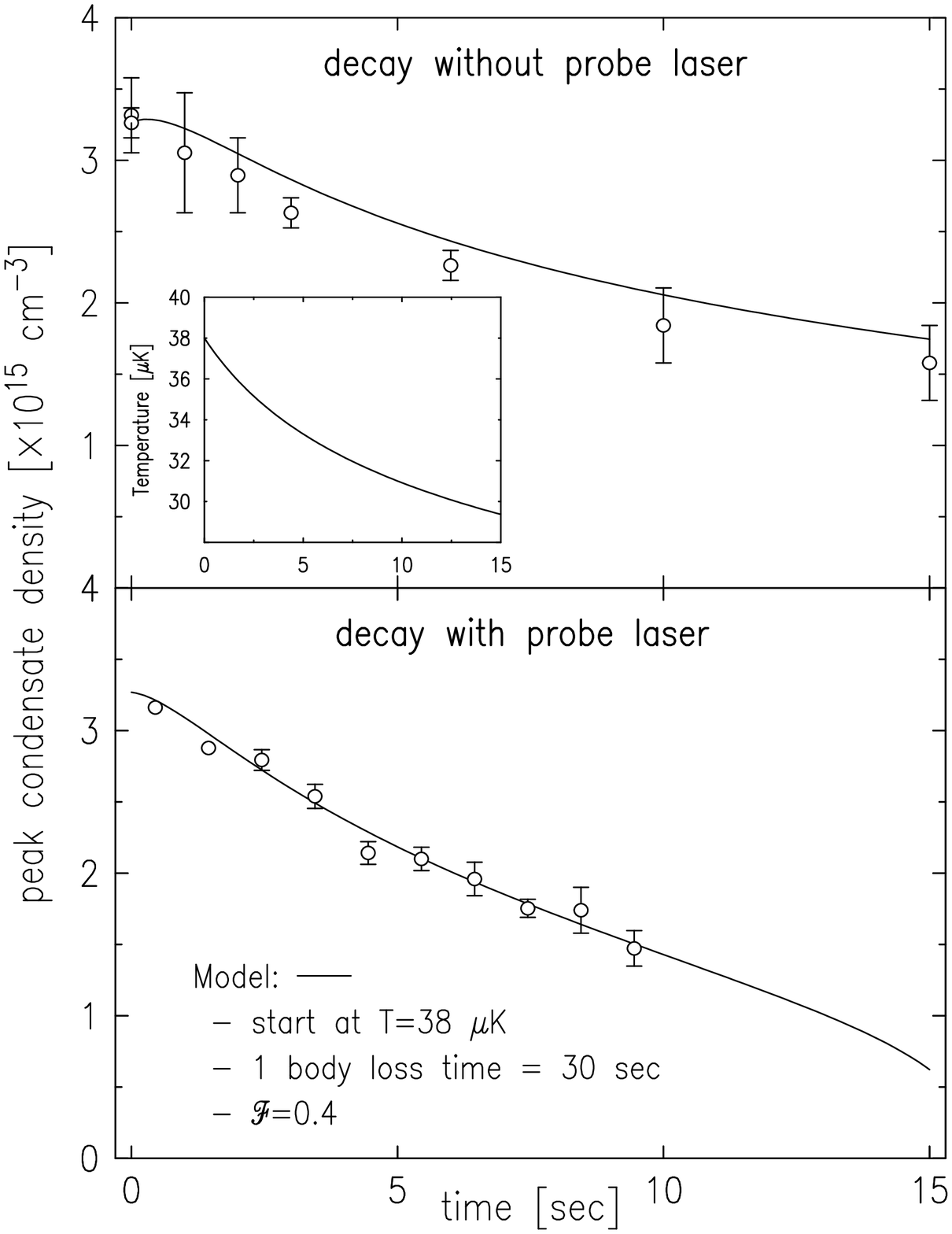}
\caption[time evolution of peak density of condensate]{Time evolution
of peak density of the condensate in trap B.  The lower panel shows
the peak condensate density extracted from a sequence of DS spectra,
some of which are shown in figure \ref{cond.DS.spec.time.evol.fig}.
The upper panel is the peak condensate measured after waiting various
times between the end of the forced evaporation cycle and the
beginning of the laser scans.  This data demonstrates the lifetime of
the condensate when the only loss mechanism is dipolar relaxation.
The solid lines are from a model of the heating and cooling processes
in the normal gas and condensate.  For the lower panel a one-body loss
rate was introduced.  The inset shows the calculated sample
temperature evolution in the absence of extraneous atom loss due to
the presence of the probe laser.}
\label{cond.time.evol.model.fig}
\end{figure}

Figure \ref{cond.time.evol.model.fig} shows the peak condensate
density as a function of time, obtained in two ways.
In the lower panel sequential spectra are taken of the same
condensate, and so losses induced by the laser beam cause the sample
to decay more quickly than it would without the extraneous one body
loss rate.  The upper panel reveals the latter situation.  Here the
sample is held for various intervals after forced rf evaporation is
finished, before laser probing is commenced.  The sequence of spectra
obtained is fit, and a peak condensate density at the beginning of the
laser scan is obtained.  This density is plotted in the figure.  We
see that the laser probe shortens the condensate life by only a small
factor.  We also observe that in the absence of losses induced by the
laser, the condensate density decays by only a factor of 2 over 15~s
(population decays by factor of $2^{5/2}\simeq6$).

The number of atoms lost from the condensate through dipolar decay
during this interval $d=15~$s can be easily estimated by integrating equation
\ref{Ndot.condensate.eqn}:
\begin{equation}
N_{c,lost}=\int_0^ddt\;\dot{N}_{2,c}
=\frac{16}{105}\:\frac{g}{2!}\varrho_0\theta U_0^{3/2} 
	\int_0^d dt \; n_p^{7/2}(t)
\label{Nc.lost.eqn}
\end{equation}
We estimate $N_{c,lost}=6(^{+0.7}_{-0.6})(^{+6}_{-3})\times10^9$
(uncertainties from present experiment and 20\% uncertainty in
$\chi_m$, respectively).  If we had used $\chi_c=\chi_m/2$ this loss
would have been $2^{7/2}\simeq11$ times larger, or
$N_{c,lost}=7(^{+0.8}_{-0.7})(^{+6}_{-4})\times10^{10}$.  For a
temperature of $50\pm10~\mu$K we estimate a normal gas population of
$N_t=5^{+3}_{-2}\times10^{10}$.  The spectra show that even after
losing $N_{c,lost}$ atoms the trap still contains a sizeable
condensate and normal gas.  It therefore appears unphysical for
$\chi_c=\chi_m/2$; however, more data should be collected to reduce
the large uncertainties in this measurement.

A computer model has been developed to study the time evolution of the
degenerate system.  As described in appendix
\ref{condensate.model.app}, this model incorporates dipolar decay
and evaporative cooling, and assumes thermal equilibrium and a
Thomas-Fermi condensate wavefunction.  The good agreement between the
model (solid line in figure \ref{cond.time.evol.model.fig}) and the
data indicates that the system is reasonably well understood.
An important result of the model is that the temperature of the sample
changes by 20-30\% during the 15~s condensate lifetime.  This behavior
is shown in the inset to figure \ref{cond.time.evol.model.fig}.

The low condensate fraction and long condensate lifetime in this
system suggest that the H condensate could be an excellent source for
a bright atom laser.  On the order of $10^8$ atoms/s could be
continuously coupled out of the condensate for 10-20 s per loading of
the trap (continuous loading schemes could be imagined to circumvent
this limitation).  The outcoupling could be accomplished by photon
recoil as atoms are excited to the $2S$ state \cite{cfk96} or by Raman
output coupling, as demonstrated in Na by Kozuma \etal\
\cite{kdh99}.  The beam thus formed should be diffraction
limited; the large condensate dimensions and short de Broglie
wavelengths achievable in this system should result in a very low
divergence beam.  The atoms in the beam could be in the metastable
$2S$ state, which could be useful because of the large energy
available for driving chemistry on a surface the atoms hits.

\section{Signature of Quantum Degeneracy in Spectrum of Normal Gas}
\label{asymetric.line.sec}

The direct condensate signatures presented so far are expected from
simple theories of BEC.  One does not expect the DF spectrum of the
normal gas to be much affected by the condensate.  Nevertheless,
a dramatic signature appears.

Simple arguments suggest that the for this experiment ($\mu\ll\kb T$)
the normal gas should be only very weakly perturbed by quantum
degeneracy.  The condensate volume is about a factor of $10^3$
smaller than the volume of the normal gas, so the high condensate
density should not affect the normal gas much.

\begin{figure}[tb]
\centering\epsfxsize=5in\epsfbox{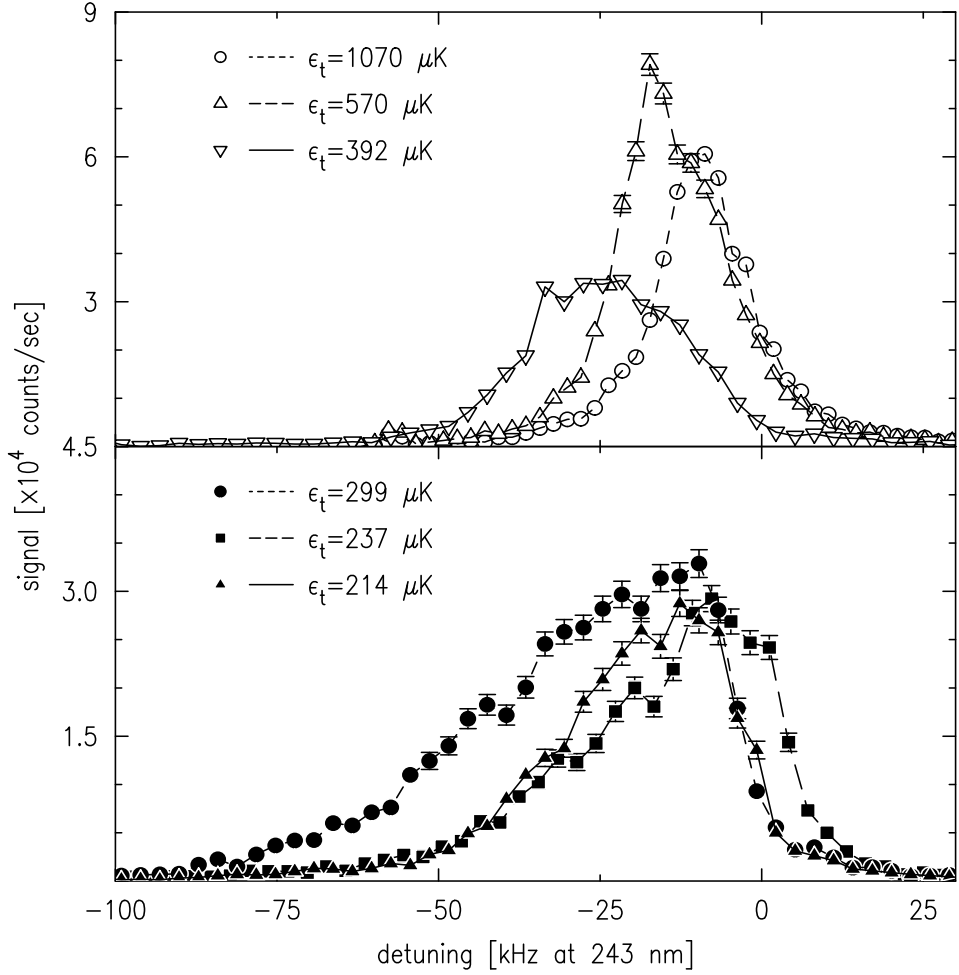}
\caption[Doppler-free spectra of normal gas in quantum degenerate
regime]{ Doppler-free spectra of normal gas showing asymmetry in
quantum degenerate regime.  The spectra in the upper panel correspond
to samples above quantum degeneracy; those in the lower panel are for
samples that are degenerate.  The trap shapes are identical, and the
trap depths are indicated.  Note the difference in vertical scale.
The lines connecting the points are a guide to the eye. The error bars
are statistical, and do not reflect systematic errors such as
fluctuations in beam alignment and laser power. }
\label{asymetric.spec.fig}
\end{figure}

In fact, measurements of the DF spectrum of the normal gas in the
degenerate regime reveal a surprise.  The lineshape is markedly
asymmetric below the transition, but largely symmetric above, as shown
in figure \ref{asymetric.spec.fig}.
Below the transition much of the spectral support is red-shifted to
large detunings.  These detunings, if simply interpreted as
cold-collision frequency shift with the measured shift-to-density
ratio $\chi_m$, correspond to sample densities a factor of two higher
than expected for a degenerate Bose gas in equilibrium at a
temperature which is reasonable for the trap.  Note that $\chi_m$ was
measured in samples far from quantum degeneracy.  If $\chi$ were
larger by a factor of two for the part of the cloud in the bottom of
the trap where the effective chemical potential is small, then the
asymmetric lineshape would make qualitative sense.  The atoms in the
highest density (red-shifted) part of the spectrum would tend to be
shifted even further (because their $\chi$ is larger) than their low
density counterparts.  However, no quantitative theory has been
formulated predicting that $\chi$ should vary for the normal gas as
quantum degeneracy is approached.

\section{Density Shift for Excitation out of the Condensate and the Normal Gas}
\label{question.of.chi.sec}

A thorough treatment of the relation between $\chi_c$ and $\chi_m$ has
been undertaken by Killian \cite{killian99}.  He
concludes that $\chi_c=\chi_m/2$ because exchange effects should be
present in the normal gas, but absent in a condensate.  Here we
discuss experimental evidence which suggests that perhaps
$\chi_c=\chi_m$.

\paragraph{Time Evolution}
As explained in section \ref{cond.time.evol.sec}, the condensate is
observed to persist for many seconds.  Integration, over the lifetime
of the condensate, of the rate of atom loss through dipolar decay in
the condensate is very sensitive to $\chi_c$ because the loss rate is
proportional to $n_p^{7/2}$; a factor of two difference in $\chi_c$
translates to a factor of 11 change in the loss rate.  If we take
$\chi_c=\chi_m/2$ then the number of atoms lost, $N_{c,lost}$ from
equation \ref{Nc.lost.eqn}, is as large as the total original
population of the trap!  The estimate here would be flawed if the trap
parameters ($\alpha,\beta,\theta$) applicable to the condensate were
somehow different from those for the thermal gas.  This is unlikely,
as noted in discussion of the trap shape in appendix
\ref{trap.shape.uncertainty.app}.

\paragraph{DS-DF Comparison}
Doppler-sensitive excitation results in excited atoms moving at the
recoil velocity $v_{recoil}\sim 3$~m/s.  These motional states are
very different from the low energy motional states inhabited during
Doppler-free excitation.  Furthermore, the high velocity of the
recoiling atoms implies that motional coherence between the ground and
excited states of the atom is very quickly lost during excitation, a
situation conceivably different for DF excitation.  Nevertheless, the
condensate spectra obtained using the two different techniques are
identical to within the (small) statistical uncertainty.

\paragraph{Spectroscopic Measurement of Condensate Fraction} 
The condensate fraction measured spectroscopically agrees well with
the fraction one computes from the population in the condensate,
assuming $\chi_c=\chi_m$.  If $\chi_c=\chi_m/2$ then the calculated
condensate fraction is $2^{5/2}\simeq 6$ times larger than indicated
by the spectroscopic technique.

\paragraph{}
The possibility should be explored that
$\chi=2\hbar(a_{2S-1S}-a_{1S-1S})/m$ (instead of
$4\hbar(a_{2S-1S}-a_{1S-1S})/m$) for excitation out of the thermal
gas.  Then it would be easy to understand why $\chi_c=\chi_m$.
Furthermore, the asymmetric spectra presented in section
\ref{asymetric.line.sec} could be understood if $\chi$ changed from
the first to the second value in the normal gas near quantum
degeneracy.  If this were true all ambiguity would be relieved.
Experiments designed to guarantee that the $1S$ and $2S$ atoms are
distinguishable could conceivably provide insight into this issue.
This possibility is addressed in section
\ref{distinguishable.normal.gas.sec}.

The Bose condensates reported in this chapter open possibilities for
many new experiments.  Some of these ideas are presented in the
following chapter.

%% file: conclusion.tex

\chapter{Conclusion}
\label{conclusions.chap}

\section{Summary and Significance of this Work}

We have created Bose-Einstein condensates of atomic hydrogen,
culminating a long research effort.  In the process we have confirmed
that, as explained by Surkov \etal\ \cite{sws96}, the reduced
dimensionality of the evaporation process was the crippling factor in
previous efforts to reach this goal in our lab \cite{dsy91} and in
Amsterdam \cite{pmw98}.  Our condensates, while huge compared to other
condensates, are only a small fraction of the total trap population, a
prediction by Hijmans \etal\ \cite{hks93} for spin-polarized hydrogen
in equilibrium.  We have this huge number of atoms in the trap because
of the cryogenic loading technique we use, which can be scaled to
arbitrarily large trap sizes (consistent with the scaling arguments in
section \ref{ultimate.condensate.pop.sec}).  This cryogenic loading
method has been generalized for arbitrary paramagnetic atoms and
molecules by Doyle and coworkers \cite{dfk95,kfk97,wck98,wcg98}, and
therefore it should be possible to create big condensates of arbitrary
bosons.  The high condensation flux we observe ($10^9$ atoms condensed
per second for 10~s) should be useful in creating a bright atom laser.
In addition, we have added another atom to the list of Bose-condensed
species.  In section \ref{hbec.props.sec} we explored many differences
between the hydrogen system and the alkali metal gases that have been
condensed so far.  Finally, we have developed and applied spectroscopy
as a new probe of condensates.

\section{Improvements to the Apparatus}

The experiments described in this thesis were frustrated by several
factors.  Changes in the apparatus could allow us to investigate
several new areas.

One of the most straightforward improvements is to reduce the stray
electric fields in the cell.  Large fields reduce the lifetime of the
$2S$ atomic state by Stark quenching; the lifetime in a static
electric field $E$ (neglecting the natural lifetime) is \cite{bes77p287}
\begin{equation}
\tau_{SQ}=\left(\frac{475~{\rm V/cm}}{E}\right)^2 \;1.6~{\rm ns}.
\end{equation}
In the current cell, with its non-conducting walls, static charges
reduce the $2S$ lifetime to 0.1-1~ms, but in previous experiments in a
metal cell \cite{cfk96} we have observed lifetimes on the order of the
natural lifetime, $\tau_0=120~$ms.  It should be possible to reduce
stray fields by more than an order of magnitude by coating the inside
of the plastic cell with a thin metallic film.  Longer $2S$ lifetimes
would allow us to address questions such as how interactions with the
condensate affect the $2S$ lifetime.  Also, low stray fields would
allow atoms ejected from the condensate in an atomic beam to travel
many centimeters toward the detector before decay (see section
\ref{atoms.laser.proposal.sec}).  Reduced stray fields would allow
optical excitation to remain coherent for many milliseconds, allowing
interesting studies of the excitation process.  The large stray fields
increase indirectly the background counts in our data: because of the
short lifetime, fewer atoms can be accumulated in the $2S$ level
before the detection pulse, and so more detection pulses must be used
to accumulate the same signal.  Each detection pulse increases the
number of background counts.  Reduction of stray fields would thus
increase our sensitivity.

Another straightforward improvement is to reduce the heat load on the
refrigerator caused by high power in the rf coils. The rf evaporation could
then begin at higher sample temperatures, and more efficient cooling
would undoubtedly result.  More atoms in the condensate would increase
the signal rates, and more detailed studies of the condensate would be
possible.  The heat load created by the rf coils appears to {\em not}
be a consequence of the fields themselves, but rather of the system
that conveys the power to the coils.  If the coils were thermally
anchored to the 4.2~K $^4$He bath the heating of the refrigerator should be
dramatically reduced.

Better knowledge of the trap shape and laser alignment could allow
more detailed studies of the influence of quantum degeneracy on the
momentum distribution of the sample, as developed in section
\ref{normal.gas.degeneracy.DS.effects.sec}.  As noted by Killian
\cite{killian99}, better knowledge of trap and laser geometry would
also allow more accurate measurements of $\chi$.  We could better understand
the sizes of the condensates we produce (recall how in section
\ref{cond.time.evol.sec} the 20\% error on $\chi_m$ translates into
nearly a factor of two uncertainty in the number of condensate atoms
lost during the lifetime of the condensate).  The evidence for
$\chi_c=\chi_m$ presented in section \ref{question.of.chi.sec} could then
be either bolstered or refuted, resulting in better understanding of
the system.

The experiments probing the condensate, described in chapter
\ref{results.chap}, suffer from low signal rate.  Straightforward
improvements to the apparatus should allow at least a factor of 10
increase in detection solid angle, thus allowing studies at higher
time resolution to be feasible.  Condensate formation and feeding
could be studied in more detail.  The detection solid angle could be
improved by moving the cold atoms into a short trap formed by small
currents in thin coils.  This small trap could be surrounded by
Lyman-$\alpha$ detectors.

\section{Suggestions for Experiments}

The success of the experiments described in this thesis has fertilized
thinking about many new experiments that could be done.  In this
section we discuss several of these ideas.

\subsection{Creating Bigger Bose Condensates}

Nearly any experiment is easier with a larger condensate, and so here
we discuss strategies for putting more atoms into a hydrogen
condensate.

First, more atoms could be initially loaded into the trap.  We do not
currently understand the loading process in detail.  Investigations
are underway to explore possibilities for loading more efficiently
from the discharge.  In addition to increasing the loading efficiency,
one could also simply increase the physical volume of the trap by
making it longer and of larger diameter.  One should be able to load a
factor of four more atoms by scaling up the apparatus using available
technology.

The overall phase space compression efficiency exponent for cooling
from loading conditions to BEC is $\gamma=-1.5$, which is small (see
section \ref{forced.evap.cooling.sec}).  Improvements to the cooling
efficiency could be made using two strategies.  The first is to
commence rf evaporation earlier in the cooling cycle.  The second is
to increase the ratio of good collisions to bad collisions.  It is
probable that the admixture of an alkali ``moderator'' atom ``M'' in
the sample could dramatically increase the effective collision rate
without significantly increasing the dipolar decay rate \cite{kgk99}.
Presumably the cross-section for H-M collisions would be much greater
than the anomalously low H-H cross-section.

Finally, improvements could be made to the cooling procedure that do
not require any changes to the apparatus.  Instead of using the rf
cooling paths shown in figure \ref{evap.trajectory.fig}, the sample
could first be compressed at constant trap depth, the depth being set
by the highest rf frequency usable in the apparatus.  Phase space
compression would be accomplished at higher temperature, and thus more
efficiently.  After this compression the sample could be cooled to
BEC at high density.

\subsection{Probing the Velocity Distribution of the Normal Component of a
Degenerate Gas}
\label{normal.gas.degeneracy.DS.effects.sec}

In section \ref{degenerate.thermometry.sec} we described the influence
of quantum degeneracy on the velocity distribution of the normal gas.
The expected spectrum is narrower and more cusp-shaped than for a
classical gas at the same temperature.  Here we consider an experiment
designed to compare these spectra.

In order to ensure that the two samples (classical and quantum) being
probed have the same temperature, we propose to use a {\em single}
sample, but probe two parts of it.  The inhomogeneous trapping
potential serves to reduce the sample density (and thus the local
chemical potential) in regions far from the deepest point in the trap.
Classical behavior can be observed in a gas containing a condensate
when the probe laser is translated a distance $\rho_p\gg\kb T/\alpha$
from the trap axis (for an IP trap with $T\geq\theta/\kb$).  The beam
diameter should be small compared to the translation distance, a
condition that is easy to fulfill in an open trap (small $\alpha$).

The signal rate is dramatically reduced by the small densities far
from the middle of the trap.  In order to obtain the same signal rates
as the present experiments (in which the high density central portion
of the sample is probed), the detection efficiency must be increased
by the ratio of the densities, $n(\rho=0)/n(\rho_p)$.  For trap B in
table \ref{trap.shape.summary.tab}, and for $\rho_p=3\kb
T/\alpha=160~\mu$m, the required improvement in detection efficiency
is about 
\begin{equation}
\frac{n(\rho=0)}{n(\rho_p)}\simeq
\frac{g_{3/2}(1)}{g_{3/2}(e^{-3})}
\frac{\Upsilon_{1/2}(\eta,0)}{\Upsilon_{1/2}(\eta-3,-3)}
\simeq 10^2 .
\end{equation}
Here we have used equation \ref{trunc.dens.func.eqn} and figure
\ref{upsilon.5.fig}.  A redesign of the apparatus for better optical
access is required.

\subsection{Creation of an Atom Laser}
\label{atoms.laser.proposal.sec}

Atoms excited to the $2S$ state by co-propagating photons receive a
momentum kick $p_r=2h/\lambda$ where $\lambda=243$~nm is the
excitation wavelength.  The atoms leave the excitation region with the
recoil velocity $v_r=3.28$~m/s added vectorially to their initial
velocity.  In the normal gas there is a wide distribution of
velocities, so only a tiny fraction are Doppler-shifted into resonance
for a given detuning.  In the condensate, however, the velocity
distribution is very narrow, and one might expect to interact
optically with all the condensate atoms at the same laser tuning.
Atoms excited out of the (coherent) condensate by a laser beam with a
long coherence time should constitute a coherent atomic beam.  The
divergence of this beam of atoms should be diffraction limited by the
size of the excitation region.  Such a scheme has been demonstrated by
Kozuma \etal\ \cite{kdh99} using Raman pulses.

Unfortunately, several complications arise.  First, the condensate
spectrum is broadened by the cold-collision frequency shifts arising
from the high and inhomogeneous condensate density.  Since these
shifts are much greater than the laser linewidth, the laser only
interacts with a small fraction of the condensate for a given
detuning.  (One can imagine efficient methods of broadening the laser
linewidth to match the condensate linewidth \cite{woh92}, resulting in
excitation of all the condensate atoms, but this broadening would
write decoherence onto the ejected atoms.)  Since the optical
radiation only excites atoms from regions of a given density
(corresponding to the detuning), the excitation region can be quite
small, and the divergence of the beam is enlarged.  For example, atoms
excited by a laser beam of linewidth 2~kHz that is tuned to the peak
condensate density will come from a region of diameter on the order of
$\omega_{exc}=1~\mu$m.  The resulting divergence angle is
$h/p_r\pi\omega_{exc}=4\times10^{-2}$ \cite{guenther90}.  The
situation is more complicated when the laser is detuned to interact
with atoms in regions of intermediate density.  Here the excitation
occurs on a shell; interesting diffraction effects should occur.

As the $2S$ atoms propagate with recoil velocity $v_r$, the potential
surface they experience is strongly perturbed by the strong attractive
mean field potential created by the high condensate density.  The
depth of this potential for the peak condensate densities listed in
table \ref{trap.shape.summary.tab} is on the order of $50~\mu$K\@. The
depth is small compared to the $650~\mu$K of recoil kinetic energy, so
the atom has enough energy to leave the condensate.  However, the
depth is much larger than the energy in the radial coordinates,
$\mu/\kb= 2~\mu$K\@.  One expects focusing effects as the recoiling
atoms leave the condensate.

The \oneStwoS\ excitation process does not effect the hyperfine state
of the atom; the recoiling atoms are still in the $d$-state, and are
trapped.  As the atoms move toward the magnetic barrier that forms the
end of the trap they slow down.  Dephasing also occurs because the
integrated action for slightly different atomic trajectories quickly
becomes quite different in the mountainous potential surface the atoms
explore as they leave the trapping region.  In order to circumvent
these problems of slowing and dephasing we propose to flatten the
potential surface: we put the atoms in a magnetic field insensitive
state by driving the $d$-$a$ transition for the $2S$ atoms.  The
transition can be driven selectively for only the $2S$ atoms because
the hyperfine splitting of the $2S$ state is $2^3=8$ times smaller
than the $1S$ hyperfine splitting of 1420~MHz.  One might be concerned
that the rf field would drive $d$-$c$ state transitions in the cold
cloud of $1S$ atoms, and thus change the effective trap depth.  This
does not occur because the frequency is too large (178~MHz corresponds
to a trap depth of 8.5~mK).  The $a$-state $2S$ atoms are free to fall
out of the trapping region.  In the current apparatus the atoms can
fall a height $h=20$~cm before hitting a window.  This should take a
time $\tau_{fall}=(-v_r/g)(1-\sqrt{1+2gh/v_r^2})=56$~ms for atoms
with downward directed recoil velocity, and
$\tau_{fall}=(v_r/g)(1+\sqrt{1+2gh/v_r^2})=790$~ms for those with
upward directed recoil velocity.  Here $g$ is the acceleration of
gravity.  We do not expect to detect the atoms launched upward because
the natural lifetime (120~ms) is too short, and because, in the
current apparatus, the atoms will hit a surface before reaching the
peak of their 55~cm high trajectory.

Detection of the atomic beam could be accomplished by quenching the
atoms with an applied electric field pulse when the atoms are at the
bottom of the cell near the detector.  To estimate the signal size, we
compare to the signal obtained when the atoms are not allowed to move
very far before detection, the experimental procedure for figure
\ref{DS.degenerate.fig}.  We expect a reduction in the number of
fluorescing atoms by $e^{-\tau_{fall}/\tau_0}\simeq 0.63$ because of
decay during the transit time $\tau_{fall}$.  We expect a factor of
two further reduction because only $1/2$ the atoms have
downward-directed recoil velocity.  We expect an increase in signal
because the detection solid angle is greater near the bottom of the
cell.  With the current cell design we should be able to achieve a
factor of seven increase in detection efficiency, but with
straightforward extensions of the cell closer to the detector this
factor could be another factor of ten larger.  Neglecting this
improvement, we expect the atoms in the atomic beam to yield about the
same signal size as observed in figure \ref{DS.degenerate.fig}.  The
experiment is thus feasible in a cell identical to the current one,
but a conductive coating to reduce stray electric fields is essential.

In addition to observing an atomic beam ejected from the sample,
another interesting experiment would be a study of collisional
quenching of the $2S$ atoms as they move through the high density
condensate.  One might expect collisional de-excitation to occur on a
time scale comparable to the characteristic collision time:
$\tau_{col}=1/n_p \sigma_{2S-1S} v_{recoil}\simeq 700~\mu$s.  A brief study
indicated no collisional quenching on the sub-millisecond time scale,
but the study should be repeated for longer times.  Reduction of the
stray fields would allow this.

\subsection{Measuring the Condensate Density through Dipolar Decay}
\label{cond.dens.measure.proposal.sec}

In the experiments described in chapter \ref{results.chap} the
condensate density was measured spectroscopically, but confusion
arises around a factor of two in the conversion constant $\chi_c$
that relates frequency shift to sample density.  We propose an
independent measure of condensate density which should resolve the
ambiguity.  The proposed measurement exploits two-body dipolar decay,
which proceeds at a well known rate in the condensate \cite{bgm97}.
If the condensate density distribution is well approximated by the
Thomas-Fermi profile in equation \ref{tf.cond.profile.eqn}, then the
characteristic loss rate is given by equation
\ref{Ndot.condensate.eqn}.  Using equation \ref{Ncond.eqn} we find
that the peak shift (frequency shift of maximally shifted part of the
spectrum) should decay in time as
\begin{equation}
\Delta_p(t)=\frac{\Delta_{p,0}}{1+\frac{4g\Delta_{p,0}}{35\chi_c}t}
\label{peak.shift.time.evol.eqn}
\end{equation}
where
$\Delta_{p,0}$ is the initial peak shift.  The only free parameters
used to fit this equation to measured data would be $\Delta_{p,0}$ and
$\chi_c$.  The factor of two ambiguity in $\chi_c$ should be easy to
resolve.  If the Thomas-Fermi density distribution is a poor
approximation, then the factor $4/35$ should be different.  The
quality of the fit of the Thomas-Fermi lineshape to the data in
figures \ref{DS.degenerate.fig}, \ref{DF.condensate.spec.fig},
\ref{cond.DS.spec.time.evol.fig}, and \ref{cond.DF.spec.time.evol.fig}
indicates that this approximation is reasonable.

To measure the condensate decay time cleanly, we require that the
condensate is not being replenished by the normal gas. The most
straightforward way to accomplish this is a complete removal of the
normal gas by rf ejection.  The trap depth would be quickly swept down
to the edge of the condensate, using high rf power to efficiently
outcouple all the normal atoms.  DF or DS spectroscopy of the
condensate would then probe the time evolution of the peak shift,
$\Delta_p(t)$.

Since the lifetime of the condensate would be short, there would not
be much signal accumulation time.  The detection efficiency in the
current apparatus is too low, but a factor of six improvement should
make this experiment feasible (a factor of six can be obtained by
replacing the Lyman-$\alpha$ filter over the MCP with a window).
Currently, sufficient signal can be obtained by summing 0.8~s sweeps
from several identically prepared samples.  A factor of six
improvement would allow the sweeps to be shortened to 0.13~s, short
compared to the decay time $35 \chi_c/4g\Delta_{p,0}\simeq 1$~s
expected for $\chi_c=\chi_m/2$ and $\Delta_{p,0}\sim 600$~kHz.

\subsection{Exchange Effects in Cold-Collision Frequency Shift}
\label{distinguishable.normal.gas.sec}
In section \ref{question.of.chi.sec} we proposed that perhaps
$\chi_c=\chi_m$, which would make sense if $\chi_m$ did {\em not}
include exchange effects for some reason.  This could be tested by
measuring the cold-collision frequency shift in the classical regime
for atoms that are known to be distinguishable.  To ensure
distinguishability, the atoms could be put into different hyperfine
states, $c$ and $d$.  A $c$-state atom interacting with a $d$-state
atom should follow an interaction potential that is very similar to
the identical particle case, but the mean-field interaction energy
would be smaller by a factor of two because the atoms are
distinguishable and exchange effects are absent.  One expects two
peaks in the DF spectrum: one from collisions between identical
particles, which is shifted twice as much as the other, which is from
collisions between distinguishable particles.  If the observed
spectrum contains only a single peak, then we conclude that the atoms
are distinguishable in all collision processes.  In that case we
expect $\chi_c=\chi_m$.

We propose to do this experiment in a very deep trap ($\ethr/\kb\gg
(\mu_B/\kb) 506~{\rm G}=34$~mK) so that the $c$-state atoms will be
confined (see figure \ref{intro.zeeman.fig}).  We propose to use a
sample typical of what exists immediately after loading the trap
($T=40$~mK, $n_0=2\times10^{13}~{\rm cm^{-3}}$).  If half the atoms
are transferred from the $d$ to the $c$ state, the expected decay time
of the $c$-state atoms is $\tau\simeq 2$~s.  The dominant terms in the
loss rate are $G_{cc\rightarrow aa}=1.4\times10^{-13}$,
$G_{cc\rightarrow bd}=7\times10^{-14}$, and $G_{cc\rightarrow
ac}=6\times10^{-14}~{\rm cm^3/s}$ for fields around 600~G
\cite{skv88}, which corresponds to the thermal energy.  We must
measure the DF spectrum quickly compared to this time $\tau$ to
observe how the spectrum changes.

Time-of-flight broadening would be significant in this spectrum, on
the 40~kHz level for the current laser beam waist (100~$\mu$m
diameter).  This is large compared to the expected cold-collision
frequency shifts of 2 and 4~kHz.  To reduce the broadening, the beam
diameter could be expanded by a factor of ten.  The excitation rate
(proportional to the square of the radiation intensity) would drop by
$10^4$, but the interaction volume would increase by $10^2$ (the
sample diameter is larger than the beam diameter).  The linewidth
would narrow by a factor of ten because of the increased interaction
time, and so the overall signal rate would drop by only a factor of
ten.  Samples can be created every two minutes and systematics are
easy to control, so it would be straightforward to accumulate signal.

\subsection{Measuring $a_{1S-1S}$}

The ground state $s$-wave scattering length has been calculated to the
level of $10^{-2}$ \cite{jdk95}, but has not been measured nearly as
well.  One could use the interaction energy $U_0n_p=4\pi\hbar^2an_p/m$
in the condensate to measure $a$, given the peak density $n_p$.  We
propose to measure the chemical potential $\mu$ directly in an
experiment similar to the measurement of $\chi_c$ in section
\ref{cond.dens.measure.proposal.sec}.  Using $\mu=U_0n_p$ we obtain
$a=\mu m/4\pi\hbar^2n_p$.  To get $\mu$ we measure the difference
between the potential energies at the center ($\theta$) and edge
($\varepsilon_e$) of the condensate.  We measure $\theta$ by sweeping the
rf frequency upward until atoms begin to fall out of the trap, as
described in section \ref{trap.bias.measurement.sec}.  We measure
$\varepsilon_e$ in a series of measurements of the sample density; the
normal atoms are removed by sweeping the rf frequency down to some
ending frequency $f_{end}$, which is varied.  For $f_{end}$ slightly
larger than $\varepsilon_e/h$ the density measurements should give
identical results.  But for $f_{end}<\varepsilon_e/h$ the condensate is
depleted, and the measured density decreases.

A measurement of $a$ to the 10\% level seems feasible.  Consider trap
B and a peak condensate density $n_p=3\times10^{15}~{\rm cm^{-3}}$.
The bias field $\theta$ must not drift more than 1~mG between
measurement of $\theta$ and $\varepsilon_e$; this stability should be
possible.  To measure $\varepsilon_e$ we vary $f_{end}$ and look for
changes in the maximum frequency shift, as described above.  Since the
maximum frequency shift scales linearly with the chemical potential,
we must measure $\Delta_{p,0}$ to better than 10\%.  This should be
possible with the factor of six detection efficiency improvement
outlined above.  Many spectra would be obtained as the condensate
decays, and so a group of many data points would be fit to the two
parameters in equation \ref{peak.shift.time.evol.eqn}.  Precision of
10\% seems feasible.

\subsection{Dynamics of $2S$ atoms in $1S$ condensate}

As described in section \ref{peak.cond.dens.sec} and in the thesis of
Killian \cite{killian99}, condensate atoms excited to the $2S$ state
are confined radially by the deep interaction potential created by the
large condensate density.  The motional eigenstates in the $2S$
potential are rather different from the $1S$ eigenfunction, the
condensate wavefunction.  We therefore expect the $2S$ atoms to exist
in superpositions of several of these excited state eigenfunctions,
and to thus oscillate across the condensate.

The initial position of the $2S$ atoms is tuned by the detuning of the
excitation radiation field.  The position after some delay time $\tau$
could be measured using Balmer-$\alpha$ absorption imaging.  One
expects breathing modes.  Coherence effects in the excitation process
could be explored with this technique.

\section{Outlook}

The studies of Bose condensed hydrogen presented here constitute
simply the appetizer in the feast of experiments to be carried out.
Once ambiguities about $\chi_c$ are straightened out and the
fascinating asymmetric lines in section \ref{asymetric.line.sec} are
understood, the creation of the high-flux coherent atomic seems to be
the next step.  The beautiful atom laser experiments of Mewes \etal\
\cite{mak97}, Kozuma \etal\ \cite{kdh99}, and Anderson and Kasevich
\cite{ank98} should provide solid stepping stones.

%% file: statmech_MB_app.tex

\chapter{Theory of Trapped Classical Gas}
\label{classical.ideal.gas.app}

In this appendix we present the statistical mechanical treatment of a
trapped ideal gas, consider decay of the sample through relaxation of
the spin-polarization, and calculate the effects of evaporation.  The
appendix exists to familiarize the reader with concepts and scaling
laws useful for hydrogen trapping and cooling experiments.  Simple
relations describe samples trapped in deep potentials. Here we address
the effects of the finite trap depth, following the work of Luiten
\etal\ \cite{lrw96}.  We find that the truncation corrections are only
important for shallow traps.  

\section{Ideal Gas}
\label{ideal.gas.props.sec}
We begin our discussion by describing a trapped ideal gas in the
classical regime.  ``Trapped'' implies confinement by an inhomogeneous
potential with a depth significantly greater than the characteristic
energy of the gas particles.  ``Ideal'' means that collisions among
the atoms, although crucial for maintaining thermal equilibrium, are
not included in the Hamiltonian.  The interaction energy for the
densities studied in this thesis will turn out to be very small
compared to the thermal energy, except in the case of when a
condensate, as described in section \ref{condensate.qm.sec}.
``Classical'' means that the gas is far from quantum degeneracy, and
the Maxwell-Boltzmann distribution adequately describes the correct
quantum distribution.  These approximations are appropriate for the
trapped hydrogen gas studied in this thesis as it is cooled from the
initial temperature of about 40~mK to near the onset of quantum
degeneracy around $50~\mu$K.

In sections \ref{energy.distrib.func.sec} and
\ref{density.of.states.sec} we describe the occupation function for
the gas and the density of states functions for traps of experimental
interest.  In section \ref{ideal.gas.thermo.sec} these concepts are
employed to calculate various thermodynamic properties.  In section
\ref{dipolar.decay.sec} we consider trap losses arising from decay of
the spin-polarization of the trapped gas.  Collisions and evaporative
cooling are the topics of section \ref{collision.sec}.

\subsection{Occupation Function}
\label{energy.distrib.func.sec}

We assume that the gas is well described by the Maxwell-Boltzmann
distribution, except that the energy distribution is truncated at some
trap depth energy $\ethr$\label{ethr.def.page}.  The gas is said to be quasi-thermal.
This is a reasonable assumption if two conditions hold \cite{lrw96}.
First, the elastic collision rate must be high enough to redistribute
particle energies quickly compared to the rate at which processes
occur which would take the gas out of equilibrium, such as changes in
the trap shape or depth.  This is called the ``sufficient ergodicity''
condition.  (Elastic collisions are discussed in section
\ref{collision.sec} below.)  Second, particles with energies greater
than the trap depth must be removed promptly compared to the rate at
which they are promoted to the energetic states.  Situations in which
this condition is not satisfied have been discussed in section
\ref{evaporation.dimensionality.sec}.

The phase space occupation function, $f(\epsilon)$, is the number of
particles per unit volume of phase space in a region of phase space
with total energy $\epsilon={\cal H}({\bf p},{\bf
r})$\label{epsilon.def.page}. (${\cal H}$ is the Hamiltonian of the
particle).  We postulate an occupation function $f(\epsilon)$ that is
the Boltzmann factor \cite{hua63} below the trap depth, and zero
above:
\begin{equation}
f(\epsilon) = \left\{ \begin{array}{lll}
	n_0 \Lambda^3 e^{-\epsilon/\kb T} &;& \epsilon \leq \ethr \\
	0 & ;& \epsilon>\ethr
\end{array}
\right.
\label{classical.occupation.func.eqn}
\end{equation}
where $\Lambda\equiv h/\sqrt{2 \pi m \kb T}$ is the thermal de Broglie
wavelength\label{lamda.def.page} and $n_0$ is a reference density; the normalization is the
total number of trapped atoms.  Evidence supporting this postulate was
obtained by numerical integration of the Boltzmann transport equation
\cite{lrw96}.  The occupation function of a simulated gas, described
initially by $f(\epsilon)=$const.\ below threshold $\ethr$ and zero
above, was observed to transform to the Boltzmann form in equation
\ref{classical.occupation.func.eqn} in a few characteristic collision
times.  The trapped gas is described by two parameters, $T$ and $n_0$,
which correspond to sample temperature\label{n0T.def.page} and maximum sample density in
an infinitely deep trap \cite{lrw96}.  For finite trap depths,
temperature is poorly defined because the distribution is truncated,
and $n_0$ is larger than the actual maximum sample density, as
discussed in section \ref{truncdenssec}.

\subsection{Density of States}
\label{density.of.states.sec}

A useful concept is the density of states, the volume of parameter
space corresponding to the fulfillment of some condition.  We use two
density of states functions.  The ``total energy density of states''
function, $\rho(\epsilon)$, is the differential volume of phase space
corresponding to a total particle energy $\epsilon$\label{rho.def.page}.  (The number of
particles $N(\epsilon)d\epsilon$ within an energy $d\epsilon$ of
$\epsilon$ is the occupation per unit volume multiplied by the volume,
$N(\epsilon)d\epsilon=f(\epsilon)\rho(\epsilon)d\epsilon$.)  The
``potential energy density of states'' function,
$\varrho(\varepsilon)$, is the differential volume of real space with
potential energy $\varepsilon$\label{varrho.def.page}.  In this thesis
$\varepsilon$ will always denote a potential energy, and $\epsilon$
will always denote a total energy.  Likewise, $\varrho(\varepsilon)$
will always be a potential energy density of states, and
$\rho(\epsilon)$ will always be a total energy density of states.

The total energy density of states function is
\begin{equation}
\rho(\epsilon)=\frac{1}{h^3} \int d^3{\bf p} \, d^3{\bf r} \, 
	\delta({\cal H}({\bf p},{\bf r})-\epsilon)
\end{equation}
where the Hamiltonian is ${\cal H}=p^2/2m + V({\bf r})$, and $V({\bf
r})$\label{Vr.def} describes the trap potential\footnote{For a particle with
magnetic moment \vecmu, the potential energy is $V({\bf r})=\vecmu
\cdot{\bf B}({\bf r})$.  If the particle is moving slowly enough so
that the Larmor precession frequency changes only slowly
($\dot{\omega}/\omega\ll 1$), then the magnetic moment will
adiabatically follow the direction of ${\bf B}({\bf r})$, and $V({\bf
r})=\mu B({\bf r})$.}.  Using $\rho(\epsilon)$, integrals over phase
space of a function $F({\cal H}({\bf r},{\bf p}))$, which depends only
on the total particle energy, can be converted from the six
dimensional integral $\int d^3{\bf r} d^3{\bf p} F({\cal H}({\bf
r},{\bf p}))$ to a one dimensional integral over energy $\int
d\epsilon \rho(\epsilon) F(\epsilon)$.

The potential energy density of states function is
\begin{equation}
\varrho(\varepsilon) \equiv \int d^3{\bf r} \, \delta(V({\bf r})-\varepsilon)
\end{equation}

The trap shapes, $V({\bf r})$, in our experiments are well
approximated by either of two model potentials.  Both potentials are
cylindrically symmetric, and we use $z$ for the axial coordinate and
$\rho$ for the radial coordinate\label{trap.coords.def.page}.  The aspect ratio of the confined
clouds (length/diameter) is typically between 10 and 400.  The
``quadrupole trap with hard wall ends'' \cite{hes86,doyle91}
approximates traps used at higher temperatures ($T\geq 1$~mK), and was treated by Doyle \cite{doyle91}.
The second model trap shape is often called the
``Ioffe-Pritchard'' \cite{pri83,lrw96} (labeled ``IP'').  It corresponds to
the potential we use at the lowest temperatures and has the form
\begin{equation}
V_{IP}(\rho,z)=\sqrt{(\alpha\rho)^2 + (\beta z^2 + \theta)^2}-\theta
\end{equation}
with radial potential energy gradient $\alpha$, axial potential energy
curvature $2\beta$ (units of energy/distance$^2$), and bias potential
energy $\theta$\label{IP.trap.def.page}.  The potential energy density
of states for this trap is
\begin{equation}
\varrho(\varepsilon)=
\frac{4\pi}{\alpha^2\sqrt{\beta}}\;\sqrt{\varepsilon}\:(\varepsilon+\theta).
\end{equation}
We define the prefactor ${\cal A}_{IP}=4\pi/\alpha^2\sqrt{\beta}$,
with units (volume)(energy)$^{-5/2}$.  The total energy density of
states for this trap is
\begin{equation}
\rho(\epsilon)=
\frac{(2\pi m)^{3/2}}{h^3}
\frac{\pi^{3/2}}{2\alpha^2\sqrt{\beta}}\;\epsilon^2(\epsilon+2\theta).
\end{equation}
We define the prefactor $A_{IP}=\pi^2m^{3/2}{\cal A}_{IP}/h^32^{3/2}$, which has units of (energy)$^{-4}$.

\subsection{Properties of the Gas}
\label{ideal.gas.thermo.sec}

Properties of the gas, such as population, energy, and density, may be
calculated using the phase space occupation function and density of states
functions \cite{hua63}.  The number of trapped atoms, $N$, is
\begin{eqnarray}
N & = & \int_0^{\ethr} d\epsilon \, f(\epsilon) \rho(\epsilon) \nonumber \\
 & = & n_0 \Lambda^3 \zz
\label{totalN:eq}
\end{eqnarray}
where the single particle partition function is
\begin{equation}
\zz = \int_0^{\ethr} d\epsilon\; \rho(\epsilon) \exp(-\epsilon/\kb T)
\label{classical.partition.func.eqn}
\end{equation}
The $N$-particle partition function for a non-interacting classical
gas is $\zz_N=\zz^N/N!$\@.  

For a power law density of states given as
$\rho(\epsilon)=\rho_0\:\epsilon^{\delta+3/2}$ we obtain
\begin{equation} \zz = \zzinf P(\delta + 5/2, \eta)
\end{equation}
where $\eta\equiv \epsilon_t/\kb T$ is a dimensionless measure of the
trap depth (typically $4<\eta <10$)\label{eta.def.page} and
\begin{equation}\zzinf\equiv\rho_0 (\kb T)^{\delta+5/2}\Gamma(\delta+5/2)
\end{equation}
is the partition function for an infinitely deep trap.  The truncation
correction factor $P(q,\eta)$ is defined in equation \ref{P.defn.eqn}.

\begin{figure}[tb]
\centering \epsfxsize=5in \epsfbox{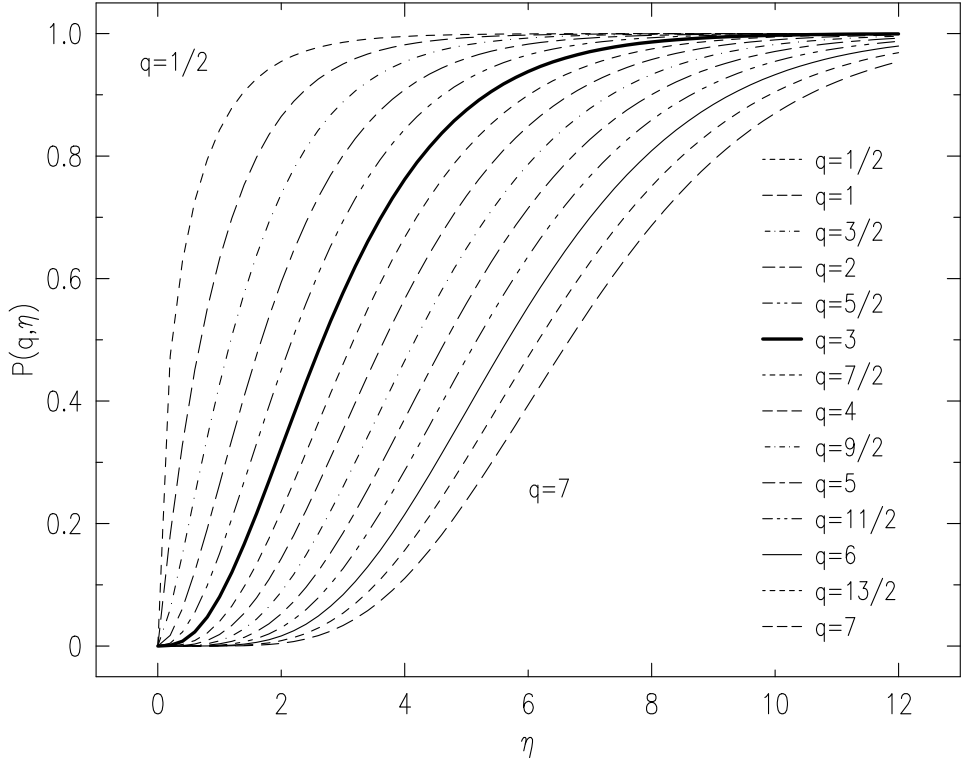}
\caption[behavior of $P(q,\eta)$]{ The behavior of the truncation
correction factor $P(q,\eta)$, plotted for various exponents $q$.  The
values $q=3$ and $q=4$ are used in finding the population of the IP
trap.  The value $q=3$ (heavy solid line) is for a harmonic trap.}
\label{pfuncfig}
\end{figure}

The partition function for the IP potential is\footnote{For this and
many other non-trivial integrals we use the computer program for
symbolic mathematics called Mathematica.}
\begin{equation}
{\cal Z}_{IP}  =  \frac{(\kb T)^{5/2} 3 \pi^{3/2}}{\Lambda^3\alpha^2
\sqrt{\beta}}\left(P(4,\eta)+\frac{2 \phi}{3} P(3,\eta)\right) 
\label{zip.eqn}
\end{equation}
where $\phi\equiv\theta/\kb T$ is a dimensionless measure of the trap
bias energy\label{phi.def.page} and $\eta\equiv\ethr/\kb T$ is a dimensionless measure of
the trap depth.  We have introduced the function $P(q,\eta)$, which is
a correction factor for the finite trap depth \cite{lrw96}.  It is
related to the incomplete gamma function
$\Gamma(q,\eta)\equiv\int_\eta^\infty t^{q-1} e^{-t} dt$ by
\begin{eqnarray}
P(q,\eta)&\equiv&\frac{1}{\Gamma(q)}\int_0^\eta t^{q-1}e^{-t} dt \nonumber \\ 
\nonumber \\
 &=&\frac{\Gamma(q)-\Gamma(q,\eta)}{\Gamma(q)}
\label{P.defn.eqn}
\end{eqnarray}
For large $\eta$ the correction is negligible, and $P(q,\eta)=1$.  The
behavior of $P(q,\eta)$ is shown in figure \ref{pfuncfig} for various
values of $q$.  
The parameter $q$ describes the energy exponents appearing in the
trapping potential.  As stated in equation \ref{zip.eqn}, the values
$q=3$ and $q=4$ are of interest for the IP trap.  Note that for
$\eta\geq 7$ there is only a small deviation from the infinitely deep
trap (corresponding to $\eta\rightarrow\infty$) for the exponents
$\delta$ of interest in this thesis.

The effective volume of the trap is 
\begin{equation}
\veff \equiv N/n_0 = \zz \Lambda^3 .
\label{veff:def}
\end{equation}

The energy  of the trapped gas is
\begin{eqnarray}
E & = & \int_0^{\ethr} d\epsilon \, \epsilon \rho(\epsilon)
f(\epsilon)
\label{E.defn.eqn}
\end{eqnarray}
For the IP trap
\begin{equation}
E_{IP} =  4 \kb T N_{IP} \frac{P(5,\eta)+\frac{\phi}{2}P(4,\eta)}
{P(4,\eta)+\frac{2\phi}{3}P(3,\eta)}.
\end{equation}
This result will be used in section \ref{equilibrium.temp.sec}.
The average energy per particle is $\overline{E}=E/N$.

\begin{figure}[tb]
\centering \epsfxsize=5in \epsfbox{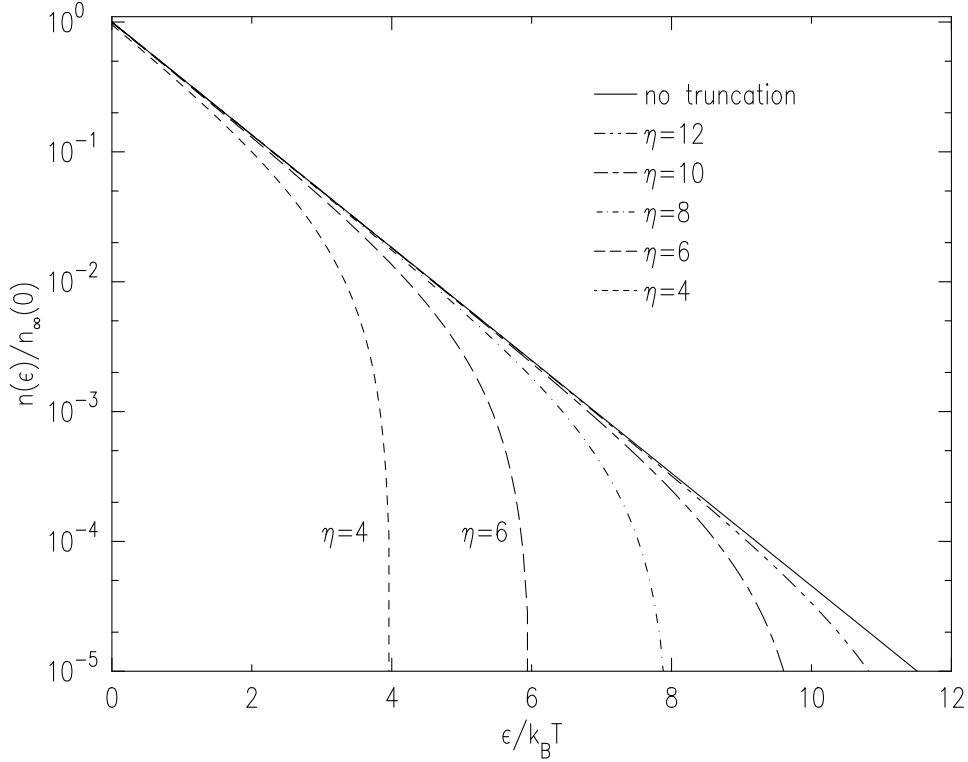}
\caption[effects of truncation on density]{ The effect of truncation
of the energy distribution on the density in the trap in a region of
potential energy $\varepsilon$, plotted for various trap depths
$\eta=\ethr/\kb T$.}
\label{densratiofig}
\end{figure}

The density \label{truncdenssec} at a position ${\bf r}$ (with
potential energy $\varepsilon=V({\bf r})$) may be obtained as
follows. Using the phase space occupation function
\begin{equation}
f({\bf r}, {\bf p})=n_0 \Lambda^3 \exp\left(-\frac{V({\bf r})+p^2/2m}{\kb T} \right)
\end{equation}
we integrate over trapped momentum states to obtain \cite{lrw96}
\begin{equation}
n({\bf r}) = n_0 e^{-V({\bf r})/\kb T} \left( \erf(t) - 2 t
e^{-t^2}/\sqrt{\pi}\right)
\label{classical.n.def.eqn}
\end{equation}
where $t=\sqrt{(\ethr-V({\bf r}))/\kb T}$ and $\erf(t)\equiv 2\int_0^t
e^{-s^2}ds/\sqrt{\pi}$ is the error function.  Note that for an
infinitely deep trap
\begin{equation}
n({\bf r}) =n_{\infty}({\bf r}) \equiv n_0 e^{-V({\bf r})/\kb T}.
\end{equation}
The ratio of the density $n({\bf r})$ in a region of potential energy
$\varepsilon=V({\bf r})$ to the density $n_{\infty}(0)$ at the bottom
of an infinitely deep trap is shown in figure \ref{densratiofig}.  We
see that, for $\eta >6$, truncation does not significantly effect the
density near the bottom of the trap where $\epsilon$ is less than a
few $\kb T$.

\subsection{Dipolar Decay}
\label{dipolar.decay.sec}

In this section we calculate the effect of spin relaxation on the
population and energy of the trapped gas.  Spin relaxation is the
dominant loss process in our trap.  In this analysis we properly treat
truncation of the energy distribution, following Luiten \etal\
\cite{lrw96}.

Dipolar spin relaxation is a process in which the magnetic dipoles of
two atoms interact, leading to a spin flip of one or both atoms
\cite{lsv86}.  If one of the interacting atoms makes a transition
from the trapped $d$-state to the $a$ state (see figure
\ref{intro.zeeman.fig}), the hyperfine energy, corresponding to 68~mK,
is released and both atoms are ejected from the trap.  An atom that
makes a transition to the $c$-state is weakly trapped in the radial
direction, but is pulled by gravity out of the trap through the weak
fields along the trap axis.  An atom that makes a transition to the
high field seeking $b$ state is expelled from the trap.  The rate
constant for loss from the trapped $d$ state has been calculated
\cite{skv88} and measured to the 20\% level \cite{rbj88,dsm89}.  The
calculated rate constant for loss from the $d$ state through all
channels at low temperatures is 
\begin{eqnarray}
g&=&2(G_{dd\rightarrow aa}+G_{dd\rightarrow
ad}+G_{dd\rightarrow ac}+G_{dd\rightarrow cc})+ G_{dd\rightarrow
cd} \nonumber \\
&=& 1.2 \times 10^{-15}~{\rm cm}^{3}~{\rm s}^{-1}, 
\label{dipolar.decay.const.eqn}
\end{eqnarray}
where $G_{\sigma\lambda\rightarrow\phi\psi}$\label{Gabcd.def} is the
event rate for the process in which two atoms in the $\sigma$ and
$\lambda$ states make transitions to the $\phi$ and $\psi$ states.
The rate depends very weakly on magnetic field, but we ignore the
dependence for the low temperature experiments which are the focus of
this thesis.

\begin{figure}[tb]
\centering \epsfxsize=5in \epsfbox{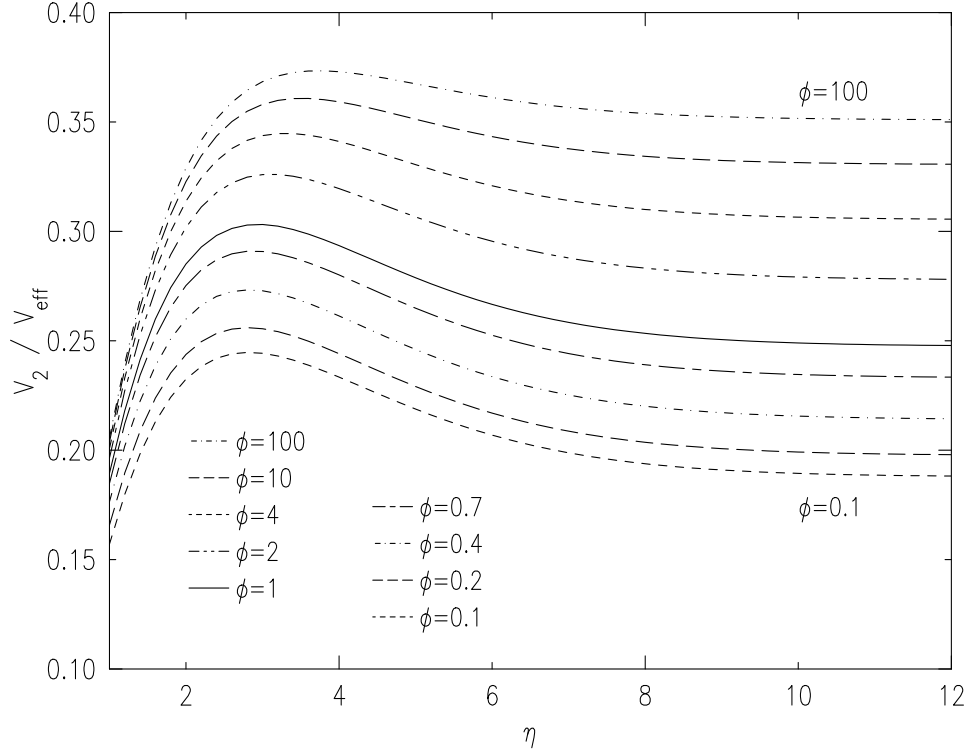}
\caption[the volume for dipolar decay]{ The volume for dipolar decay
compared to the effective volume, plotted for various bias fields
$\phi=\theta/\kb T$.  The difference between large $\phi$ (harmonic)
and small $\phi$ (linear radial confinement) arises because in the
harmonic trap the density distribution is flatter in the middle.  For
a box potential the density distribution is completely flat, and the
two volumes are identical.}
\label{vdipveffratfig}
\end{figure}

The dipolar decay rate from a region of density $n$ is
$\dot{n}=-gn^2$.  Consequently, loss occurs most rapidly from regions
of high density near the bottom of the trap.  The rate at which atoms
are lost from the trap through dipolar decay is
\begin{equation}
\dot{N}_2=-g \int_0^{\ethr} d\varepsilon \,
\varrho(\varepsilon) n^2(\varepsilon)
\label{classical.Ndot2.eqn}
\end{equation}
where $n(\varepsilon)$ is the gas density in regions with potential
energy $\varepsilon$.  An effective volume for dipolar decay, $V_2$, is
defined by
\begin{equation}
\dot{N}_2  =  -g n_0^2\:V_2 
= -g n_0 N \left(\frac{V_2}{V_{eff}}\right)
\end{equation}
The ratio $V_2/V_{eff}$ for the IP trap is plotted in figure
\ref{vdipveffratfig}.  The volume for dipolar decay is smaller 
than the effective volume because dipolar decay occurs preferentially
in regions of highest density, which are restricted to the bottom of the trap.

\begin{figure}[tb]
\centering \epsfxsize=5in \epsfbox{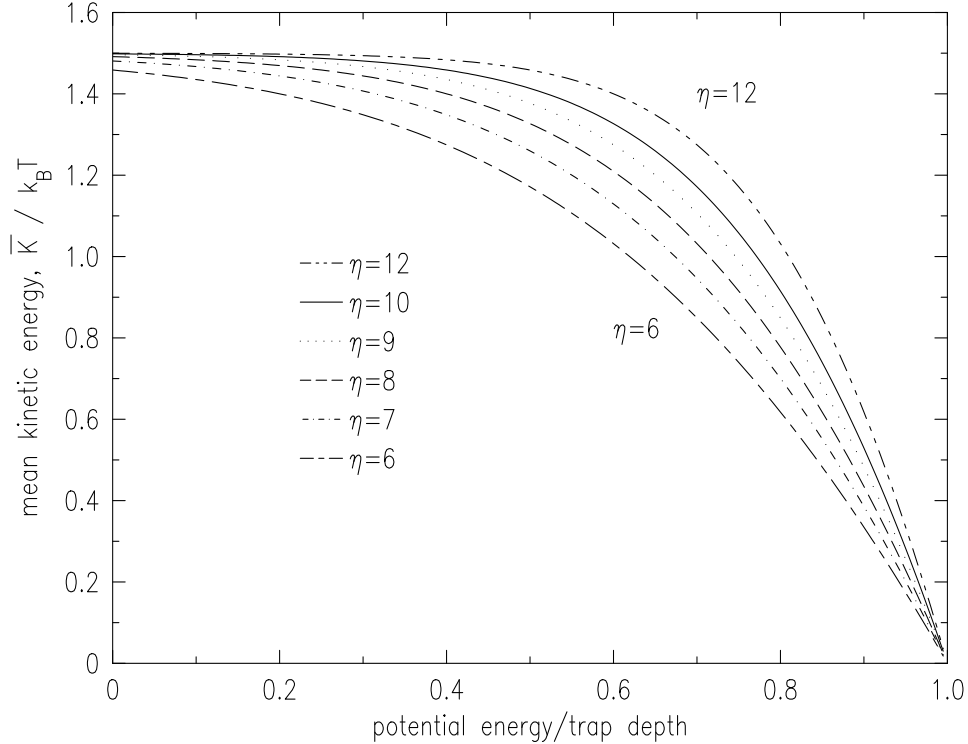}
\caption[mean kinetic energy of atoms in a trap of finite depth]{ The
mean kinetic energy of atoms in a region of given potential energy,
plotted for samples in traps of various depths.  The trap appears
shallower for atoms at higher potential energies.}
\label{gammafig}
\end{figure}

\begin{figure}[tb]
\centering \epsfxsize=5in \epsfbox{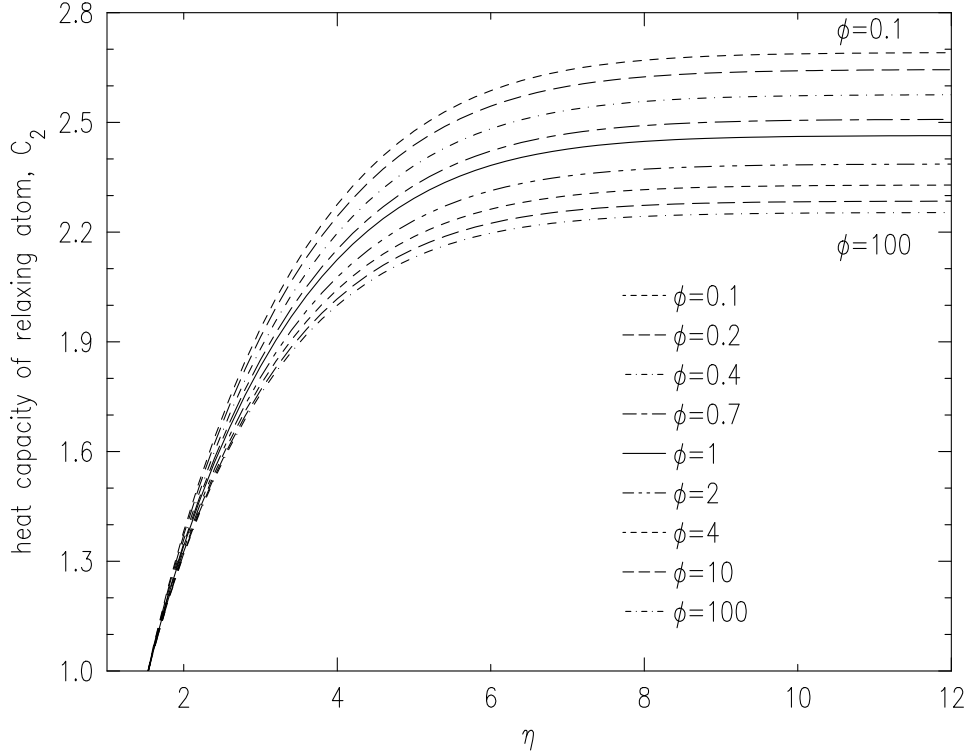}
\caption[average energy of a relaxing atom]{ The average total energy
per atom that is lost from a Ioffe-Pritchard trap through dipolar
relaxation, in units of $\kb T$.  The parameter $\phi\equiv \theta/\kb
T$ is the strength of the bias field, and $\eta\equiv\epsilon_t/\kb T$
is the trap depth.  We see that we can ignore truncation effects for
$\eta\geq 6$.  For the harmonic trap (large $\phi$) the particles have
less potential energy because the potential is quadratic; in the
linear radial potential (small $\phi$) there is more potential energy
per particle. }
\label{decayenergyfig}
\end{figure}

Since relaxation occurs predominantly near the bottom of the trap, the
escaping atoms typically have less energy than the sample average
energy per particle, $\overline{E}$.  Dipolar decay is thus a heating
mechanism.  The energy loss rate is the integral over the trap of the
local loss rate multiplied by the local average of the total energy per
particle:
\begin{equation}
\dot{E}_2 = -g \int_0^{\ethr} d\varepsilon \,
\varrho(\varepsilon) n^2(\varepsilon) \left(\varepsilon +
\overline{K}(\varepsilon)\right).
\label{classical.Edot2.eqn}
\end{equation}
The average kinetic energy per particle is $\overline{K}=3\kb T/2$ in an
untruncated Maxwell-Boltzmann distribution of velocities, but here we
include the finite trap depth in $\overline{K}$.   
We calculate the average kinetic energy per particle for
particles with potential energy $\varepsilon$ in a trap of depth
$\ethr$:
\begin{eqnarray}
\overline{K}(\varepsilon) & = &
\frac{\displaystyle \int_0^{\sqrt{2m(\ethr-\varepsilon)}}d^3p\,
\frac{p^2}{2m}e^{-p^2/2m \kb
T}}
{\displaystyle \int_0^{\sqrt{2m(\ethr-\varepsilon)}}d^3p \, e^{-p^2/2m \kb T}}
\nonumber \\
& = & \kb T \frac{6 t + 4t^3 -3 s}{4 t - 2 s}
\end{eqnarray}
where $s=\sqrt{\pi} \erf(t) \exp(t^2)$ and
$t=\sqrt{(\ethr-\varepsilon)/\kb T}$. 
Figure \ref{gammafig} shows the behavior of
$\overline{K}$.
The average energy carried away per atom is $C_2\kb
T\equiv\dot{E}_2/\dot{N}_2$\label{C2.def.page}.  This is plotted in
figure \ref{decayenergyfig} for the IP trap, for various values of
$\phi$.  We see that for $\eta\geq 6$ we can reasonably ignore
truncation effects.

\section{Collisions and Evaporative Cooling}
\label{collision.sec}

So far we have not addressed elastic collisions, but they play a
crucial role in the dynamics of the trapped gas.  In this section we
examine we examine the role of collisions in maintaining thermal
equilibrium and in evaporatively cooling the sample.  We calculate the
equilibrium temperature of the gas.  And we conclude the section with
a discussion of an impediment to efficient evaporative cooling, the
impediment that prevented the creation of BEC in hydrogen before this
work.

Only low energy collisions occur in our cold samples, so the elastic
scattering process is described by quantum mechanical $s$-wave
scattering theory\footnote{$s$-wave scattering is allowed for
identical bosons, but is prohibited for identical fermions}
\cite{sak85,wal94}.  The result of this theory is that elastic
interactions are parameterized by a single quantity, the $s$-wave
scattering length $a$\label{swave.a.def}.  The cross section for
collisions between identical particles is $\sigma=8\pi
a^2$\label{sigma.def.page}.  In a gas of density $n$ and mean particle
speed $\bar{v}=\sqrt{8\kb T/\pi m}$\label{vbar.def.page}, the
collision rate is $\Gamma_{col}(n)=n \sigma \bar{v}
\sqrt{2}$\label{Gamma.def.page}.  The collision rate determines the
time scale for energy redistribution in the gas.

A sample in a trap of finite depth is never in complete equilibrium
since collisions constantly tend to populate that portion of the
equilibrium energy distribution which lies above the trap depth.  The
removal of these energetic atoms is called evaporation.  Its use as a
cooling mechanism for a trapped atomic gas was first proposed by Hess
\cite{hes86} and was first demonstrated in the MIT spin-polarized
hydrogen group \cite{mds88}.  Evaporation occurs when a collision of
two atoms transfers enough energy to one atom that it may escape the
trap.  This atom leaves the trap with an energy $\epsilon>\ethr$, much
greater than the average energy per atom in the sample, $\overline{E}$.
The average energy per remaining particle is lower, and after
rethermalization of the trapped gas, the temperature is lower.
Cooling has occurred.  Note that rethermalization is given a headstart
by the preferential evaporation of atoms from the upper end of the
energy distribution, closer in energy to the trap depth, $\ethr$, as
described in appendix \ref{evapcoll:appendix}.

\subsection{Evaporative Cooling Rate}
\label{evap.cooling.rate.sec}

The evaporative cooling rate for a sample described by a truncated
Boltzmann occupation function has been calculated by Luiten \etal\
\cite{lrw96} using the Boltzmann transport equation.  We quote their
results here.

\begin{figure}[tb]
\centering \epsfxsize=5in \epsfbox{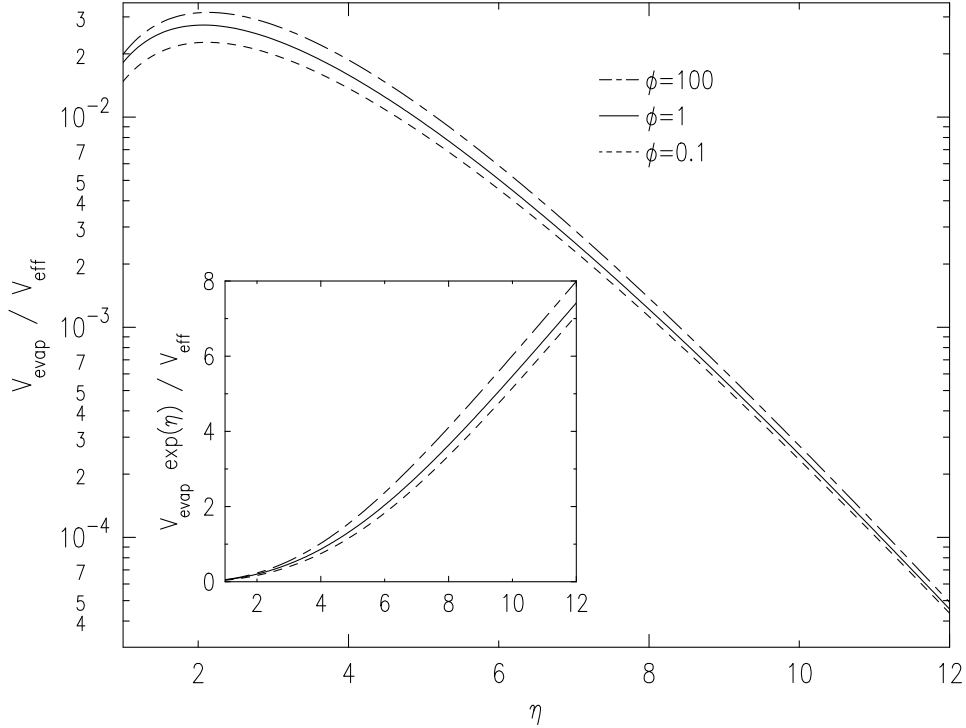}
\caption[evaporation volume $V_{evap}$]{ The quantity
$V_{evap}/V_{eff}$ for an IP trap as a function of $\eta$, plotted for
various bias fields $\phi=\theta/\kb T$.  The inset shows
$V_{evap}e^{\eta}$, which becomes linear in $\eta$ for large $\eta$.
This linear relation is useful when estimating evaporation rates.}
\label{vevapvefffig}
\end{figure}

The particle loss rate due to evaporation is
\begin{equation}
\dot{N}_{evap}=-n_0^2 \, \sigma \, \bar{v} \, V_{evap} 
= - N \: \frac{\Gamma_{col}(n_0)}{\sqrt{2}}\: \frac{V_{evap}}{V_{eff}},
\end{equation}
a product of the total population, a characteristic collision rate,
and the fraction of collisions which produce an atom with enough
energy to escape.  The ratio $V_{evap}/V_{eff}$ for an IP trap is
plotted in figure \ref{vevapvefffig} for various $\phi=\theta/\kb T$
(we use equation 43a from \cite{lrw96}).

\begin{figure}[tb]
\centering \epsfxsize=5in \epsfbox{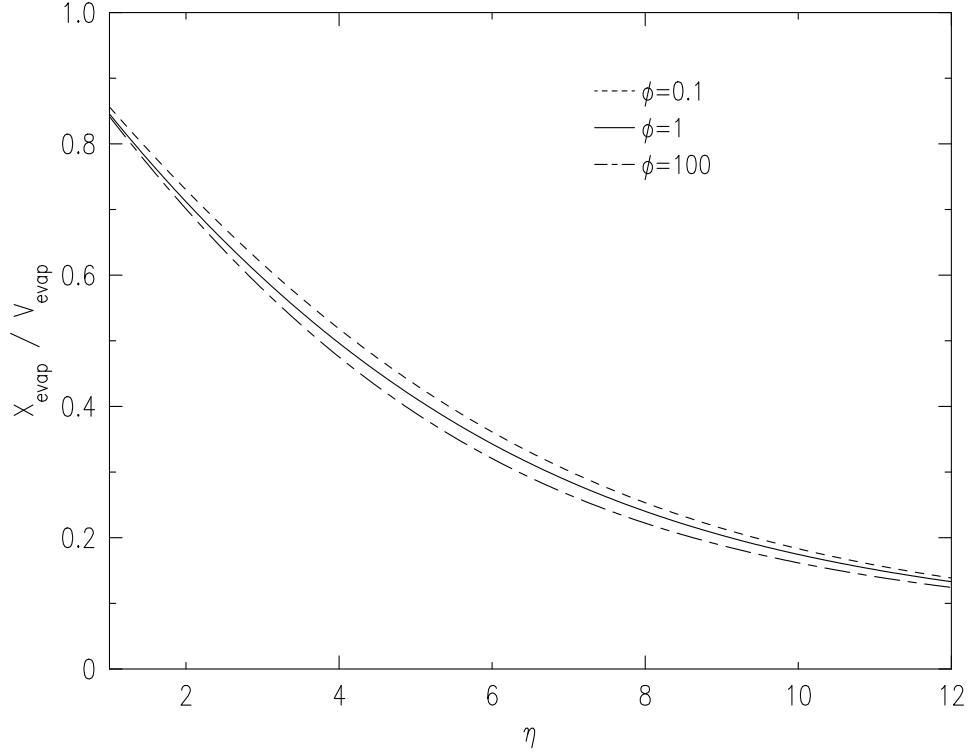}
\caption[evaporation volume $X_{evap}$]{ The quantity
$X_{evap}/V_{evap}$ for an IP trap as a function of $\eta=\ethr/\kb
T$, plotted for various bias fields $\phi=\theta/\kb T$.  The volume
$X_{evap}$ is a small contribution to the evaporation volumes for
$\eta\geq 6$}
\label{xevapvevapfig}
\end{figure}

The energy loss rate due to evaporation is given by equation 40 in
\cite{lrw96}:
\begin{equation}
\dot{E}_{evap}=\dot{N}_{evap} \kb T \left\{ \eta+1 - X_{evap}/V_{evap}
\right\}
\label{u.evapeqn}
\end{equation}
where $X_{evap}$ is a truncation correction factor.  Figure
\ref{xevapvevapfig} shows the behavior of $X_{evap}/V_{evap}$ for the
IP trap (we use equation 43b from \cite{lrw96}).  
For large $\eta$,
the quantity in curly braces in equation \ref{u.evapeqn} is nearly
$\eta+1$.

\subsection{Equilibrium Temperature}
\label{equilibrium.temp.sec}

The temperature of the trapped gas is set by a competition between
heating due to dipolar relaxation and cooling due to evaporation.  The
sample is in dynamic equilibrium when the temperature is no longer changing,
meaning that the average energy per particle is stationary.  Setting
$d(E/N)/dt=0$ dictates that
\begin{equation}
\frac{E}{N} 
=\frac{\dot{E}}{\dot{N}}
=\frac{\dot{E}_{evap}+\dot{E}_2}{\dot{N}_{evap}+\dot{N}_2}
\label{energy.loss.balance.MB.eqn}
\end{equation}
If the temperature drops below equilibrium, the evaporation rate drops
exponentially.  In addition, the effective volume decreases, the
density increases, and the heating rate due to dipolar decay
increases.  On the other hand, if the temperature increases above
equilibrium, the heating rate decreases and the evaporative cooling
rate increases.  A dynamic equilibrium is thus maintained.

\begin{figure}[tb]
\centering\epsfxsize=5in \epsfbox{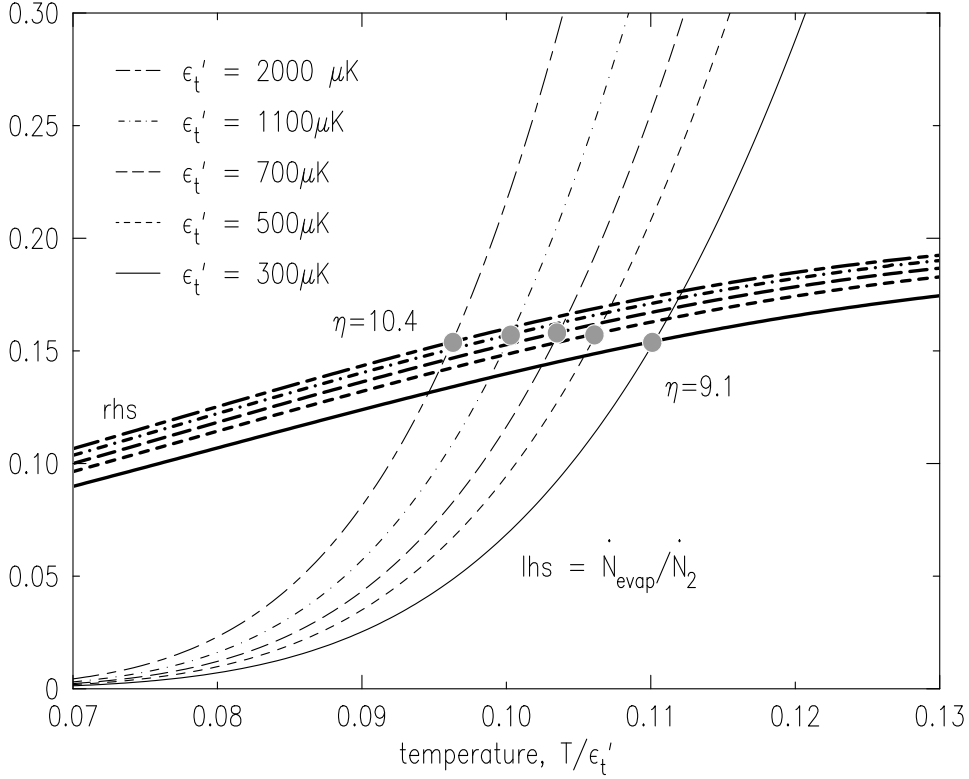}
\caption[graphical solution of heating/cooling balance equation]{
Graphical solution of the heating/cooling balance equation, plotted
for traps of various depths $\ethr$.  The bias energy is
$\theta/\kb=34~\mu$K.  The heavy lines are the right hand side (rhs)
of equation \ref{heating.cooling.eqn}.  The light lines are the left
hand side.  The solid dots indicate the equilibrium temperature for
the various trap depths.  At lower temperatures the equilibrium $\eta$
is smaller, indicating less efficient evaporation. }
\label{heating.cooling.balance.fig}
\end{figure}

The equilibrium temperature is found by numerically solving equation
\ref{energy.loss.balance.MB.eqn}.  For notational simplicity we define
the factor $B$ which indicates whether the IP trap is primarily linear
or harmonic in the radial direction,
\begin{equation}
B(\eta,\phi)\equiv 
\frac{P(5,\eta)+\phi P(4,\eta)/2}{P(4,\eta)+2\phi P(3,\eta)/3}.
\label{Bfunc.MB.def.eqn}
\end{equation}
Then equation \ref{energy.loss.balance.MB.eqn} may be written
\begin{equation}
\frac{\dot{N}_{evap}}{\dot{N}_2}=
\frac{4B(\eta,\phi)-C_2}{\eta+1+\frac{X_{evap}}{V_{evap}}-4
B(\eta,\phi)}.
\label{heating.cooling.eqn}
\end{equation}
The left and right hand sides are plotted in figure
\ref{heating.cooling.balance.fig} for typical trap depths and a bias
field $\theta/\kb=34~\mu$K\@.  
The equilibrium $\eta$ is larger at higher temperatures because the
mean particle speed, and thus collision rate, is higher.  The higher
$\eta$ at higher temperatures indicates that evaporative cooling is
more efficient there; fewer atoms are expended in the battle against
heating caused by dipolar decay.

\subsection{Forced Evaporative Cooling}
\label{forced.evap.cooling.sec}

The analysis above indicates that, for a given trap depth, a minimum
temperature will be achieved which is set by the competition between
heating and cooling.  In order to cool the sample further the trap
depth must be reduced.  A comprehensive review by Ketterle and van
Druten \cite{ked96} of forced evaporation analyses the process in
detail.  For the experiments described in this thesis the trap depth
is reduced by either lowering the magnetic confinement field at one
end of the trap, or by reducing the frequency of an RF magnetic field
that selectively removes the most energetic atoms.  This second method
is the focus of chapter \ref{rfcellchapter}.

A computer model (described in appendix \ref{evap.model.app}) 
has been
used to evolve in time the total energy and total number of particles
in the sample as it is cooled.  Temperature and density are calculated
along the way.  The effects of changing the trap depth and changing
the functional form of the trapping potential have been included.
This model agrees well with experiments.

The rate at which the temperature of the gas can be changed is an
important parameter.  Luiten \etal\ \cite{lrw96} have shown that a
classical gas with an energy distribution far from equilibrium will
reasonably approximate a Maxwell-Boltzmann distribution after about
three collision times.  One therefore expects fluctuations of the
energy distribution to dissipate on this time scale.  To maintain
quasi-equilibrium the trap depth and the trapping potential should
thus be changed slowly compared to this collision time.

Since trapped atoms are a scarce commodity, a figure of merit for a
cooling process is the amount of phase space compression achieved per
atom lost.  Far from quantum degeneracy the phase space density is the
number of atoms per cubic thermal de Broglie wavelength, $D\equiv n
\Lambda^3$\label{D.def.page}.  A useful measure of the cooling efficiency is the
cooling exponent \cite{ked96}, which is the differential fractional
change in phase space density per differential fractional change in
population,
\begin{equation}
\gamma \equiv \frac{d\log D}{d\log N}.
\label{gamma.def.eqn}
\end{equation}
A large, negative value indicates that the phase space density
increases by a large factor for a decrease in population by only a small
factor.  Cooling exponents as large as $\gamma=-3$ have been observed
\cite{mbg97}.

In cooling clouds of alkali metal atoms, a process called ``runaway
evaporation'' is observed, in which cooling of the sample leads to an
increase in sample density because the atoms are more localized in the
bottom of the trap.  In turn, the increased density leads to an increased
collision rate and faster evaporation.  The cooling accelerates, a
desirable phenomena because the number of atoms lost through
background gas collisions during the time used to cool the sample is
reduced when the cooling time is shorter.  The conditions required for
runaway evaporation are carefully explained by Ketterle \etal\
\cite{ked96}.  Runaway evaporation does not occur in clouds of trapped
hydrogen because the dominant loss rate, due to dipolar relaxation,
increases with increasing density with the same exponent as the
evaporation rate.  The ratio of the evaporation loss rate (``good''
losses) to the dipolar decay loss rate (``bad'' losses),
\begin{equation}
\frac{\dot{N}_{evap}}{\dot{N}_2}=
\frac{\sigma \bar{v} V_{evap}}{g V_2}\sim
50 e^{-\eta}(\eta-4.5) \sqrt{\frac{T}{1~\mu{\rm K}}},
\label{dependence.of.r.eqn}
\end{equation}
is independent of density, depending only on $\sqrt{T}$ for a given
$\eta$.  The effective volumes for dipolar decay and evaporation used
in equation \ref{dependence.of.r.eqn} are for $\phi\sim 1$ and
$\eta>7$.  From this equation we see that evaporation is more
efficient at higher temperatures, as noted above.

%% file: evap_collision.tex
\chapter{Fraction of Collisions Which Produce an Energetic Atom}
\label{evapcoll:appendix}

In a collision between two atoms with a medium energy, what is the
probability of creating an ``energetic atom'', i.e. an atom with
enough energy to escape the trap?  Consider a collision between two
distinguishable particles of identical mass $m$ and initial momenta
${\bf p}_1$ and ${\bf p}_2$.  After the collision the momenta are
labeled ${\bf q}_1$ and ${\bf q}_2$.  The total and relative momenta
are ${\bf P}=({\bf p}_1+{\bf p}_2)$ and ${\bf p}={\bf p}_1-{\bf p}_2$,
and similarly for ${\bf Q}$ and ${\bf q}$.

Conservation of momentum leads to the equation ${\bf P}={\bf Q}$.  We
orient the axes of the coordinate system so that ${\bf Q}=Q\hat{z}$
and we work in spherical coordinates.  The total energy is
\begin{equation}
E_0=(P^2 + p^2)/4m.
\label{E0eqn}
\end{equation}
Conservation of energy demands that $p^2=q^2$.  We thus
have four equations and six unknowns, leaving two undetermined
coordinates, which will be labeled $\theta$ and $\phi$ in the
description of
\[{\bf q}=q\left\{\,(\hat{x}\cos\phi+\hat{y}\sin\phi)\,\sin\theta +
\hat{z}\cos\theta\right\}.\]

After aligning the coordinate system we see that the there are two
conserved scalar quantites, $P=Q$ and $p=q$.  We can compute the
fraction of this allowable phase space corresponding to various
conditions by assuming that, in a collision, $\theta$ and $\phi$ may
take any value with equal probability.  We define the function
$f(\theta,\phi)\;d\Omega$ as the volume of phase space within a
differential solid angle $d\Omega$ of the given angles.  Since the
collisions are postulated to be independent of angle, we take
$f(\theta,\phi)=f_0=1/4\pi$.

In the calculations to follow we scale the energies to the trap depth,
$E_t$.  We use the dimensionless variables $a\equiv q/\sqrt{4E_t m}$ and
$b=Q/\sqrt{4E_tm}$.  The total energy is then $E_0=E_t(a^2+b^2)$.

We first find the fraction of phase space corresponding to one atom
with energy greater than the trap depth.  We call this an ``energetic
atom''.  The condition is expressed as $E_1=q_1^2/2m>E_t$ or
$E_2=q_2^2/2m>E_t$.  Using ${\bf q}\cdot{\bf Q}=qQ\cos\theta$ and
defining the ``energy distribution'' parameter $\xi\equiv
ab\cos\theta$ we translate the conditon above to
\[|\xi|>1 - E_0/2E_t.\]
The fraction of phase space (consistent with the given parameters $a$
and $b$) that corresponds to the existence of at least one energetic
atom is then
\begin{eqnarray}
G(Q,q)&=&\int_{|\xi|>1-E_0/2E_t} d\theta \sin\theta d\phi \;
f(\theta,\phi) \nonumber \\ &=& \int_{\alpha}^1 du
\end{eqnarray}
where $u$ has been substituted for $\cos\theta$ and the lower limit on
the integral is $\alpha=0$ if $E_0\geq 2E_t$ or, if $E_0<2E_t$, the
smaller of $u_0\equiv (2-a^2-b^2)/2ab$ and $1$.  We obtain
\begin{equation}
G(a,b)=\left\{ \begin{array}{ll}
	0 & u_0 \geq 1 \\
	1-u_0 & 0<u_0<1 \\
	1 & u_0\leq 0 \\
\end{array}
\right.
\end{equation}

This quantity may now be averaged over all the initial conditions of
interest to obtain the average fraction of collisions which produce an
energetic atom.  The allowable initial conditions are that $E_1<E_t$,
$E_2<E_t$, and $E_0>E_t$ since we assume both particles are initially
trapped but there is enough total energy for one particle to escape.
As we sum over all the possible initial states with the given $a$ and
$b$, we must weight the average by the fraction of initial states
which have both particles trapped.  This factor is simply the volume
of phase space left over from the above computation applied to the
initial conditions, $1-G$.  Since $P=Q$ and $p=q$, the average
fraction is then
\[F=F_0\int_{E_t<E_0<2E_t} d^3{\bf Q}\:d^3{\bf q}\; G\, (1-G)\]
where $F_0$ is a normalization factor to be computed below.
Exploiting the spherical symmetry and switching to scaled variables we
obtain
\begin{equation}
F=F_0\: (4\pi)^2 \: 4 E_t m \int_0^{\sqrt{2}} b^2 db
\int_{a_{min}}^{a_{max}} a^2 da \; G(a,b) (1-G(a,b))
\end{equation}
where $a_{min}=-b+\sqrt{2}$ and $q_{max}=\sqrt{2-b^2}$ are obtained
from the condition that $0\leq u_0 \leq 1$.  The limits on the $b$
integral are chosen to keep the limits on the $a$ integral
reasonable.
The result is
\[F=F_0 \: 16 \pi^2 \: E_t m\: \frac{20-6\pi}{9}.\]

The normalization constant $F_0$ is simply the weighted integral over
the allowable initial states,
\[F_0 = \left(\int_{E_t<E_0<2E_t}  d^3{\bf Q}\: d^3{\bf q}\;(1-G)\right)^{-1}.\]
We follow the same scheme, but note that when $u_0>1$ the integrand,
$1-G$, remains at unity. We obtain
\[F_0 = \left(16\pi^2 E_t m \,\frac{4}{9} \right)^{-1}.\]
We conclude that the fraction of the collisions which will produce an
energetic atom is
\[F=5-\frac{3\pi}{2}\approx 0.288.\]

\begin{figure}[tb]
\centering \epsfxsize=5in \epsfbox{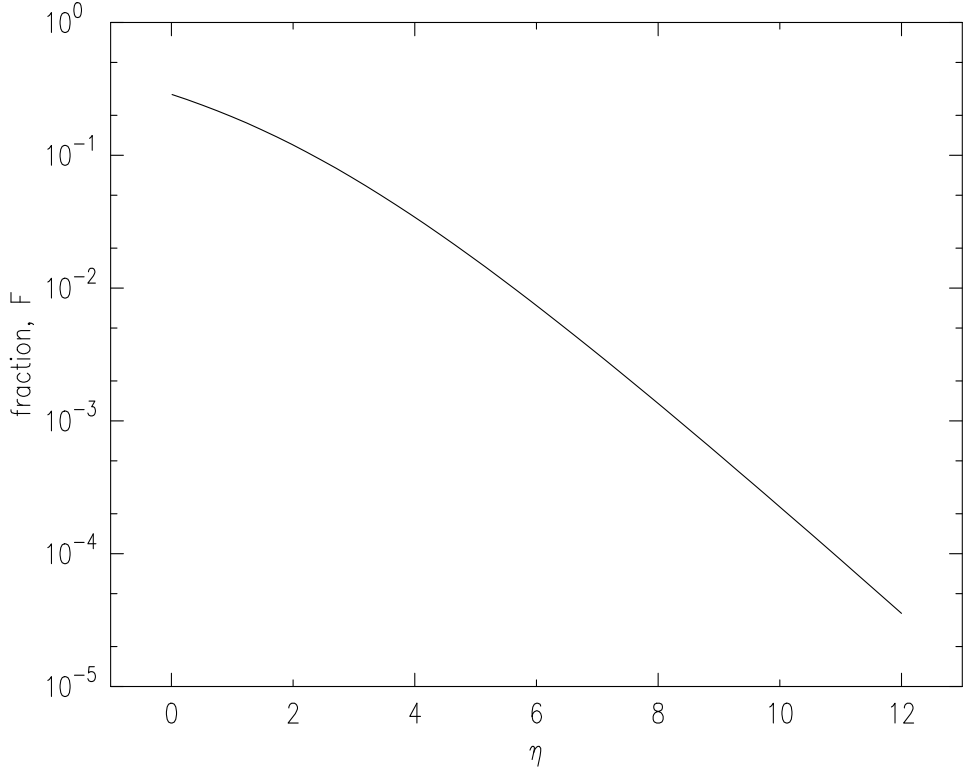}
\caption[fraction of collisions which produce an energetic atom]{
The fraction of collisions which produce an atom with enough energy to
escape the trap.  A truncated Maxwell-Boltzmann distribution of
velocities is assumed.  The parameter $\eta$ is the truncation energy
divided by $\kb T$.  }
\label{evapfracfig}
\end{figure}

The above calculation assumed that the probability of having any of
the initial states was identical.  In fact, the higher energy states
are less likely than the lower energy states, as dictated by the
Boltzmann factor.  Into the above integrals, then, must be inserted
the factor $\exp(-E_0/\kb T)=\exp(-\eta(a^2+b^2))$ where $\eta\equiv
E_t/\kb T$.  The $b$ integral must be computed numerically.  We obtain
behavior as indicated in Fig \ref{evapfracfig}.
We see that for typical values of the parameter $\eta$, a collision
which produces an energetic atom is an extremely rare event.

Another quantity of interest is the probability that an energetic atom
which collides with another atom will becoming trapped.  This is
essentially the reverse of the process calculated above.  We conclude
that in the vast majority of collisions between a thermal atom and an
energetic atom, neither atom will be energetic afterward.

%% file: statmech_bose_app.tex

\chapter{Behavior of Non-Condensed Fraction of a Degenerate Gas}
\label{normal.degenerate.app}

In this appendix we consider a weakly interacting Bose gas of
temperature $T$ and chemical potential $\mu\leq0$ confined in a trap
of depth $\ethr$. A truncated Bose-Einstein energy distribution is
assumed.  Here we derive results that will be used in section
\ref{hbec.props.sec}.  Truncation effects are important because of the
shallow traps used for the experiments in chapter \ref{results.chap}.

As discussed in section \ref{condensate.qm.sec}, the interactions in
the condensate increase the chemical potential slightly above zero.
The mean-field repulsion energy of the condensate deforms the
effective trap shape slightly, moving the energy at the bottom of the
trap from $\varepsilon=0$ to $\varepsilon=\mu$ and flattening the
potential there.  However, we ignore these effects on the thermal
gas because for our experiments $\kb T\gg \mu$; the magnitude of the
deformation is much smaller than the characteristic energy of the
atoms.

\section{Ideal Degenerate Bose Gas}

\subsection{Density}

\begin{figure}[tb]
\centering \epsfxsize=5in \epsfbox{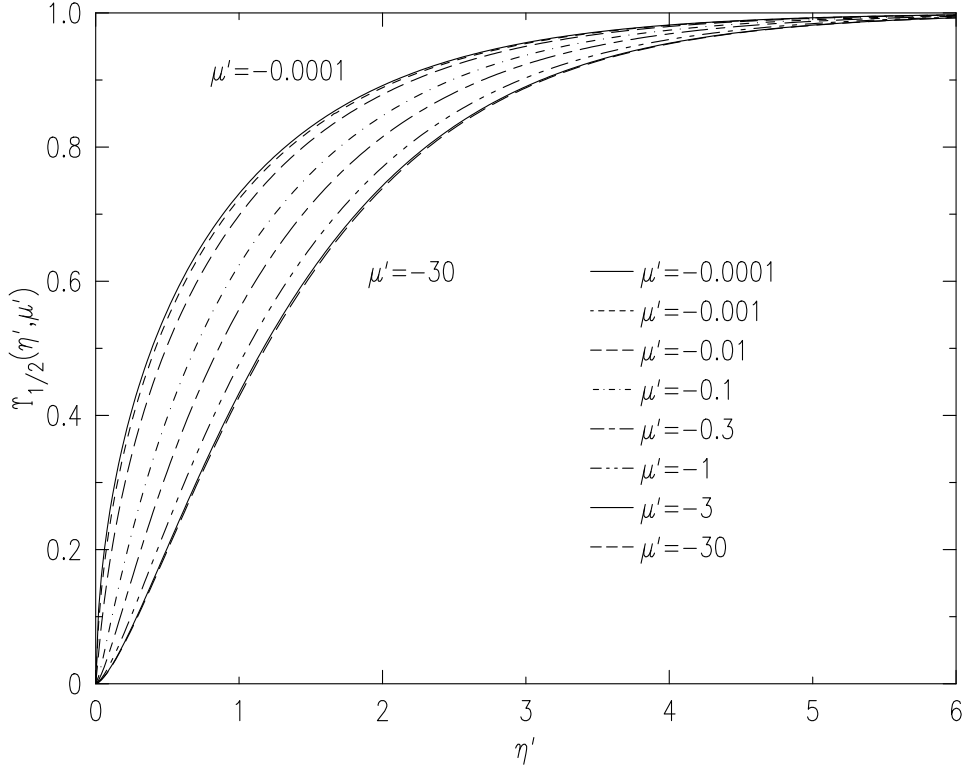}
\caption[behavior of $\Upsilon_{1/2}$]{ Behavior of the truncation
correction factor $\Upsilon_{1/2}$, plotted for various effective
chemical potentials.  The largest deviations from unity occur for
small $\eta^\prime$, corresponding to potential energies within a few
$\kb T$ of the energy truncation point $\ethr$.  The range of
$\mu^\prime$ shows how the correction factor changes between the
classical ($\mu^\prime\ll -1$) and the degenerate quantum
($|\mu^\prime| \ll 1$) regimes.  The asymptotic functional form for
low and high $\mu^\prime$ can be seen.}
\label{upsilon.5.fig}
\end{figure}

Here we calculate the density of the gas in a region of the trap with
potential energy $\varepsilon$.  We include truncation effects
because, as will be shown in the experimental data in chapter
\ref{results.chap}, the typical $\eta$ is between 4 and 6 when the
condensate is present.  Integrating the occupation function over all
the trapped momenta we obtain a density
\begin{equation}
n(\varepsilon; T, \mu, \epsilon_t)=\frac{1}{h^3}\int_0^{p_{max}} 
\frac{4 \pi p^2}{e^{(p^2/2m + \varepsilon-\mu)/\kb T}-1}\; dp
\end{equation}
where $p_{max}^2/2m=\epsilon_t-\varepsilon$ is the largest allowable
kinetic energy for atoms with potential energy $\varepsilon$.  We
separate the result of the integration into an expression for the
density in an infinitely deep trap and a unitless correction factor
$\Upsilon_{1/2}$ which accounts for truncation of the distribution.  We
obtain
\begin{equation}
n(\varepsilon; T, \mu, \epsilon_t)= 
\frac{g_{3/2}(e^{(\mu-\varepsilon)/\kb T})}{\Lambda^3(T)} \;
\Upsilon_{1/2}\left(\frac{\epsilon_t-\varepsilon}{\kb T}, 
\frac{\mu-\varepsilon}{\kb T}\right).
\label{trunc.dens.func.eqn}
\end{equation}
The correction factor is
\begin{equation}
\Upsilon_k(\eta^\prime,\mu^\prime)\equiv
\left\{\begin{array}{lll}
\frac{\displaystyle\int_0^{\eta^\prime} \frac{x^k}{e^{x-\mu^\prime}-1}\;dx}
{\displaystyle\int_0^{\infty} \frac{x^k}{e^{x-\mu^\prime}-1}\;dx} 
& ; & \eta^\prime>0 \\
0 & ; & \eta^\prime\leq 0 \\
\end{array}
\right.
\label{upsilon.def.eqn}
\end{equation}
where $\eta^\prime$ is an effective unitless trap depth and
$\mu^\prime$ is an effective chemical potential.  In the
limit of large $\eta^\prime$, $\Upsilon_k(\eta^\prime, \mu^\prime)=1$.
The behavior of $\Upsilon_{1/2}$ is displayed in figure \ref{upsilon.5.fig}.
As expected, the truncation correction factor is most pronounced at
potential energies within a few $\kb T$ of the energy threshold
$\epsilon_t$, for which $\eta^\prime$ is small.  Also, for a nearly
degenerate gas ($\mu^\prime$ small) the population is concentrated
more at low energies so the effects of truncation are less important.
For trap depths greater than $4\kb T$ the truncations effects can be
neglected.

\subsection{Kinetic Energy}

The mean kinetic energy of the particles in a region of potential
energy $\varepsilon$ is
\begin{equation}
\overline{K}(\varepsilon; T, \mu, \epsilon_t)=
\frac{1}{n(\varepsilon;T,\mu,\epsilon_t)h^3}\int_0^{p_{max}} 
\frac{4 \pi p^2}{e^{(p^2/2m + \varepsilon-\mu)/\kb T}-1}\; \frac{p^2}{2m}\;dp.
\end{equation}
Again we break the result into an expression for $\overline{K}$ in an
infinitely deep trap and a correction factor.  We obtain
\begin{equation}
\overline{K}(\varepsilon;T,\mu,\epsilon_t) =
\left(\frac{3 \kb T}{2}\right)\;
\left(\frac{g_{5/2}(e^{\mu^\prime})}{g_{3/2}(e^{\mu^\prime})}\right)\;
\left(\frac{\Upsilon_{3/2}(\eta^\prime,\mu^\prime)}
	{\Upsilon_{1/2}(\eta^\prime,\mu^\prime)}\right),
\label{BE.mean.KE.eqn}
\end{equation}
where $\eta^\prime=(\epsilon_t-\varepsilon)/\kb T$ and
$\mu^\prime=(\mu-\varepsilon)/\kb T$.  The first term is the classical
term, obtained for a Maxwell-Boltzmann distribution.  The middle term
is the difference between classical and Bose statistics.  The third
term is the truncation correction, which is unity for deep traps.  For
$\mu^\prime=0$ the middle term is 0.513.  This result is used in the
calculation of the energy loss rate due to dipolar decay in section
\ref{degenerate.dipolar.decay.sec}.  This result tells us that the
velocity distribution of a degenerate Bose gas should be narrower than
its classical counterpart, an effect observed in section
\ref{degenerate.thermometry.sec}.

\subsection{Population}

\begin{figure}[tb]
\centering \epsfxsize=5in \epsfbox{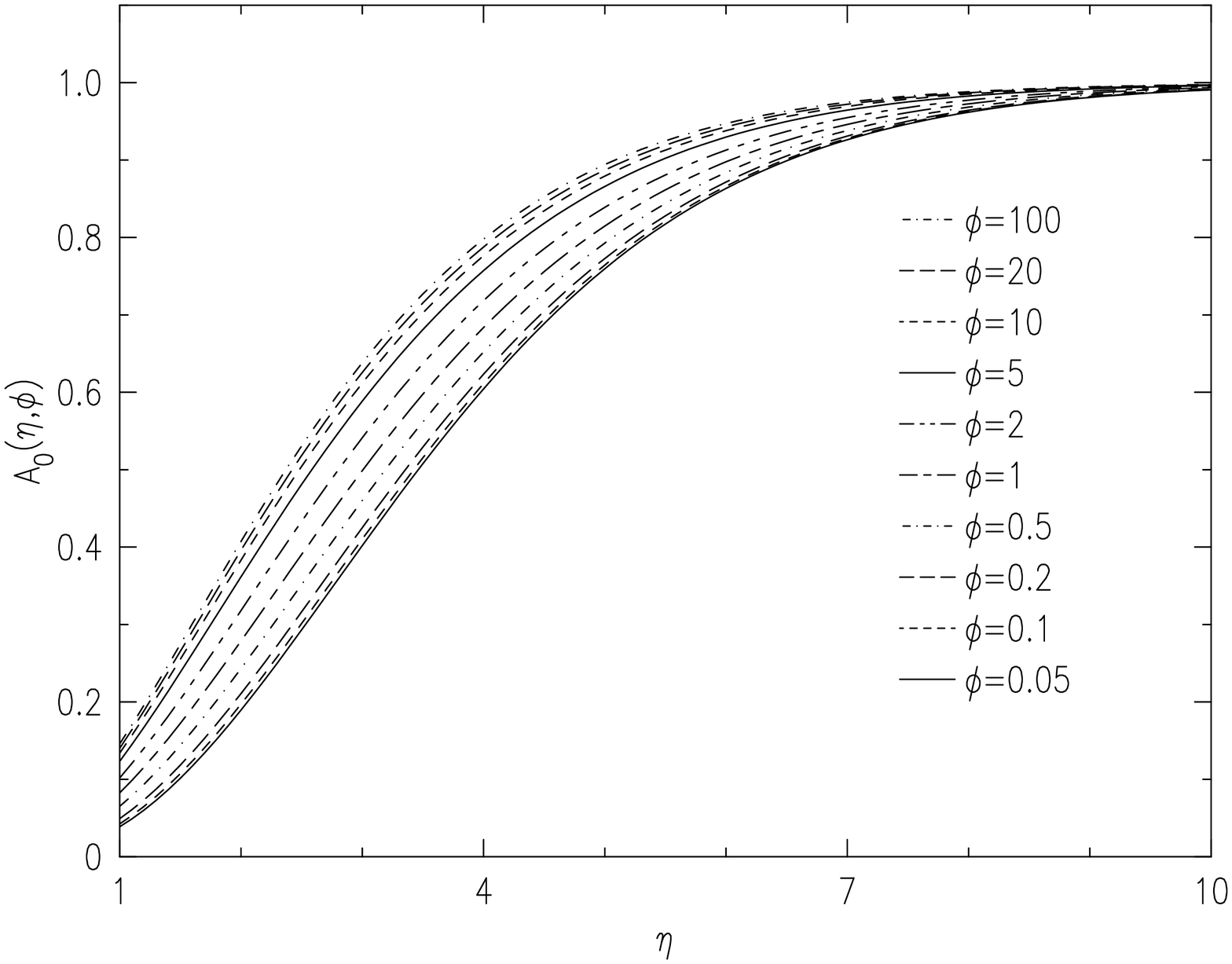}
\caption[behavior of $A_0$]{ Behavior of the truncation correction
factor $A_0$ as a function of scaled trap depth $\eta=\epsilon_t/\kb
T$, plotted for various bias field energies $\phi=\theta/\kb T$.  The
asymptotic functional form for low and high $\phi$ can be seen.  }
\label{A0.fig}
\end{figure}

The number of atoms in the thermal gas is obtained by integrating
the Bose occupation function, weighted by the density of states, over
the trapped energy states:
\begin{equation}
N_t(T,\mu,\epsilon_t)=
\int_0^{\epsilon_t} \frac{\rho(\epsilon)}{e^{(\epsilon-\mu)/\kb T}-1}\; 
d\epsilon.
\label{thermal.population.simple.eqn}
\end{equation}
The lower bound on the integral is the bottom of the trap.  For the
analysis which follows the thermal gas will exist in the presence of a
condensate, and so we set $\mu=0$.  Specializing to the
Ioffe-Pritchard trap, we obtain
\begin{equation}
N_t(T,\mu,\epsilon_t)=
A_{IP}(\kb T)^4 \frac{\pi^4 + 60\zeta(3)\phi}{15}\; A_{0}(\eta,\phi)
\label{thermal.population.eqn}
\end{equation}
where $\zeta$ is the Riemann zeta function ($\zeta(3)=1.202$) and
$A_k(\eta,\phi)$ is a correction factor for the truncation of the
energy distribution:
\begin{equation}
A_k(\eta,\phi)=
\frac{\displaystyle\int_0^\eta \frac{x^{2+k}(x+2\phi)}{e^x-1}\;dx}
{\displaystyle\int_0^\infty \frac{x^{2+k}(x+2\phi)}{e^x-1}\;dx}.
\end{equation}
The behavior of $A_0$ is shown in figure \ref{A0.fig}.

\subsection{Total Energy}

\begin{figure}[tb]
\centering \epsfxsize=5in \epsfbox{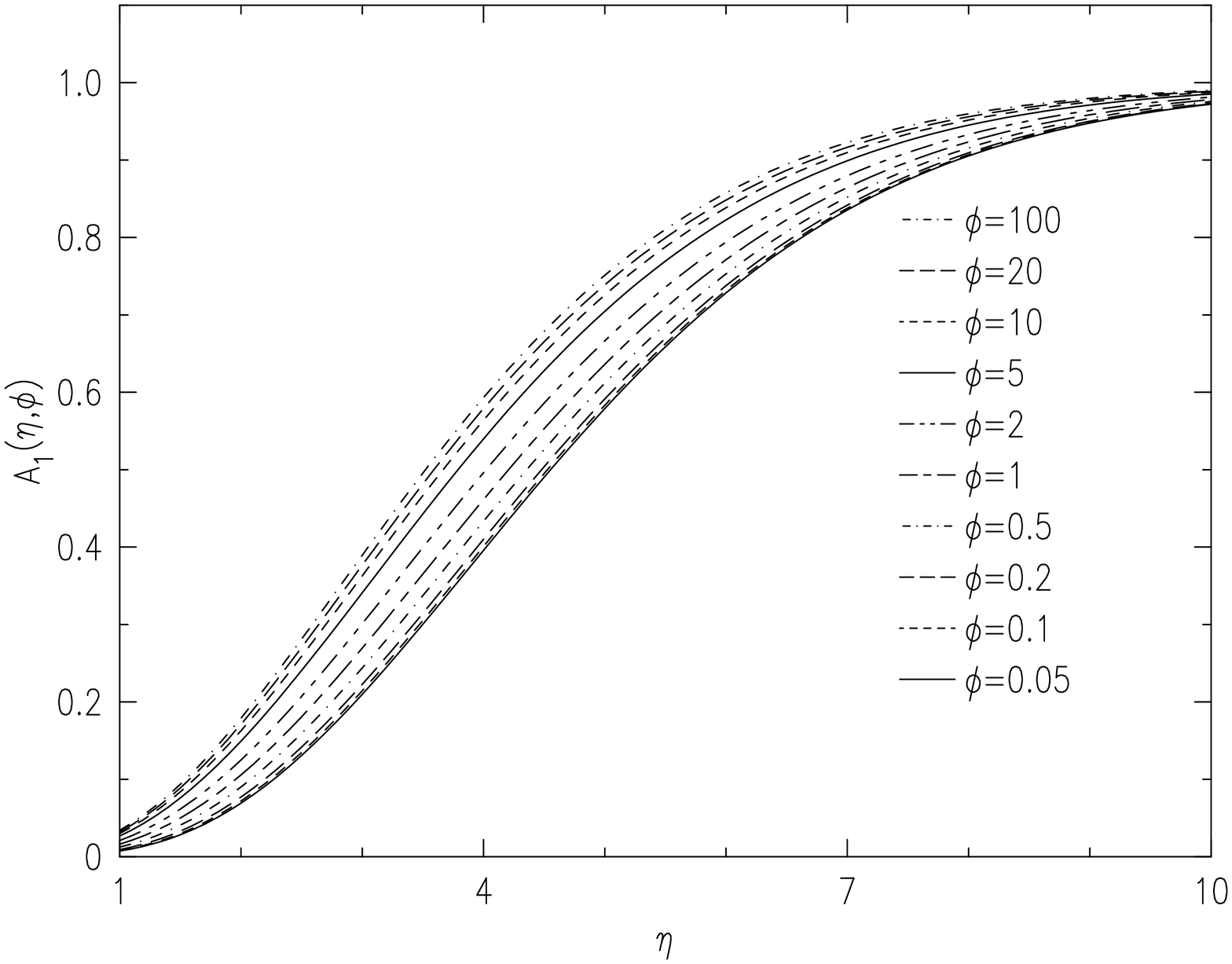}
\caption[behavior of $A_1$]{ Behavior of the truncation correction
factor $A_1$ as a function of scaled trap depth $\eta=\epsilon_t/\kb
T$, plotted for various bias field energies $\phi=\theta/\kb T$.  The
asymptotic functional form for low and high $\phi$ can be seen. }
\label{A1.fig}
\end{figure}

\begin{figure}[tb]
\centering \epsfxsize=5in \epsfbox{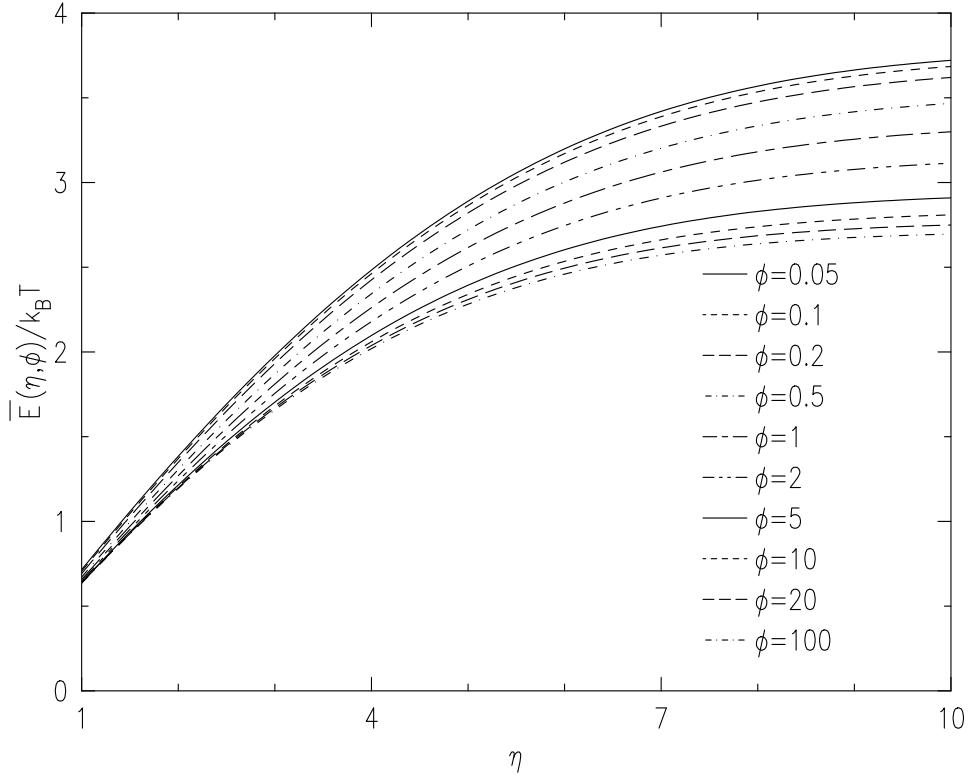}
\caption[behavior of $\overline{E}$ for a degenerate Bose gas]{ The
average energy per particle in the thermal portion of a degenerate
Bose gas as a function of scaled trap depth $\eta=\epsilon_t/\kb T$,
plotted for various bias field energies $\phi=\theta/\kb T$.  For low
$\phi$ the potential is linear in the radial direction.  More energy
can be carried as radial potential energy for low $\phi$ than for
large $\phi$, which corresponds to harmonic potentials. }
\label{bose.Ebar.fig}
\end{figure}

The total energy of the thermal gas is derived in a manner analogous
to the occupation:
\begin{equation}
E_t(T,\mu,\epsilon_t)=
\int_0^{\epsilon_t} \frac{\rho(\epsilon)}{e^{(\epsilon-\mu)/\kb T}-1}\; 
\epsilon \;d\epsilon.
\label{thermal.energy.simple.eqn}
\end{equation}
We obtain, for a degenerate gas ($\mu=0$) in a Ioffe-Pritchard trap,
\begin{equation}
E_t(T,\epsilon_t)= A_{IP}(\kb T)^5 \left(\frac{2 \phi
\pi^4}{15}+24\zeta(5)\right)\; A_1(\eta,\phi)
\end{equation}
where $\zeta(5)=1.037$.  The behavior of $A_1$ is shown in figure
\ref{A1.fig}.  
The average energy per particle in the thermal gas, $\overline{E}_t$, is then
\begin{equation}
\overline{E}_t(T,\epsilon_t)=
\frac{E_t}{N_t}=\kb T\; B(\phi)
\frac{A_1(\eta,\phi)}{A_0(\eta,\phi)}
\end{equation}
where 
\begin{eqnarray}
B(\phi) & \equiv & 
2\frac{\phi\pi^4+180\zeta(5)}{\pi^4+60\zeta(3)\phi} \nonumber \\
& \simeq & 2.701\frac{\phi + 1.916}{\phi+1.351}
\label{b.phi.defn.eqn}
\end{eqnarray}
is a factor that indicates whether the trap is predominantly linear or harmonic in the radial direction.
This average energy is shown in figure \ref{bose.Ebar.fig}.
For a classical gas the average energy per quadratic degree of freedom
is $\kb T/2$, but for a degenerate Bose gas the energy is less.  For
large $\phi$ the potential is harmonic and there are six quadratic
degrees of freedom, implying a mean energy of $3\kb T$ for a classical
gas; for a degenerate Bose gas the mean energy is 2.7~$\kb T$\@.  The
difference between the classical and degenerate Bose results (3-2.7)
is not as great as the factor $\sim 0.5$ mentioned for $\mu^\prime=0$
in the discussion of equation \ref{BE.mean.KE.eqn}.  This is because
in calculating the mean kinetic energy an average is made over all the
atoms in the trap with their many effective chemical potentials.
Atoms far from the center of the gas have $\mu^\prime\ll -1$, and are
essentially in the classical regime.

\section{Dipolar Decay}
\label{degenerate.dipolar.decay.sec}

\begin{figure}[tb]
\centering \epsfxsize=5in \epsfbox{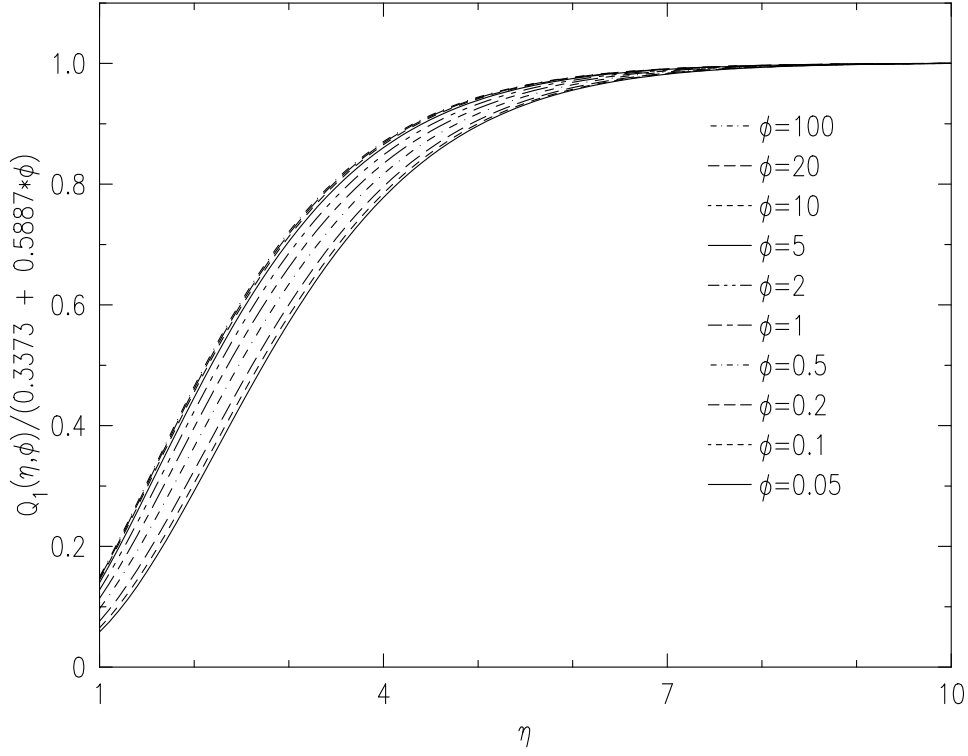}
\caption[behavior of $Q_1$ for a degenerate Bose gas]{ Behavior of
the squared density integral $Q_1$ as a function of scaled trap depth
$\eta=\epsilon_t/\kb T$, plotted for various bias field energies
$\phi=\theta/\kb T$.  In the figure $Q_1$ is scaled by $0.337 +
0.589\phi$, the high $\eta$ asymptote. }
\label{q1.fig}
\end{figure}

The rate at which atoms leave the thermal gas through dipolar decay is
calculated by integrating $\dot{n}=-gn^2$, over the
volume of the gas.  For a degenerate gas in the IP trap we obtain a
two-body loss rate
\begin{eqnarray}
\dot{N}_{2,t}(T,\epsilon_t)&=&
\int_0^{\epsilon_t}\varrho(\varepsilon) (-g) n^2(\varepsilon;T,\mu=0,\epsilon_t)
\;d\varepsilon
\nonumber \\
&=&
-\frac{(\kb T)^{5/2}{\cal A}_{IP} g}{\Lambda^6}\;Q_1(\phi,\eta)
\label{bose.dipolar.particle.rate.eqn}
\end{eqnarray}
where $Q_1$ is the (unitless) integral of the density squared, and includes truncation effects:
\begin{equation}
Q_1(\phi,\eta)= \int_0^\eta \sqrt{x}(x+\phi)\;
g_{3/2}^2(e^{-x})\Upsilon_{1/2}^2(\eta-x,-x)\; dx.
\label{Q1.def.eqn}
\end{equation}
The behavior of $Q_1$ is shown in figure \ref{q1.fig}.

\begin{figure}[tb]
\centering \epsfxsize=5in \epsfbox{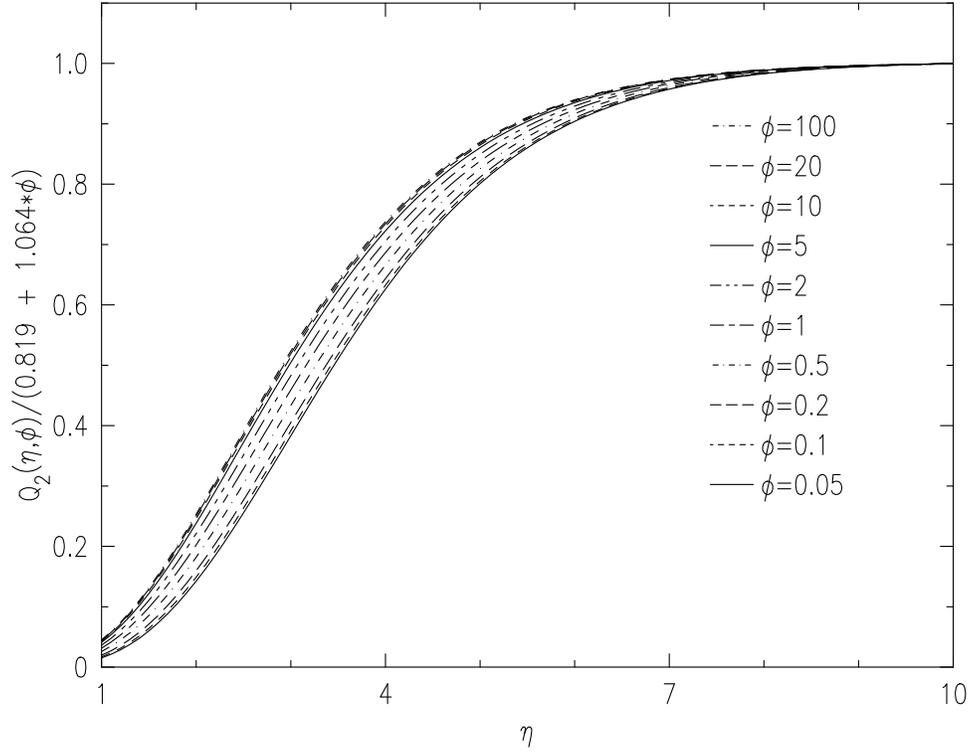}
\caption[behavior of $Q_2$ for a degenerate Bose gas]{ Behavior of
the dipolar decay energy loss integral $Q_2$ as a function of scaled
trap depth $\eta=\epsilon_t/\kb T$, plotted for various bias field
energies $\phi=\theta/\kb T$.  In the figure $Q_2$ is scaled by $0.819
+ 1.064\phi$, the high $\eta$ asymptote.  The asymptotic functional
form can be seen for high $\phi$.}
\label{q2.fig}
\end{figure}

\begin{figure}[tb]
\centering \epsfxsize=5in \epsfbox{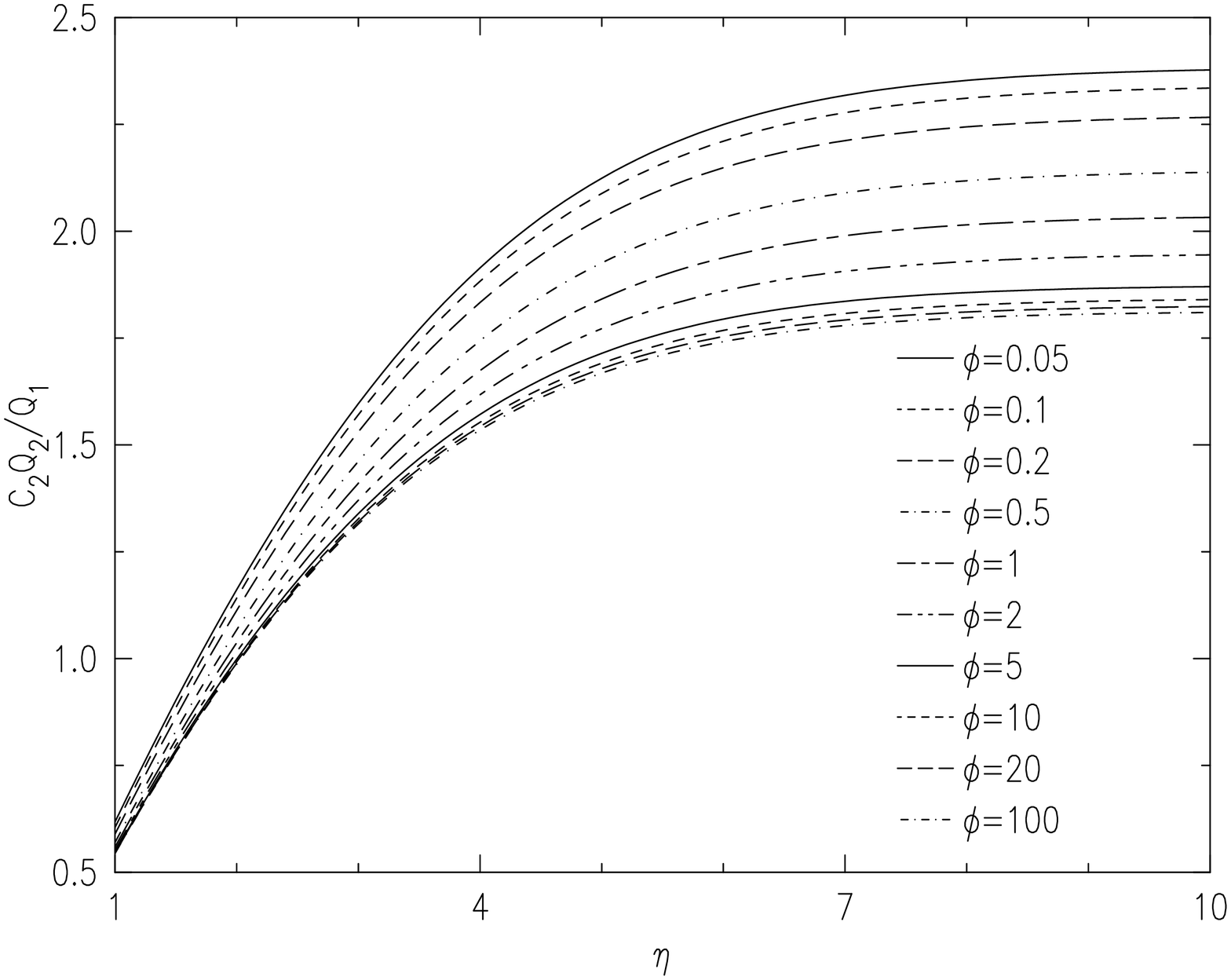}
\caption[average energy of a relaxing atom for a degenerate Bose gas]{
Average energy of atoms lost from thermal portion of a degenerate gas
through dipolar decay, in units of $\kb T$, plotted for $\mu=0$ and
various bias energies $\phi=\theta/\kb T$. The asymptotic functional
forms for small and large $\phi$ are apparent. }
\label{q2q1.fig}
\end{figure}

The rate at which energy leaves the thermal gas through dipolar
decay is
\begin{eqnarray}
\dot{E}_{2,t}(T,\epsilon_t)&=&
\int_0^{\epsilon_t}\varrho(\varepsilon) (-g) n^2(\varepsilon;T,\mu=0,\epsilon_t)
\;\left(\varepsilon+\gamma(\varepsilon)\right)\;
d\varepsilon
\nonumber \\
&=&
-\frac{(\kb T)^{7/2}{\cal A}_{IP} g}{\Lambda^6}\;Q_2(\phi,\eta)
\label{bose.dipolar.energy.rate.eqn}
\end{eqnarray}
where
\begin{eqnarray}
Q_2(\phi,\eta)&=&
\int_0^\eta \sqrt{x}(x+\phi)\; g_{3/2}^2(e^{-x})\Upsilon_{1/2}^2(\eta-x,-x)\;
\nonumber \\
& &\times \left(x+\frac{3}{2}\frac{g_{5/2}(e^{-x})}{g_{3/2}(e^{-x})}
\frac{\Upsilon_{3/2}(\eta-x,-x)}{\Upsilon_{1/2}(\eta-x,-x)}\right) dx
\label{Q2.def.eqn}
\end{eqnarray}
The behavior of $Q_2$ is shown in figure \ref{q2.fig}.
The average energy carried away by each particle that is lost through dipolar relaxation
is then $C_2\kb T=\kb T Q_2/Q_1$\label{C2.bose.def}, and is shown in figure
\ref{q2q1.fig}.
As expected, the average energy lost per particle through dipolar
relaxation is less than $\overline{E}$, the average energy per particle
throughout the sample (compare figure \ref{q2q1.fig} with figure
\ref{bose.Ebar.fig}), and thus dipolar relaxation leads to heating.
Compared to the Maxwell-Boltzmann distribution, the Bose-Einstein
distribution has more particles at lower energy.  We thus expect $C_2$
to be smaller, as is seen by comparing figure \ref{q2q1.fig} to figure
\ref{decayenergyfig}.

%% file: evap_model.tex

\chapter{Model of the Dynamics of the Non-Degenerate Gas}
\label{evap.model.app}

This appendix describes a computer model of the dynamics of the
trapped gas, including evaporation, dipolar decay, and changes of the
trap shape and depth.  The C source code is available.

\newcommand{\mytrap}{\verb1mytrap1}

\section{Overview}

We desire a general purpose computer model of the time evolution of
our trapped gas to assist in choosing paths through parameter space
toward BEC.  Previous models \cite{doyle91} employed approximations
that are not valid at the low temperatures we wish to access and are
not easily adaptable to the more complicated trap shapes we wish to
employ.

\subsection{Approximations}
The primary approximation is that the phase space distribution
function for the gas is always a Maxwell-Boltzmann truncated at the
trap depth.  Luiten \etal\ \cite{lrw96} have confirmed this
approximation for slow evaporation.  This approximation is valid as
long as two additional assumptions are valid: (a) instantaneous
ergodicity and (b) instantaneous removal of energetic atoms.  The
validity of these approximations depends on some time scales.  For
approximation (a) we need the time required for the gas to
re-equilibrate to be short compared to the speed at which the trap
shape and depth are changed.  For approximation (b) we need the
energetic atoms to be removed quickly compared to the energy
redistribution time.  These are the approximations described in
section \ref{energy.distrib.func.sec}.

\subsection{Describing the Trap}
We make no approximations about the trap shape.  Instead the trap is
described by energy and potential energy density-of-states functions
($\rho(\epsilon)$ and $\varrho(\varepsilon)$) which may involve any
number of parameters, all (possibly) time dependent.  The energy
truncation point is also given as a time-dependent parameter.

\section{Statistical Mechanical Description of the Gas}
\subsection{Variables}

We describe the trap by an evaporation energy threshold, $\ethr(t)$,
and the density of states functions and their set of (time dependent)
parameters, $\alpha_\sigma(t)$.  These are our independent variables.
Some function is given by the user to relate these trap shape
parameters to time, the ultimate independent variable.

The primary dependent variables are $E$ and $N$, the total energy and
total population of the sample.  At each time step during the
evolution of the trapped gas we calculate changes to the total energy,
$dE$, and to the total number of atoms in the trap, $dN$.  The new
value for $T$ may then be calculated from $E$.  If the form of
$\rho(\epsilon)$ is known, it is often possible to analytically relate
$T$ to $E$.  Otherwise $T$ must be determined numerically, as is done
in this model.  The density parameter $n_o$ is trivially related to
$N$ by equation \ref{veff:def}.

\subsection{Computing Properties of the Gas} 

The population, $N$, total energy, $E$, and density $n({\bf r})$ are
computed as described in section \ref{ideal.gas.props.sec}.

For notational simplicity, we introduce the functions
\begin{equation}
\psi_n \equiv \int_0^{\ethr} d\epsilon \, \rho(\epsilon)\, \epsilon^{n} \,
e^{-\epsilon/\kb T}
\end{equation}
and
\begin{equation}
\varphi_n \equiv \int_0^{\ethr} d\epsilon \, \rho(\epsilon) \, \epsilon^{n}.
\end{equation}

The Helmholtz free energy is given by $A=-\kb T \log\zz_N$, where $\zz_N$
is the $N$-particle partition function defined below equation
\ref{classical.partition.func.eqn}.  We can define a ``single
particle'' free energy
\[
A_1 \equiv -\kb T \log \zz
\]
so that $A=N A_1 + \kb T \log N!.$
The entropy is 
\begin{equation}
S=-\left. \frac{\partial A}{\partial T}\right|_V
\end{equation}
To keep $\veff$ constant, we replace $\zz$ with $\veff/\Lambda^3$
(using equation \ref{veff:def}) to obtain
\begin{equation}
S=\kb \left(N\log \zz + \frac{3N}{2} - \log N! \right).
\label{entropy:eq}
\end{equation}

\section{Treatment of Physical Effects}
The model accounts for cooling due to evaporation, heating due to
two-body spin relaxation, adiabatic trap shape changes, and the skimming
of energetic atoms as the energy threshold is lowered.  The finite
trap depth is fully integrated into the model.

At each time step we calculate
\begin{equation}
\dot{E} = \dot{E}_{evap} + \dot{E}_{2} + \dot{E}_{trap} + \dot{E}_{skim}
\end{equation}
and
\begin{equation}
\dot{N} = \dot{N}_{evap} + \dot{N}_{2} + \dot{N}_{skim}.
\end{equation}
Each of these terms is explained below.

\subsection{Evaporation}
Luiten \etal\ \cite{lrw96} have described the evaporation process,
starting from the kinetic equation.  We summarize their results here.
Following section \ref{evap.cooling.rate.sec}, the particle loss rate
due to evaporation is
\begin{equation}
\dot{N}_{evap}=-n_o^2 \, \sigma \, \bar{v} \, V_{evap}
\end{equation}
where $\sigma$ is the collision cross section, $\bar{v}=\sqrt{8 \kb
T/\pi m}$ is the mean speed,
\begin{eqnarray}
V_{evap} &= &\Lambda^3 e^{-\eta} \int_0^{\ethr} d\epsilon \,
\rho(\epsilon) \left[(\eta-1-\epsilon/\kb T)e^{-\epsilon/\kb T}
+e^{-\eta}\right] \nonumber \\
& = & \Lambda^3 e^{-\eta}\left((\eta-1)\psi_0 - \frac{\psi_1}{\kb T} + 
\varphi_0 e^{-\eta}\right)
\end{eqnarray}
is the effective volume for evaporation, and $\eta\equiv\ethr/\kb T$.
The energy loss rate due to evaporation is
\begin{equation}
\dot{E}_{evap}=-n_o^2 \sigma \bar{v} \left\{ (\ethr+\kb T)V_{evap} -
\kb T X_{evap} \right\}
\end{equation}
where
\begin{eqnarray}
X_{evap} &=& \Lambda^3 e^{-\eta} \int_0^{\ethr}d\epsilon \,
\rho(\epsilon)\, \left[ e^{-\epsilon/\kb T} - (\eta +1 - \epsilon/\kb
T)e^{-\eta} \right] \nonumber \\
&=& \Lambda^3 e^{-\eta}\left(\psi_0 - (\eta+1)\varphi_0 e^{-\eta} +\frac{\varphi_1}{\kb T} e^{-\eta}\right)
\end{eqnarray}

\subsection{Dipolar Spin Relaxation}

The particle and energy loss rates $\dot{N}_2$ and $\dot{E}_2$ due to
dipolar relaxation are given by equations \ref{classical.Ndot2.eqn}
and \ref{classical.Edot2.eqn} in section \ref{dipolar.decay.sec}.

\subsection{Trap Shape Changes}
We assume that the trap shape is changing slowly compared to the
transit time of an atom across the trap.  Trap shape changes are then
adiabatic.  We can use $dS=0$ to compute $dT$ from changes in the
parameters to $\rho(\epsilon)$, and eventually obtain $dE$.

We break $dE$ into its constituents
\begin{equation}
dE = \left. \frac{\partial E}{\partial T}\right|_{\alpha} dT \, +
\sum_{\sigma}\left. \frac{\partial E}{\partial
\alpha_{\sigma}}\right|_T d\alpha_\sigma
\end{equation}
where the first partial derivative is done at constant trap shape and
the sum in the second term is over all the trap shape parameters.
Since the process is adiabatic, $dS=0$.  From equation \ref{entropy:eq} we
see that if $dS=0$ and $N$ is fixed we must have $d\zz=0$.  Then we
may find $dT$ by writing out $d\zz$.  We obtain
\begin{equation}
dT = - \frac{\sum_\sigma \left. \frac{\partial \zz}{\partial
\alpha_\sigma}\right|_T d\alpha_\sigma}{\left. \frac{\partial
\zz}{\partial T}\right|_{\alpha_\sigma}}
\end{equation}
Note that
\begin{equation}
\left.\frac{\partial \psi_n}{\partial T}\right|_\alpha 
= \frac{\psi_{n+1}}{\kb T^2}.
\end{equation}
Remembering that $E=N \psi_1/\psi_0$ we get
\begin{equation}
\dot{E}_{trap}=\frac{E}{\psi_1}\sum_\sigma\left(\frac{\partial
\psi_1}{\partial \alpha_\sigma} - \frac{\psi_2}{\psi_1} \frac{\partial
\psi_0}{\partial \alpha_\sigma} \right) \dot{\alpha}_\sigma .
\end{equation}
In practice the partial derivatives are evaluated numerically.

\subsection{Skimming Energetic Atoms}
As the trap evaporation threshold is lowered, atoms are skimmed from
the top of the energy distribution.  Using equation \ref{totalN:eq}
we calculate
\begin{eqnarray}
dN_{skim} & = & \left. \frac{\partial N}{\partial \ethr}\right|_{T,
\alpha_\sigma} d\ethr \nonumber \\
\dot{N}_{skim} & = & \frac{N \rho(\ethr) e^{-\eta}}{\psi_0} \dot{\epsilon}_t
\end{eqnarray}
and
\begin{equation}
\dot{E}_{skim} = \ethr \, \dot{N}_{skim}.
\end{equation}

\section{Software Implementation}
The evaporation simulator is a set of routines which numerically
integrate the differential equations for $E$ and $N$ through time.  A
user of these routines must write C routines that describe the trap to
the integration routines.  The description takes the form of an energy
density of states, a potential energy density of states, and an energy
threshold above which the energy distribution is truncated.  Other
code (described below) is also expected to parse the command line and
do initialization.  Ideally, all of the information that specializes
the general evaporation routines to a specific trap and ramp function
is contained in a single file in the directory {\tt specialized}.  This
file is mentioned in the top level Makefile.  Henceforth this file
will be referred to as {\em userfile}.

Information about the trap is passed to the evaporation simulation
routines in a data structure, \mytrap, defined in {\tt prog.h}.  The
components of this data structure are pointers to functions and
arrays.  The data structure is a global variable, and is initialized
by the routine \verb1initialize_mytrap1 in {\em userfile}.  This
routine is called at the very beginning of the program.

\subsection{Program Flow}
Execution of the program flows as follows.  First,
\verb1initialize_mytrap()1 is called to set up the data structure.
Then a routine, pointed to by \mytrap, is invoked to parse the command
line and stash user-provided information into the appropriate arrays.  The
differential equation integrator is then started.  It steps $E$ and
$N$ through time using the Runge-Kutta routines from Numerical Recipes
\cite{numerical.recipes}.  At each time step the routines from {\em
userfile} are used as follows.  First, the threshold and parameters
used for calculating the density of states are calculated.  These
might correspond to magnet currents or rf field frequencies, as
computed for controlling the experiment.  These values are printed to
a log file.  Since these parameters may often be manipulated into a
form that economizes computation when calculating the density of
states, another routine, pointed to by \mytrap, is invoked to do this.
Finally, we are ready to calculate $\psi_n$ and $\varphi_n$ for this
instant.  The new temperature and density are calculated.  The effects
of evaporation, relaxation, skimming, and adiabatic trap shape changes
are determined, and we are ready for the next step.

\subsection{Data Structure}
The data structure \mytrap \ is referenced often is these calculations,
and is the link between user-supplied routines and the general
simulation machinery.  The entries in the data structure are:
\begin{list}{}{}

\item{{\tt parse}} Pointer to a function which parses the command
line, obtaining the initial density and temperature of the simulated
gas, and the duration of the simulation.  This information is returned
to the simulator.  Further information might be the beginning and
ending currents in certain magnets, shape factors for ramps, etc.,
which would be used to specify the trap shape at each instant of time.

\item{{\tt threshold}} Pointer to a function that gives the energy
threshold (atoms with more than this energy are assumed to escape from
the trap), as a function of time.

\item{\verb1calc_dos_params1} Pointer to a function which calculates,
at a given time, the parameters used to calculate the density of
states.  The values calculated here will be printed in the log file for
each time step.

\item{\verb1n_dos_params1} An integer, giving the number of density of
states parameters calculated by \verb1calc_dos_params1.  These are
printed in the log file.

\item{\verb1precalculate1} Pointer to a function which manipulates the
density of states parameters to speed up actual calculation of the
density of states.  For example, in a Ioffe trap the density of states
is proportional to the square root of the radial confinement field.
This square root may be calculated here in this routine.

\item{{\tt edos} } A pointer to a routine which returns the energy
density of states at a given energy.

\item{{\tt pedos}} A pointer to a routine which returns the potential
energy density of states at a given energy.

\item{{\tt trapspecs}} An array of information used (possibly) by
\verb1calc_dos_params1 and {\tt threshold}.  These data are not
touched by the simulator.  The only C file that looks at these is {\em
userfile}.

\item{{\tt ntrapspecs}} Size of the array {\tt trapspecs}.  Only used
in file {\em userfile}. 

\end{list}

Several examples of {\em userfile} exist, and may be copied and
hacked.

%% file: vel_distribs.tex

%

\chapter{Bose-Einstein Velocity Distribution Function}
\label{be.velocity.distrib.app}

To find the one-dimensional velocity distribution function which is
used in calculating the Doppler-sensitive lineshape observed in
section \ref{degenerate.thermometry.sec} one should use the
Bose-Einstein (BE) occupation function instead of the more convenient
Maxwell-Boltzmann (MB) occupation function.  For a gas at temperature
$T$ and chemical potential $\mu$ the occupation function is
\begin{equation}
f(\epsilon;T,\mu)=\frac{1}{e^{(\epsilon-\mu)/\kb T}-1}
\label{BE.energy.distrib.func}
\end{equation}
In this appendix we derive the BE one-dimensional velocity
distribution function and use it to calculate characteristic spectral
lineshapes.  Truncation effects for a trap of depth $\ethr$ are
included, but shown to be small when the laser beam intersects the
middle of the trap and preferentially excites atoms from the deepest
part of the trap.  In this case, quantum degeneracy is shown to narrow
the spectrum.

\section{Velocity Distribution}
\begin{figure}[tb]
\centering \epsfxsize=5in \epsfbox{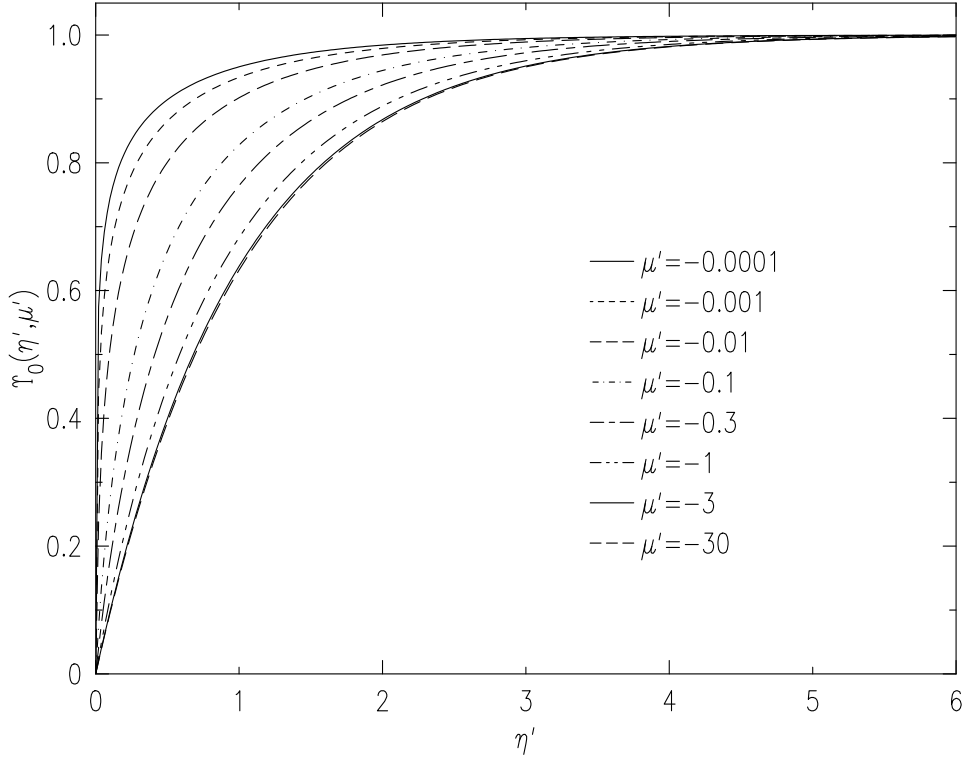}
\caption[behavior of $\Upsilon_0$]{ Behavior of the truncation
correction factor $\Upsilon_0$ as a function of effective trap depth
$\eta^\prime$, plotted for various effective chemical potentials
$\mu^\prime$.  }
\label{upsilon0.fig}
\end{figure}

The one-dimensional momentum distribution function is obtained in a
way analogous to the density, in equation \ref{trunc.dens.func.eqn},
but the momentum integral is over only two dimensions.  We define
$f_p(p_z)\:dp_z$ as the probability of an atom having a momentum along
the $\hat{z}$ axis within a range $dp_z$ of $p_z$; the normalization
is 1.  The distribution is
\begin{equation}
f_p(p_z)=\frac{\displaystyle
\int_0^{p_{\perp,max}} \frac{2\pi p_\perp}
{e^{(p_\perp^2/2m + p_z^2/2m +\varepsilon-\mu)/\kb T}-1}
\;dp_\perp}
{\displaystyle \int_0^{p_{max}} 
\frac{4 \pi p^2}{e^{(p^2/2m + \varepsilon-\mu)/\kb T}-1}\; dp
}
\end{equation}
where $p_\perp$ is the momentum perpendicular to $\hat{z}$ and
$p_{\perp,max}^2/2m=\epsilon_t-\varepsilon-p_z^2/2m$ is the maximum
allowable kinetic energy perpendicular to $\hat{z}$.  The denominator
is $h^3 n(\varepsilon; T, \mu, \epsilon_t)$, calculated above.  The
numerator can be broken into an expression which neglects truncation
and a correction factor $\Upsilon_0$ defined in equation
\ref{upsilon.def.eqn}.  We obtain
\begin{eqnarray}
f_p(p_z; T, \mu, \epsilon_t, \varepsilon) & = & \frac{-2\pi m \kb T}{h^3
n(\varepsilon; T, \mu, \epsilon_t)}
\log\left(1-\exp\left(\frac{\mu-\varepsilon-p_z^2/2m}{\kb T}\right)\right)
\nonumber \\
& & \times \Upsilon_0\left(\frac{\epsilon_t-\varepsilon-p_z^2/2m}{\kb T},
\frac{\mu-\varepsilon-p_z^2/2m}{\kb T}\right)
\end{eqnarray}
The behavior of $\Upsilon_0$ is shown in figure \ref{upsilon0.fig}.
For large $\eta^\prime$ truncation is unimportant and $\Upsilon_0=1$.

\begin{figure}[tb]
\centering \epsfxsize=5in \epsfbox{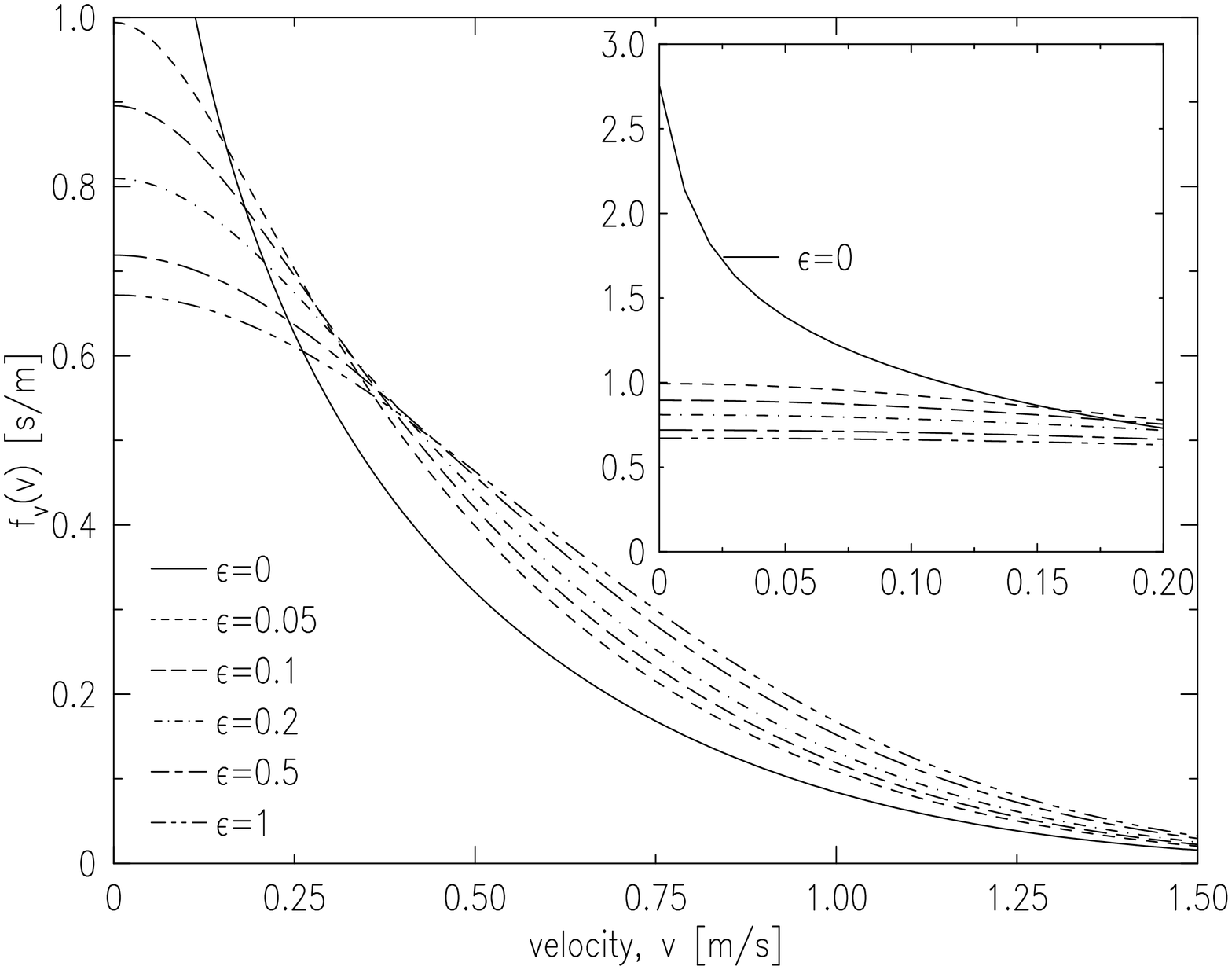}
\caption[distribution of velocities for various effective trap depths]{
One dimensional velocity distribution function, plotted for regions
with various potential energies $\varepsilon$ (listed in units of $\kb
T$).  The sample temperature is $T=50~\mu$K, the chemical potential is
$\mu=-2\times10^{-7}\kb T$, and the trap depth is $6\kb T$. The inset
reveals part of the $\varepsilon=0$ distribution that is not visible
in the main figure.  }
\label{vel.distrib.vary.PE.fig}
\end{figure}

The velocity distribution function is $f_v(v_z)=mf_p(mv_z)$.  Figure
\ref{vel.distrib.vary.PE.fig} shows this distribution function for various
potential energies in the trap.
At higher potentials the effective chemical potential becomes more
negative, and the distributions look more classical.  Only very near
the trap minimum is there significant clumping of the velocities
around zero.

\begin{figure}[tb]
\centering \epsfxsize=5in \epsfbox{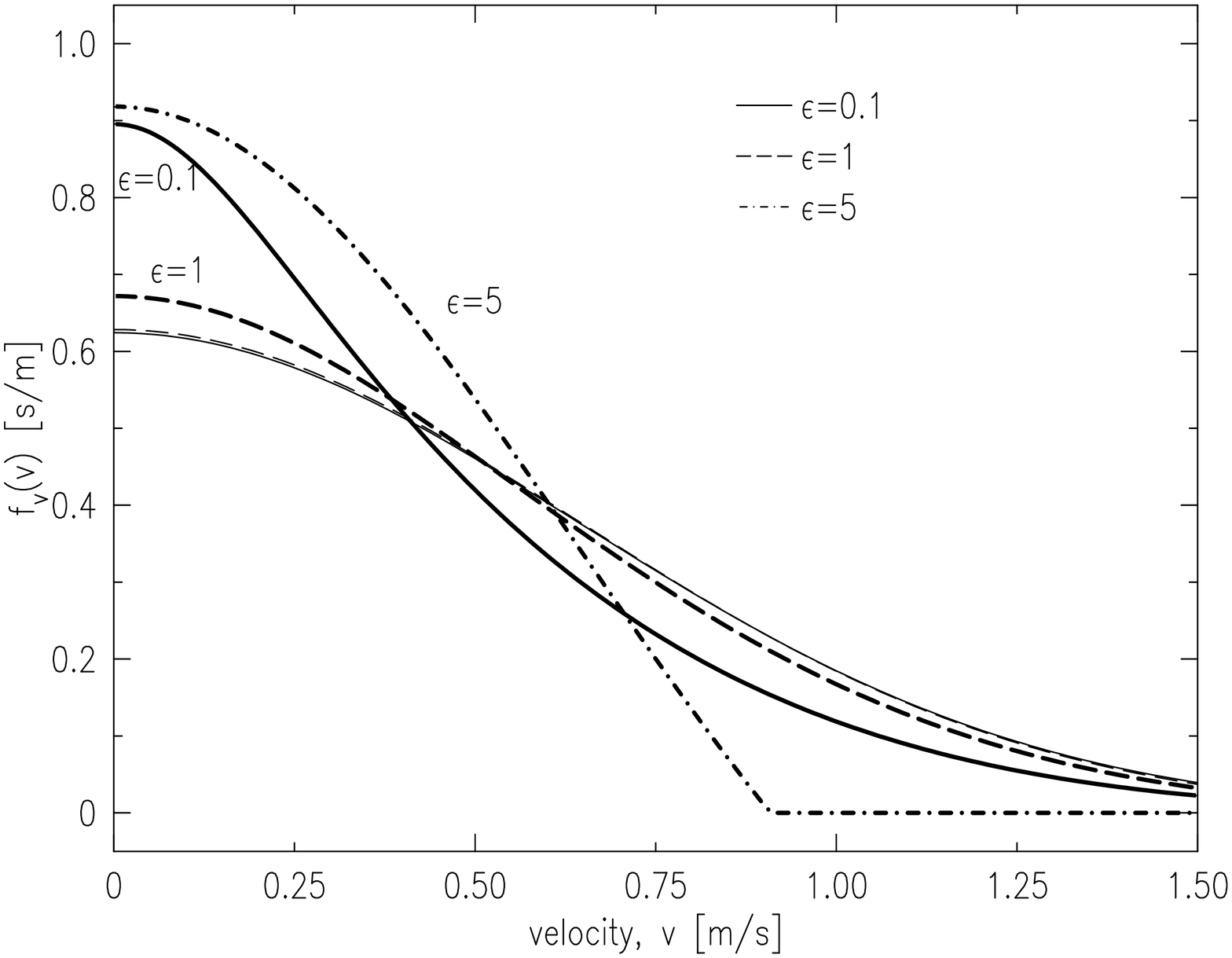}
\caption[comparison of velocity distributions for classical and
degenerate quantum samples]{ Comparison of one dimensional velocity
distributions for samples near to (far from) quantum degeneracy,
having chemical potential $\mu/\kb T=-2\times10^{-7}$ ($\mu/\kb
T=-20$), plotted with heavy (light) lines.  The sample temperature
is $T=50~\mu$K\@.  The trap depth is $6\kb T$\@. The velocity
distribution for each chemical potential is plotted for three
different potential energies, listed in units of $\kb T$.  For the
$\varepsilon=5$ case the two lines overlap; also, truncation effects
are clear because of the shallow effective depth.  }
\label{vel.distrib.vary.mu.fig}
\end{figure}

Figure \ref{vel.distrib.vary.mu.fig} directly compares the quantum velocity
distribution (small $\mu$) with the classical distribution (large
$\mu$).  
As expected, the classical velocity distribution at
$\varepsilon=0.1\kb T$ and $\varepsilon=1\kb T$ are nearly identical,
but differ substantially from their quantum counterparts.  For
$\varepsilon=5\kb T$ the classical and quantum distributions are
nearly identical because even the degenerate gas becomes dilute at
this high potential energy.  Truncation effects are important, as
evidenced by the sharp cutoff of the velocity distribution.

\section{Spectral Signature of Velocity Distribution}
The Doppler-sensitive spectrum is a direct measure of the
one-dimensional velocity distribution function $f_v(v_z)$ through the
(radian) frequency shift $\Delta={\bf k}\cdot{\bf v}$, where ${\bf
k}=2\pi\hat{z}/\lambda$ is the wavevector of the radiation and
$\lambda=243.2$~nm.  The spectrum, $S(\Delta)$, is given by
considering the spectrum of the atoms at each potential energy in the trap:
\begin{equation}
S(\Delta) = S_0 \int_0^{\epsilon_t} d\varepsilon\;
\varrho(\varepsilon)n(\varepsilon) \;
\int_{-v_{max}}^{v_{max}}dv_z f_v(v_z) \delta(\Delta-kv_z)
\end{equation}
where $mv_{max}^2/2\equiv\epsilon_t-\varepsilon$ is the maximum
kinetic energy allowable for atoms with potential energy
$\varepsilon$.  The inner integral vanishes if $|\Delta/k|>v_{max}$.
The upper integration limit on the outer integral thus becomes $\varepsilon_{max}=\epsilon_t-m\Delta^2/2k^2$.  The spectrum is
\begin{eqnarray}
\displaystyle
S(\Delta)=&\displaystyle
\frac{S_0}{h\Lambda^2}
\int_0^{\varepsilon_{max}}d\varepsilon\varrho(\varepsilon)
\log\left(1-e^{(\mu-\varepsilon-m\Delta^2/2k^2)/\kb T}\right) \nonumber \\
&\displaystyle \times 
\Upsilon_0\left(\frac{\epsilon_t-\varepsilon-m\Delta^2/2k^2}{\kb T},
\frac{\mu-\varepsilon-m\Delta^2/2k^2}{\kb T}\right)
\label{spectrum.be.ve.distrib.eq}
\end{eqnarray}
This calculation has assumed uniform laser illumination of the sample,
has ignored coherence effects for atoms that pass repeatedly through
the laser beam during excitation, has assumed all atoms are driven for
identical durations, and has neglected the cold-collision frequency
shift.  

\begin{figure}[tb]
\centering \epsfxsize=5in \epsfbox{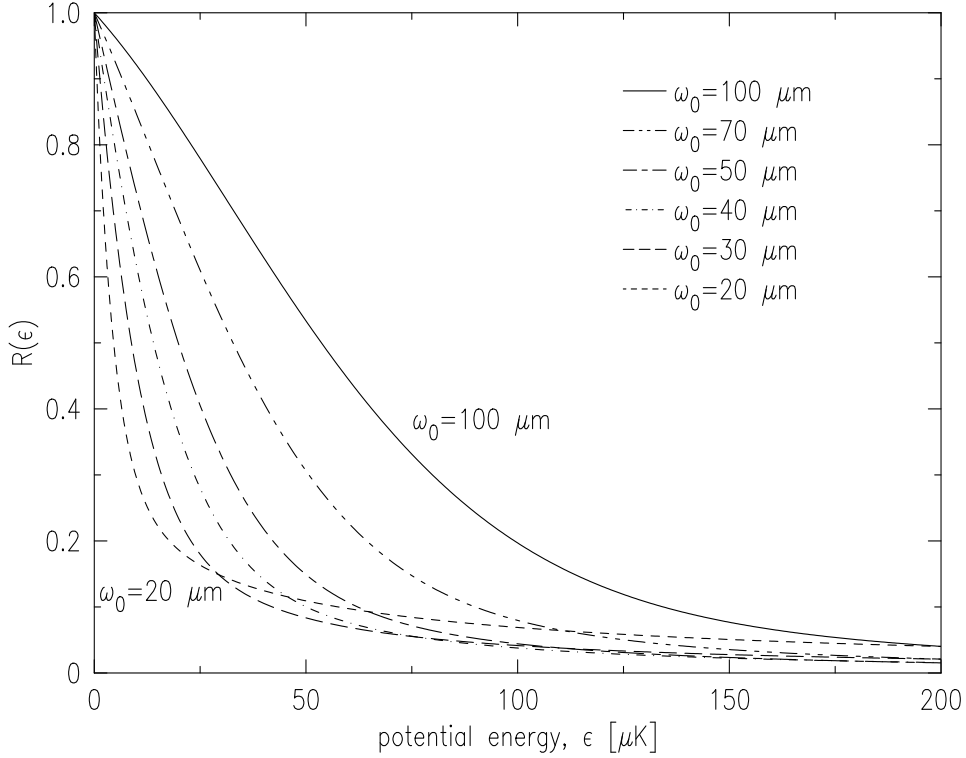}
\caption[how the small laser beam preferentially excites low energy
atoms]{ The average of the square of the laser beam intensity on an
equipotential surface of energy $\varepsilon$, plotted for various
beam waist radii $\omega_0$.  The trap has radial energy gradient
$\alpha/\kb=16$~mK/cm, axial curvature $\beta/\kb = 25~\mu{\rm
K/cm^2}$, and bias $\theta/\kb=34~\mu$K\@. At large $\varepsilon$ the
average vanishes very slowly because the beam diverges and illuminates
more of the surface. }
\label{intensity.avg.fig}
\end{figure}

The non-uniform laser illumination can be inserted into equation
\ref{spectrum.be.ve.distrib.eq} as a  factor $R(\varepsilon )$, 
an average of the square of the laser intensity over the surface of
constant potential energy $\varepsilon$ in a trap potential $V({\bf
r})$,
\begin{equation}
R(\varepsilon) I_0^2=\frac{1}{\varrho(\varepsilon)}
\int d^3{\bf r} \; I^2(\rho,z)\;\delta(\varepsilon-V({\bf r}))
\end{equation}
where $I(\rho,z)=I_0 \exp(-2\rho^2/\omega^2(z))$ describes the spatial
dependence of the intensity of a Gaussian beam \cite{guenther90}; the
beam radius a distance $z$ from the focus is
$\omega(z)=\omega_0\sqrt{1+(z/d)^2}$ where $\omega_0$ is the waist
radius and $d=\pi \omega_0^2/\lambda$ is the divergence length.  We
assume the beam focus is at the center of the trap and the beam is
oriented parallel to the cylindrical symmetry axis of the trap.  For a
Ioffe-Pritchard trap potential $V(\rho,z)=\sqrt{(\alpha
\rho)^2 + (\beta z^2 + \theta)^2}-\theta$, we obtain
\begin{equation}
R(\varepsilon)= \sqrt{\frac{\beta}{\varepsilon}}\;
\int_0^{\sqrt{\varepsilon/\beta}} dz
\exp\left\{-\frac{4}{\alpha^2 \omega_0^2}
\frac{(\varepsilon+\theta)^2-(\beta z^2+\theta)^2}{1+(z/d)^2}\right\}
\end{equation}
This average has been calculated for the trap shape of immediate
interest for the experiments described in this thesis.  See figure
\ref{intensity.avg.fig}.  
The laser beam primarily excites the atoms at the bottom of the trap
where the beam intensity is largest.  The energy distribution
truncation effects are thus not very important, and the shape of the
velocity distribution is well approximated by
$-\log(1-\exp((\mu-\varepsilon-mv_z^2/2)/\kb T))$.  Because the the
effective chemical potential is small for these atoms, however,
quantum effects can be important.  

\begin{figure}[tb]
\centering \epsfxsize=5in \epsfbox{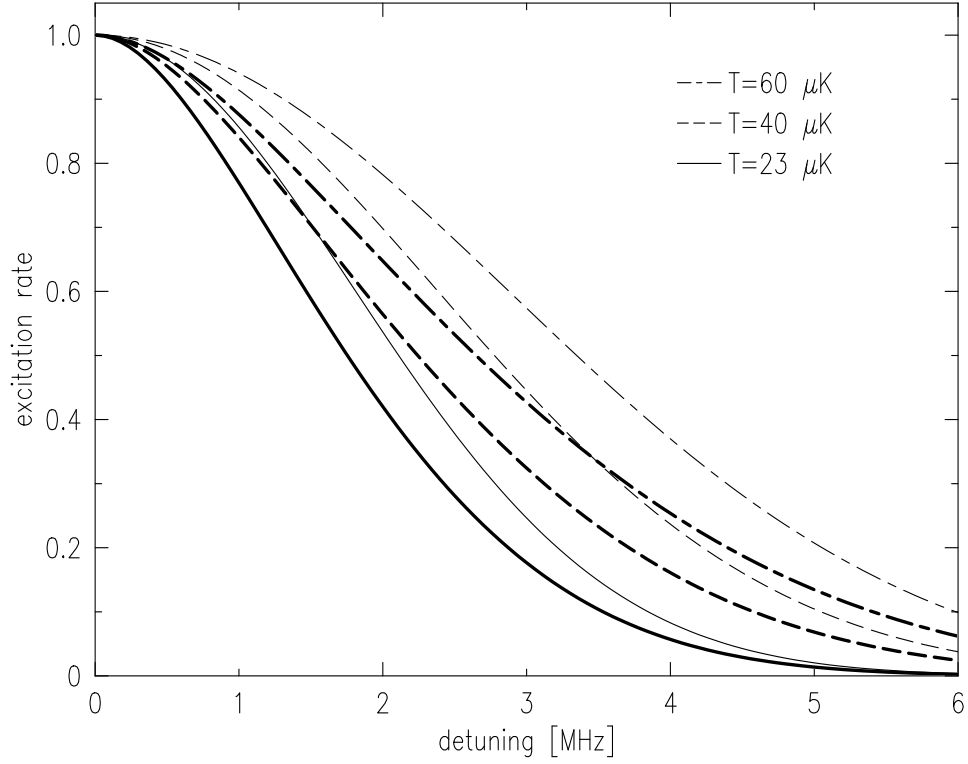}
\caption[Doppler-sensitive lineshape for a quantum degenerate gas]{
Expected Doppler-sensitive lineshape for a gas near to (far from) the
quantum degenerate regime, plotted for samples at various
temperatures.  The thick (thin) lines are for samples with chemical
potential $\mu/\kb=-1\times10^{-4} (-5\times10^{2})~\mu$K,
corresponding to the degenerate (non-degenerate) regime.  For these
calculations $\ethr=300~\mu$K, $\alpha=16$~mK/cm, $\beta=25~\mu{\rm
K/cm^2}$, $\theta=34~\mu$K, and $\omega_0=48~\mu$m. }
\label{calc.ds.spec.fig}
\end{figure}

Figure \ref{calc.ds.spec.fig} 
shows calculations of the lineshape, using equation
\ref{spectrum.be.ve.distrib.eq} and including $R(\varepsilon)$, for
samples in the quantum degenerate regime at various temperatures for a
set of trap parameters of immediate interest to this experiment.  We
see that the distributions are significantly narrower than those
arising from samples far from quantum degeneracy.  If the spectrum for
a degenerate sample is fit to the distribution expected for a
non-degenerate sample, the quoted temperature will be about 40\% too
low; the actual temperature is about 1.6 times higher for the
conditions of immediate interest in our experiment.

%% file: cond_loss.tex

\chapter{Model of the Dynamics of the Degenerate Gas}
\label{condensate.model.app}

\section{Overview}
This appendix describes a computer model of the dynamics of the
degenerate trapped gas, including evaporation and dipolar decay losses
from the condensate and thermal gas.  As depicted in figure
\ref{system.cartoon.fig}, the trapped gas is modeled as a system
composed of two tightly coupled components, the thermal gas and the
condensate.  We assume the loss rates from the system are slow
compared to the rates of transferring particles and energy between the
two components.  A dynamical equilibrium is thus postulated.  The
results of this model are shown in figure
\ref{cond.time.evol.model.fig}.

The trap shape is assumed to be static and of the Ioffe-Pritchard
form.  The depth is fixed.  The Bose-Einstein occupation function is
used for calculation of dipolar decay losses from the thermal gas, and
the Thomas-Fermi condensate wavefunction is the basis for calculations
of loss from the condensate.

The particle and energy loss rates due to evaporation should be
calculated using a quantum kinetic theory derived from the quantum
Boltzmann transport equation.  This approach would correctly include
the Bose statistics.  As a very simple approximation to the desired,
but complicated, formulation, we simply apply the evaporation rates
calculated for a Maxwell-Boltzmann distribution by Luiten \etal\
\cite{lrw96} (equations in section \ref{evap.cooling.rate.sec}).  One
expects a lower evaporation rate near quantum degeneracy because atom
energies are distributed to lower energies relative to the thermal
energy $\kb T$ than is typical in a classical gas.  Roughly speaking,
the effective $\eta$ is larger.  Another way to think about the
situation is to realize that collisions which lead to evaporation
occur preferentially among higher energy atoms.  As shown in figure
\ref{MB.BE.population.fig}, the Maxwell-Boltzmann distribution calls
for many more atoms in these regions of higher energy than does the
Bose-Einstein distribution in the degenerate regime.  We therefore
reduce the evaporation rate by the adjustable factor ${\cal F}$ of
order $0.4$.  This ${\cal F}$ is the factor mentioned in figure
\ref{cond.time.evol.model.fig}.  Note that the {\em rate} of
evaporation is reduced, but the energy of the escaping atoms is still
approximately $(\eta+1)\kb T$.

\section{Simulation Details}

The input to the simulation is the initial temperature of the thermal
gas and the initial condensate population (the proper thermodynamic
parameters are $T$ and $\mu$; the chemical potential is directly
linked to the peak condensate density, which is directly connected to
the condensate population).  The trap shape and depth
($\alpha,\beta,\theta,\ethr$) are additional parameters.  Finally, a
one body loss rate with time constant $\tau_1$ is specified.  Using
this input, the program calculates the total occupation of the trap,
$N_T$, and the total energy, $E_T$, using equations
\ref{thermal.population.simple.eqn}, \ref{thermal.energy.simple.eqn},
and \ref{Ncond.eqn}, and the condensate energy $E_c=\mu N_c=n_p U_0
N_c$.  It is the total energy and the total population that is evolved
in time.  At each time step the temperature is found by numerically
solving equation \ref{thermal.energy.simple.eqn} for the total energy
of the thermal gas.  We assume $\mu=0$ for the thermal gas during the
entire simulation.  If we allowed $\mu$ to vary, then {\em two}
equations would have to be consistently solved numerically, one for
$N_t$ and one for $E_t$ (possible, but more difficult).  Once the
temperature is found, the occupation of the thermal gas is calculated.
The condensate occupation is the total trap occupation minus the
thermal occupation, $N_c=N_T-N_t$ (we postulate equilibrium between
the thermal gas and the condensate, a questionable assumption as
discussed in section \ref{condensate.feeding.rate.sec}).  The particle
and energy loss rates due to dipolar relaxation from the thermal gas
and condensate are calculated using equations
\ref{bose.dipolar.particle.rate.eqn} and
\ref{bose.dipolar.energy.rate.eqn}.  Possible one-body losses are
added.  The particle and energy loss rates due to evaporation from the
thermal gas are calculated for a classical gas of the same peak
density as the Bose gas being simulated, $n_o=2.612/\Lambda^3(T)$.
These rates are multiplied by the correction factor ${\cal F}$.  The sum of
all the particle and energy loss rates for the given time step is
returned to the differential equation stepper routine.

The simulation is stepped through time using the Romberg integration
technique \cite{numerical.recipes}.  Errors in $N_T$ and $E_T$ are
maintained below the $10^{-5}$ level.

\section{Improvements}

One important improvement would be to determine the evaporation rate
using the quantum Boltzmann equation, which is based on the correct
quantum occupation function.  This has been done very recently
\cite{yki99}.

Further improvements would include treatment of the finite feeding
rate of the condensate from the thermal gas.  This bottleneck should
be playing a role for the largest condensates in this thesis, as
suggested in section \ref{condensate.feeding.rate.sec}.

%% file: trapshape.tex

\chapter{Trap Shape Uncertainties}
\label{trap.shape.uncertainty.app}

In order to calculate various properties of the thermal gas and
condensate, it is necessary to combine measurements (of density and
temperature, for example) with knowledge of the trap shape.  In
principle the trap shape can be calculated simply from the Biot-Savart
law.  In practice various complications arise.  In this appendix we
discuss these complications and estimate their effect on the trap shape.  We
conclude that the trap parameters (radial gradient $\alpha$ and axial
curvature $\beta$) should have uncertainties of less than 20\%.

The trap shapes used for the experiments in this thesis are created by
currents in 17 superconducting coils.  Various compensation schemes
are employed to produce steep gradients and large curvatures along the
$z$ axis, and  shim coils are used to adjust the bias field.  One
source of uncertainty is knowledge of the precise geometry of the
coils.  Additional uncertainties exist because fields of unknown
strength and direction arise from trapped fluxes \cite{doyle91}
and magnetized materials in the cryostat.

We first address the stray fields: how can these stray fields
influence the trap shape?  Let us consider the influence of stray
fields on the three Ioffe-Pritchard trap parameters, $\alpha$,
$\beta$, and $\theta$.  We label the three space coordinates $x,y,z$,
and assume the trap exhibits cylindrical symmetry about the $z$ axis.
A uniform field perpendicular to the $z$ axis will shift the location
of the trap minimum in the plane, but will not affect $\alpha$,
$\beta$, or $\theta$.  A uniform field parallel to the $z$ axis will
shift $\theta$, but not affect $\alpha$ or $\beta$.  A gradient field
with strength $B=s\:x$ ($s$ is the gradient strength) aligned
perpendicular to the $z$ axis will add linearly with $\alpha$, but
will not affect $\beta$ or $\theta$.  A field with strength $B=s\:z$
aligned parallel to the $z$ axis will move the trap minimum along the
axis and shift $\theta$, but will not affect $\beta$.  A field with
strength $B=s\:z$ aligned perpendicular to the $z$ axis will tip the
trap symmetry axis away from the $z$ axis, but not affect any
parameters directly.  Finally, a field parallel to the $z$ axis and
with strength $B=s\:x$ will bow the trap and directly perturb all
three parameters.  Defining the unitless parameter
$\epsilon=\sqrt{1+(\mu_B s/\alpha)^2}$, we find that
$\alpha\rightarrow \epsilon\alpha$, $\beta\rightarrow\beta/\epsilon$,
and $\theta\rightarrow\theta/\epsilon$.

In the traps of immediate interest in this thesis, the samples are
short (a few cm) and very narrow (less than 1~mm in diameter).  The
radial gradient fields are large ($\sim 10^2$~G/cm), and sources of
stray fields are far (several cm) from the sample.  Thus, we expect
only very weak perturbations to the gradient parameter $\alpha$ and
the curvature $\beta$.  The parameter $\theta$ is easily perturbed,
but we can measure it directly, as described in section
\ref{trap.bias.measurement.sec}.  Trapped fluxes are expected to exist
on the 1~G level, but the distance from the sources (superconducting
wires) to the trap should cause these fields to appear as simple
gradients, which have been shown above to only slightly perturb
$\beta$ and $\alpha$.  Magnetic materials in the cryostat could
produce large fields ($\sim 10-100$~G), but they are very far away and
the field should therefore appear uniform across the trap (i.e. not
appear as gradients).  We conclude that stray fields should not
perturb the trap parameters $\alpha$ and $\beta$ significantly.  The
parameter $\theta$ may be perturbed, but it can be measured directly.

Measurements of $\theta$ can be complicated by the effects of gravity:
for very cold samples gravity can pull the atom cloud away from the
region of lowest magnetic field, thus making the measurement of
$\theta$ ambiguous.  This distortion becomes important when the
gravitational potential energy varies over the height, $h$, of the
cloud by many times the thermal energy: $h\gg \kb T/mg=(0.84~{\rm
mm/\mu K}) T$.  For the present samples we are far from this limit.

Since stray fields do not strongly effect $\alpha$ or $\beta$, the
primary uncertainty in knowledge of the trap shape arises from
uncertainty of the coil winding geometry.  The field cancelation
scheme used to create tight curvature $\beta$ along the $z$ axis
amplifies the uncertainties since the net field is the small
difference between multiple large fields.  Reasonable uncertainties in the
winding geometry (elongations and translations of coils on the scale
of 5~mm) give rise to no more than 10\% changes to the calculated
$\beta$.

\begin{figure}[tb]
\centering\epsfxsize=5in\epsfbox{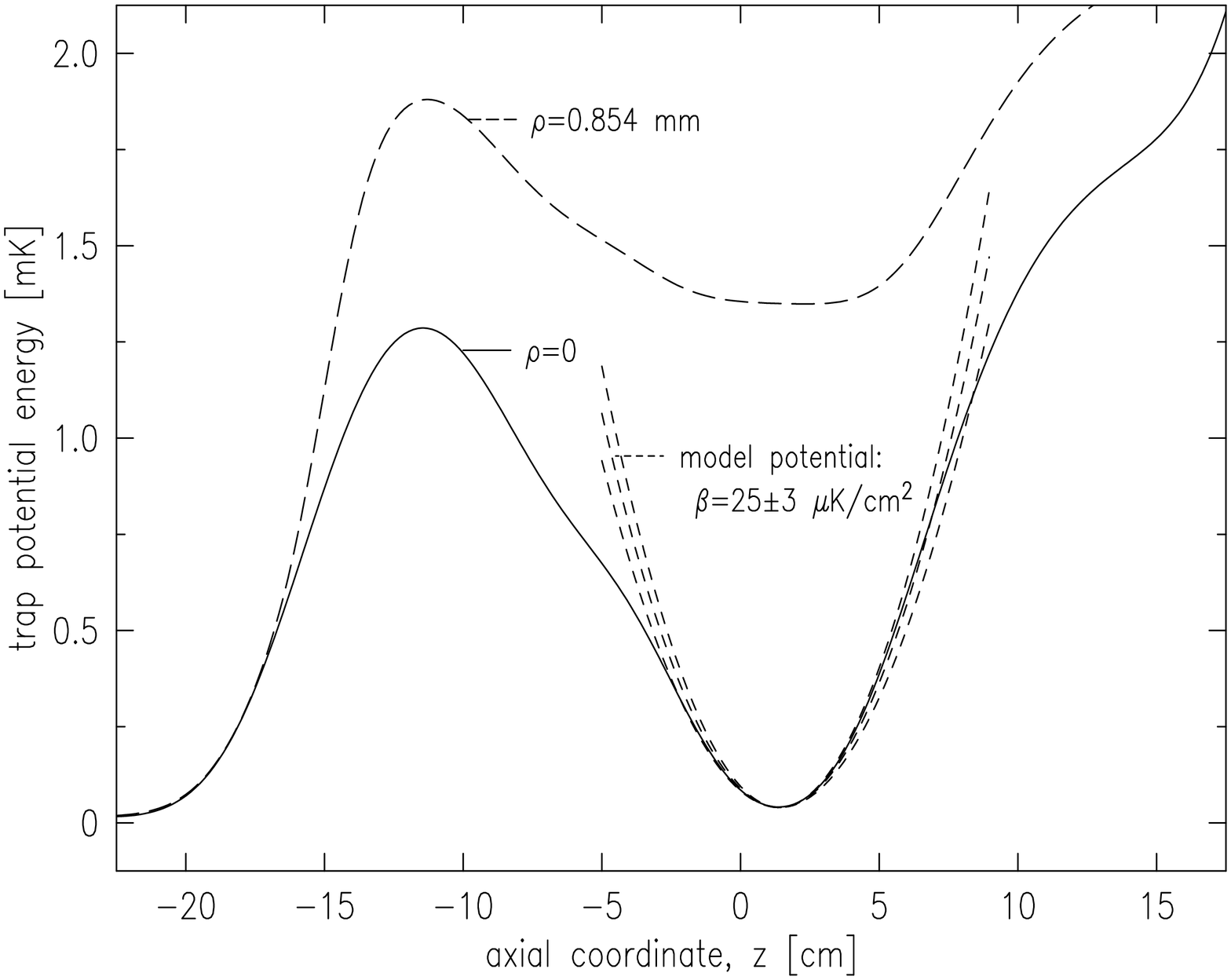}
\caption[calculation of potential energy profile in trap A]{
Calculation of potential energy contours in trap A, along the trap
axis ($\rho=0$) and offset from the axis by distance $\rho=0.854$~mm.
A harmonic potential with curvature $\beta=25\pm3~\mu{\rm K/cm^2}$ is
shown.  This curvature is used in chapter \ref{results.chap}.}
\label{trap.A.calc.fig}
\end{figure}

A calculation of one of the trap shapes used in chapter
\ref{results.chap} is shown in figure \ref{trap.A.calc.fig}, along
with analytic fits to the bottom of the potential.
The harmonic approximation of the axial curvature of the trap should
be precise to the 10\% level.

We conclude that uncertainties in the trap shape due to both stray
fields and imprecise winding geometry should be below the 20\% level.

%% file: variable2.tex


\chapter{Symbols}

Here is a compilation of the symbols used in this thesis, and the pages
on which they are introduced.

\begin{centering}
\begin{tabbing}
\underline{symbol} \hspace{0.3in} \= \underline{description} \\

$\alpha$ \> radial energy gradient in Ioffe-Pritchard trap, p\pageref{IP.trap.def.page} \\
$\beta$ \> axial energy curvature in Ioffe-Pritchard trap, p\pageref{IP.trap.def.page} \\
$\Gamma_{col}$ \> elastic collision rate, p\pageref{Gamma.def.page} \\
$\gamma$ \> phase space compression efficiency parameter, p\pageref{gamma.def.eqn} \\
$\epsilon$ \> total energy of a trapped atom, p\pageref{epsilon.def.page} \\
$\ethr$ \> trap depth (energy), p\pageref{ethr.def.page} \\
$\varepsilon$ \> potential energy, p\pageref{varrho.def.page} \\
$\eta$ \> scaled trap depth, $\epsilon_t/\kb T$, p\pageref{eta.def.page} \\
$\theta$ \> bias field energy in Ioffe-Pritchard trap, p\pageref{IP.trap.def.page} \\
$\Lambda$ \> thermal de Broglie wavelength,  $h/\sqrt{2 \pi m \kb T}$, p\pageref{lamda.def.page} \\
$\mu$ \> chemical potential of sample, p\pageref{muT.def.page} \\
$\mu_B$ \> Bohr magneton, $e\hbar/2m_e=9.274 \times 10^{-24}$~J/T , p\pageref{muB.def.page} \\
$\rho$ \> radial coordinate relative to the trap center, p\pageref{trap.coords.def.page} \\
$\rho(\epsilon)$ \> total energy density of states, p\pageref{rho.def.page} \\
$\varrho(\varepsilon)$ \> potential energy density of states (differential volume with potential energy $\varepsilon$), p\pageref{varrho.def.page} \\
$\sigma$ \> collision cross section for identical particles,  $8 \pi a^2$, p\pageref{sigma.def.page} \\
$\phi$ \> scaled bias field, $\theta/\kb T$, p\pageref{phi.def.page} \\
$\omega_\rho$ \> radial trap oscillation frequency, p\pageref{radial.osc.freq.defn.eq} \\
$\omega_z$ \> axial trap oscillation frequency, p\pageref{axial.osc.freq.defn.eq} \\

$A_{IP}$ \> total energy density of states prefactor for IP trap, p\pageref{rho.def.page} \\
${\cal A}_{IP}$ \> potential energy density of states prefactor for IP trap, p\pageref{varrho.def.page} \\
$a$ \> scattering length for collisions between two $1S$ H atoms, 0.648~\AA, p\pageref{swave.a.def} \\
$B(\phi)$ \>  indicates whether trap is predominantly linear or harmonic in radial direction, p\pageref{b.phi.defn.eqn} \\
$C(\phi,\eta,f)$ \> dependence of heating ratio $H_c/H_t$ on trap shape, p\pageref{heating.ratio.eqn} \\
$C_2$ \> average energy of atoms lost from thermal gas through dipolar decay, p\pageref{C2.def.page}, p\pageref{C2.bose.def} \\
$D$ \> phase space density when far from quantum degeneracy,  $n_0 \Lambda^3$, p\pageref{D.def.page} \\
$E$ \> total energy of sample, p\pageref{E.defn.eqn} \\
$E_c$ \> total energy of condensate, p\pageref{Ec.def.page} \\
$E_t$ \> total energy of thermal gas, p\pageref{thermal.energy.simple.eqn} \\
$\dot{E}_{2,c}$ \> energy loss rate due to dipolar decay from the condensate, p\pageref{Edot2c.def} \\
$\dot{E}_{2,t}$ \> energy loss rate due to dipolar decay from the thermal gas, p\pageref{bose.dipolar.energy.rate.eqn} \\
$F$ \>  condensate fraction, $N_c/(N_t + N_c)=f/(1+f)$ , p\pageref{condensate.fraction.def.eqn} \\
$f$ \> condensate occupation ratio, $N_c/N_t$ , p\pageref{f.def} \\
$G_{\sigma\lambda\rightarrow\phi\psi}$ \> dipolar decay event rate for a given channel, p\pageref{Gabcd.def} \\
$g$ \> dipolar decay rate for a cold gas of $d$-state atoms,  $1.2 \times 10^{-15}~{\rm cm^3/s}$, p\pageref{dipolar.decay.const.eqn} \\
$g_k(z)$ \> Bose function,  $\sum_{l=0}^\infty z^l/l^n$, p\pageref{boseg.def} \\
$H_\sigma$ \> rate at which process $\sigma$ heats the gas, p\pageref{H.def.eqn} \\
$\kb$ \> Boltzmann's constant, $1.3807 \times 10^{-29}$~J/$\mu$K \\
$m$ \> mass of H atom,  1.0078 amu \\
$N$ \> population, number of atoms, p\pageref{totalN:eq} \\
$N_c$ \> population of condensate, p\pageref{Ncond.eqn} \\
$N_t$ \> population of thermal cloud, p\pageref{thermal.population.simple.eqn} \\
$\dot{N}_{2,t}$ \> particle loss rate due to (two-body) dipolar decay from thermal gas, p\pageref{bose.dipolar.particle.rate.eqn} \\
$n$ \> sample density (atoms per unit volume), p\pageref{classical.n.def.eqn}, p\pageref{trunc.dens.func.eqn} \\
$n_0$ \> reference density of thermal gas, p\pageref{n0T.def.page} \\
$n_c$ \> critical density for BEC,  $2.612/\Lambda^3(T)$, p\pageref{bec.density.eqn} \\
$n_{cond}$ \> condensate density, p\pageref{ncond.def.page} \\
$n_p$ \> peak condensate density, p\pageref{np.def.page} \\
$Q_1$ \> unitless dipolar particle decay rate integral, p\pageref{Q1.def.eqn} \\
$Q_2$ \> unitless dipolar energy decay rate integral, p\pageref{Q2.def.eqn} \\
$T$ \> sample temperature, p\pageref{n0T.def.page}, p\pageref{muT.def.page} \\
$U_0$ \> mean field energy per unit density,  $4 \pi \hbar^2 a/m$, p\pageref{U0.def} \\
$V({\bf r})$ \> trap potential energy function, p\pageref{Vr.def} \\
$V_{cond}$ \> volume of the condensate, assuming a Thomas-Fermi wavefunction, p\pageref{Vcond.def.eqn} \\
$\veff$ \> effective volume of the sample, p\pageref{veff:def} \\
$\bar{v}$ \> mean speed of particles in a classical gas,  $\sqrt{8 \kb T/\pi m}$, p\pageref{vbar.def.page} \\
$z$ \> axial coordinate relative to the trap center, p\pageref{trap.coords.def.page} \\
$\zz$ \> classical single particle partition function, p\pageref{classical.partition.func.eqn} \\

\end{tabbing}
\end{centering}

%% file: energy_units.tex



\chapter{Laboratory Units for Energy}
\label{energy.unit.app}

The chart on the following page may prove helpful when comparing the
various ways energy is manifest in atom trapping experiments.  The
relevant constants are:

\begin{tabbing}
\underline{constant} \hspace{0.2in} \= \underline{value} \\
$k_B$ \> $1.381\times 10^{-29}$~J/$\mu$K, Boltzmann constant \\
$\mu_B$ \> $9.274 \times 10^{-24}$~J/T, Bohr magneton  \\
$h$ \> $6.626\times 10^{-34}$~J/Hz, Planck constant \\
$m g$ \> $1.6411\times 10^{-28}$~J/cm, gravitational energy for H \\
\end{tabbing}
\newpage

\begin{centering}
\setlength{\unitlength}{1mm}
\begin{picture}(120,120)(-10,-10)
\thicklines
\put(0,0){\makebox(0,0){\Huge $mgz$}}
\put(100,0){\makebox(0,0){\Huge $h\nu$}}
\put(0,100){\makebox(0,0){\Huge $\kb T$}}
\put(100,100){\makebox(0,0){\Huge $\mu_B B$}}

\put(0,0){\begin{picture}(50,50)
\put(14,0){\vector(1,0){25}}
\put(0,14){\vector(0,1){25}}
\put(14,14){\vector(1,1){25}}
\put(26.5,-5){\makebox(0,0)[t]{$\frac{mg}{h}=250$~kHz/cm}}
\put(-4,26.5){\makebox(0,0)[r]{$\frac{mg}{\kb}=12~\mu$K/cm}}
\put(26,25){\makebox(0,0)[tl]{$\frac{mg}{\mu_B}=0.177$~G/cm}}
\end{picture}}

\put(100,100){\begin{picture}(50,50)
\put(-14,0){\vector(-1,0){25}}
\put(0,-14){\vector(0,-1){25}}
\put(-14,-14){\vector(-1,-1){25}}
\put(-26.5,5){\makebox(0,0)[b]{$\frac{\mu_B}{h}=67.2~\mu$K/G}}
\put(4,-26.5){\makebox(0,0)[l]{$\frac{\mu_B}{\kb}=1.4$~MHz/G}}
\put(-26,-25){\makebox(0,0)[br]{$\frac{\mu_B}{mg}=5.7$~cm/G}}
\end{picture}}

\put(100,0){\begin{picture}(50,50)
\put(-14,0){\vector(-1,0){25}}
\put(-26.5,-5){\makebox(0,0)[t]{$\frac{h}{mg}=4.0~$cm/MHz}}
\put(-14,14){\vector(-1,1){25}}
\put(-20,17){\makebox(0,0)[tr]{$\frac{h}{\kb}=48~\mu$K/MHz}}
\put(0,14){\vector(0,1){25}}
\put(5,26.5){\makebox(0,0)[l]{$\frac{h}{\mu_B}=0.71~$G/MHz}}
\end{picture}}

\put(0,100){\begin{picture}(50,50)
\put(14,0){\vector(1,0){25}}
\put(26.5,5){\makebox(0,0)[b]{$\frac{\kb}{\mu_B}=14.9$~G/mK}}
\put(14,-14){\vector(1,-1){25}}
\put(20,-17){\makebox(0,0)[bl]{$\frac{\kb}{h}=20.8~$MHz/mK}}
\put(0,-14){\vector(0,-1){25}}
\put(-5,-26.5){\makebox(0,0)[r]{$\frac{\kb}{mg}=84$~cm/mK}}
\end{picture}}

\end{picture}

\end{centering}